\pdfoutput=1
\documentclass[iop]{emulateapj}
\usepackage{graphicx}
\usepackage{color}

\shorttitle{Slipping magnetic reconnection during an X-class flare}
\shortauthors{Dud\'ik et al.}

\begin{document}

\title{Slipping magnetic reconnection during an X-class solar flare observed by SDO/AIA}

\author{J. Dud\'ik\altaffilmark{1}}
    \affil{RS Newton International Fellow, DAMTP, CMS, University of Cambridge, Wilberforce Road, Cambridge CB3 0WA, United Kingdom}
    \email{J.Dudik@damtp.cam.ac.uk}
\author{M. Janvier}
    \affil{Department of Mathematics, University of Dundee, Dundee DD1 4HN, Scotland, United Kingdom} \email{mjanvier@maths.dundee.ac.uk}
\author{G. Aulanier}
    \affil{LESIA, Observatoire de Paris, UMR 8109 (CNRS), 92195 Meudon Principal Cedex, France}
\author{G. Del Zanna}
    \affil{DAMTP, CMS, University of Cambridge, Wilberforce Road, Cambridge CB3 0WA, United Kingdom}
\author{M. Karlick\'y}
    \affil{Astronomical Institute of the Academy of Sciences of the Czech Republic, Fri\v{c}ova 298, 251 65 Ond\v{r}ejov, Czech Republic}
\author{H. E. Mason}
    \affil{DAMTP, CMS, University of Cambridge, Wilberforce Road, Cambridge CB3 0WA, United Kingdom}
\author{B. Schmieder}
    \affil{LESIA, Observatoire de Paris, UMR 8109 (CNRS), 92195 Meudon Principal Cedex, France}

\altaffiltext{1}{DAPEM, Faculty of Mathematics Physics and Computer Science, Comenius University, Mlynsk\'a Dolina F2, 842 48 Bratislava, Slovakia}

\begin{abstract}
We present SDO/AIA observations of an eruptive X-class flare of July 12, 2012, and compare its evolution with the predictions of a 3D numerical simulation. We focus on the dynamics of flare loops that are seen to undergo slipping reconnection during the flare. In the AIA 131\AA~observations, lower parts of 10\,MK flare loops exhibit an apparent motion with velocities of several tens of km\,s$^{-1}$ along the developing flare ribbons. In the early stages of the flare, flare ribbons consist of compact, localized bright transition-region emission from the footpoints of the flare loops. A DEM analysis shows that the flare loops have temperatures up to the formation of \ion{Fe}{24}. A series of very long, S-shaped loops erupt, leading to a CME observed by STEREO. The observed dynamics are compared with the evolution of magnetic structures in the ``standard solar flare model in 3D''. This model matches the observations well, reproducing both the apparently slipping flare loops, S-shaped erupting loops, and the evolution of flare ribbons. All of these processes are explained via 3D reconnection mechanisms resulting from the expansion of a torus-unstable flux rope. The AIA observations and the numerical model are complemented by radio observations showing a noise storm in the metric range. Dm-drifting pulsation structures occurring during the eruption indicate plasmoid ejection and enhancement of reconnection rate. The bursty nature of radio emission shows that the slipping reconnection is still intermittent, although it is observed to persist for more than an hour.
\end{abstract}


%
\section{Introduction}
\label{Sect:1}
Solar flares are the most energetic manifestation of solar magnetic activity. They are characterized by a rapid increase in emission over a broad range of the electromagnetic spectrum, from X-rays and extreme-ultraviolet (EUV) to radio wavelengths \citep[e.g.,][]{Kane74,Fletcher11,White11}. A typical flare encompasses a wealth of dynamical phenomena that are manifestations of the release of magnetic energy via the process of magnetic reconnection \citep{Parker57,Sweet58,Priest00,Zweibel09}. One of the most distinct signatures of these dynamical phenomena is that of the formation of flare loops, which emit strongly in X-rays and EUV \citep[e.g.,][]{Fletcher11}, and the accompanied flare ribbons prominent in EUV and up to visible wavelengths \citep[e.g.,][]{Warren01}. Eruptive flares also exhibit large-scale restructuring of the magnetic field accompanied by coronal mass ejections (CMEs) that possibly result from the expulsion of coronal magnetic flux ropes \citep[e.g.,][]{vanB89,Amari00,Moore01,Green09,Green11,Patsourakos13} and are an important driver of space weather.

Based on the observations of eruptive flares, various models have been developed to describe and interpret their main features. The standard 2D CSHKP model \citep{Carmichael64,Sturrock66,Hirayama74,Kopp76}, for example, describes the formation of flare loops and the flux rope \citep[e.g.,][]{Dere99,Cheng13}, from the reconnection of coronal magnetic field lines. Magnetic reconnection releases magnetic energy in the form of particle acceleration and thermal energy. Energetic particles guided along the magnetic field can impact the chromosphere \citep[e.g.,][]{Reid12}, where they lead to the formation of flare ribbons \citep[e.g.,][]{Schmieder96}. The chromosphere can also be heated by thermal conduction \citep{Petkaki12}. Both processes lead to chromospheric evaporation \citep{Neupert68}, which quickly fills the flare loops \citep[e.g.,][]{Raftery09,Milligan09,Brosius10,DelZanna11a,Ning11,Doschek13,Brosius13}. The high density in the flare loops makes them a prominent emission structures. Flare ribbons are the footpoints of these loops, and their emission comes from compact transition region features \citep[e.g.,][]{Graham11,Young13} and the underlying chromosphere.

While the CSHKP model provides a straightforward picture of eruptive flares, it fails at describing the intrinsically 3D features of solar flares. Examples of such features include e.g. motion of EUV or X-ray sources parallel to the flare ribbons \citep[e.g.][]{Inglis13}, or the evolution of the twist or shear of coronal loops, which is important for the dynamics of both flare loops and the flux rope \citep{Aulanier12}. Other models, for example the tether cutting model \citep{Moore97,Fan12} or torus-unstable flux rope models \citep{Torok04,Aulanier10} have been proposed to extend the standard model for eruptive flares to 3D. In particular, the 3D numerical simulation of a torus-unstable flux rope \citep{Aulanier12,Janvier13} has helped to shed light on the underlying reconnection mechanism forming both flare loops and the flux rope. The absence of 3D null-points means that the reconnection takes place in Quasi-Separatrix Layers (QSLs), and in particular in the thinning high current density region underneath the expanding flux rope.

The QSLs \citep{Priest95,Demoulin96a,Demoulin97} are locations where the mapping of the magnetic field lines exhibits strong gradients, but is not discontinuous. This means that there are no separatrices dividing the magnetic field into regions of different connectivities, as is the case with magnetic null-points. Rather, connectivity domains are bordered by regions where the connectivity is strongly changing, but is still continuous. The magnetic field distortion is analytically measured by calculating the norm $N$ of the field line mapping \citep{Demoulin96b} or the squashing degree $Q$ \citep{Titov02,Pariat12}. Because of the distortion of the magnetic field, high electric current density regions can form where the distortion is the strongest \citep[e.g.,][]{Aulanier05,Masson09,Wilmot09}. This therefore results in magnetic reconnection as ideal MHD can break down \citep{Parker57,Sweet58,Priest00}.

In the case of QSLs, magnetic field lines passing through them can undergo successive reconnection that is seen as an apparent ``flipping'', or ``slipping'' motion \citep{Priest95,Priest03,Aulanier06}. The apparent slipping motion is a consequence of the local diffusion in the reconnection region, where the neighbouring field lines continually exchange connectivities \citep{Aulanier06}. Because of the local rotation of the magnetic field vector within the coronal diffusion region, this process induces apparent field line velocities all along their length that can be different from the plasma velocity \citep{Priest03,Aulanier06}. Slipping reconnection is theoretically predicted in both eruptive and confined flares, as investigated in \citet{Masson12}. \citet{Janvier13} showed that the speed of the apparent motion of the field lines can be directly linked to the norm $N$ and the reconnection rate. This study provided an important insight on reconnection mechanisms in 3D and the link with magnetic topology. Since the QSLs are more general than true null-points which require discontinuities, slipping reconnection is a more general mechanism of energy release not only during solar flares, but also in active regions \citep{Aulanier07}.

At first, current sheets required for reconnection can be formed dynamically, both in 2D \citep[e.g.,][]{Magara96,Lin00,Barta11a} and also in 3D \citep[e.g.,][]{Antiochos99,Lynch08,Aulanier10,Kliem10,Aurass11}. Then, the current layer itself can be destabilized, leading to the formation of several plasmoids and null-points moving along the sheet \citep[e.g.,][]{Loureiro12}. Merging and fragmentation of plasmoids then results in smaller current sheet systems that can be associated with fast reconnection regimes, providing the necessary energy release rate and acceleration of particles \citep{Shibata01,Karlicky04,Barta08,Uzdensky10,Barta11a,KarlickyBarta11,Karlicky12}. These processes can be observed in X-rays as the formation, ejection and interaction of plasmoids \citep{Ohyama98,Kolomanski07,Milligan10} and in the dm-radio range as drifting pulsation structures \citep[DPS,][]{Kliem00,Karlicky02,Karlicky04,KarlickyBarta07,Karlicky10}. Magnetic reconnection at high coronal altitudes can also manifest itself as a radio noise storm \citep{Elgaroy77,DelZanna11b}.

Based on the above, the general expectation is that during eruptive flares, slipping motion of the field lines should manifest itself as an apparent slipping motion of flare loops, and plasmoid emission in the form of DPS should be observed at radio wavelengths as a consequence of both the flux rope eruption and the reconnection regime involving dynamics of the current layer. To the author's knowledge, these processes have not previously both been reported in the same flare. Moreover, reports of the slipping reconnection itself are rare \citep{Aulanier07,Testa13} and none are associated with flares.

In this paper, we report on the X1.4 flare of July 12, 2012 that manifests both the slipping reconnection and in later phase
the fast reconnection with plasmoids (DPSs). AIA observations of the flare are described in Sect. \ref{Sect:2}. This section contains an overview of the events during the flare together with analysis of the apparent slipping motion of the flare loops. Individual contributions to flare EUV emission as observed by AIA are obtained from DEM analysis in Sect. \ref{Sect:3}. Section \ref{Sect:4} presents the results of an earlier numerical simulation reproducing the expansion of an unstable asymmetric flux rope. With the simulated region having similar features as the observed active region 11520, the simulation is directly compared to observations. Evolution of the QSLs and the flux rope are also described along with the slipping reconnection process. In Sect. \ref{Sect:5}, we report on radio emission associated with the flare. A Summary and Conclusions are given in Sect. \ref{Sect:6}.

%
%
   \begin{figure}
       \centering
    \vspace{0.3cm}
       \includegraphics[width=8.8cm,bb=0  0 498 255]{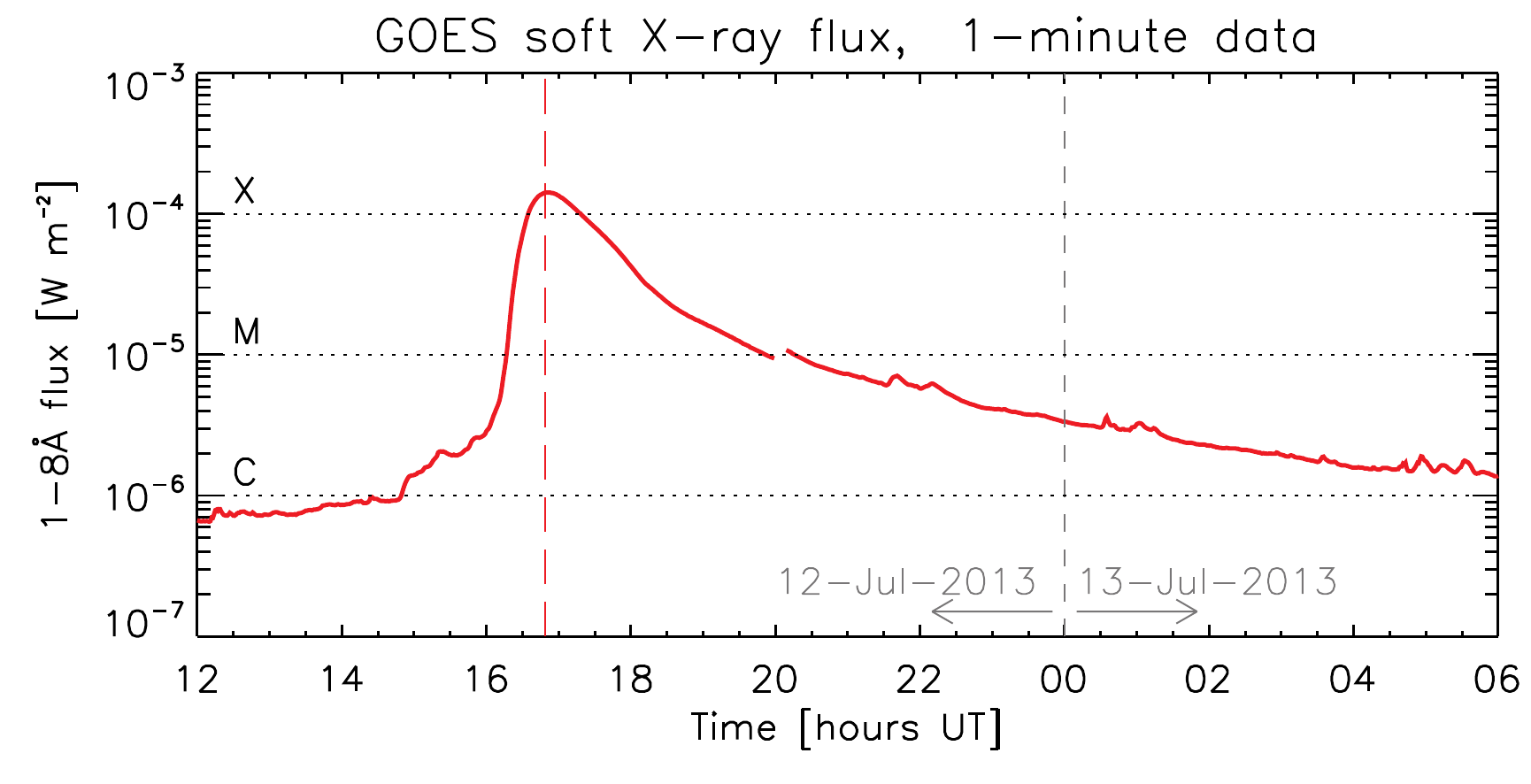}
       \includegraphics[width=8.8cm,bb=0 40 498 255]{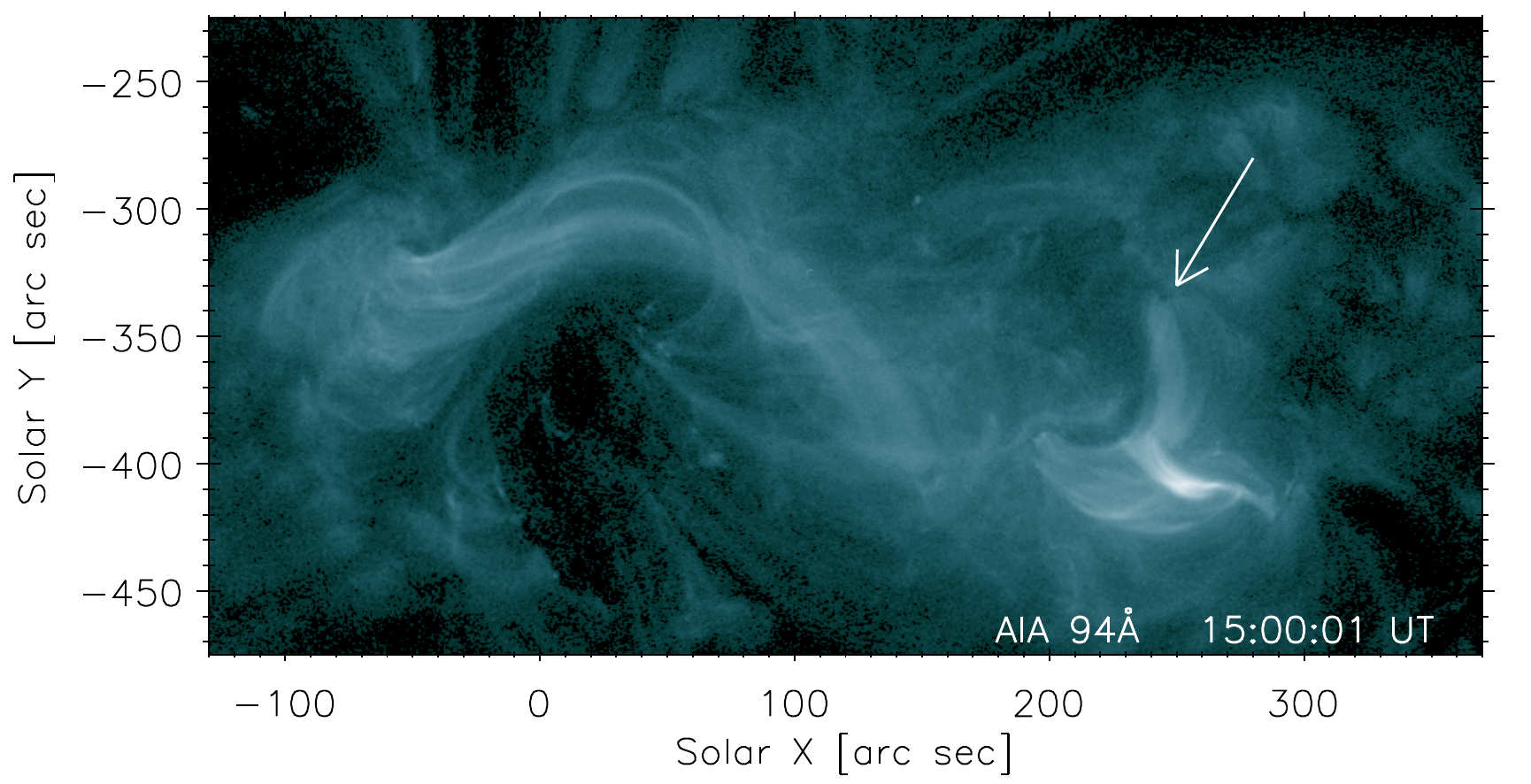}
       \includegraphics[width=8.8cm,bb=0 40 498 255]{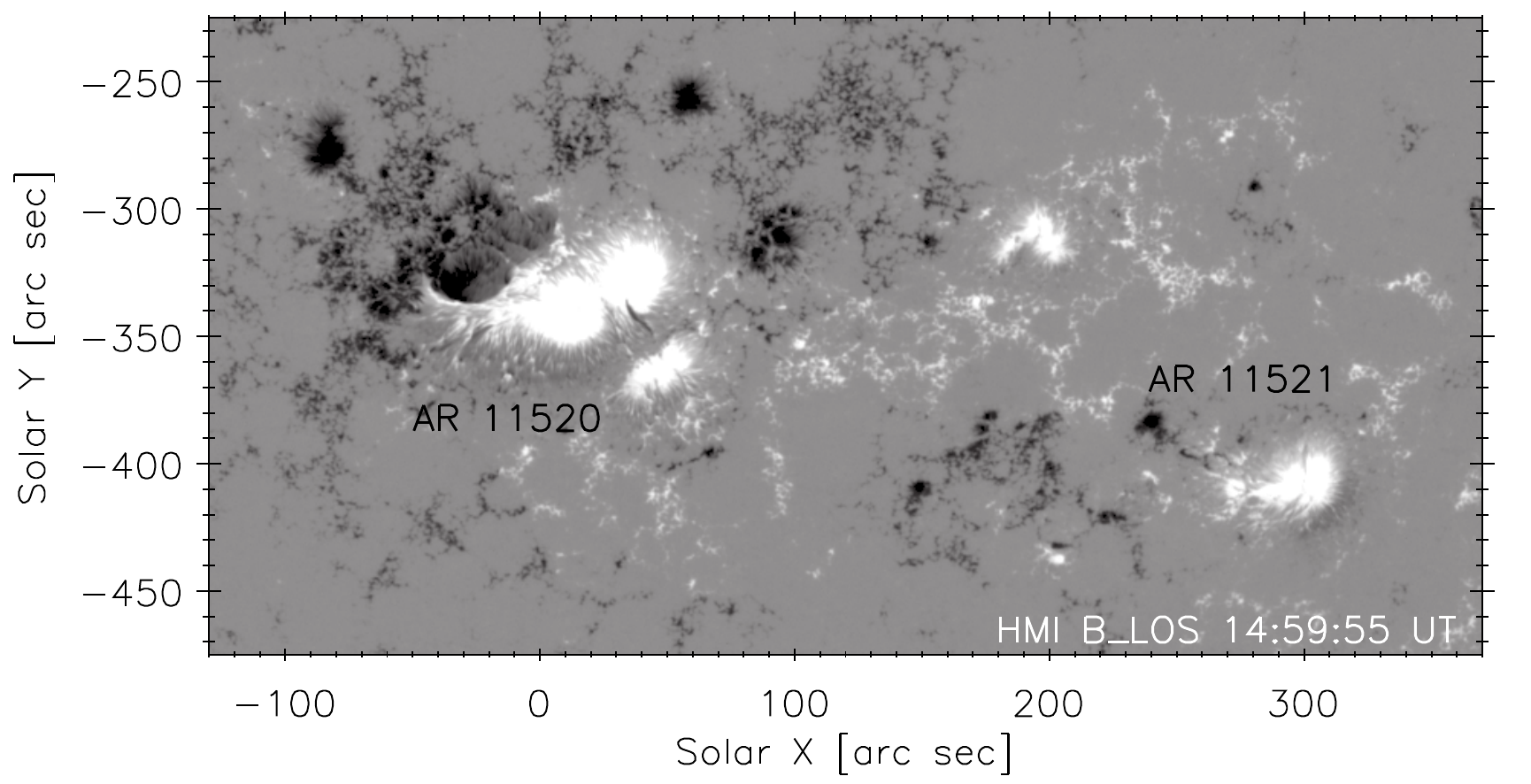}
       \includegraphics[width=8.8cm,bb=0  0 498 255]{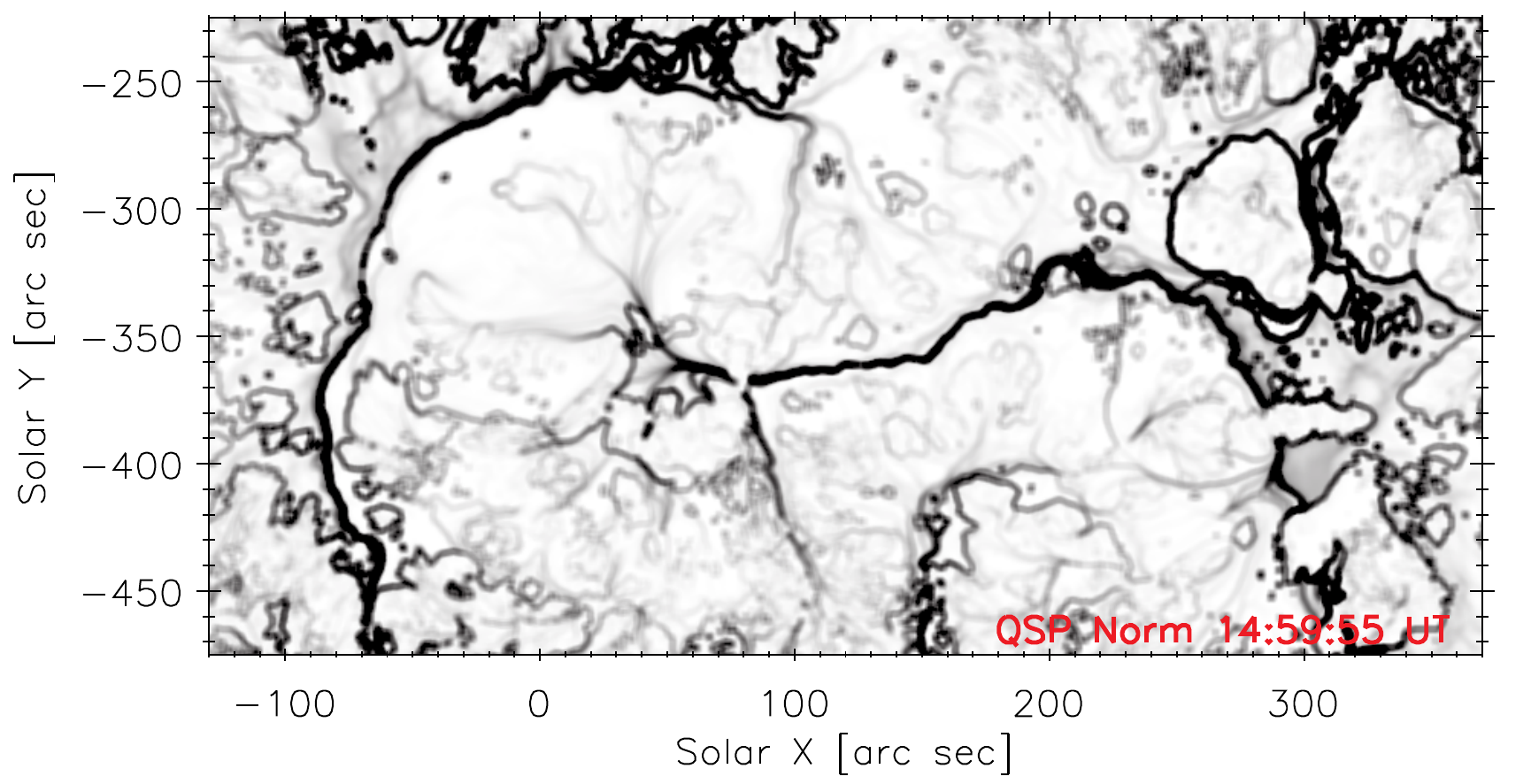}
    \caption{\textit{Top}: Evolution of the X-ray flux in the 1--8\AA\,channel observed by GOES. \textit{Second row}: Pre-flare coronal configuration in the SDO/AIA 94\AA\,filter showing a large sigmoid in AR 11520 and a brightening in AR 11521. Footpoints of one of the loop systems participating in the brightening are denoted by a white arrow. \textit{Third row}: Portion of the longitudinal SDO/HMI magnetogram showing the active region complex. The AR 11519 is not shown as it lies further to the West. \textit{Bottom}: Intersections of the large-scale quasi-separatrix layers with the photosphere, calculated from a potential extrapolation of the magnetic field.}
       \label{Fig:1500UT}
   \end{figure}
   \begin{figure}
       \centering
       \includegraphics[width=8.8cm]{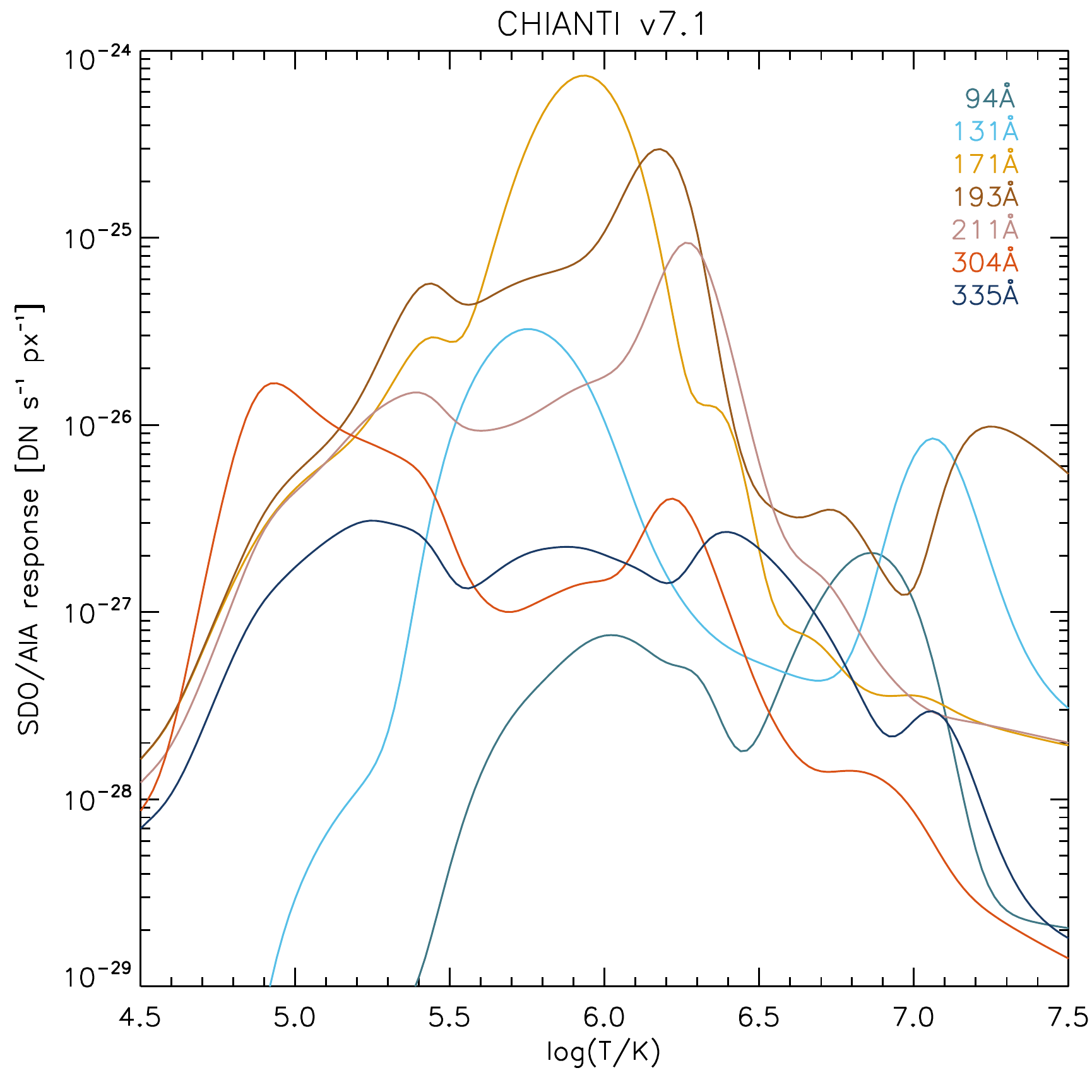}
    \caption{Temperature responses of the AIA EUV filters, calculated using log$(n_\mathrm{e}/\mathrm{cm}^3)$\,=\,9 and the abundances from \citet{Schmelz12}.}
       \label{Fig:AIA_resp}
   \end{figure}
%
%
%
\begin{table*}
\begin{center}
\scriptsize
\caption{Summary of individual events during the X1.4 flare. Time and locations are approximate, as the events are dynamical.
\label{Table:1}}
\tabletypesize{\normalsize}
\begin{tabular}{llll}
\tableline\tableline
Approx. Time [UT]   & Event description					& Location [Solar $X$, $Y$]		& Notes                                        				\\
\tableline
14:48--15:15 UT	& Brightening of loop systems in AR 11521		& $[+300\arcsec, -400\arcsec]$		& Fig. \ref{Fig:1500UT}, \textit{second row}                            \\
15:00 UT	& Flare start in AR 11520 (first flare loop)		& $[0\arcsec, -300\arcsec]$, above F1   & Arrow 1 in Fig. \ref{Fig:Overview} \textit{top left}			\\
15:00 UT        & Radio noise storm beginning at 200--500 MHz		&					& Fig. \ref{Fig:Radio}, type III--like bursts				\\
15:01--15:34 UT & Flare loops undergo slipping motion and gradual brightening   & $[-20\arcsec, -320\arcsec]$	& Sect. \ref{Sect:2.2.1}, Figs. \ref{Fig:Slip1}--\ref{Fig:Slip1_stackplots} \\
15:07--16:10 UT & Growing system of hot loops in 131\AA~(erupts later)	& $[+100\arcsec, -330\arcsec]$		& Arrow 2 in Fig. \ref{Fig:Overview} (\textit{rows 2--4})		\\
15:43--16:07 UT	& Second slipping event, gradual developing of NR       & $[-20\arcsec, -320\arcsec]$,~~\,NR	& Sect. \ref{Sect:2.2.2}, Figs. \ref{Fig:Slip2}--\ref{Fig:Slip2_stackplots} \\
16:02--16:17 UT	& Hot 131\AA~loop appearing at the end of NRH           & $[-100\arcsec, -320\arcsec]$		& Arrow 3 in Fig. \ref{Fig:Overview} \textit{fourth row}		\\
16:05--16:25 UT	& Brightening and development of the NRH		& $[-100\arcsec, -320\arcsec]$, NRH	& Fig. \ref{Fig:Overview} \textit{fifth row}				\\
16:05--16:35 UT	& Third slipping event along developing NRH		& $[-50\arcsec, -320\arcsec]$		& Sect. \ref{Sect:2.2.3}, Figs. \ref{Fig:Slip3}--\ref{Fig:Slip3_stackplots} \\
16:13--16:33 UT	& Large intensity enhancement along NR moving westward	& $[-30\arcsec, -270\arcsec]$,~~\,NR	& Fig. \ref{Fig:Overview} \textit{fifth row}                            \\
16:14--16:27 UT	& PRH brightening, slipping assoc. with eruption in 131\AA & $[+150\arcsec, -360\arcsec]$, PRH	& Fig. \ref{Fig:Overview} \textit{fifth row}, Figs. \ref{Fig:Slipe}--\ref{Fig:Slipe_stackplots} \\
16:16 UT	& Start of a strong radio flare				&					& Sect. \ref{Sect:2.2}, Figs. \ref{Fig:Radio}--\ref{Fig:Radio}		\\
16:30--17:00 UT	& PRH widens, with associated coronal dimming		& $[+200\arcsec, -350\arcsec]$		& Fig. \ref{Fig:Overview} \textit{sixth row}                            \\
16:34--16:40 UT	& Deformation of the NRH				& $[-90\arcsec, -320\arcsec]$		&									\\
16:40--17:00 UT	& NRH extension in N-S direction, assoc. coronal dimming & $[-100\arcsec, -350\arcsec]$		& Fig. \ref{Fig:Overview} \textit{sixth row}				\\
16:49 UT	& Peak of the GOES 1--8\AA\,X-ray flux			&					& Fig. \ref{Fig:1500UT} \textit{top}					\\
\tableline\tableline
\end{tabular}
\end{center}
\end{table*}

%
   \begin{figure*}[!ht]
       \centering
       \includegraphics[width=8.8cm,bb=0 43 498 255]{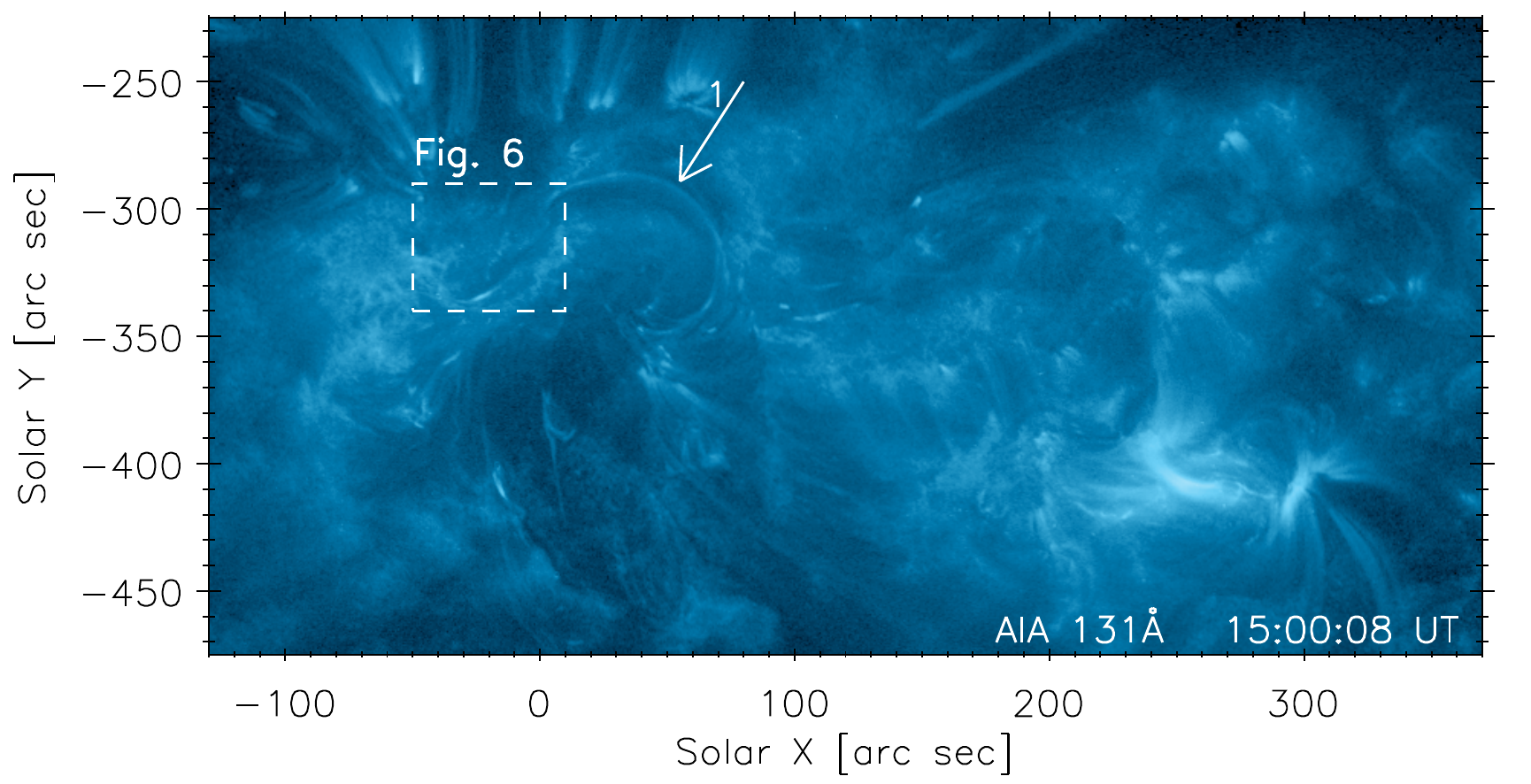}
       \includegraphics[width=8.8cm,bb=0 43 498 255]{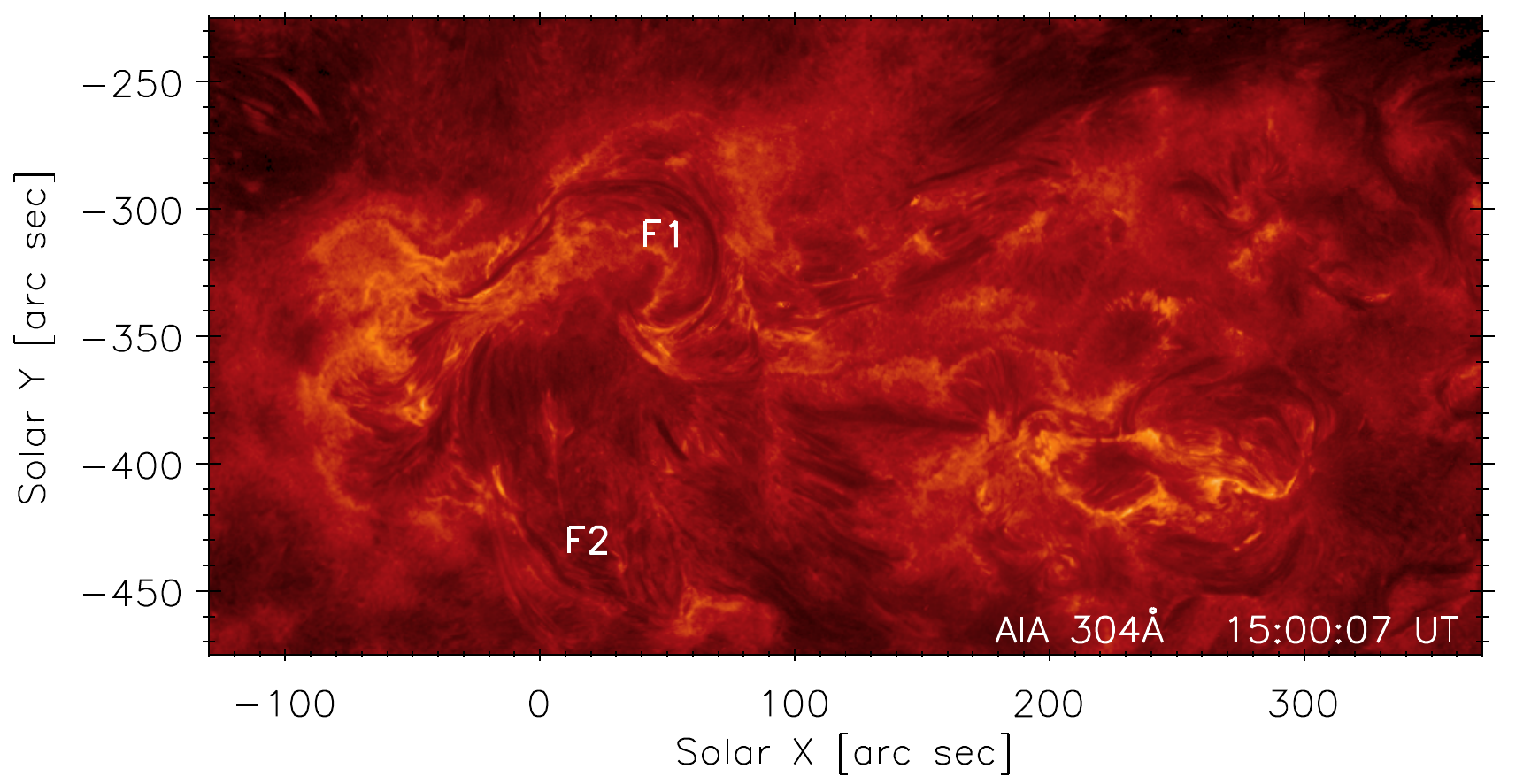}
       \includegraphics[width=8.8cm,bb=0 43 498 255]{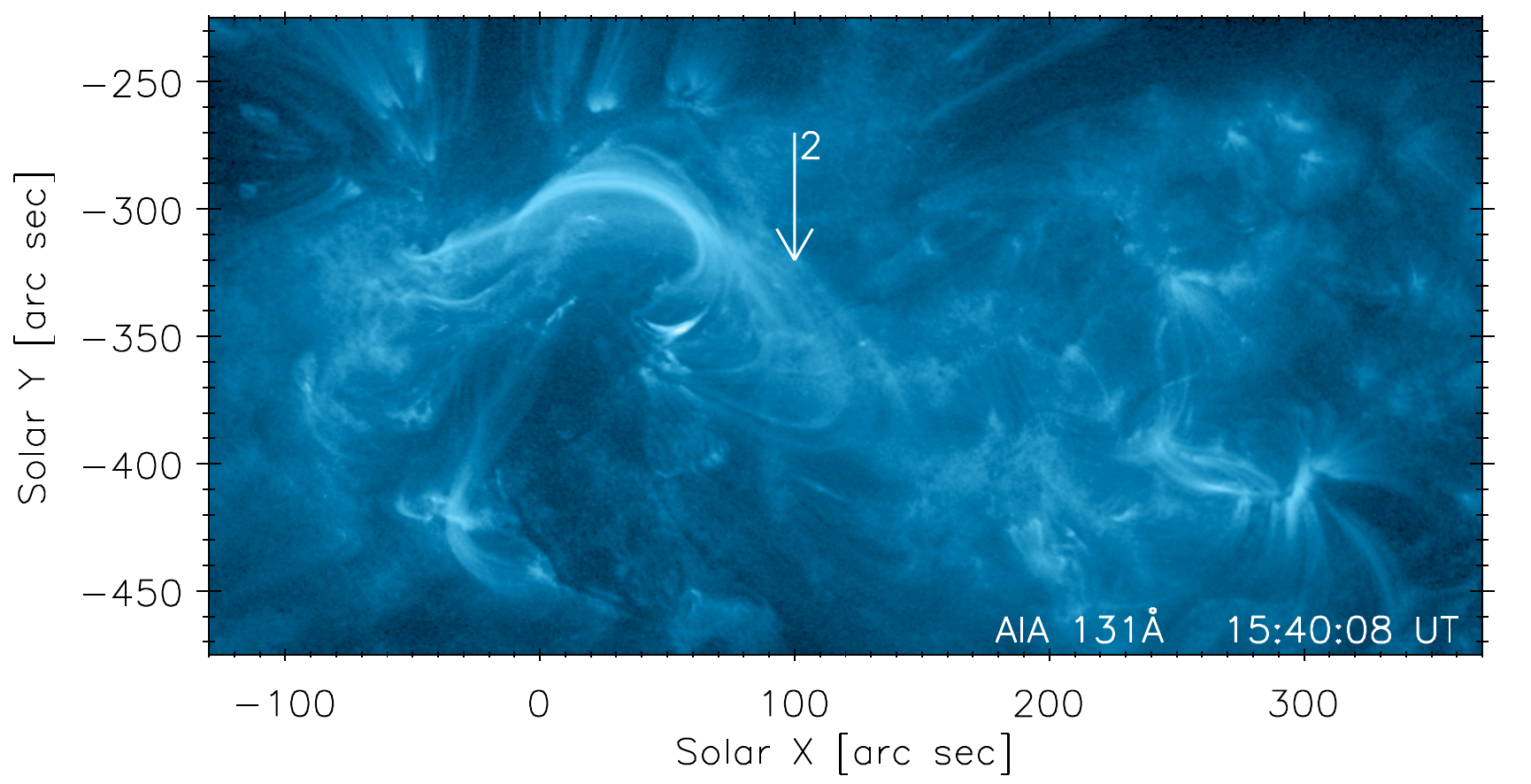}
       \includegraphics[width=8.8cm,bb=0 43 498 255]{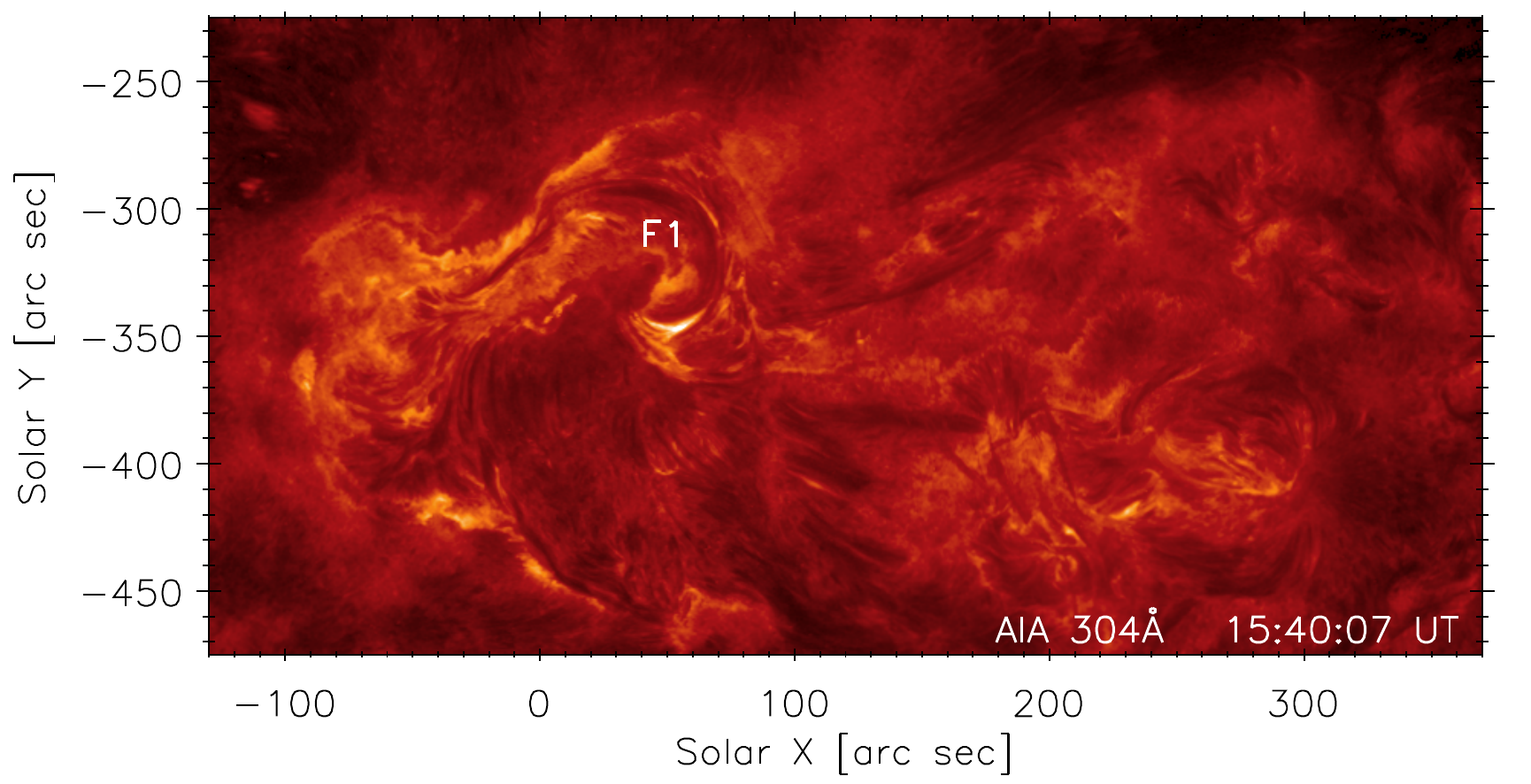}
       \includegraphics[width=8.8cm,bb=0 43 498 255]{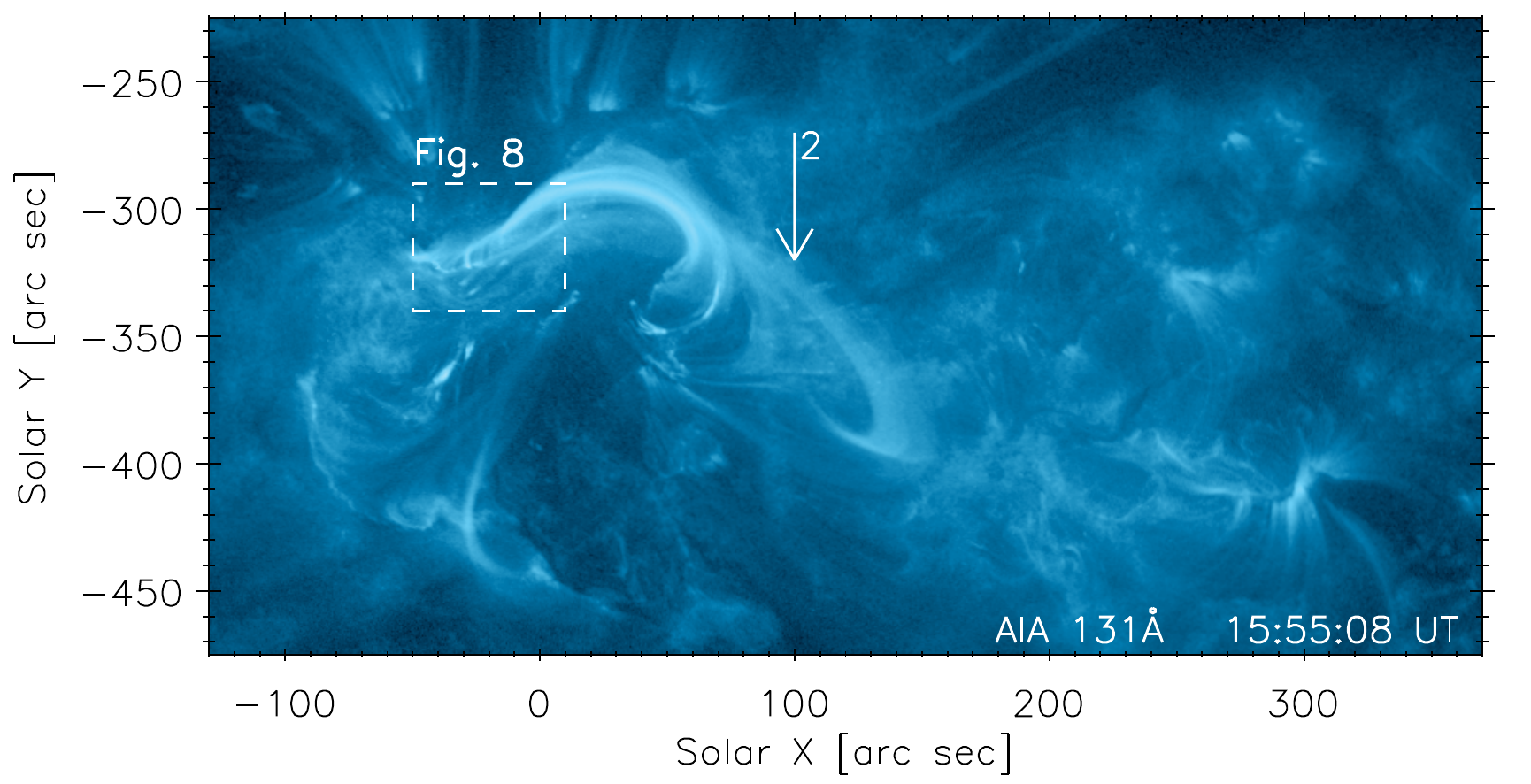}
       \includegraphics[width=8.8cm,bb=0 43 498 255]{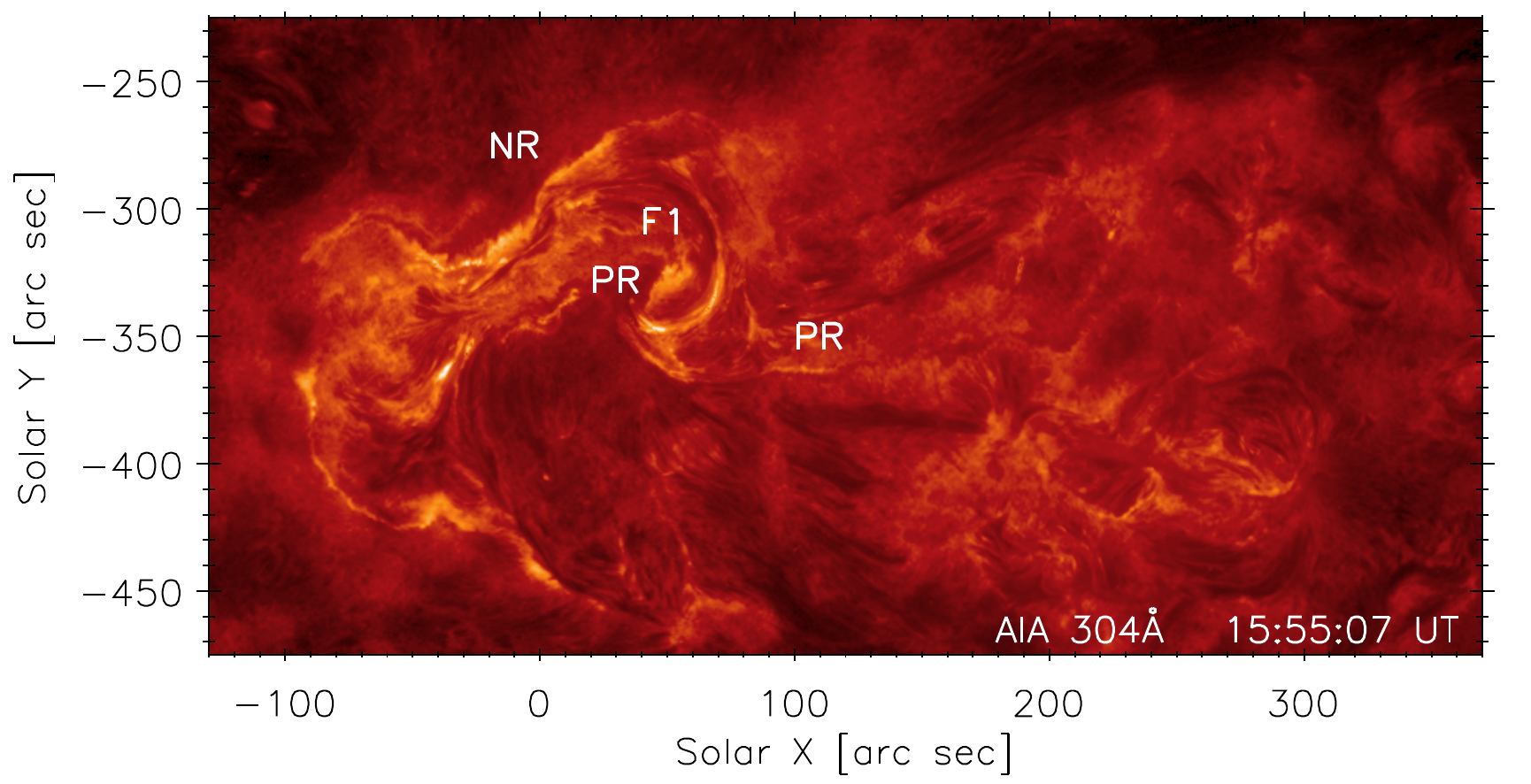}
       \includegraphics[width=8.8cm,bb=0 43 498 255]{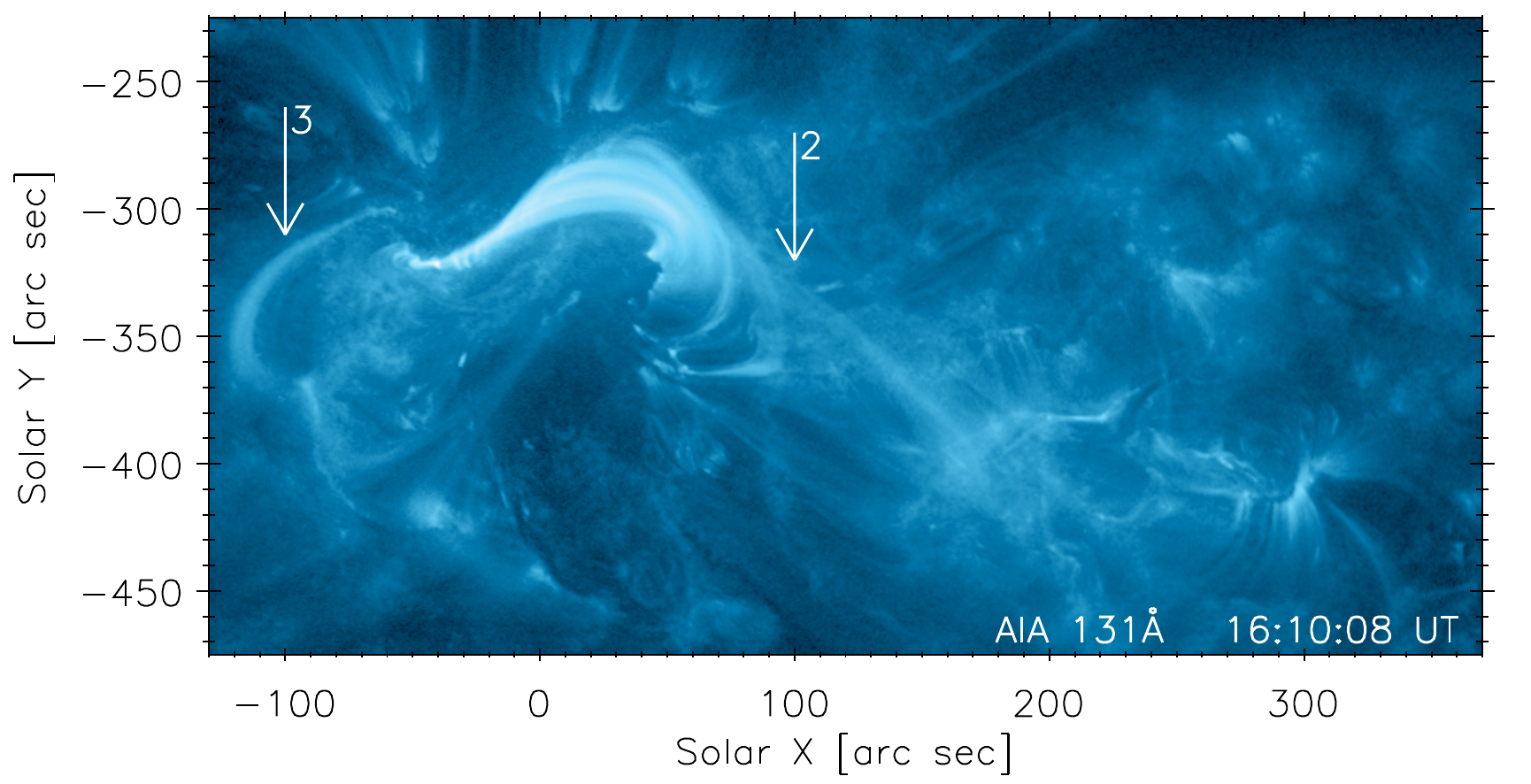}
       \includegraphics[width=8.8cm,bb=0 43 498 255]{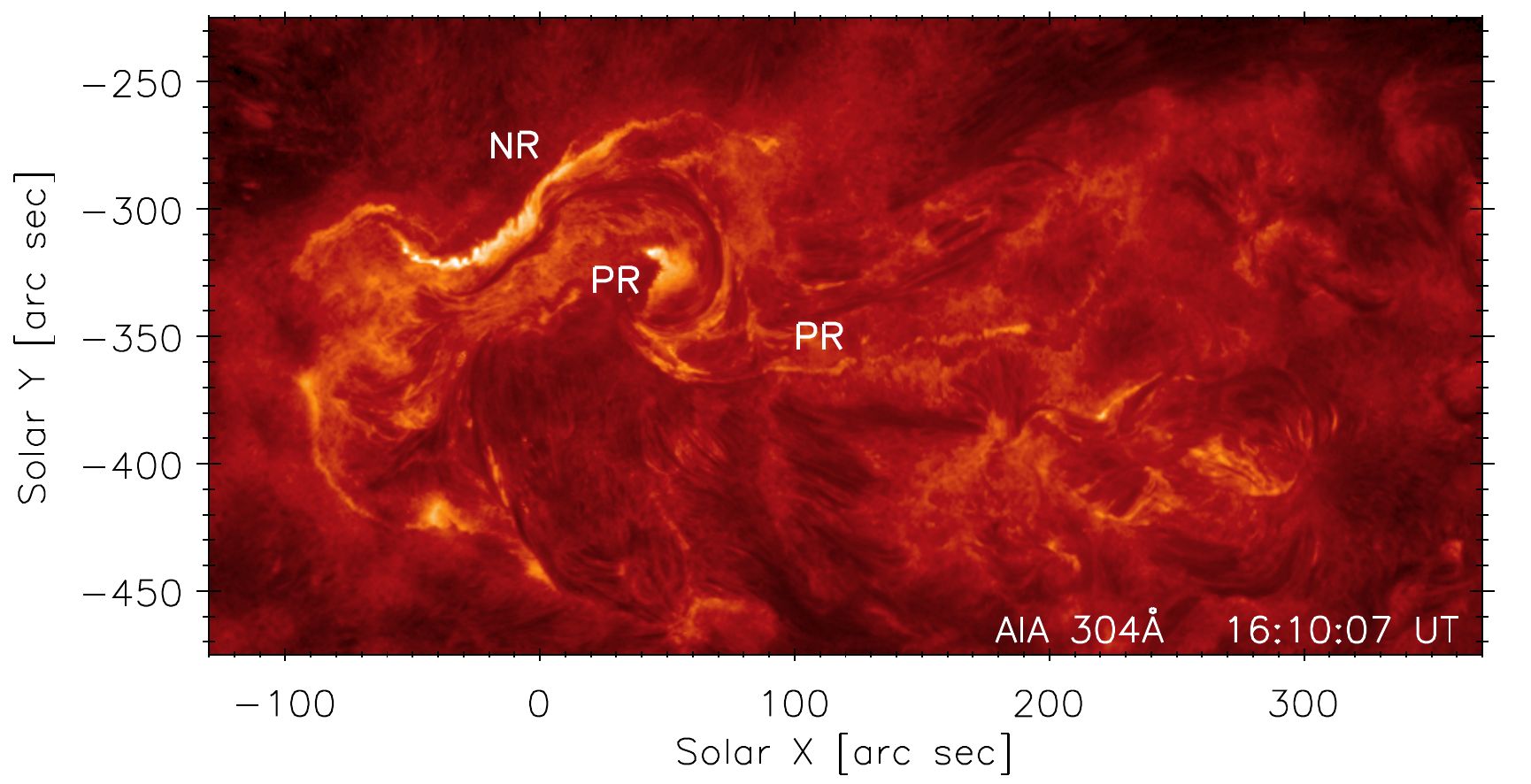}
       \includegraphics[width=8.8cm,bb=0 43 498 255]{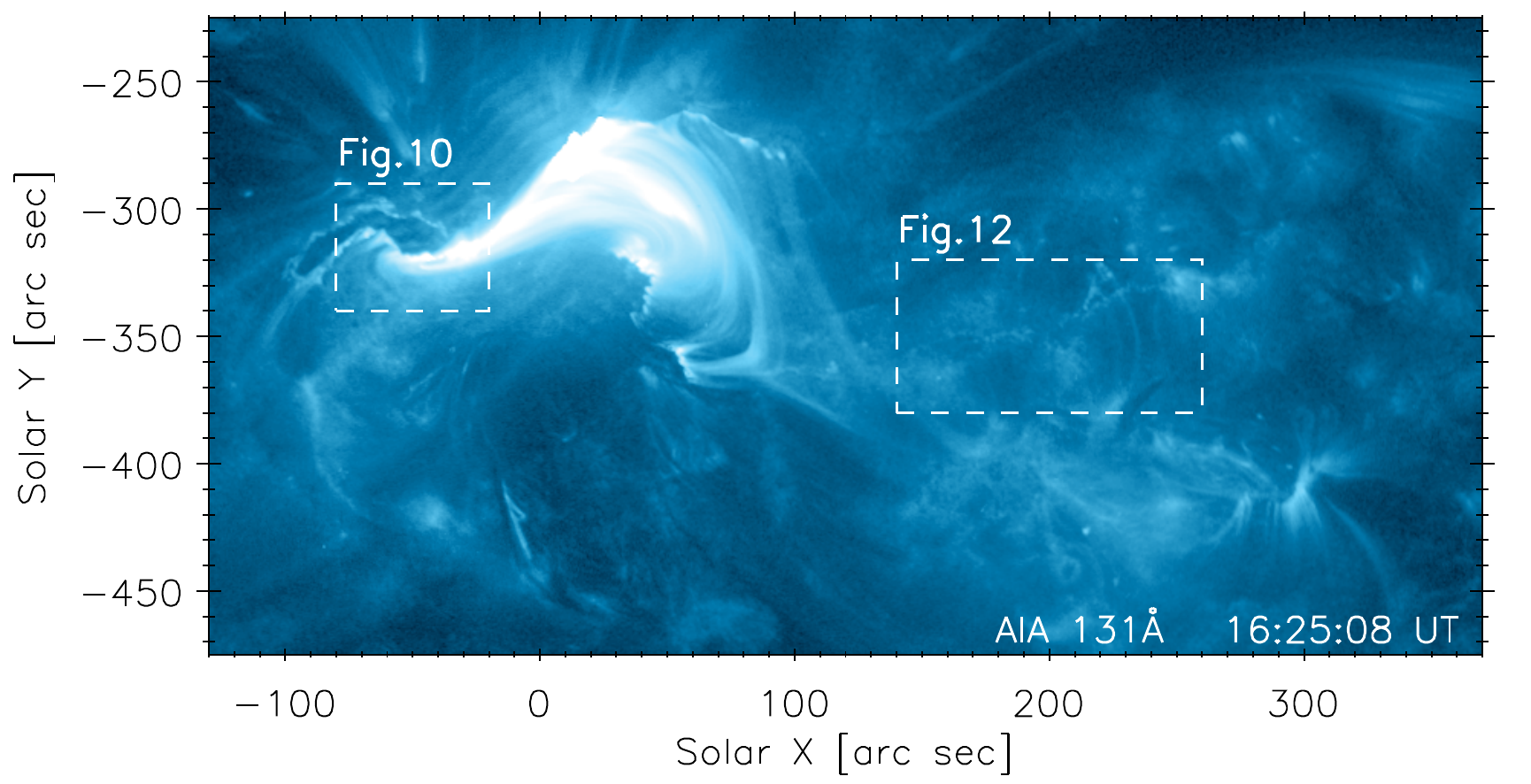}
       \includegraphics[width=8.8cm,bb=0 43 498 255]{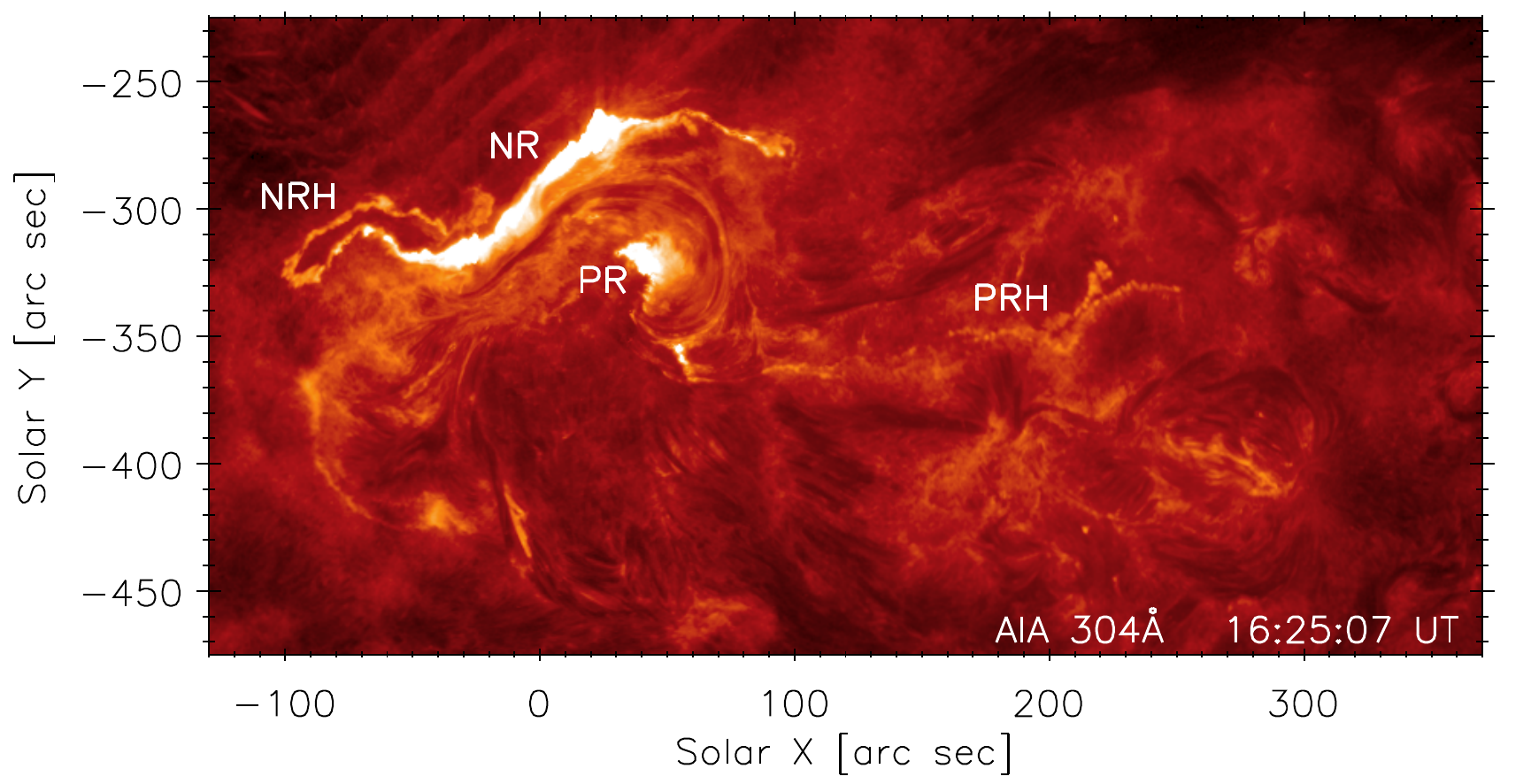}
       \includegraphics[width=8.8cm,bb=0  0 498 255]{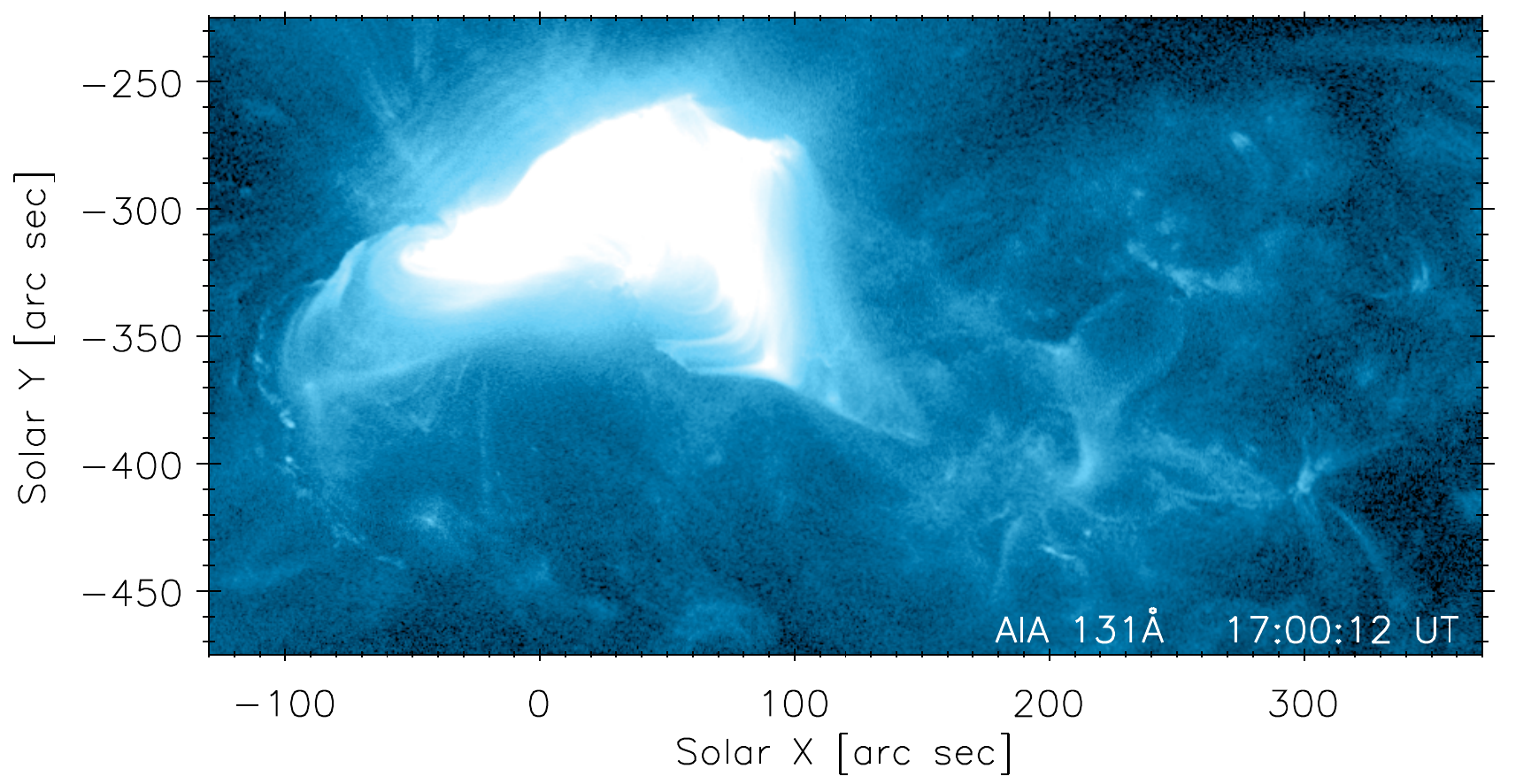}
       \includegraphics[width=8.8cm,bb=0  0 498 255]{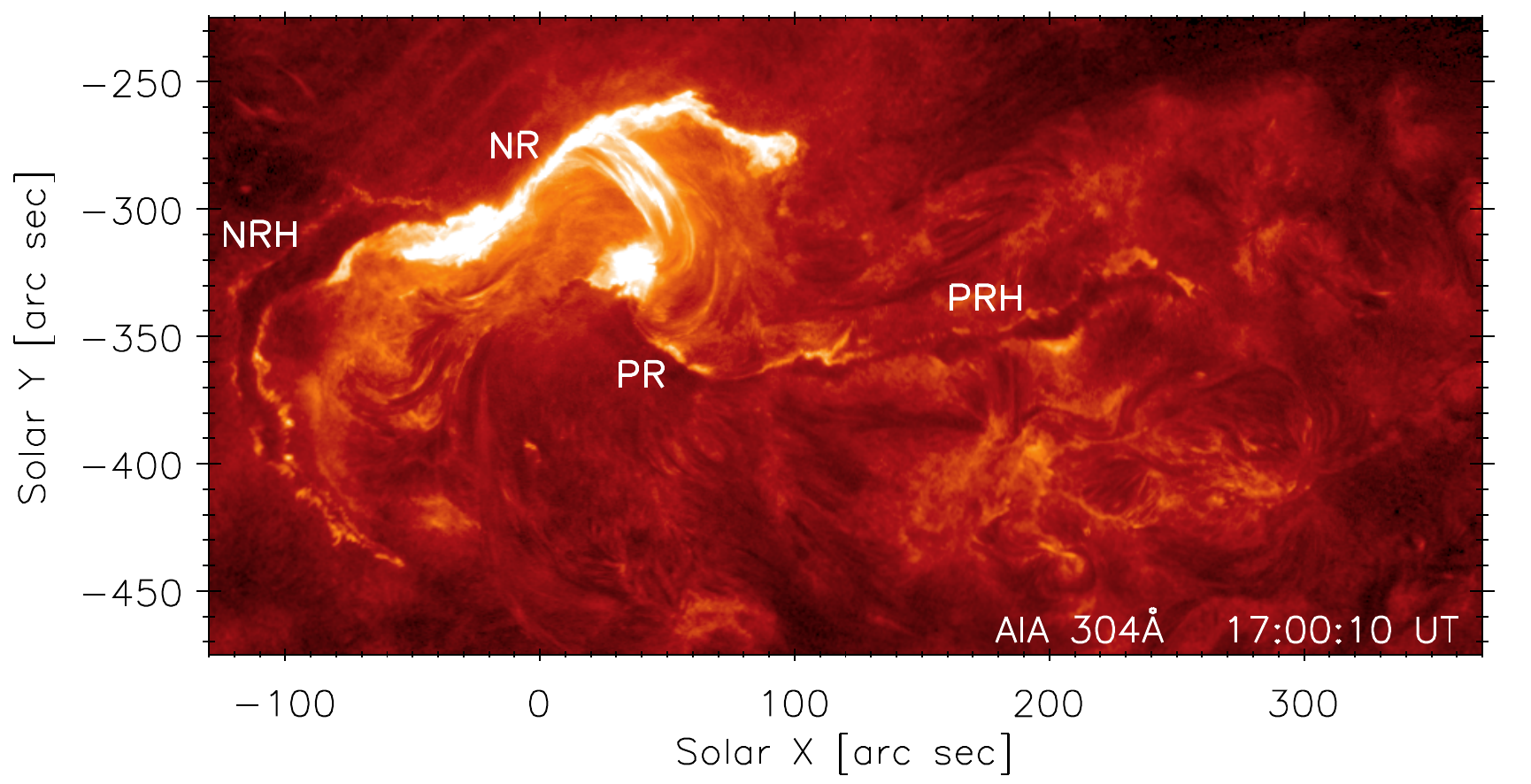}
     \caption{Overview of the evolution of the X1.4 flare on July 12, 2012. Arrow 1 denotes the first flare loop visible. Arrow 2 points to the erupting hot loops, while Arrow 3 denotes the loop appearing at the end of the NRH. Individual features involved in the flare are marked. See the text for details. Boxes enclose areas for more detailed time-series shown in the figures indicated. The intensities are scaled logarithmically. The images are available as mpeg animations (Movies 1 and 2) in the online version.
        }
       \label{Fig:Overview}
   \end{figure*}
%
   \begin{figure*}[!ht]
       \centering
       \includegraphics[width=3.50cm,clip]{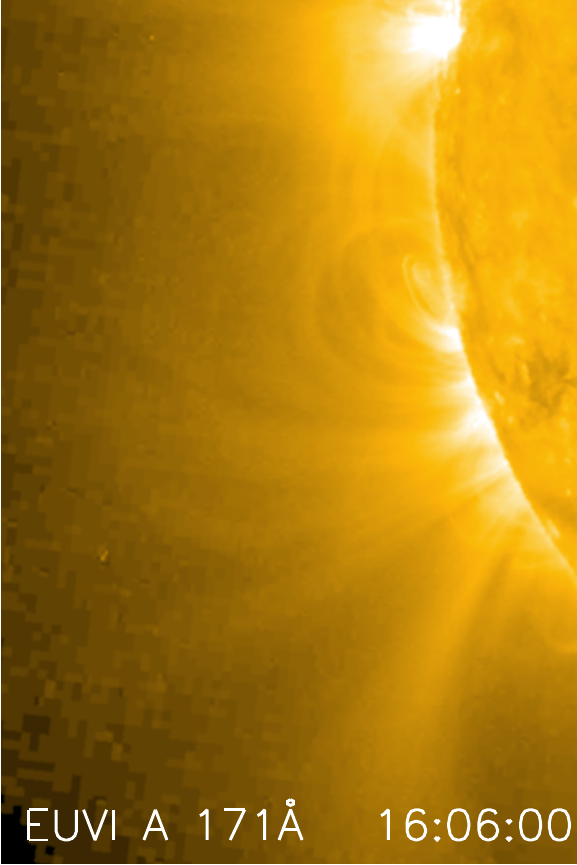}
       \includegraphics[width=3.50cm,clip]{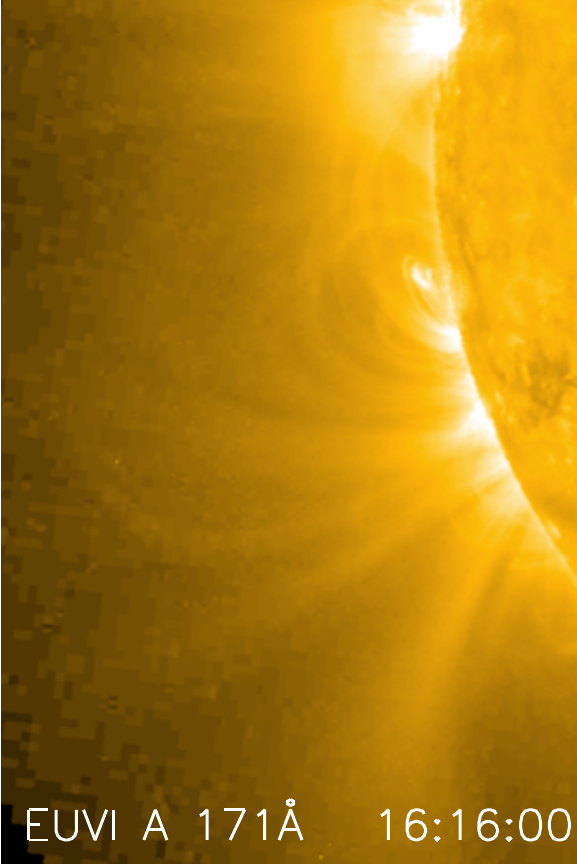}
       \includegraphics[width=3.50cm,clip]{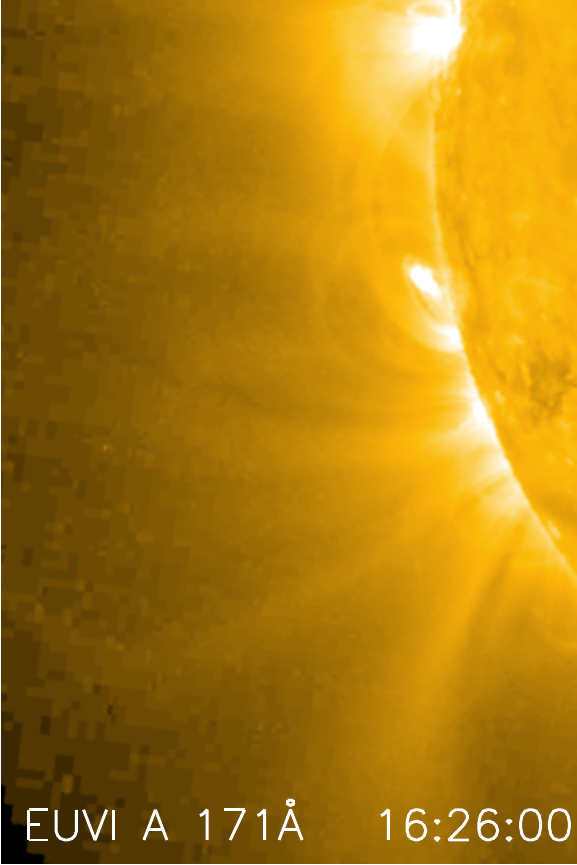}
       \includegraphics[width=3.50cm,clip]{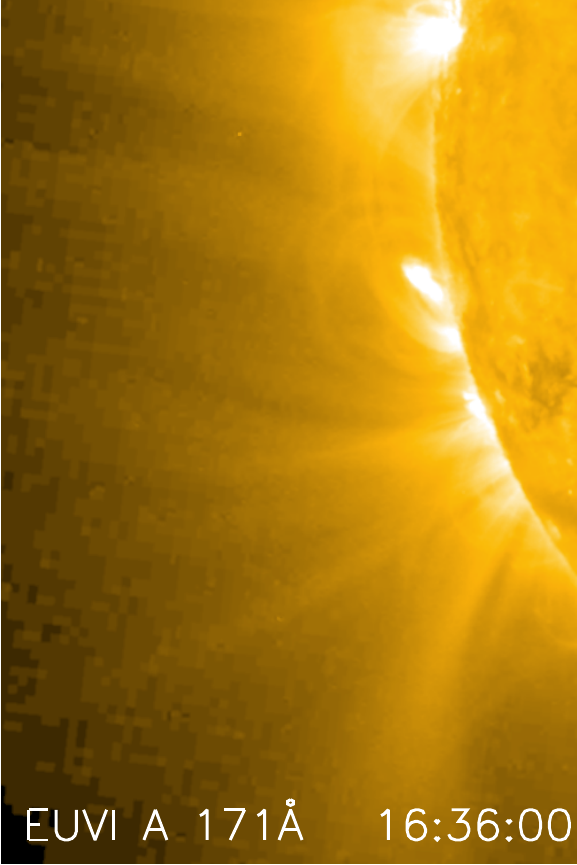}
       \includegraphics[width=3.50cm,clip]{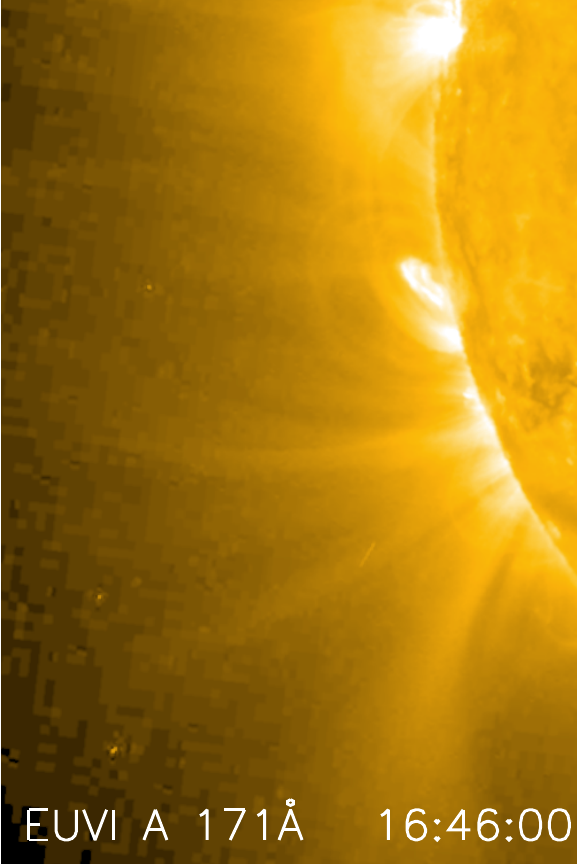}
     \caption{CME observed by the STEREO-A SECCHI/EUVI telescope in the 171\AA\,channel. The separation angle between STEREO-A spacecraft and Earth was $+120.0^\circ$.}
       \label{Fig:CME}
   \end{figure*}
%
%
%
   \begin{figure*}
       \centering
       \includegraphics[width=8.8cm,bb=0 40 498 255,clip]{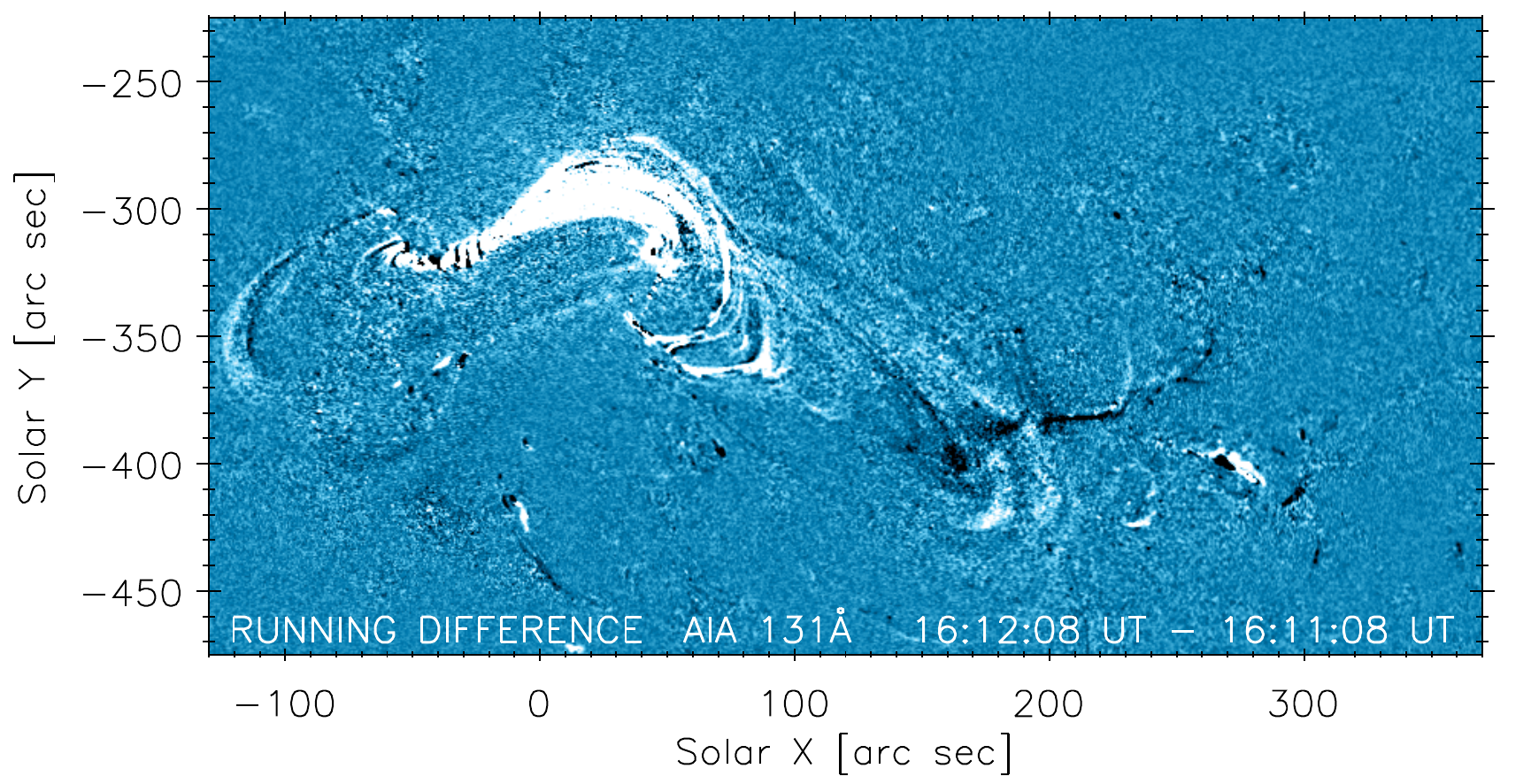}
       \includegraphics[width=8.8cm,bb=0 40 498 255,clip]{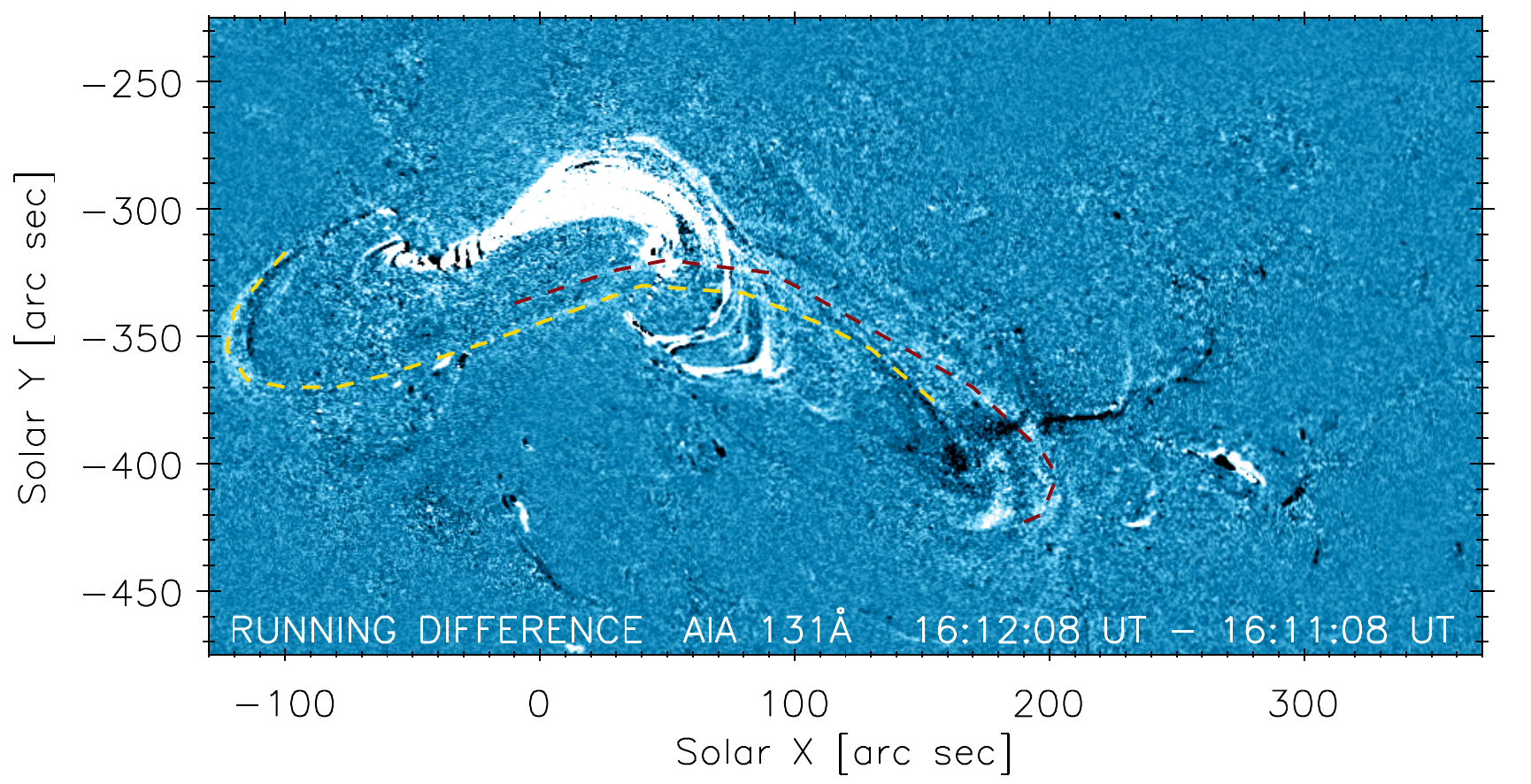}
       \includegraphics[width=8.8cm,bb=0 40 498 255,clip]{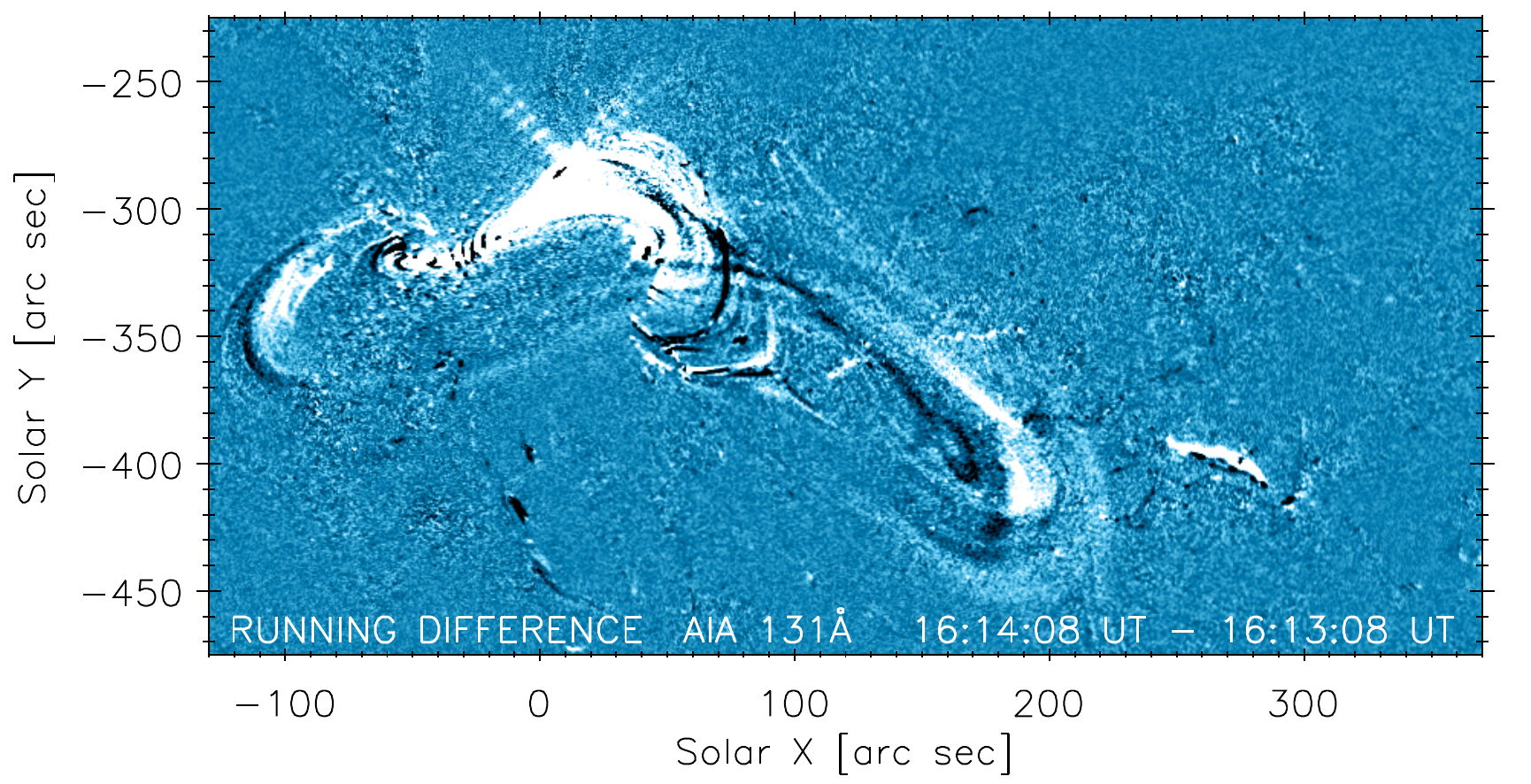}
       \includegraphics[width=8.8cm,bb=0 40 498 255,clip]{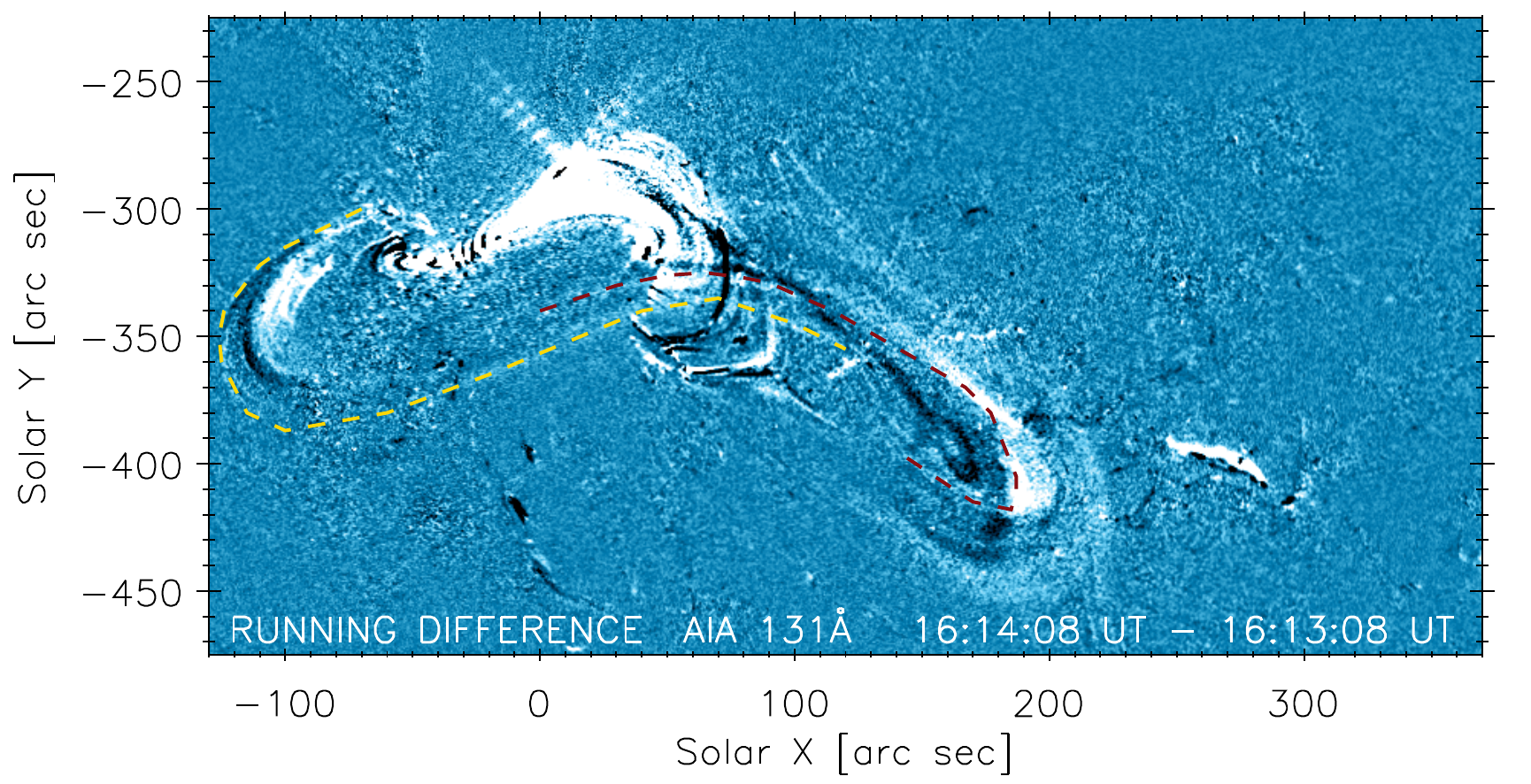}
       \includegraphics[width=8.8cm,bb=0  0 498 255,clip]{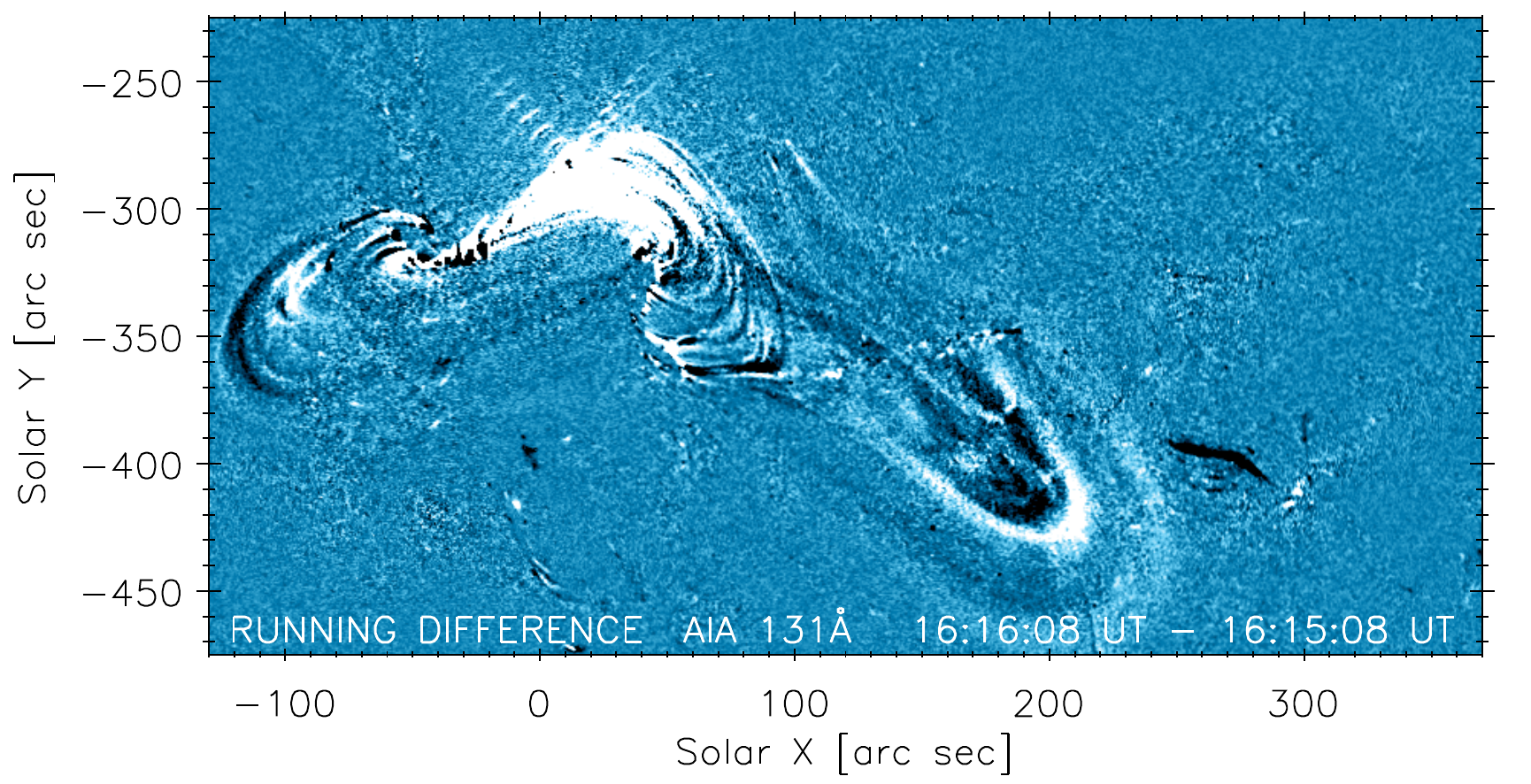}
       \includegraphics[width=8.8cm,bb=0  0 498 255,clip]{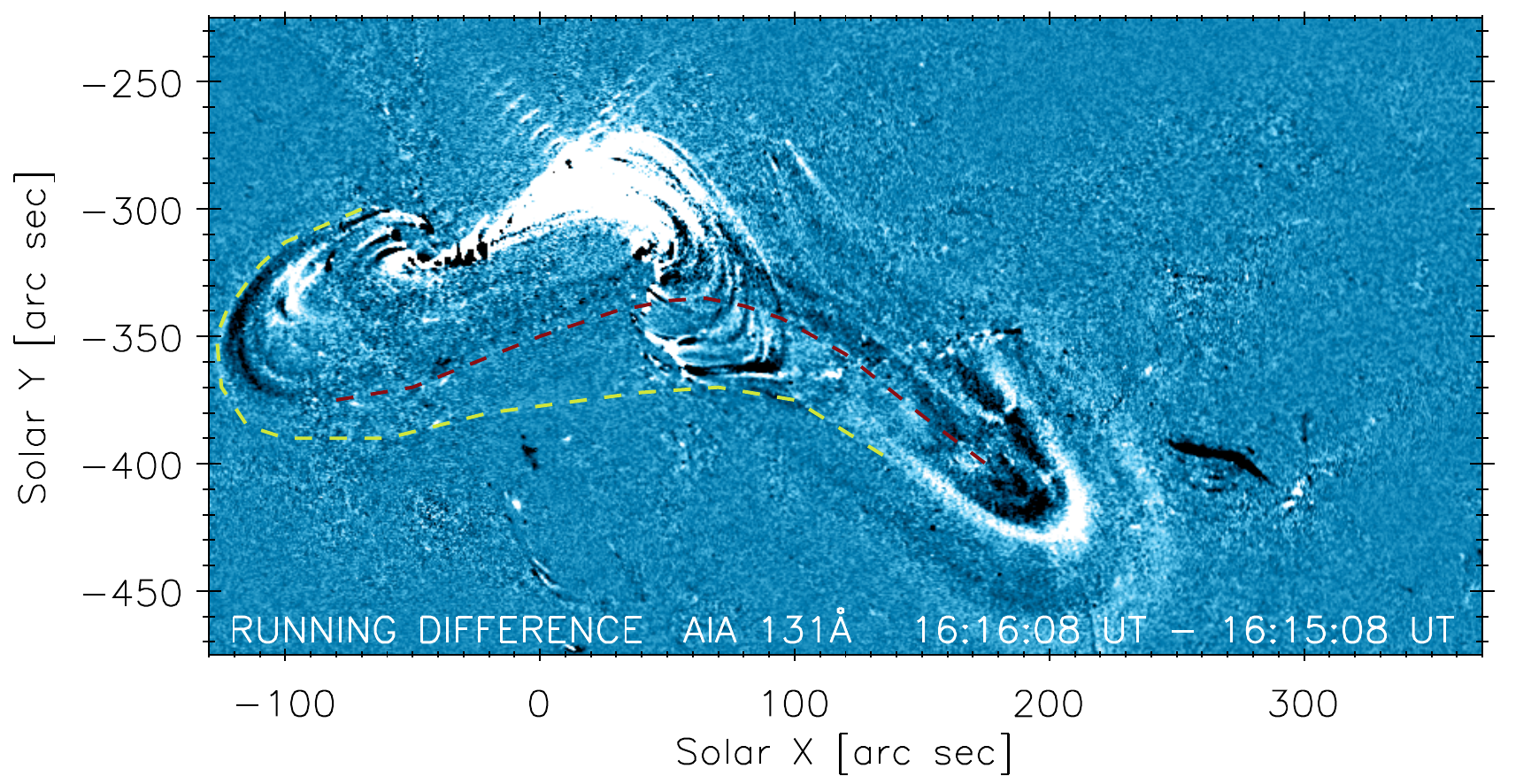}
    \caption{AIA 131\AA~images running-difference images with time delay of 1 min, saturated to $\pm$20\,DN\,s$^{-1}$\,px$^{-1}$. Portions of two long erupting loops are outlined by colored lines in the \textit{right} images.}
       \label{Fig:Eruption}
   \end{figure*}
%
%
   \begin{figure*}[!ht]
       \centering
       \includegraphics[height=1.15cm, bb=0   0 220 56, clip]{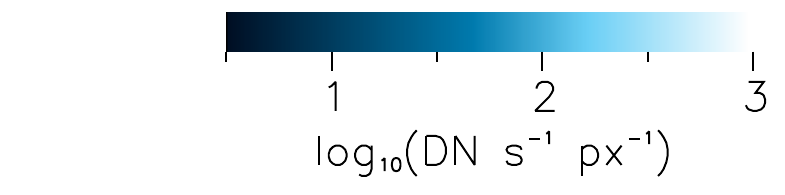}
       \includegraphics[height=1.15cm, bb=60  0 220 56, clip]{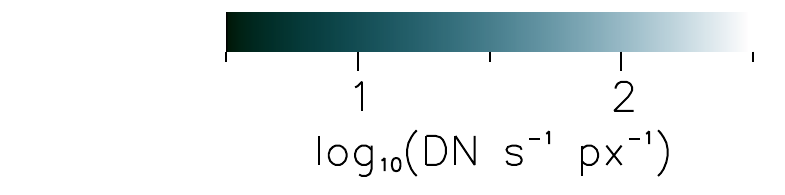}
       \includegraphics[height=1.15cm, bb=60  0 220 56, clip]{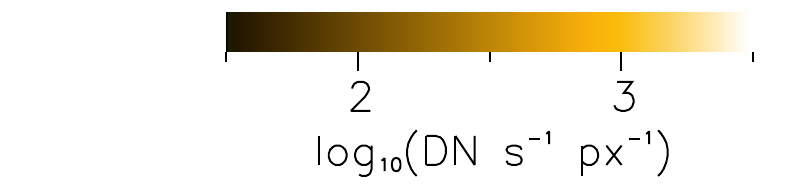}
       \includegraphics[height=1.15cm, bb=60  0 220 56, clip]{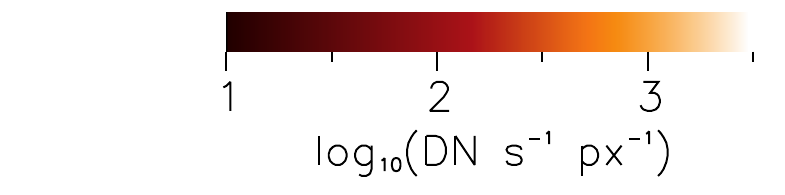}
       \includegraphics[height=1.15cm, bb=60  0 220 56, clip]{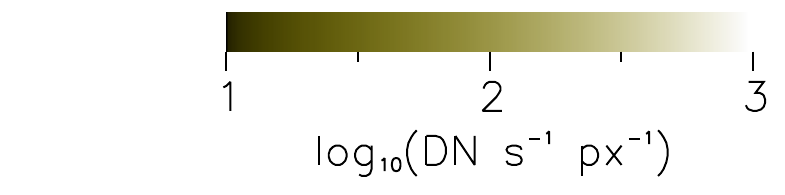}

       \includegraphics[height=2.76cm, bb=0  40 220 175, clip]{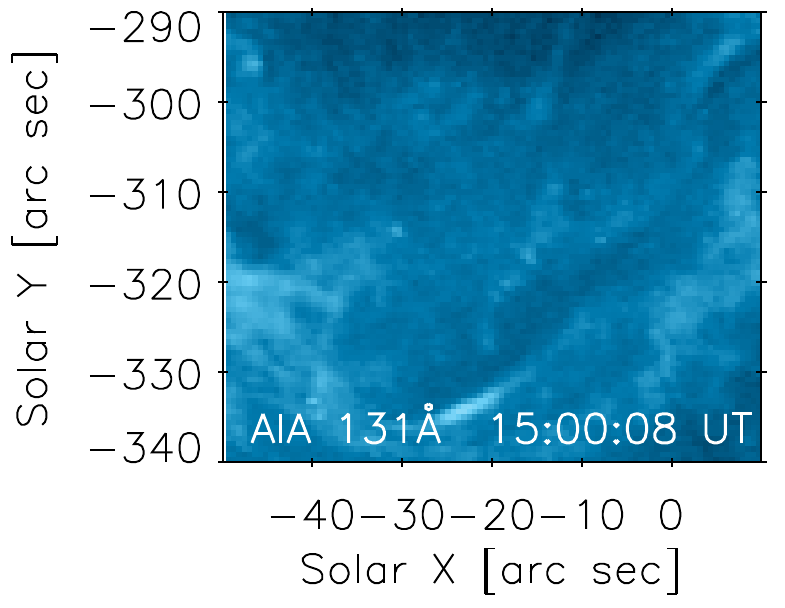}
       \includegraphics[height=2.76cm, bb=60 40 220 175, clip]{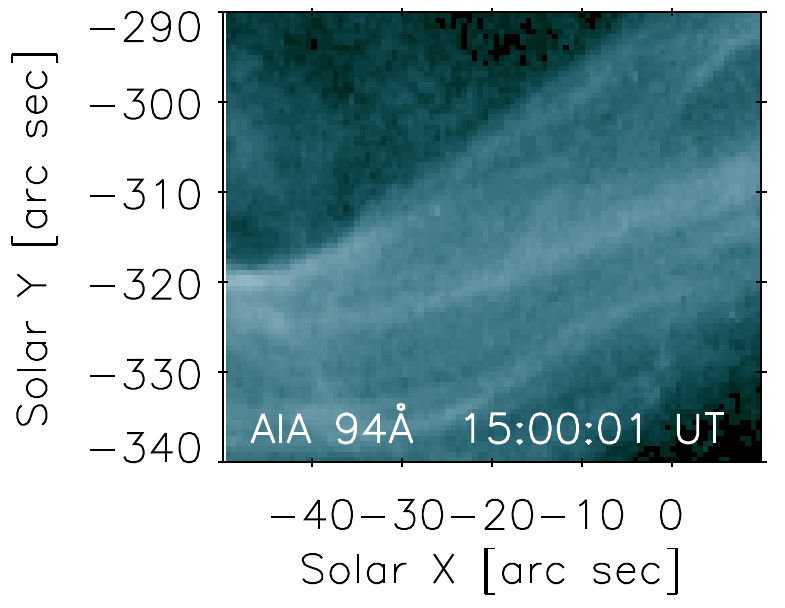}
       \includegraphics[height=2.76cm, bb=60 40 220 175, clip]{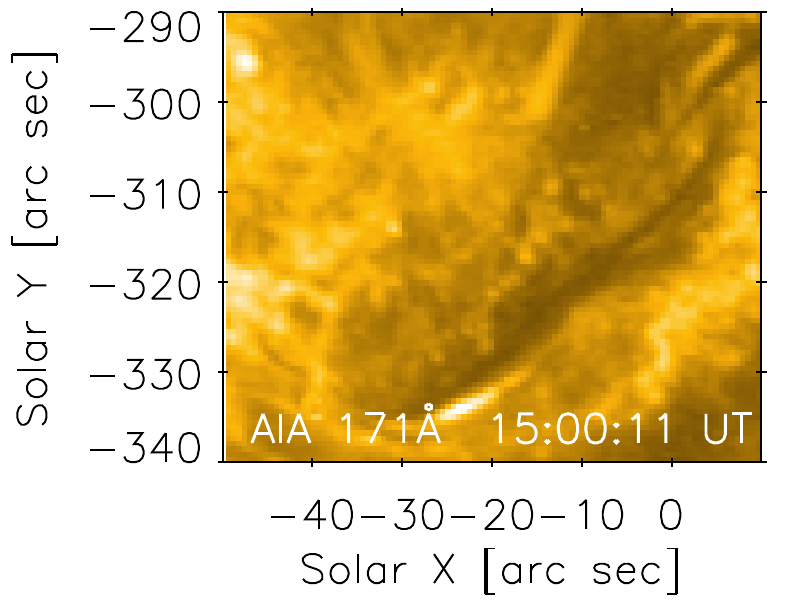}
       \includegraphics[height=2.76cm, bb=60 40 220 175, clip]{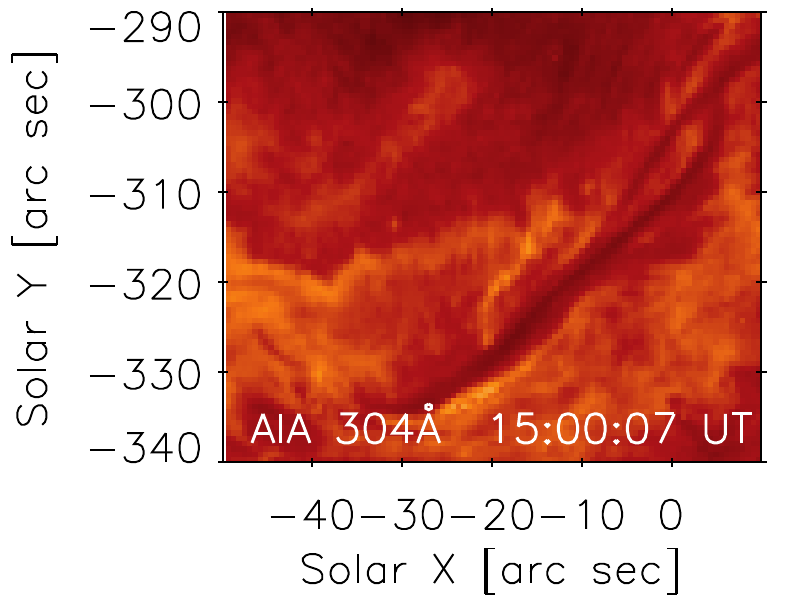}
       \includegraphics[height=2.76cm, bb=60 40 220 175, clip]{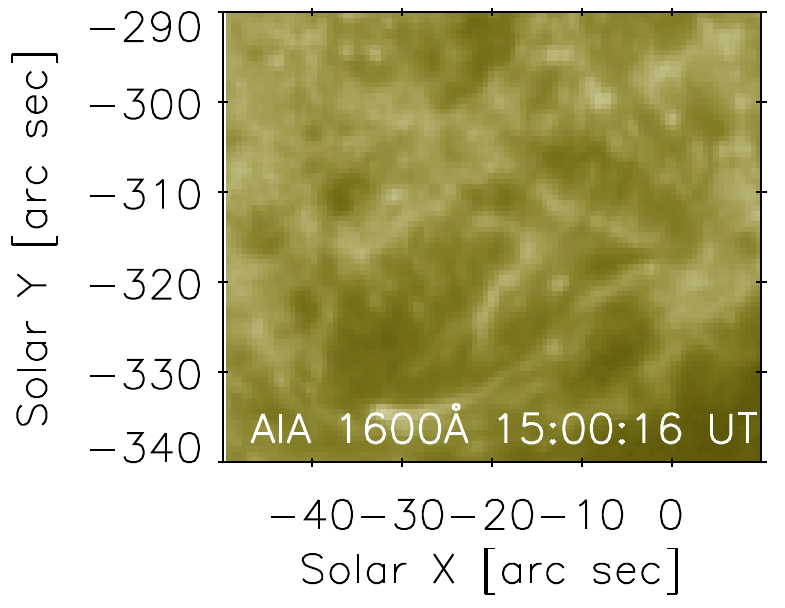}

       \includegraphics[height=2.76cm, bb=0  40 220 175, clip]{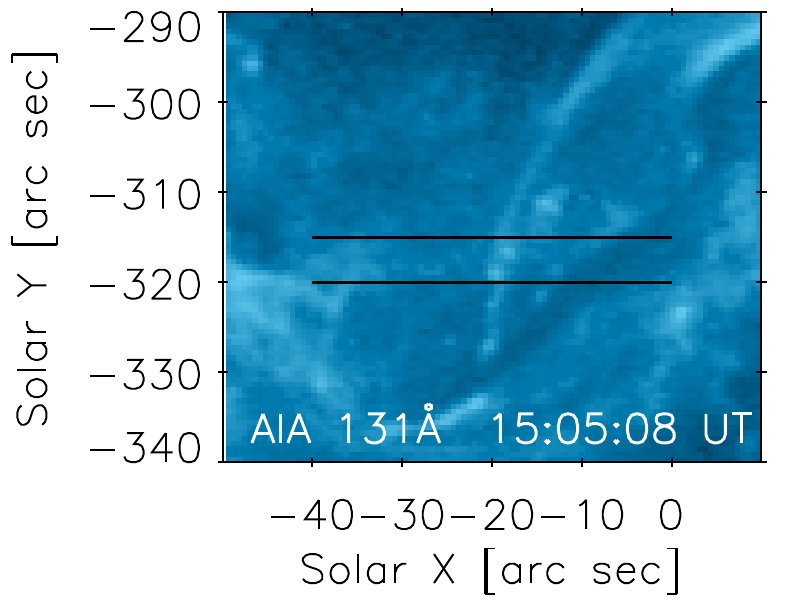}
       \includegraphics[height=2.76cm, bb=60 40 220 175, clip]{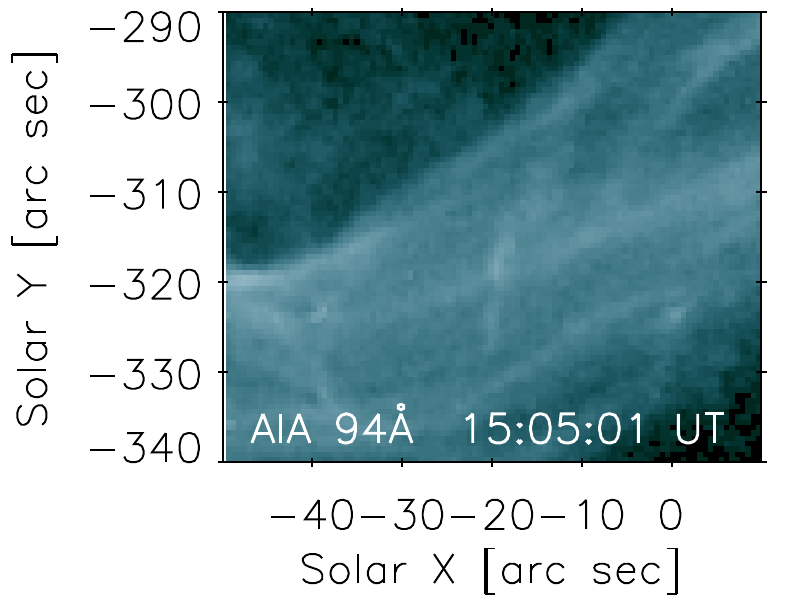}
       \includegraphics[height=2.76cm, bb=60 40 220 175, clip]{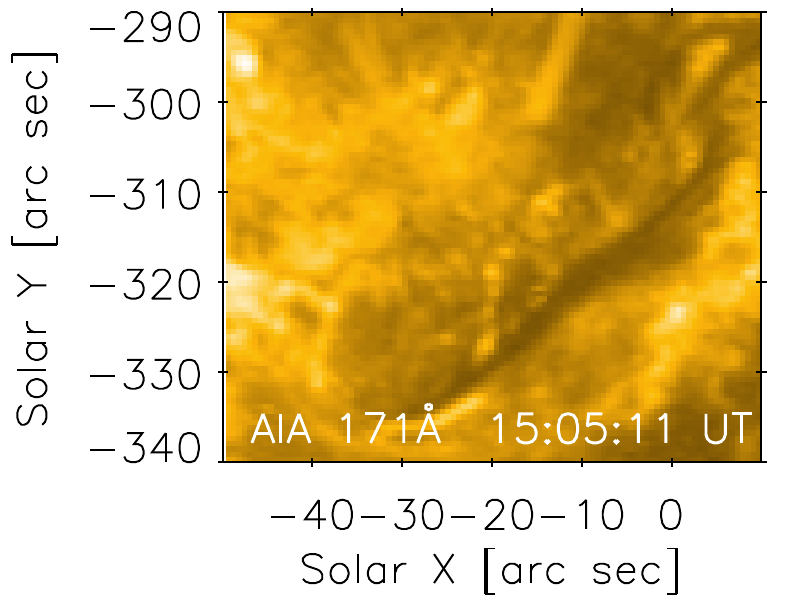}
       \includegraphics[height=2.76cm, bb=60 40 220 175, clip]{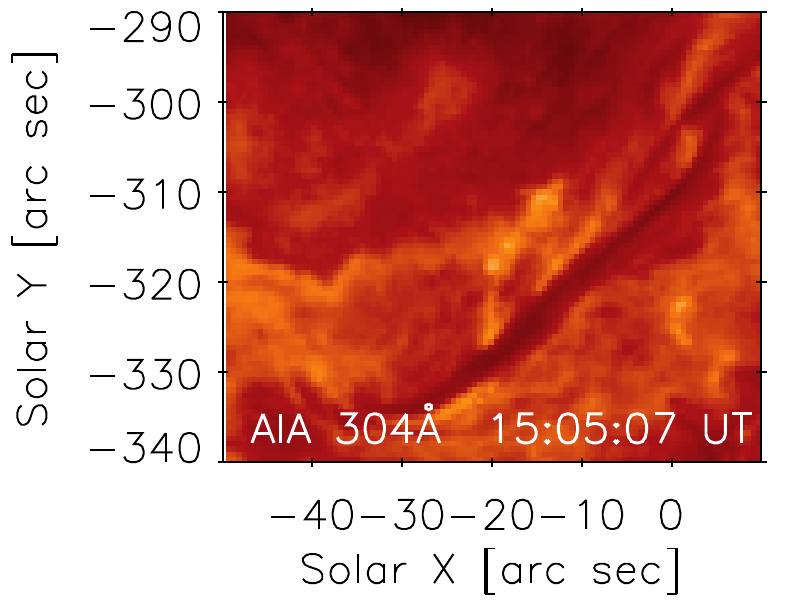}
       \includegraphics[height=2.76cm, bb=60 40 220 175, clip]{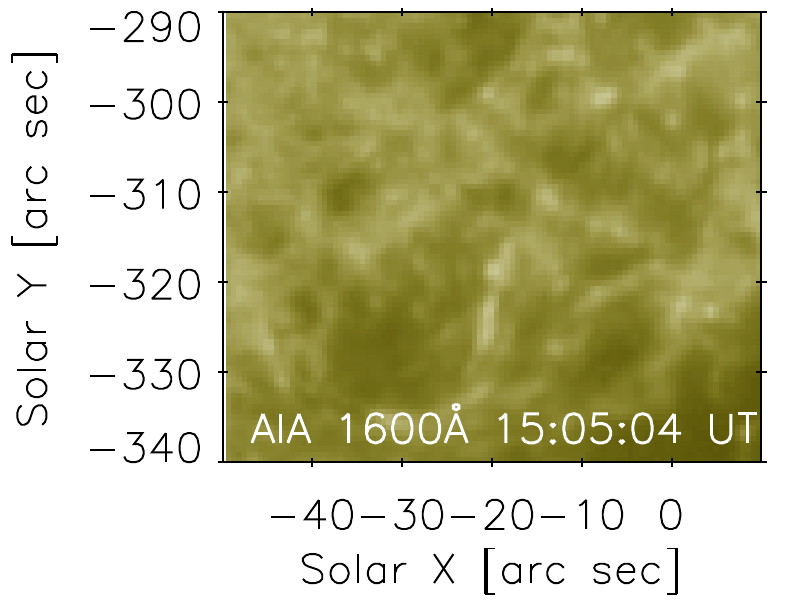}

       \includegraphics[height=2.76cm, bb=0  40 220 175, clip]{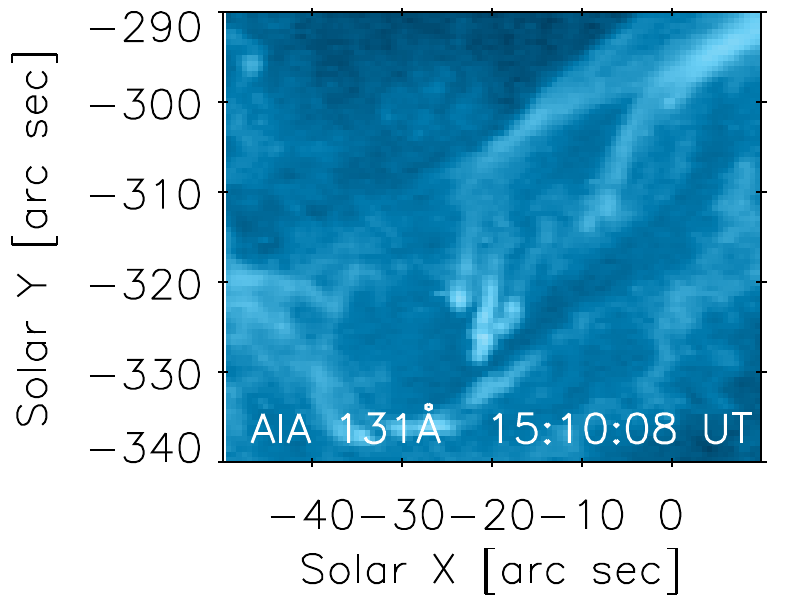}
       \includegraphics[height=2.76cm, bb=60 40 220 175, clip]{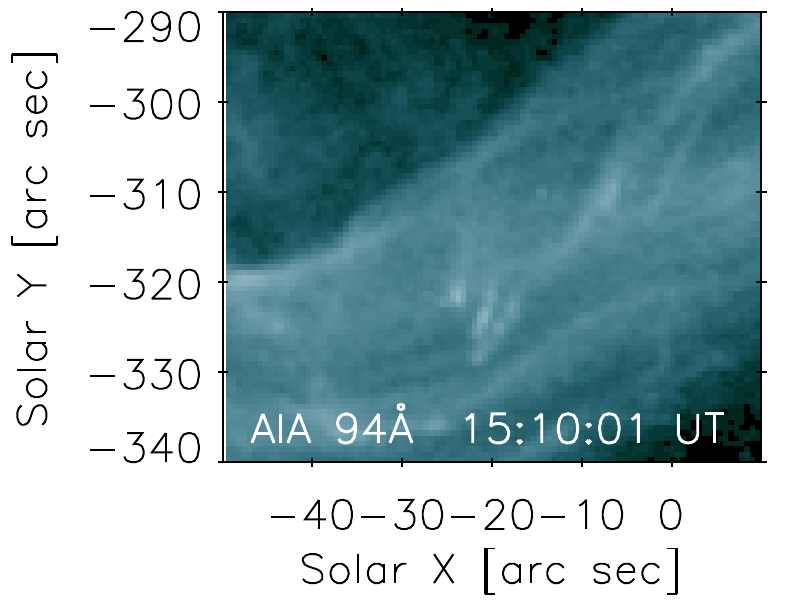}
       \includegraphics[height=2.76cm, bb=60 40 220 175, clip]{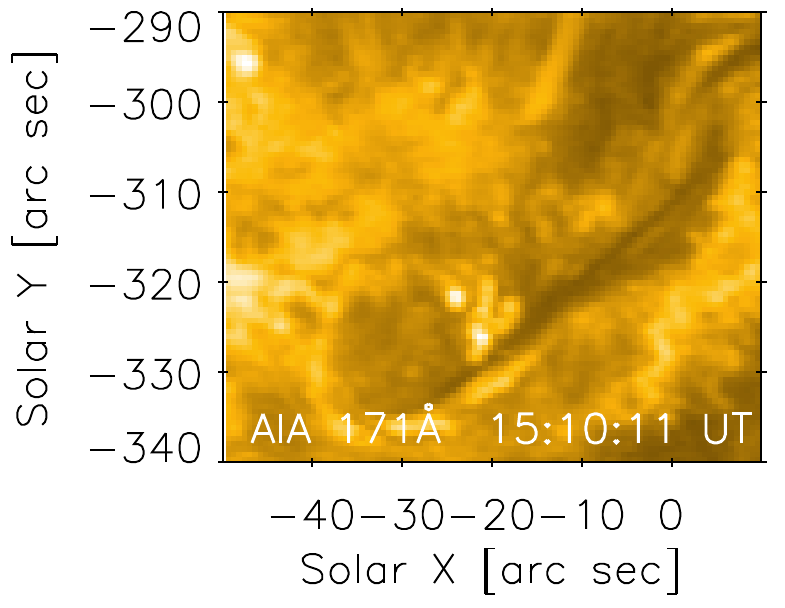}
       \includegraphics[height=2.76cm, bb=60 40 220 175, clip]{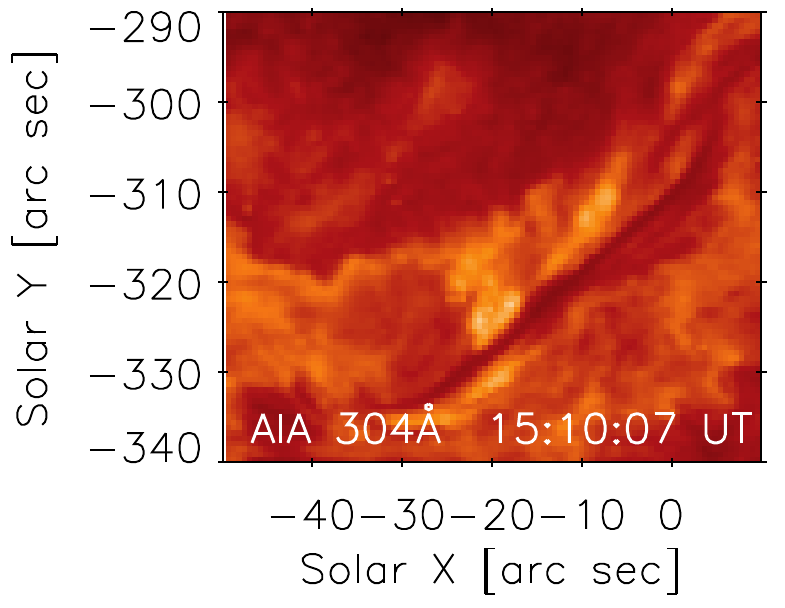}
       \includegraphics[height=2.76cm, bb=60 40 220 175, clip]{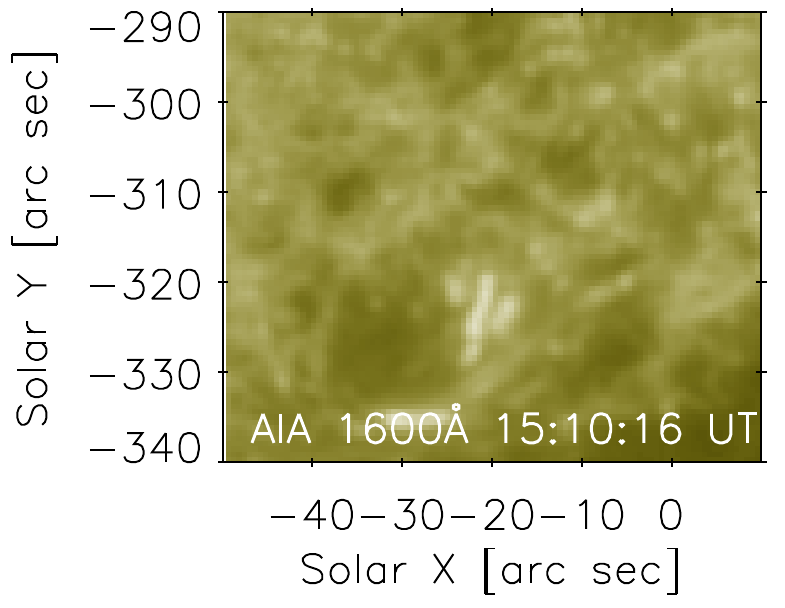}

       \includegraphics[height=2.76cm, bb=0  40 220 175, clip]{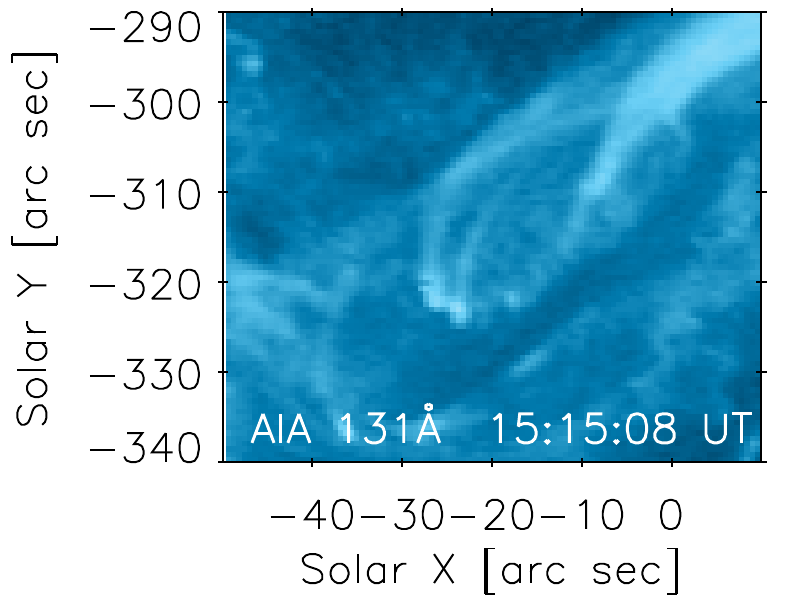}
       \includegraphics[height=2.76cm, bb=60 40 220 175, clip]{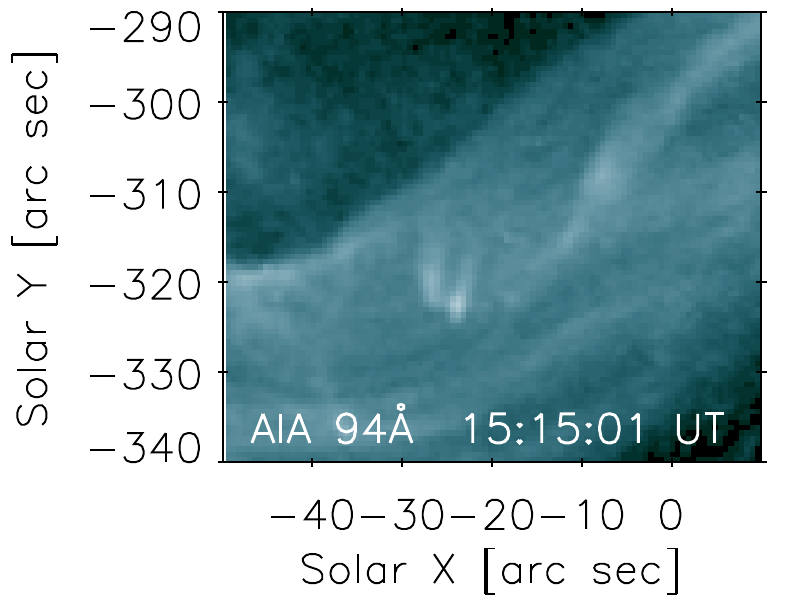}
       \includegraphics[height=2.76cm, bb=60 40 220 175, clip]{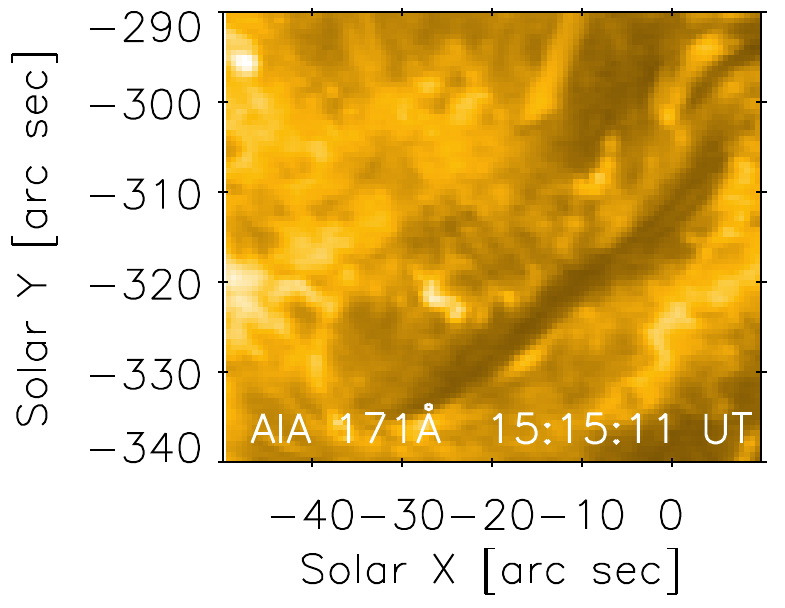}
       \includegraphics[height=2.76cm, bb=60 40 220 175, clip]{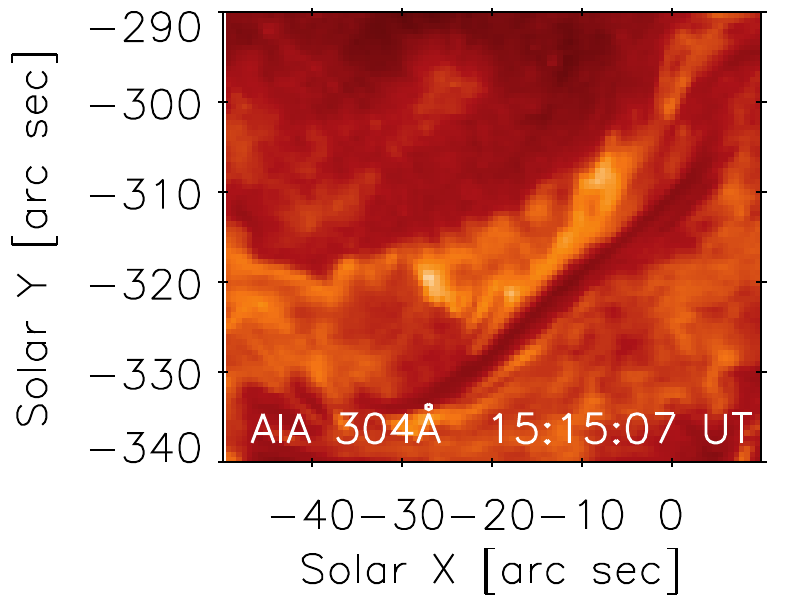}
       \includegraphics[height=2.76cm, bb=60 40 220 175, clip]{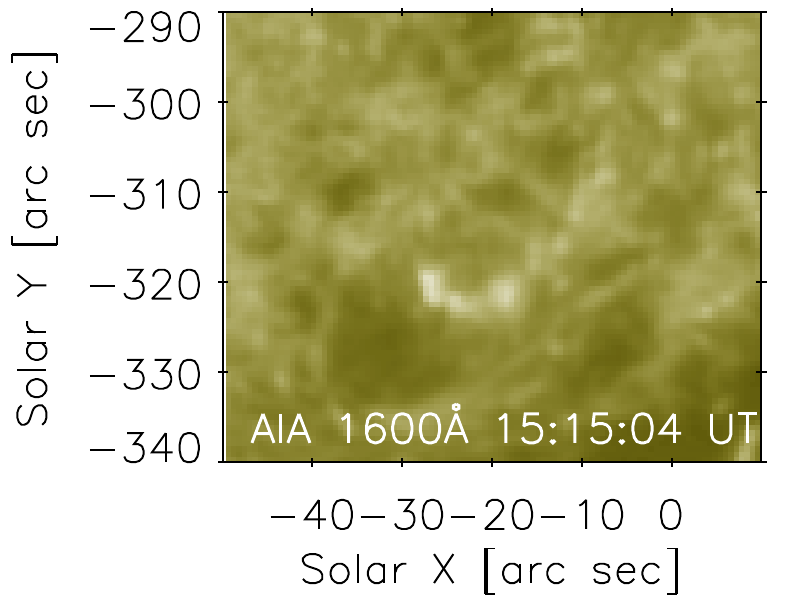}

       \includegraphics[height=3.58cm, bb=0   0 220 175, clip]{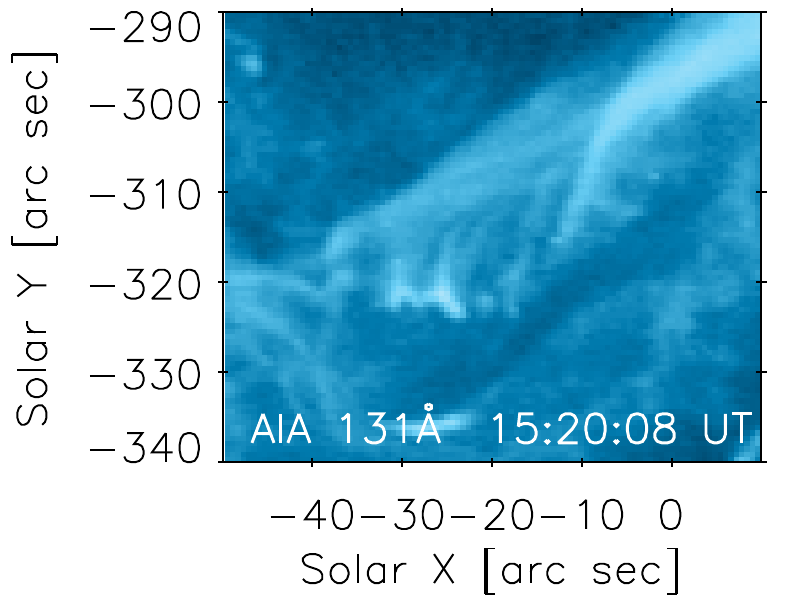}
       \includegraphics[height=3.58cm, bb=60  0 220 175, clip]{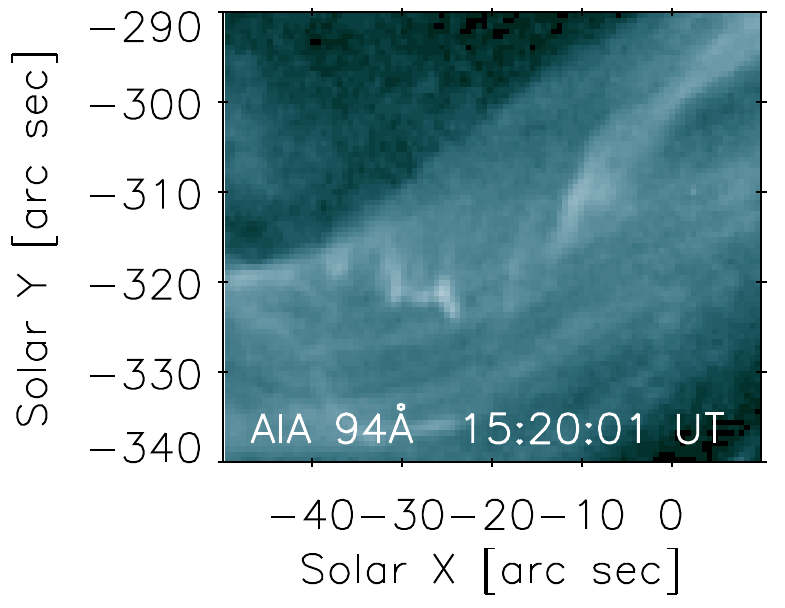}
       \includegraphics[height=3.58cm, bb=60  0 220 175, clip]{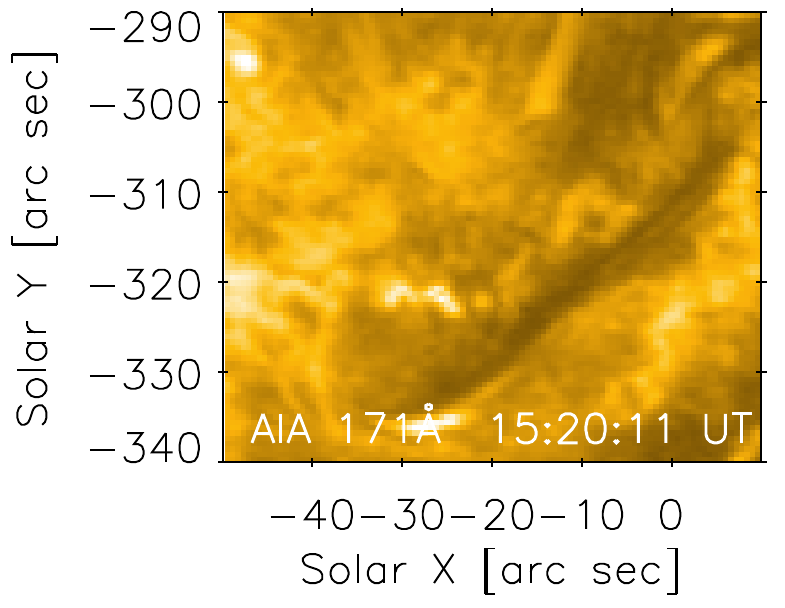}
       \includegraphics[height=3.58cm, bb=60  0 220 175, clip]{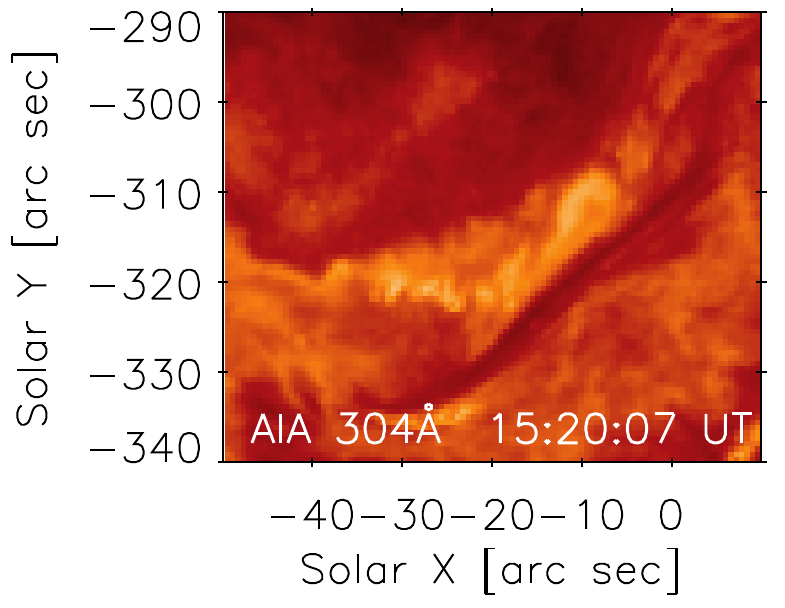}
       \includegraphics[height=3.58cm, bb=60  0 220 175, clip]{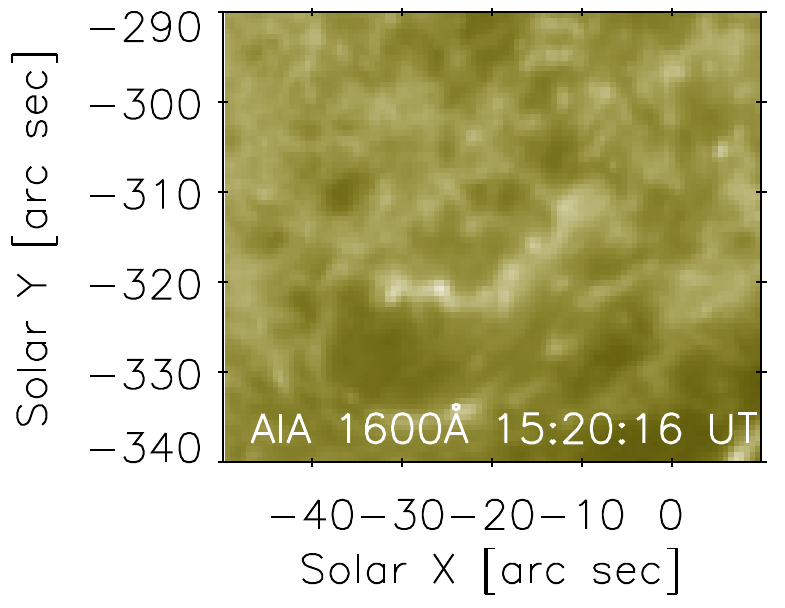}

      \caption{Slipping magnetic loops at the beginning of the flare. Dark lines in the \textit{top left} panel show positions of the cuts used to construct $X$-$t$ plots (stackplots) shown in Fig. \ref{Fig:Slip1_stackplots}. The intensities are scaled logarithmically, with units of DN\,s$^{-1}$\,px$^{-1}$. An animation of the AIA 131\AA~observations (\textit{left column}) is available as the online Movie 4. 
        }
       \label{Fig:Slip1}
   \end{figure*}
%
%
   \begin{figure}
    \centering
    \includegraphics[height=4.7cm,bb=5  0 249 249,clip]{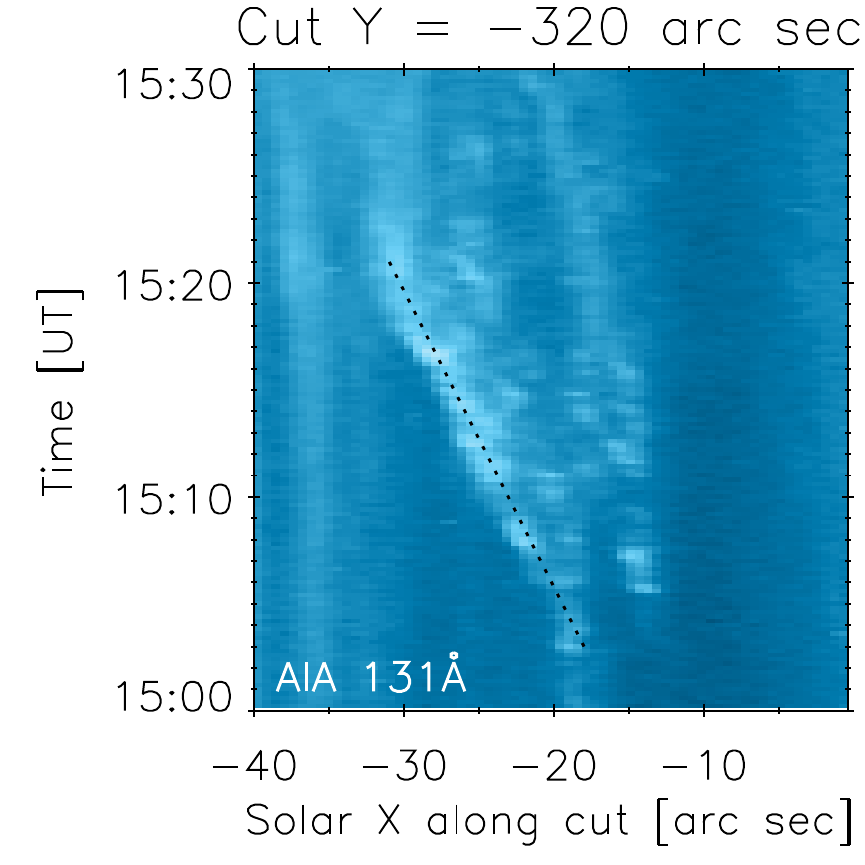}
    \includegraphics[height=4.7cm,bb=67 0 249 249,clip]{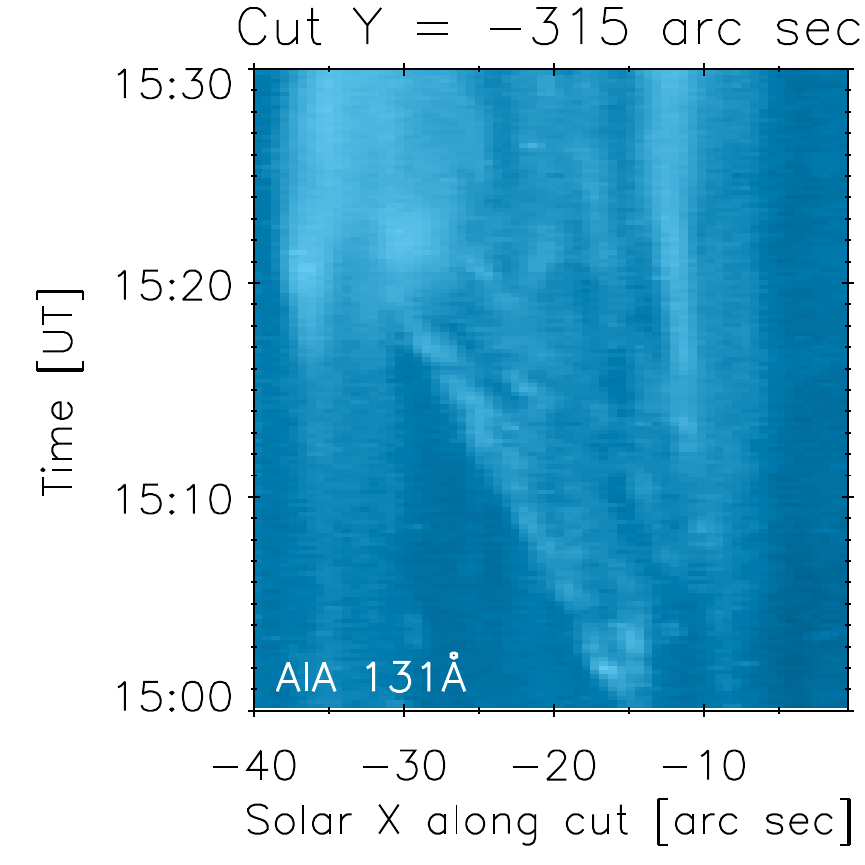}
    \caption{Stackplots along the two cuts plotted in Fig.\,\ref{Fig:Slip1} showing slipping loops in the AIA 131\AA filter. The black line on the \textit{left} image corresponds to the velocity of 8.7 km\,s$^{-1}$.
        }
       \label{Fig:Slip1_stackplots}
   \end{figure}
%
%
   \begin{figure*}[!ht]
       \centering
       \includegraphics[height=1.15cm, bb=0   0 220 56, clip]{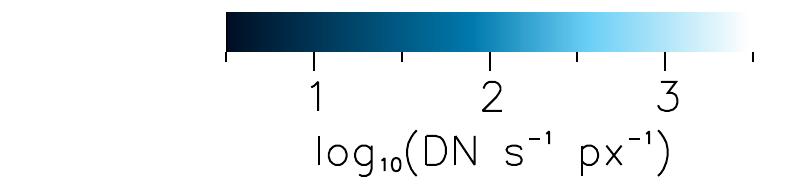}
       \includegraphics[height=1.15cm, bb=60  0 220 56, clip]{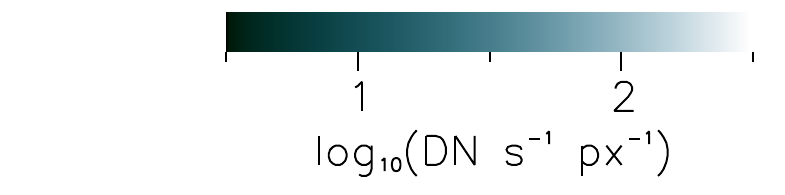}
       \includegraphics[height=1.15cm, bb=60  0 220 56, clip]{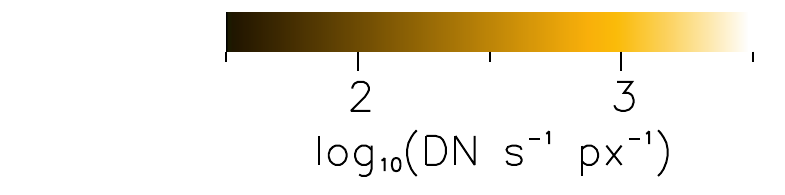}
       \includegraphics[height=1.15cm, bb=60  0 220 56, clip]{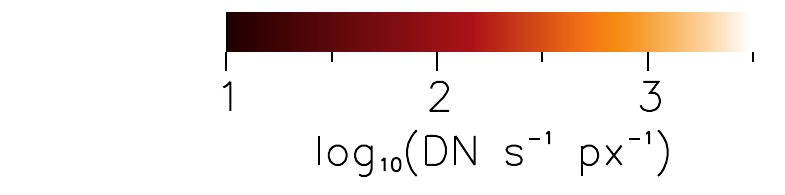}
       \includegraphics[height=1.15cm, bb=60  0 220 56, clip]{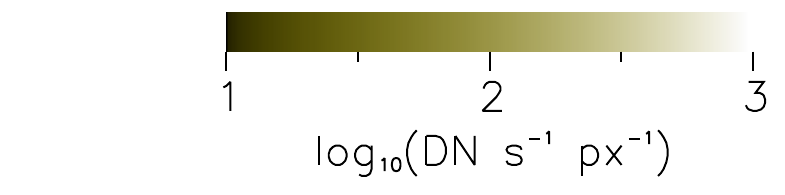}

       \includegraphics[height=2.76cm, bb=0  40 220 175, clip]{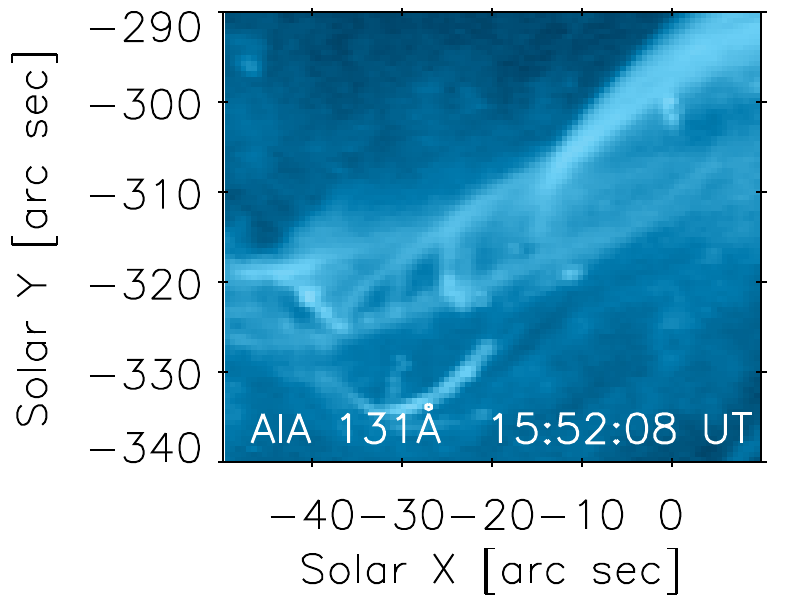}
       \includegraphics[height=2.76cm, bb=60 40 220 175, clip]{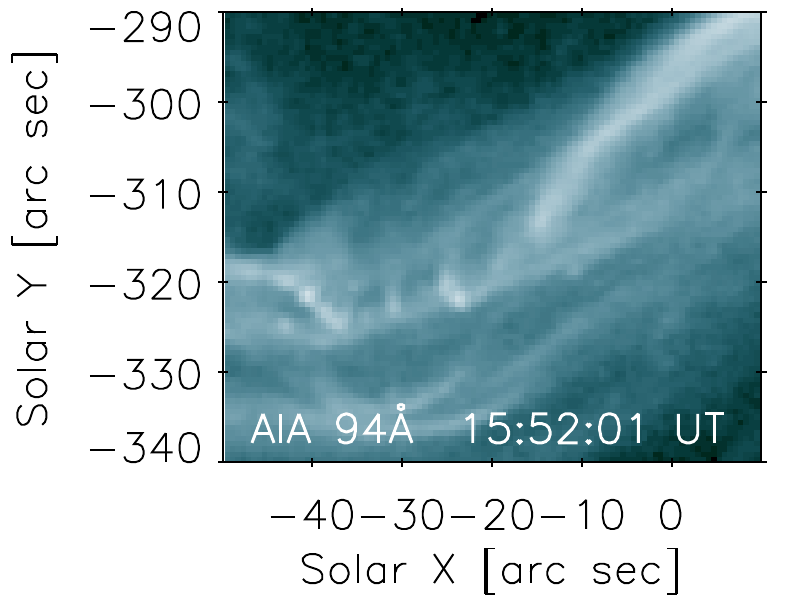}
       \includegraphics[height=2.76cm, bb=60 40 220 175, clip]{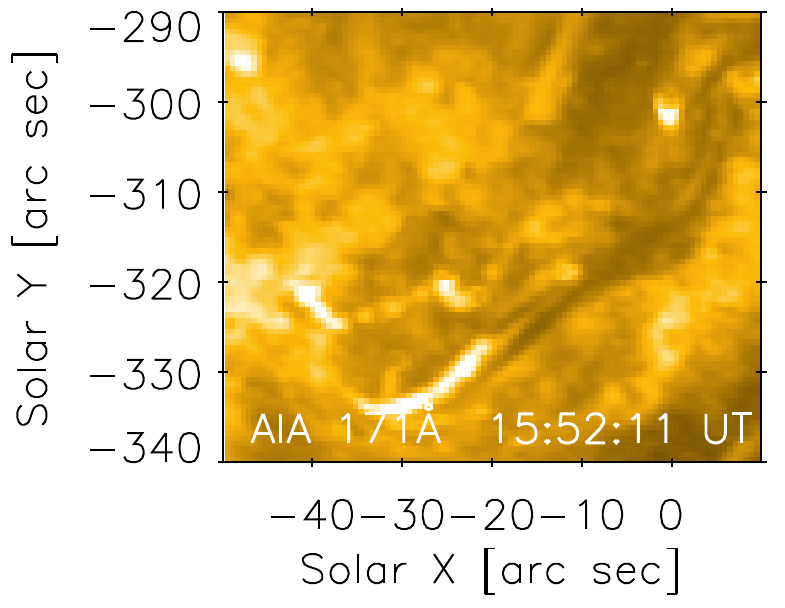}
       \includegraphics[height=2.76cm, bb=60 40 220 175, clip]{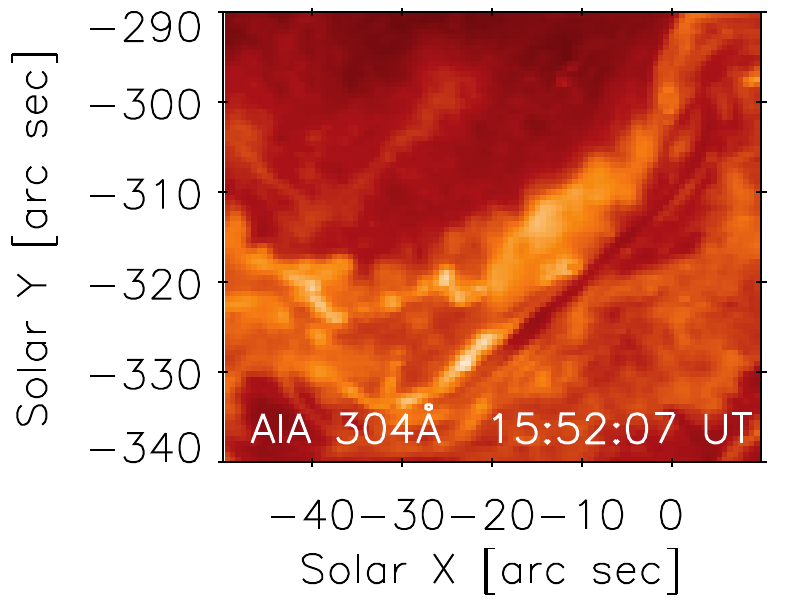}
       \includegraphics[height=2.76cm, bb=60 40 220 175, clip]{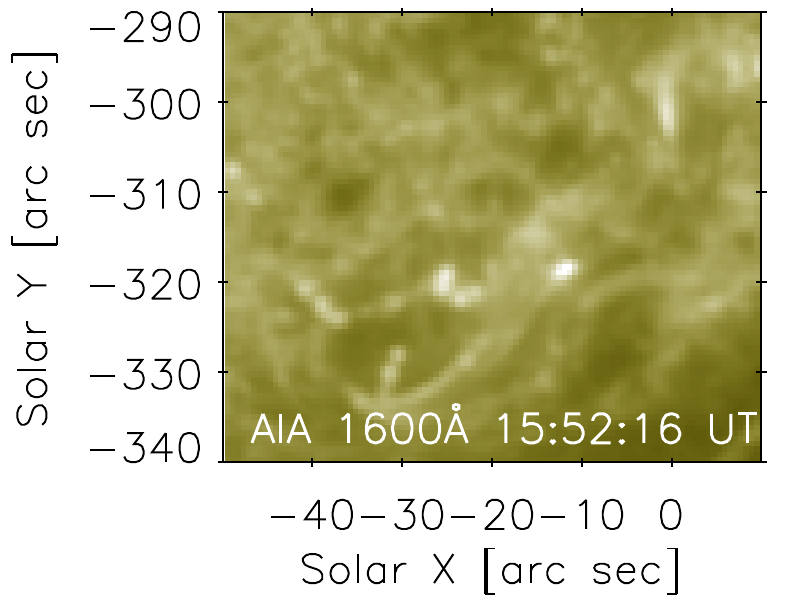}

       \includegraphics[height=2.76cm, bb=0  40 220 175, clip]{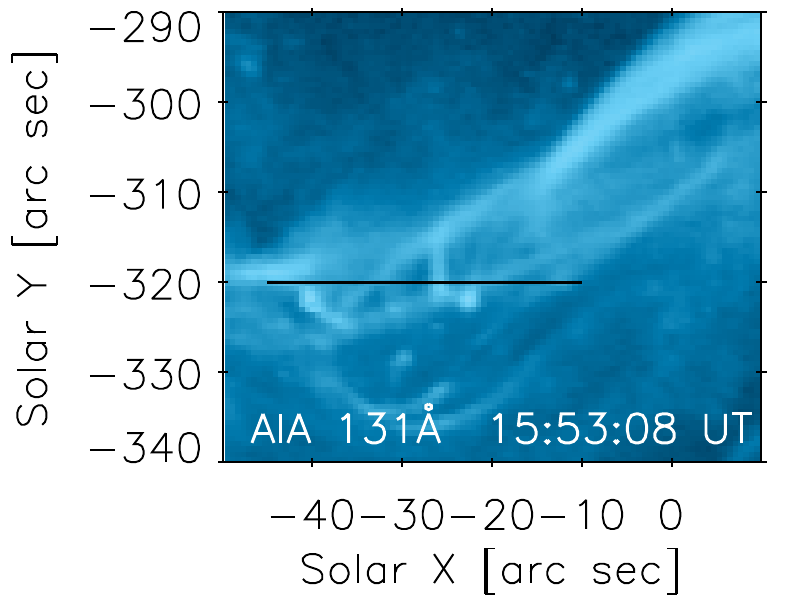}
       \includegraphics[height=2.76cm, bb=60 40 220 175, clip]{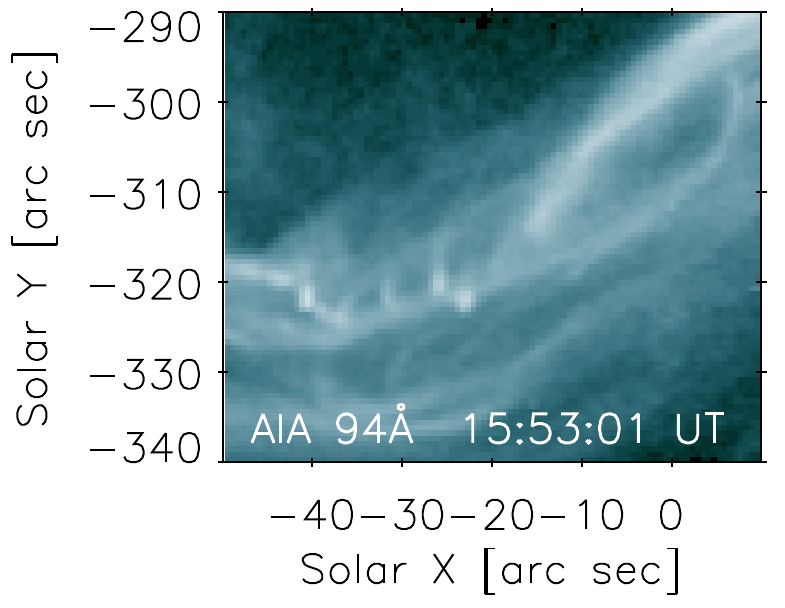}
       \includegraphics[height=2.76cm, bb=60 40 220 175, clip]{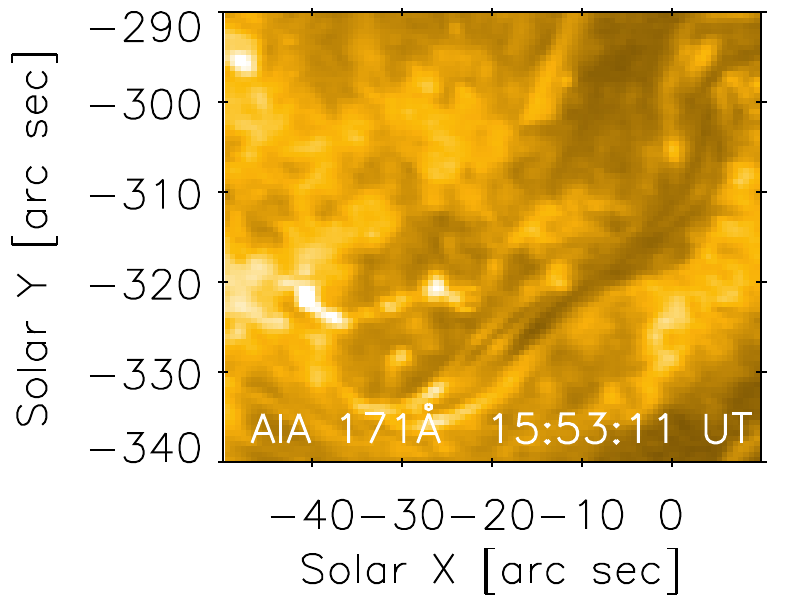}
       \includegraphics[height=2.76cm, bb=60 40 220 175, clip]{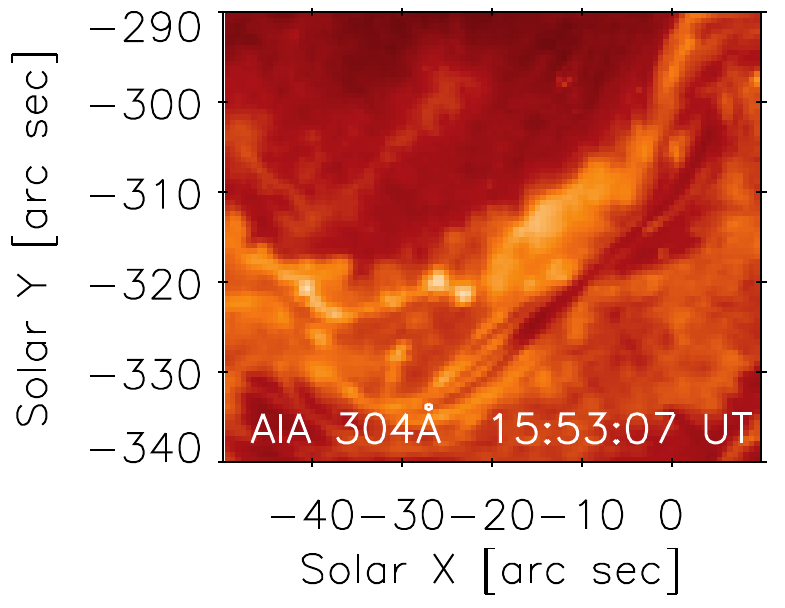}
       \includegraphics[height=2.76cm, bb=60 40 220 175, clip]{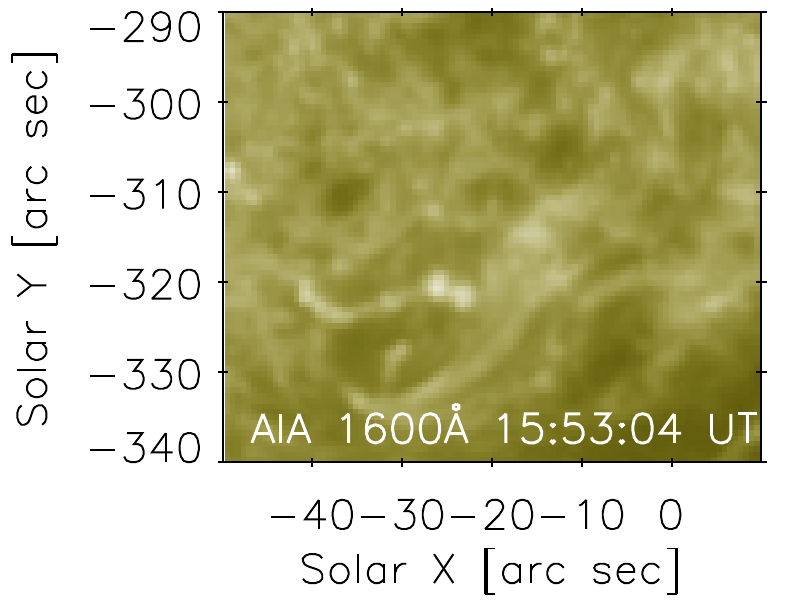}

       \includegraphics[height=2.76cm, bb=0  40 220 175, clip]{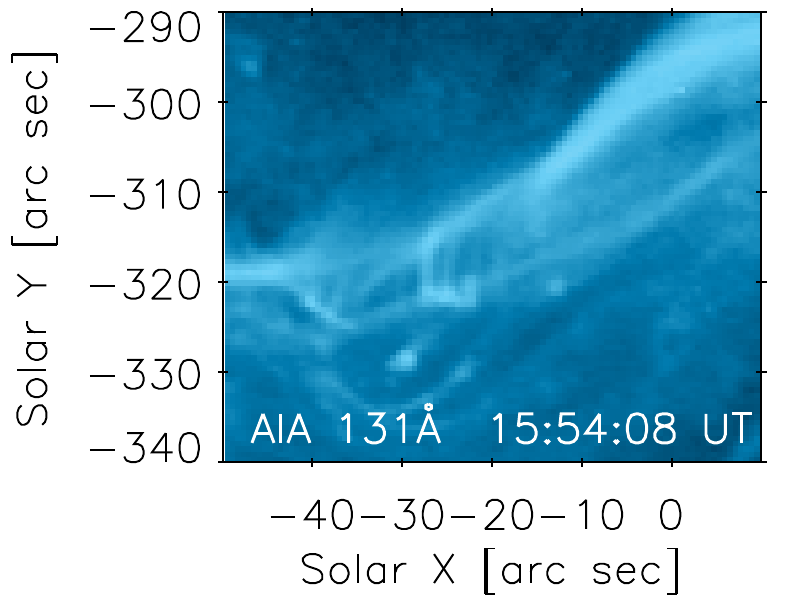}
       \includegraphics[height=2.76cm, bb=60 40 220 175, clip]{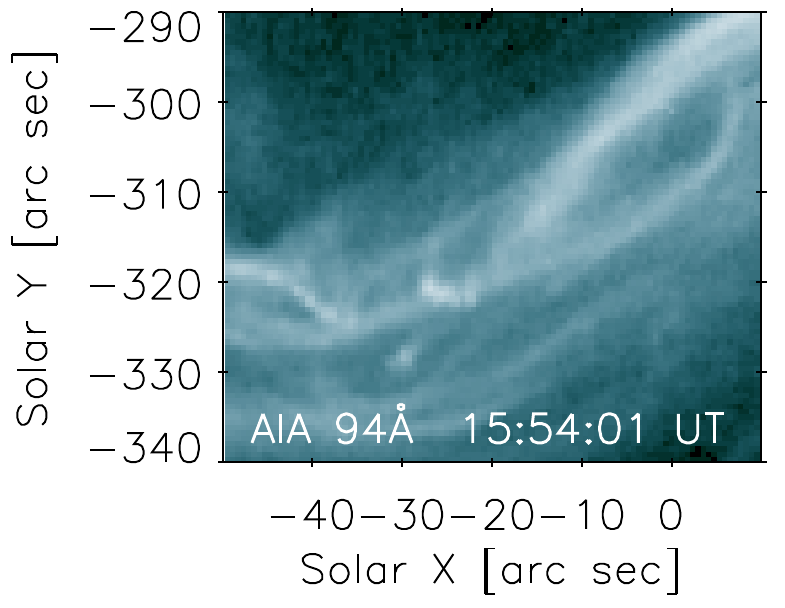}
       \includegraphics[height=2.76cm, bb=60 40 220 175, clip]{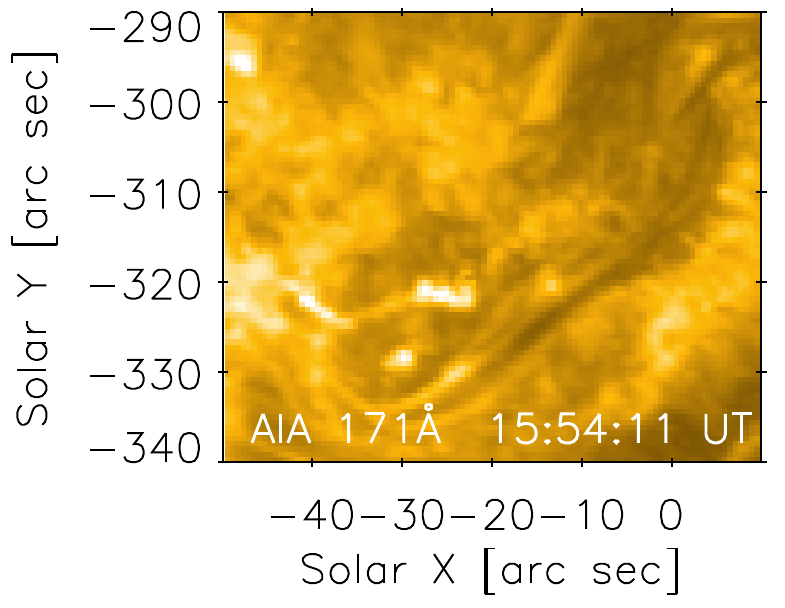}
       \includegraphics[height=2.76cm, bb=60 40 220 175, clip]{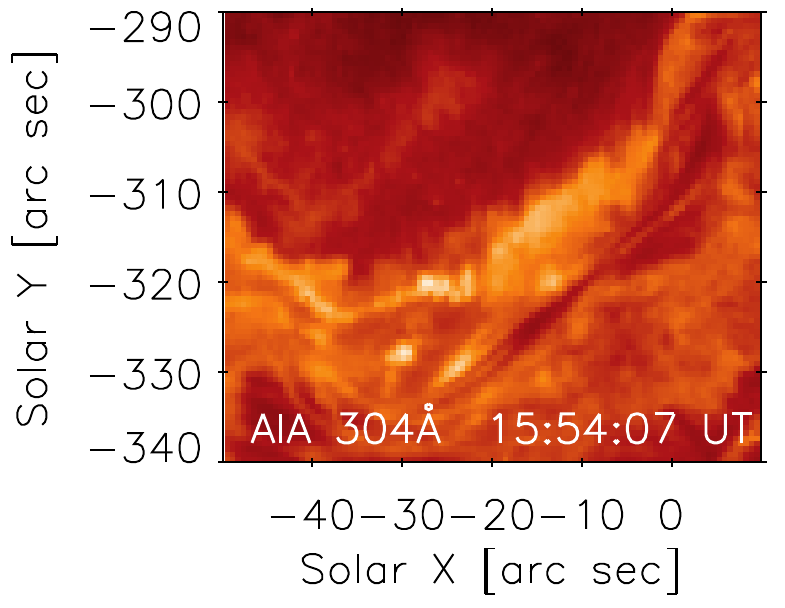}
       \includegraphics[height=2.76cm, bb=60 40 220 175, clip]{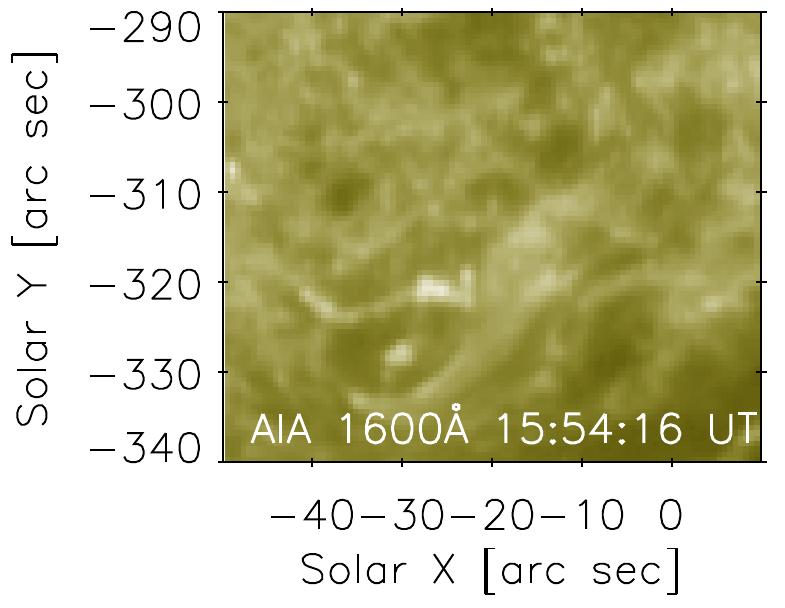}

       \includegraphics[height=2.76cm, bb=0  40 220 175, clip]{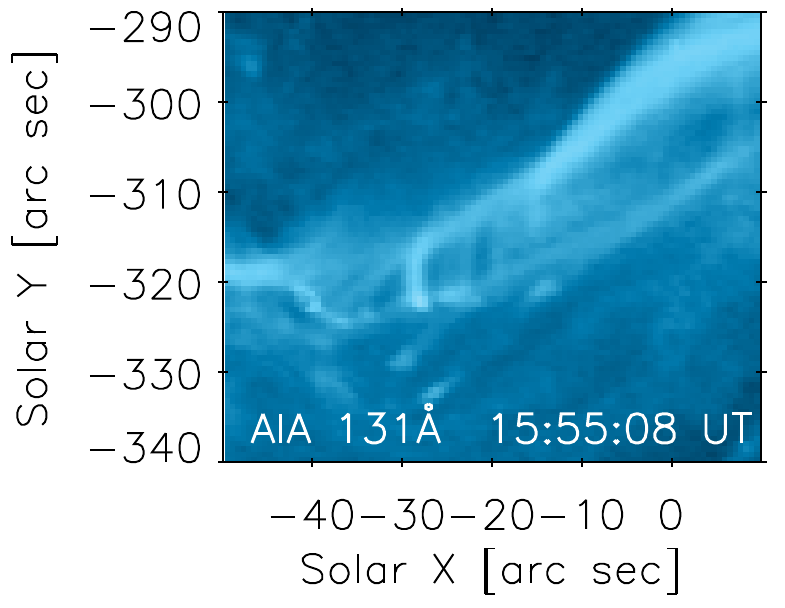}
       \includegraphics[height=2.76cm, bb=60 40 220 175, clip]{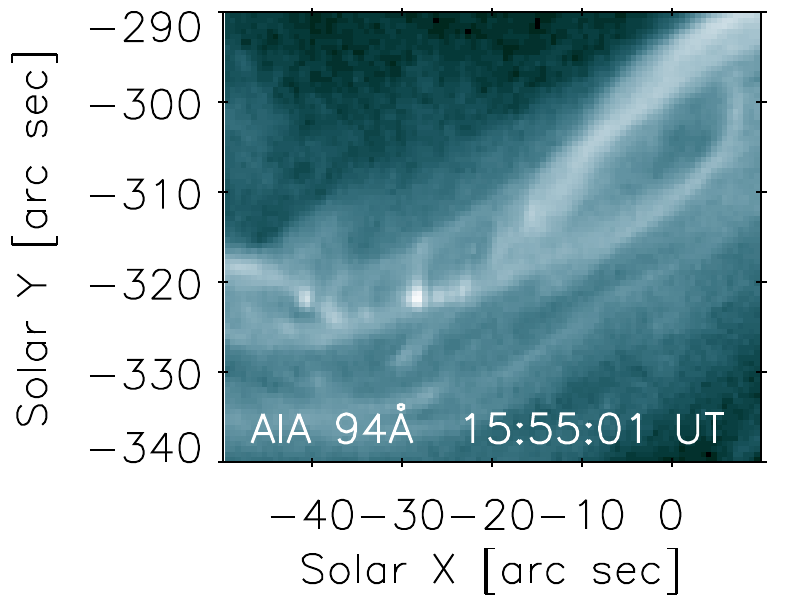}
       \includegraphics[height=2.76cm, bb=60 40 220 175, clip]{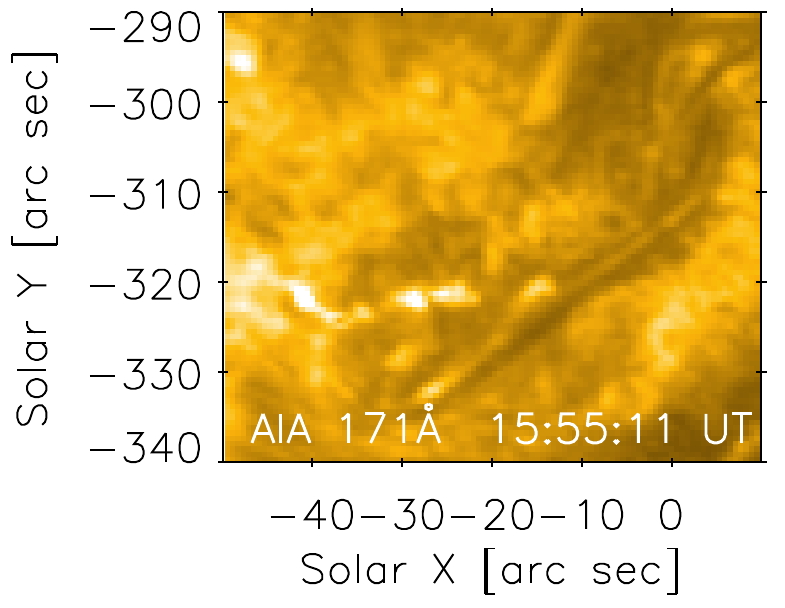}
       \includegraphics[height=2.76cm, bb=60 40 220 175, clip]{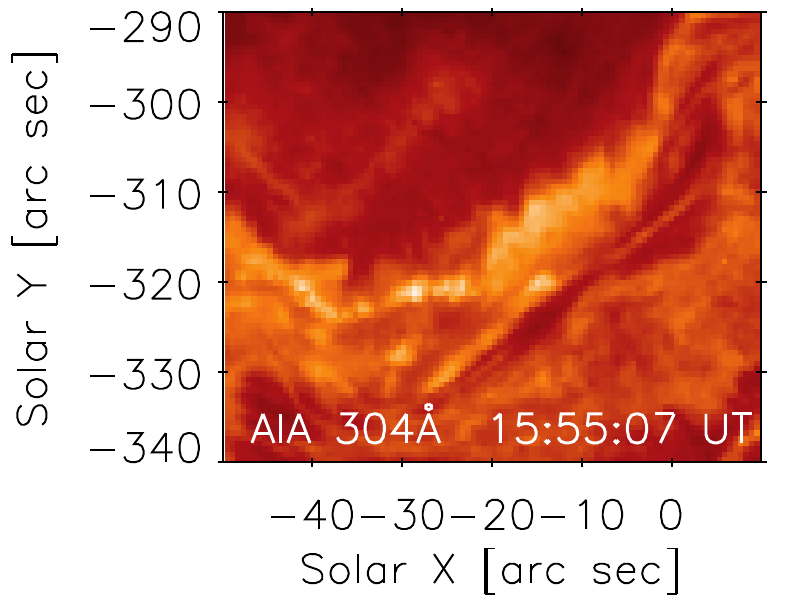}
       \includegraphics[height=2.76cm, bb=60 40 220 175, clip]{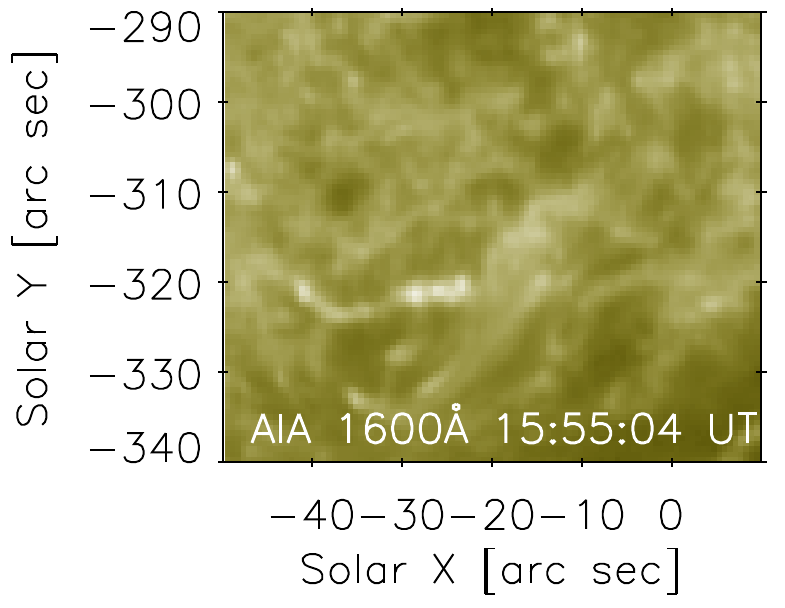}

       \includegraphics[height=2.76cm, bb=0  40 220 175, clip]{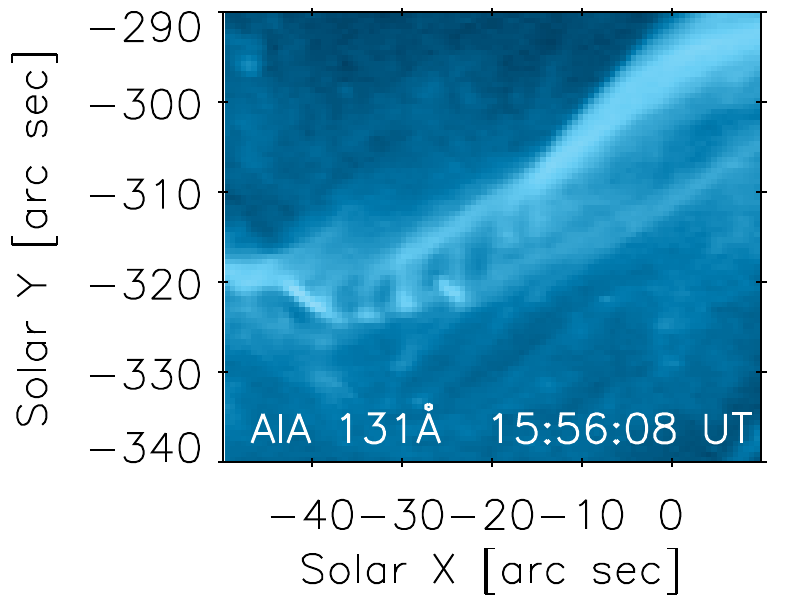}
       \includegraphics[height=2.76cm, bb=60 40 220 175, clip]{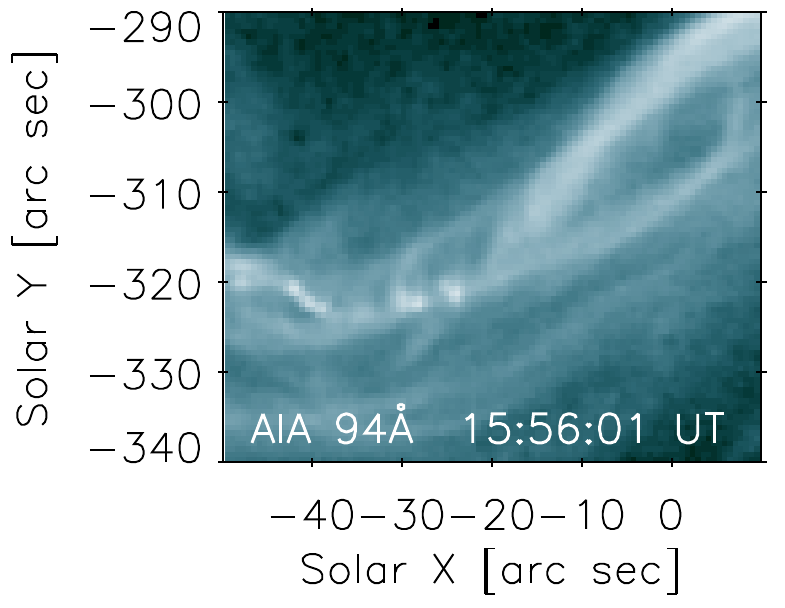}
       \includegraphics[height=2.76cm, bb=60 40 220 175, clip]{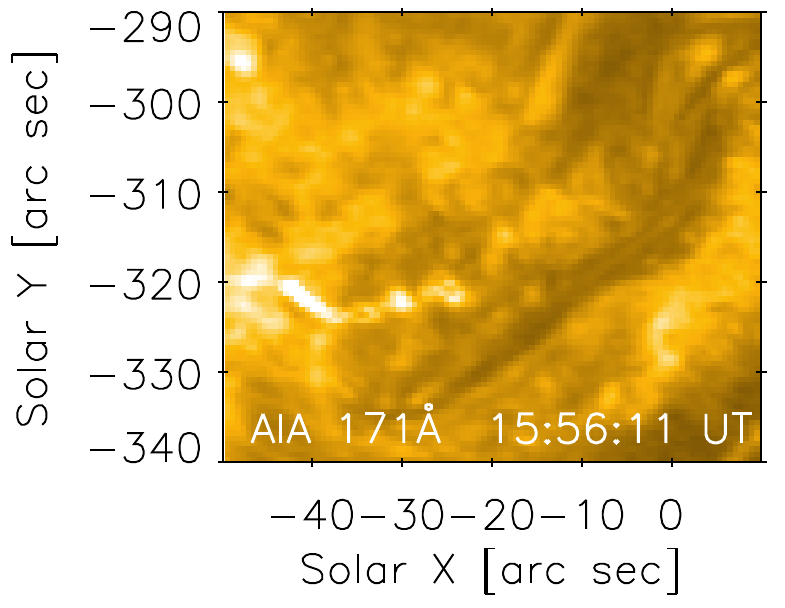}
       \includegraphics[height=2.76cm, bb=60 40 220 175, clip]{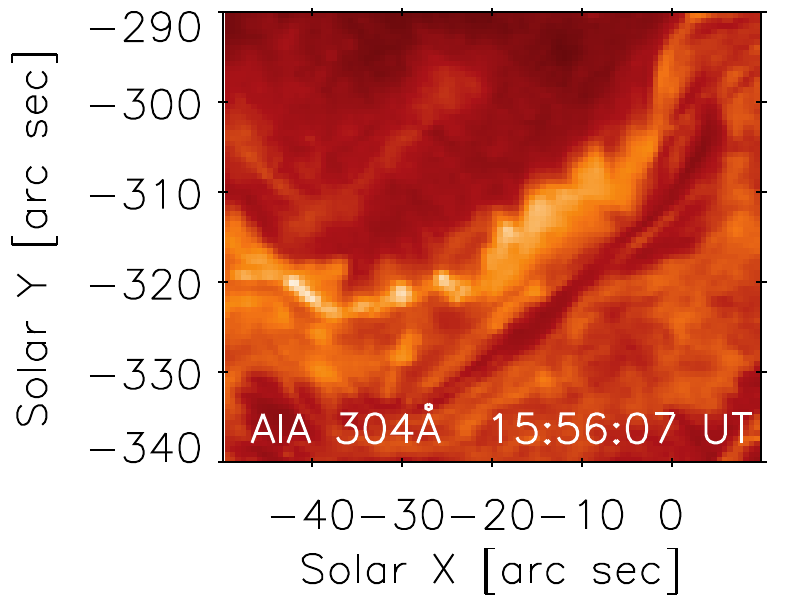}
       \includegraphics[height=2.76cm, bb=60 40 220 175, clip]{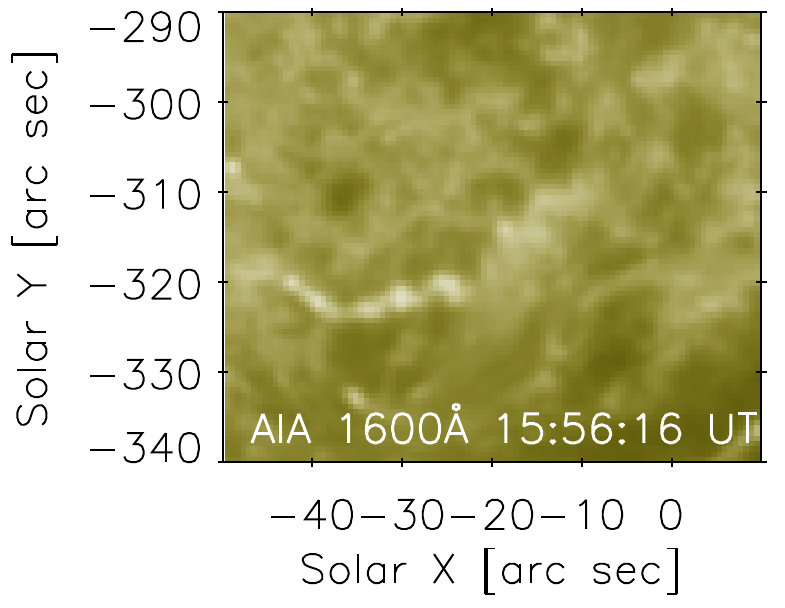}

       \includegraphics[height=3.58cm, bb=0   0 220 175, clip]{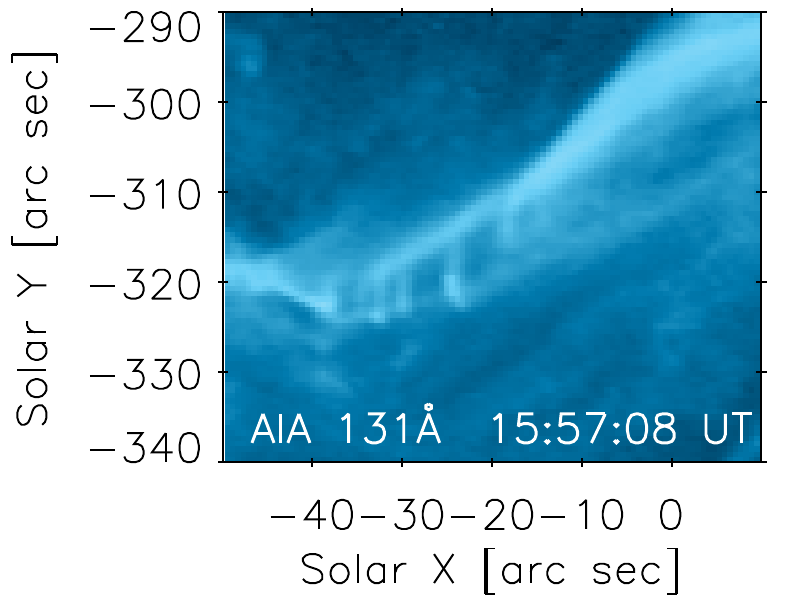}
       \includegraphics[height=3.58cm, bb=60  0 220 175, clip]{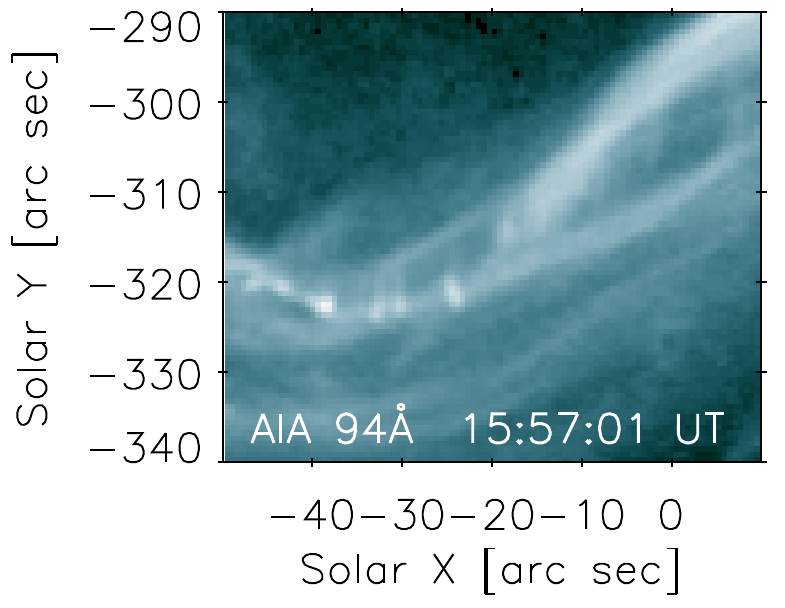}
       \includegraphics[height=3.58cm, bb=60  0 220 175, clip]{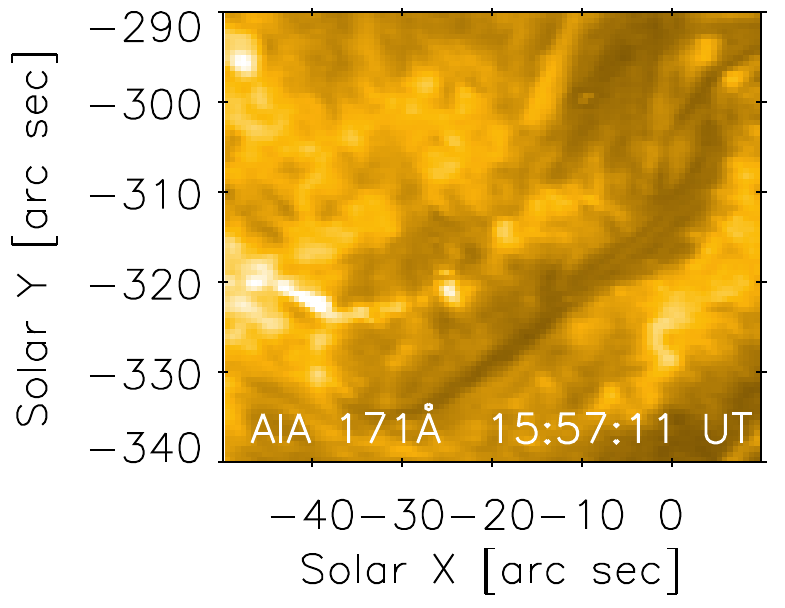}
       \includegraphics[height=3.58cm, bb=60  0 220 175, clip]{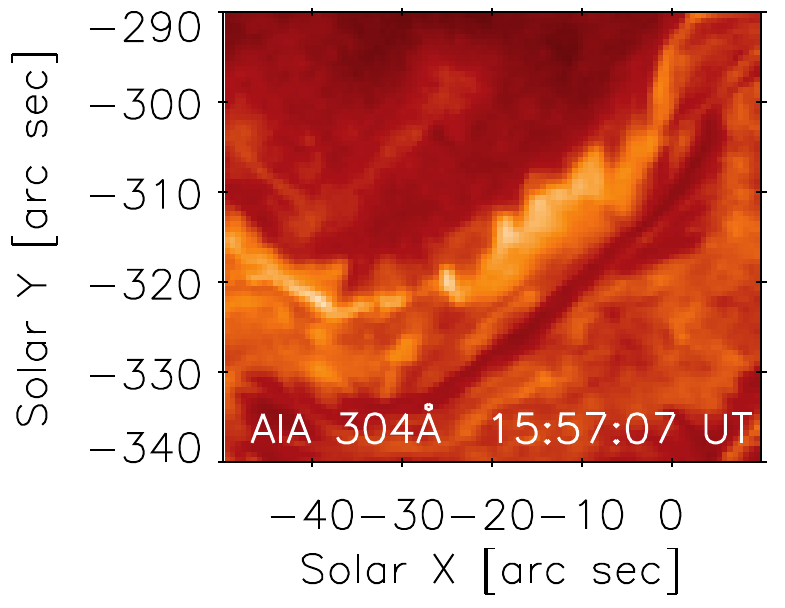}
       \includegraphics[height=3.58cm, bb=60  0 220 175, clip]{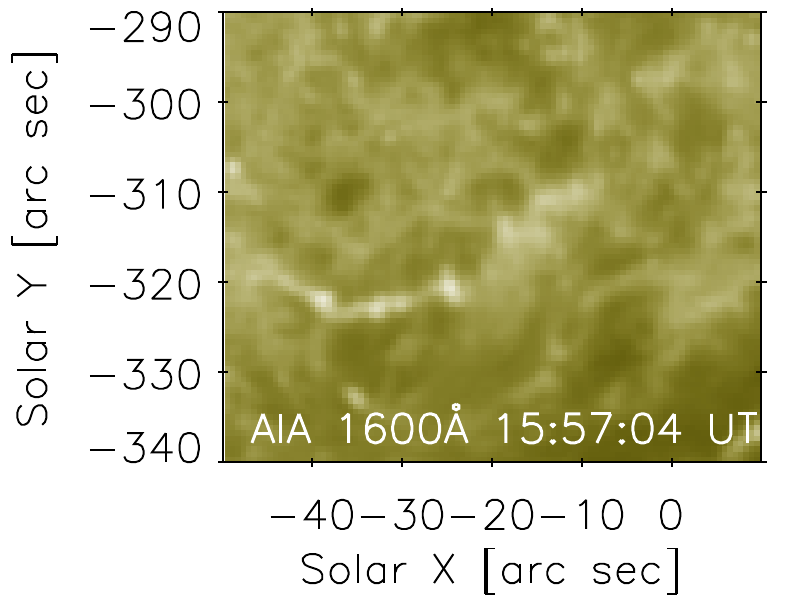}

       \caption{Slipping magnetic loops during the second slipping event. Dark lines in the \textit{top left} panel show positions of the cuts used to construct $X$-$t$ plots (stackplots) shown in Fig. \ref{Fig:Slip2_stackplots}. The intensities are scaled logarithmically, with units of DN\,s$^{-1}$\,px$^{-1}$. An animation of the AIA 131\AA~observations (\textit{left column}) is available as the online Movie 5. 
        }
       \label{Fig:Slip2}
   \end{figure*}
%
   \begin{figure}
    \centering
    \includegraphics[height=4.7cm,bb=5  0 249 249,clip]{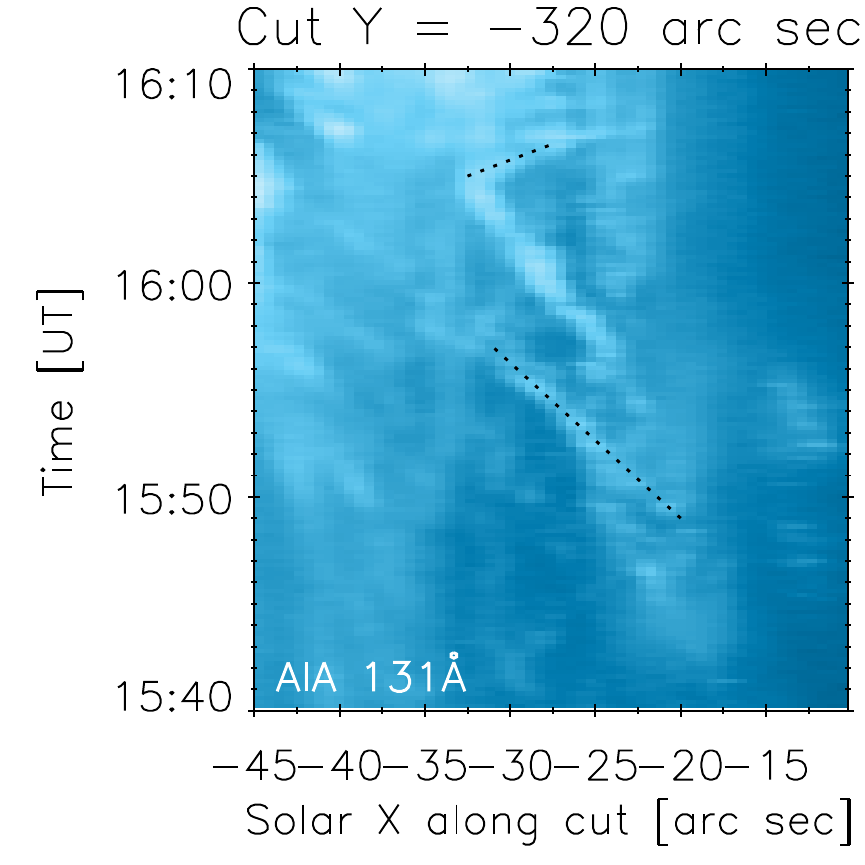}
    \includegraphics[height=4.7cm,bb=67 0 249 249,clip]{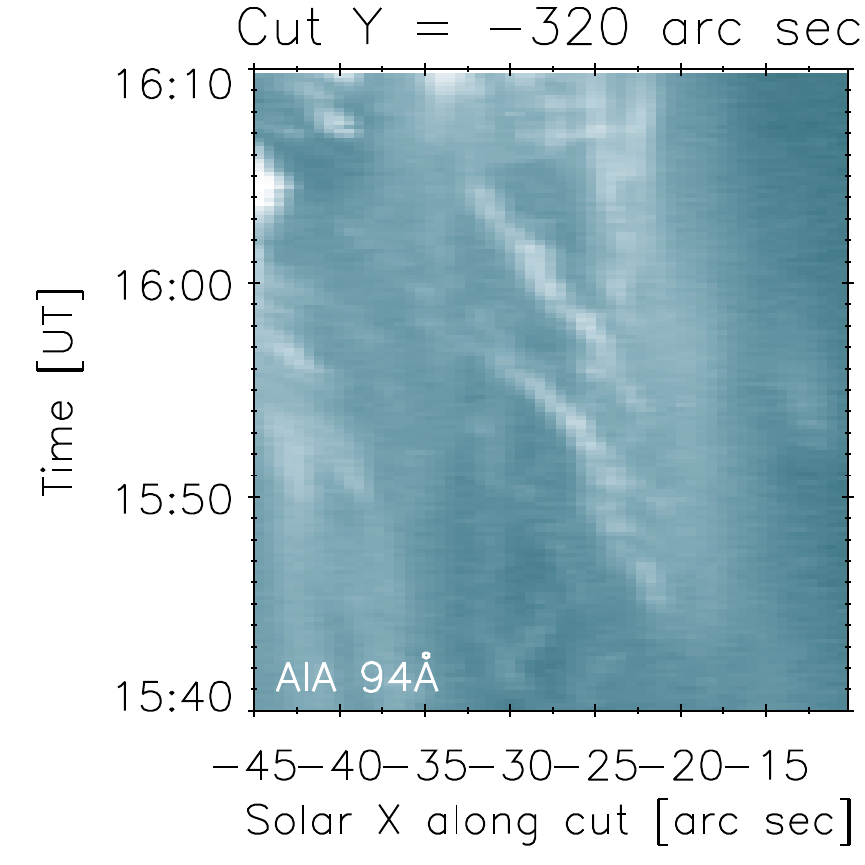}
        \includegraphics[height=3.58cm, bb=0   0 220 175, clip]{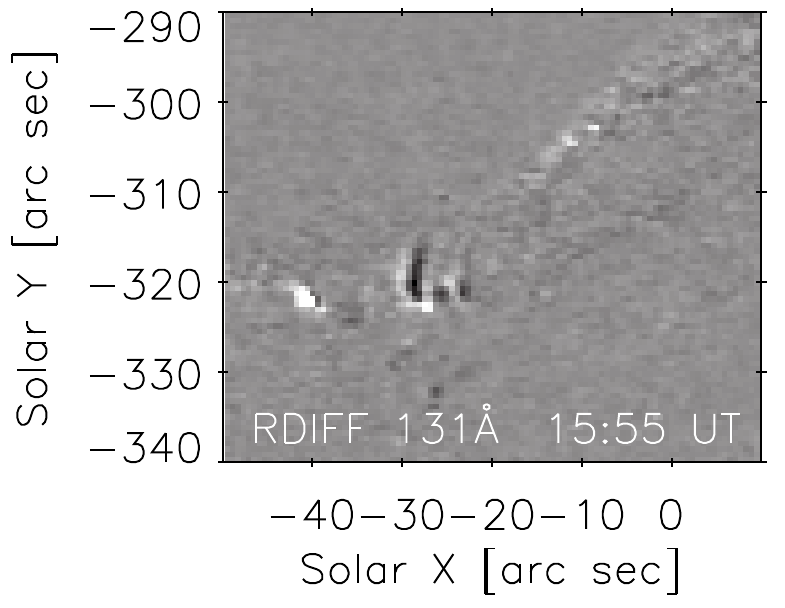}
        \includegraphics[height=3.58cm, bb=60  0 220 175, clip]{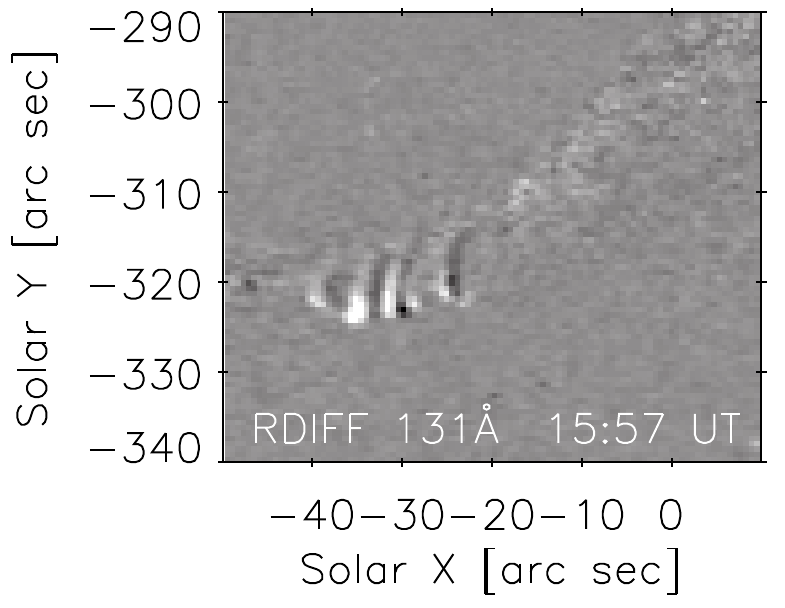}
    \caption{\textit{Top}: $X$-$t$ stackplots along the cut plotted in Fig. \ref{Fig:Slip2} showing apparently slipping loops in the AIA 131\AA~and 94\AA~filters. The long, dotted black line on the \textit{top left} image corresponds to the velocity of 16.6\,km\,s$^{-1}$, while the short dotted line corresponds to velocity of 4.5\,km\,s$^{-1}$. \textit{Bottom}: Examples of running-difference images at 15:55:32 and 15:57:32 UT showing one (\textit{bottom left}) and multiple (\textit{bottom right}) slipping loops.
        }
       \label{Fig:Slip2_stackplots}
   \end{figure}
%
%
%
   \begin{figure*}[!ht]
       \centering
       \includegraphics[height=1.15cm, bb=0   0 220 56, clip]{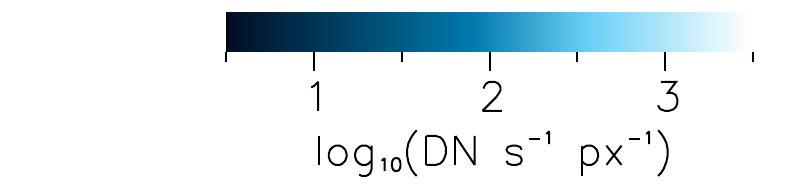}
       \includegraphics[height=1.15cm, bb=60  0 220 56, clip]{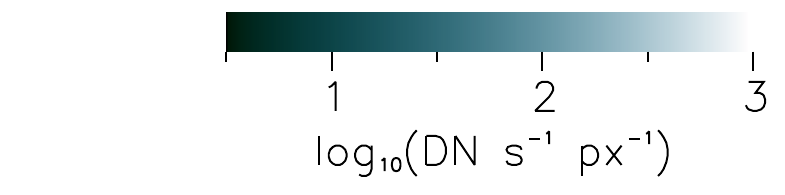}
       \includegraphics[height=1.15cm, bb=60  0 220 56, clip]{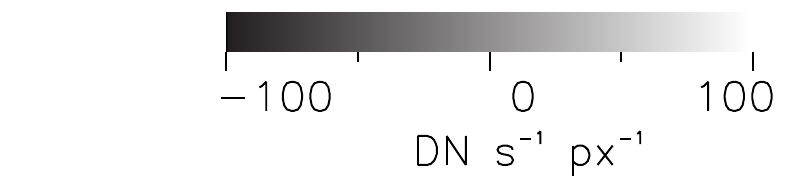}
       \includegraphics[height=1.15cm, bb=60  0 220 56, clip]{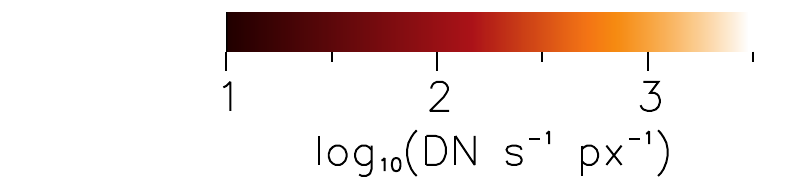}
       \includegraphics[height=1.15cm, bb=60  0 220 56, clip]{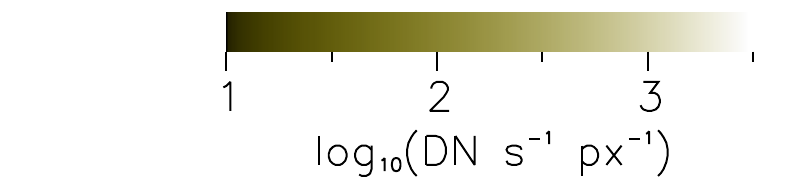}

       \includegraphics[height=2.76cm, bb=0  40 220 175, clip]{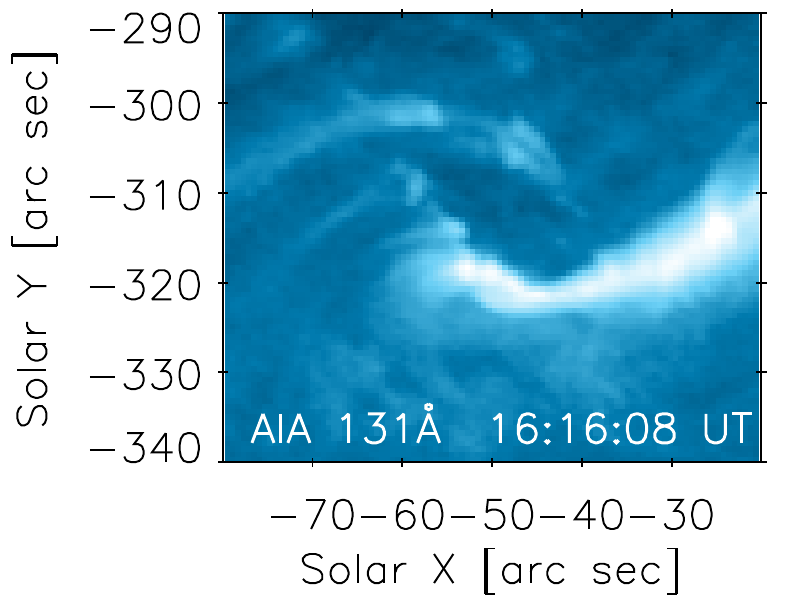}
       \includegraphics[height=2.76cm, bb=60 40 220 175, clip]{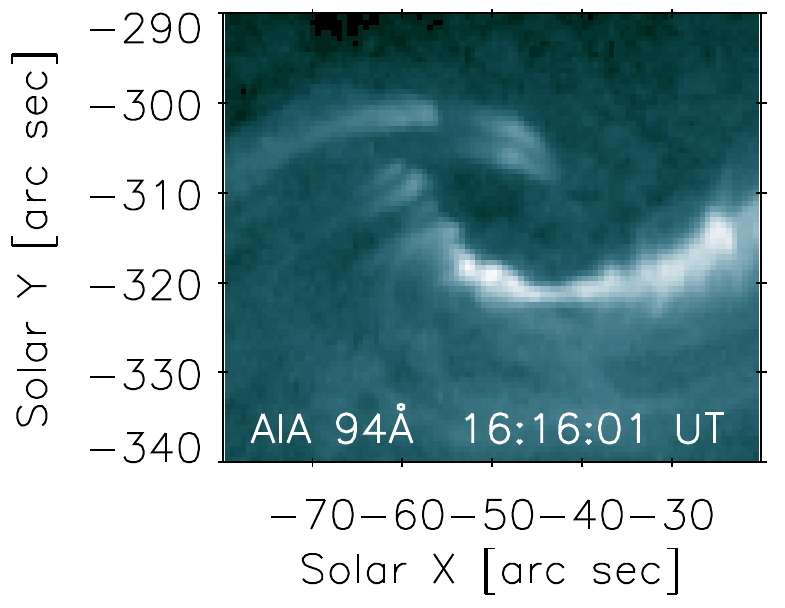}
       \includegraphics[height=2.76cm, bb=60 40 220 175, clip]{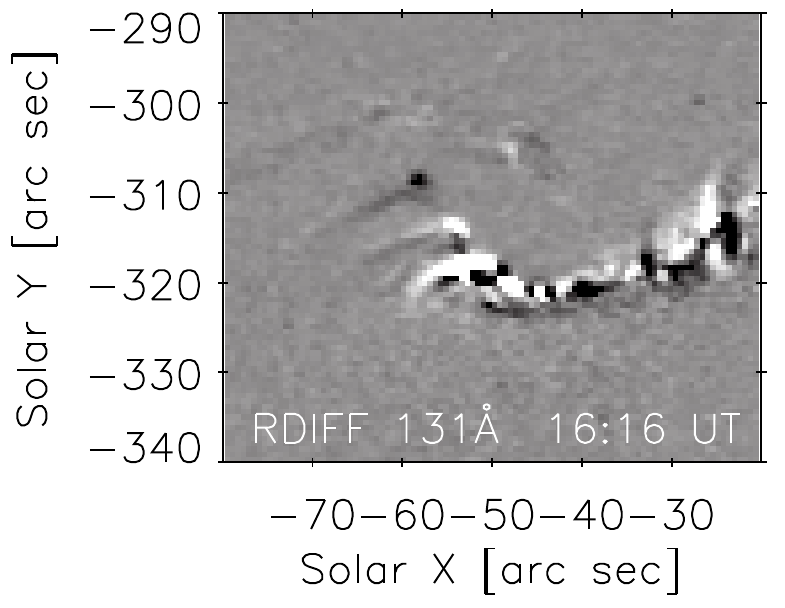}
       \includegraphics[height=2.76cm, bb=60 40 220 175, clip]{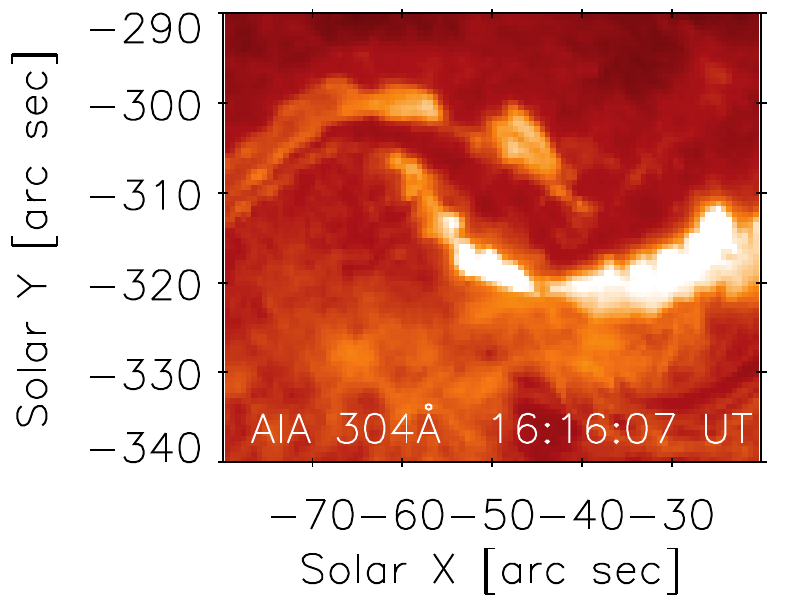}
       \includegraphics[height=2.76cm, bb=60 40 220 175, clip]{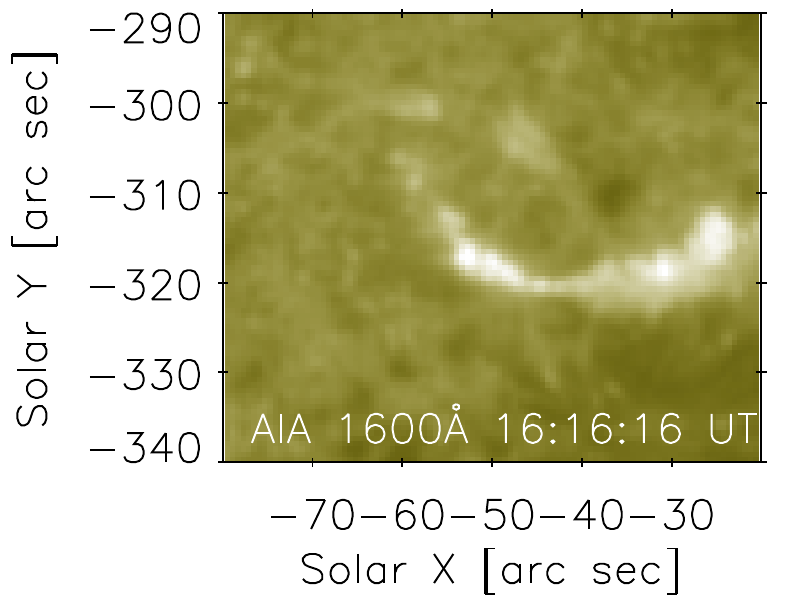}

       \includegraphics[height=2.76cm, bb=0  40 220 175, clip]{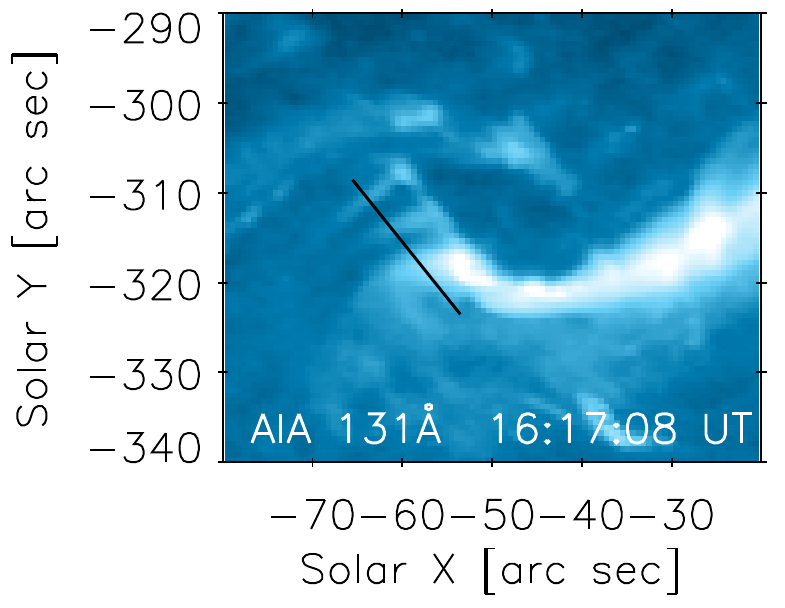}
       \includegraphics[height=2.76cm, bb=60 40 220 175, clip]{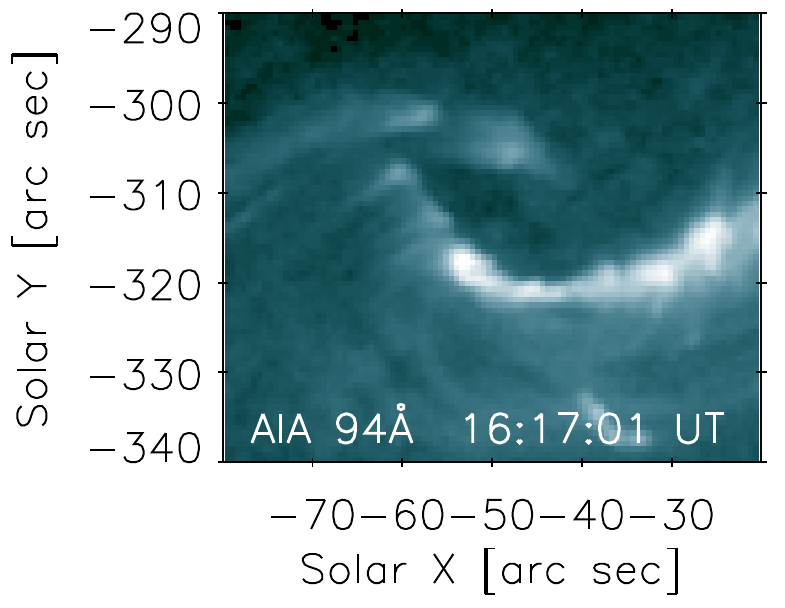}
       \includegraphics[height=2.76cm, bb=60 40 220 175, clip]{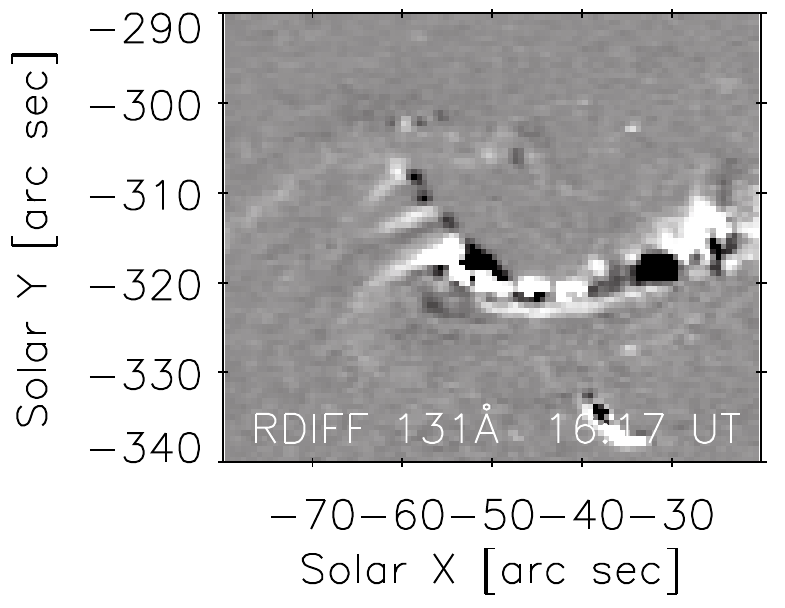}
       \includegraphics[height=2.76cm, bb=60 40 220 175, clip]{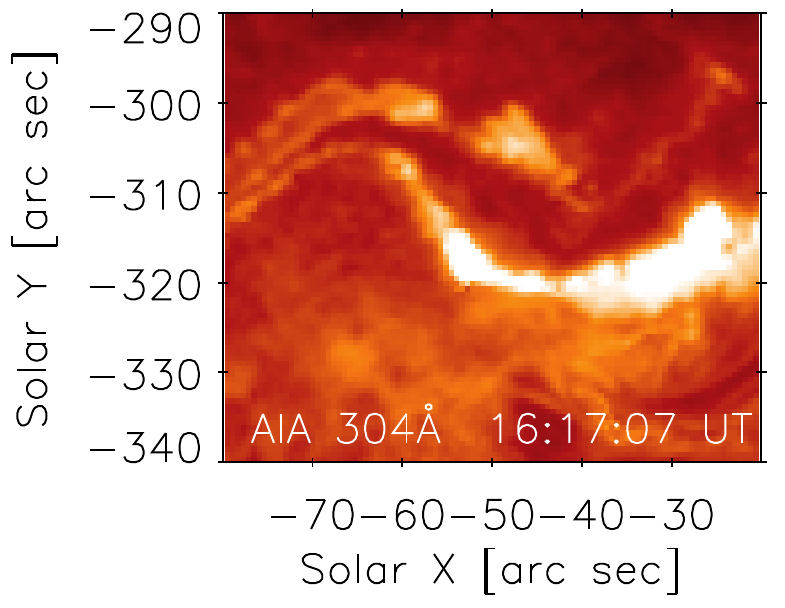}
       \includegraphics[height=2.76cm, bb=60 40 220 175, clip]{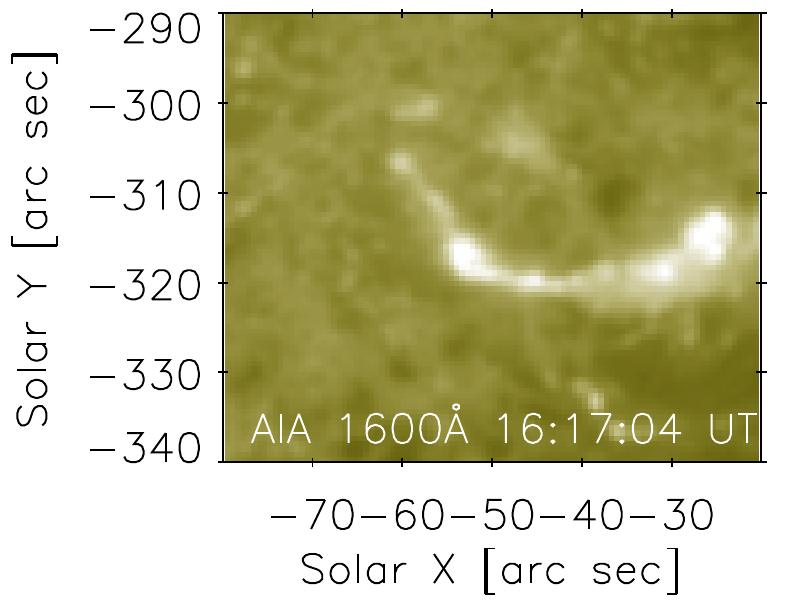}

       \includegraphics[height=2.76cm, bb=0  40 220 175, clip]{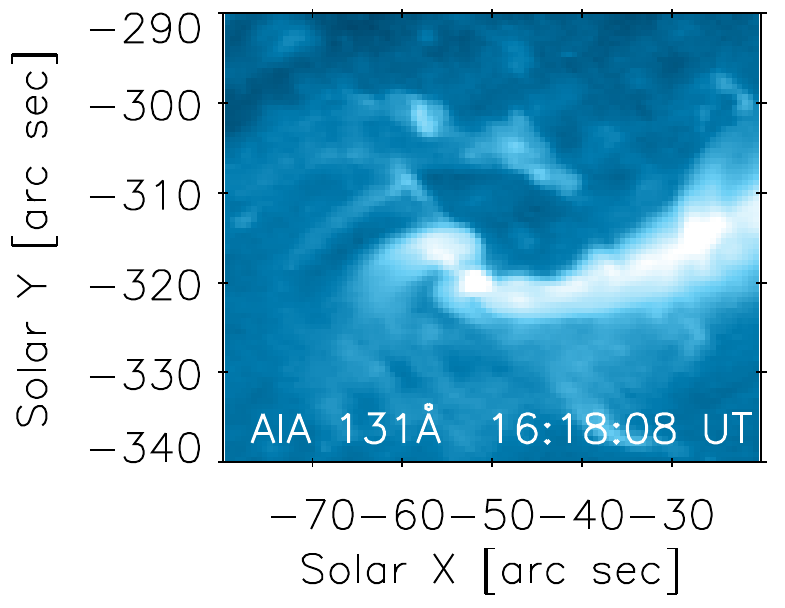}
       \includegraphics[height=2.76cm, bb=60 40 220 175, clip]{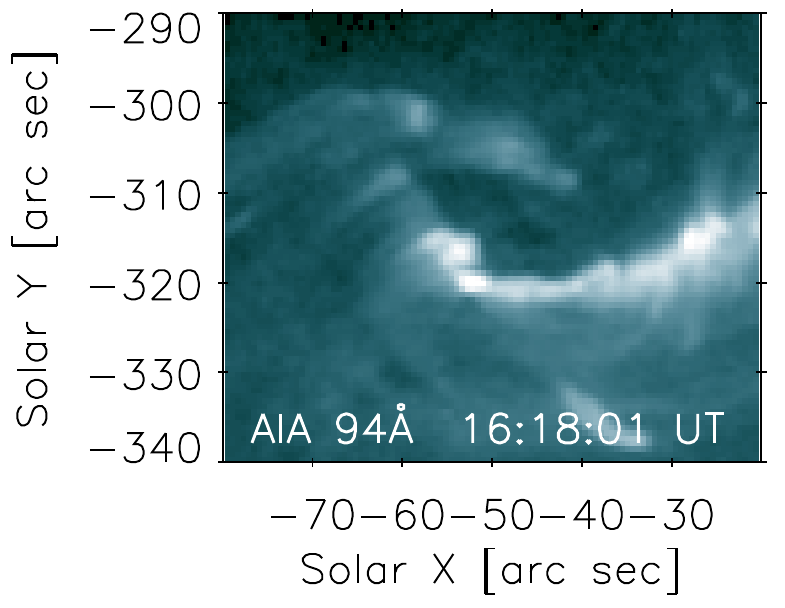}
       \includegraphics[height=2.76cm, bb=60 40 220 175, clip]{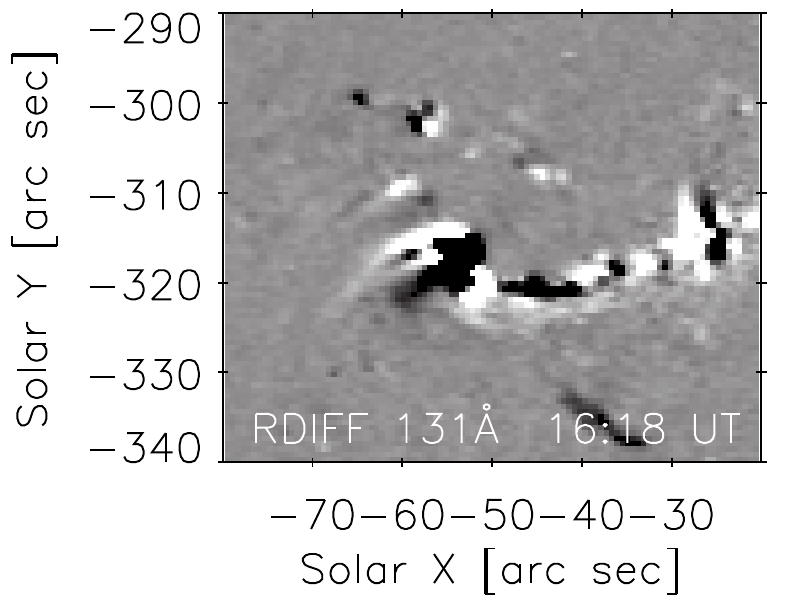}
       \includegraphics[height=2.76cm, bb=60 40 220 175, clip]{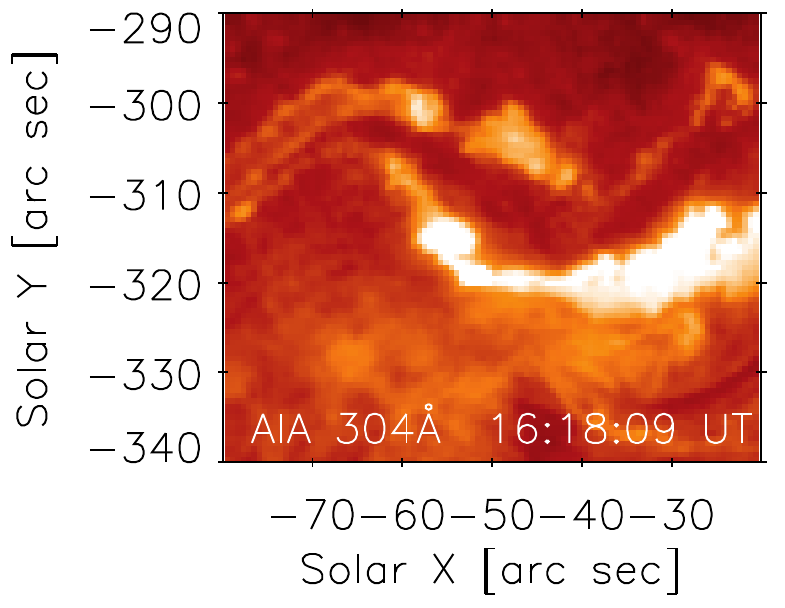}
       \includegraphics[height=2.76cm, bb=60 40 220 175, clip]{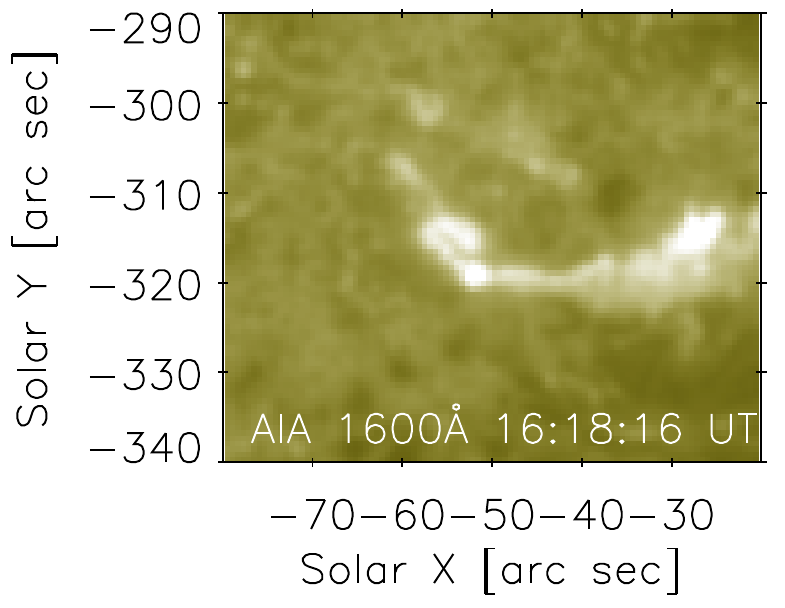}

       \includegraphics[height=2.76cm, bb=0  40 220 175, clip]{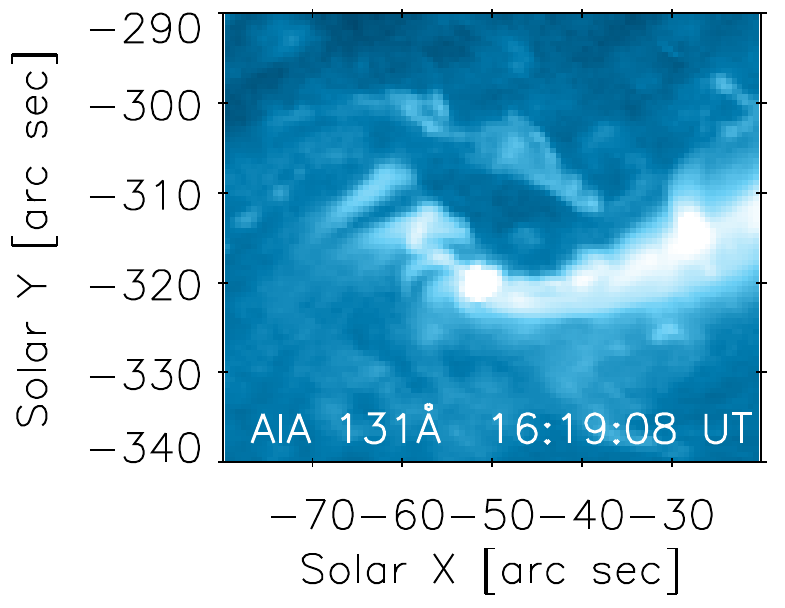}
       \includegraphics[height=2.76cm, bb=60 40 220 175, clip]{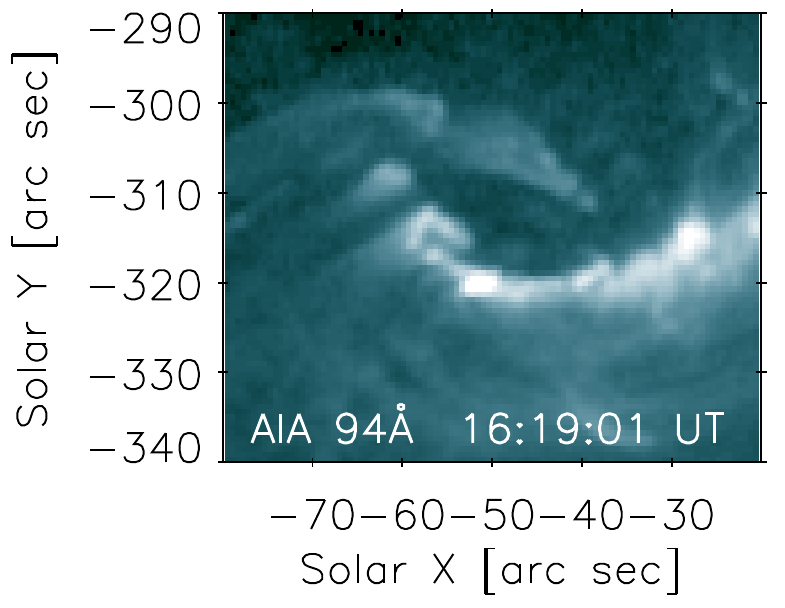}
       \includegraphics[height=2.76cm, bb=60 40 220 175, clip]{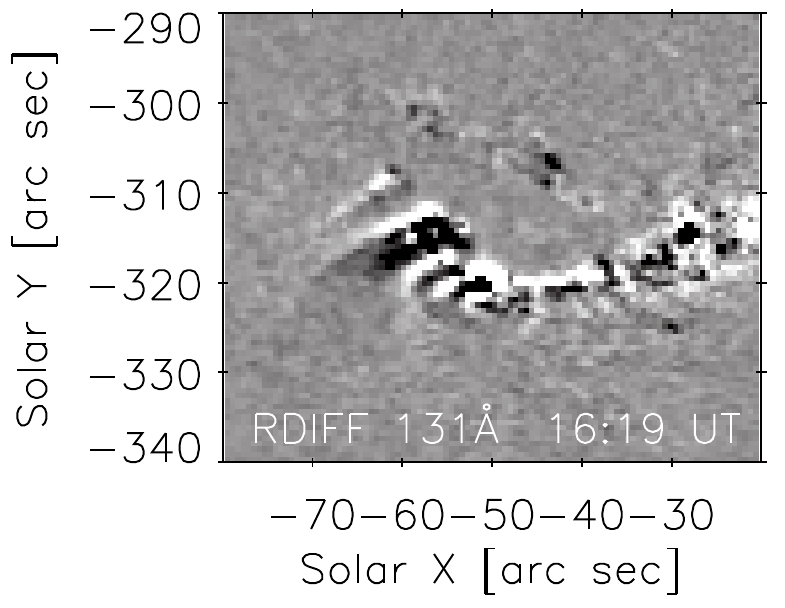}
       \includegraphics[height=2.76cm, bb=60 40 220 175, clip]{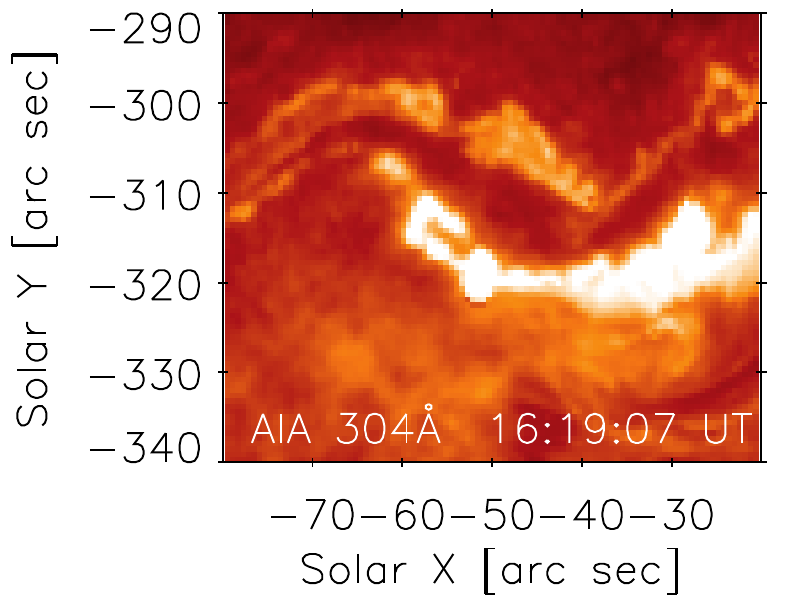}
       \includegraphics[height=2.76cm, bb=60 40 220 175, clip]{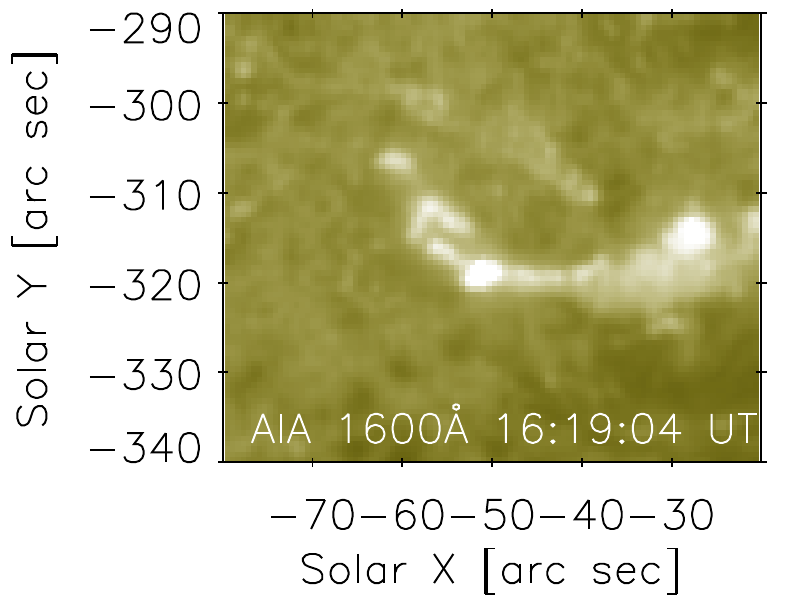}

       \includegraphics[height=3.58cm, bb=0   0 220 175, clip]{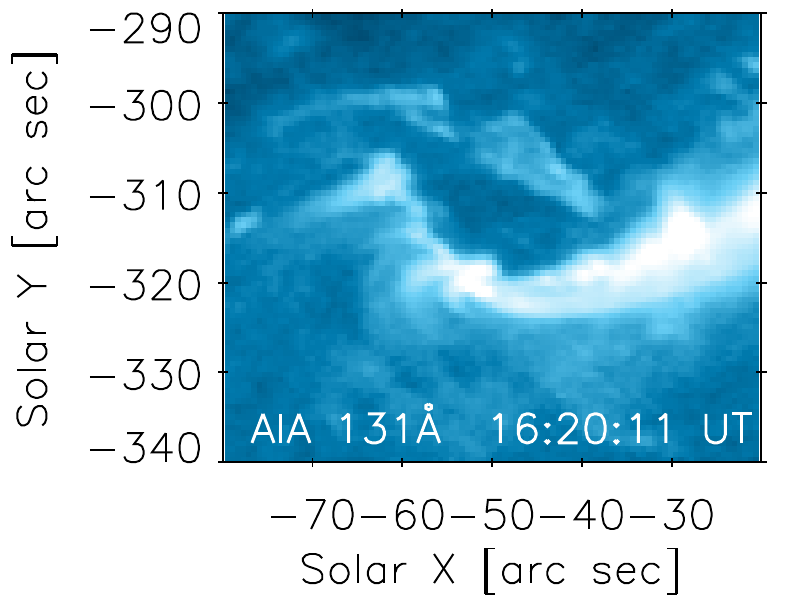}
       \includegraphics[height=3.58cm, bb=60  0 220 175, clip]{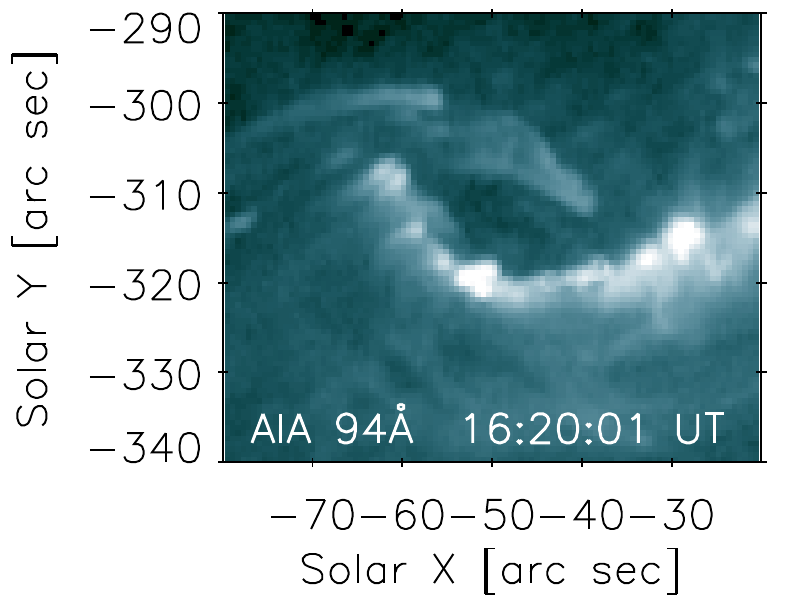}
       \includegraphics[height=3.58cm, bb=60  0 220 175, clip]{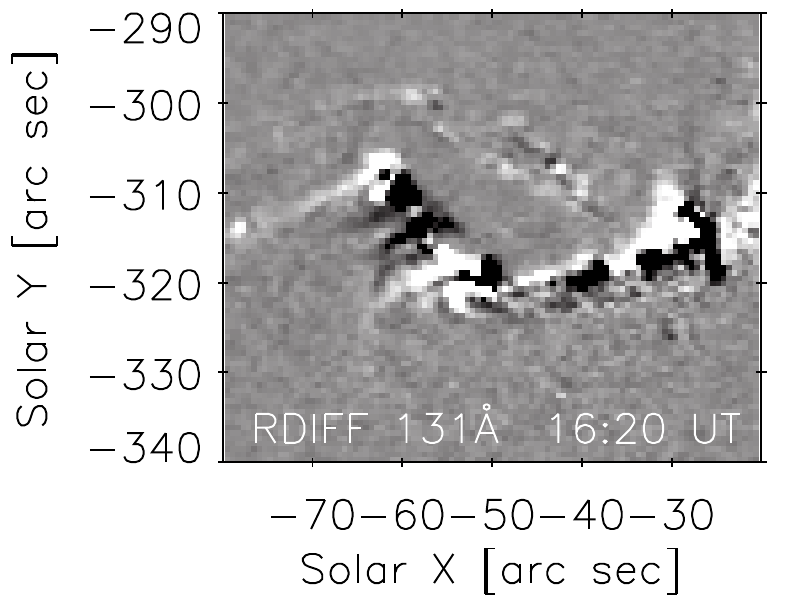}
       \includegraphics[height=3.58cm, bb=60  0 220 175, clip]{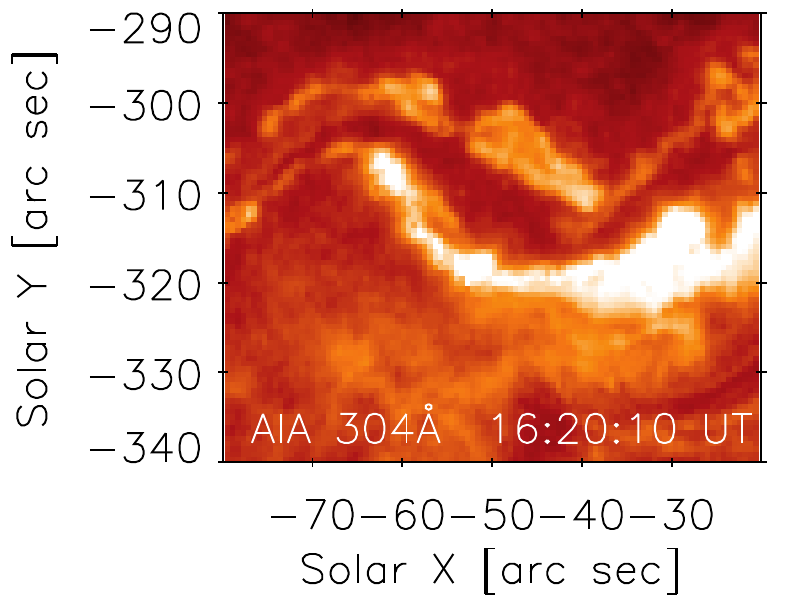}
       \includegraphics[height=3.58cm, bb=60  0 220 175, clip]{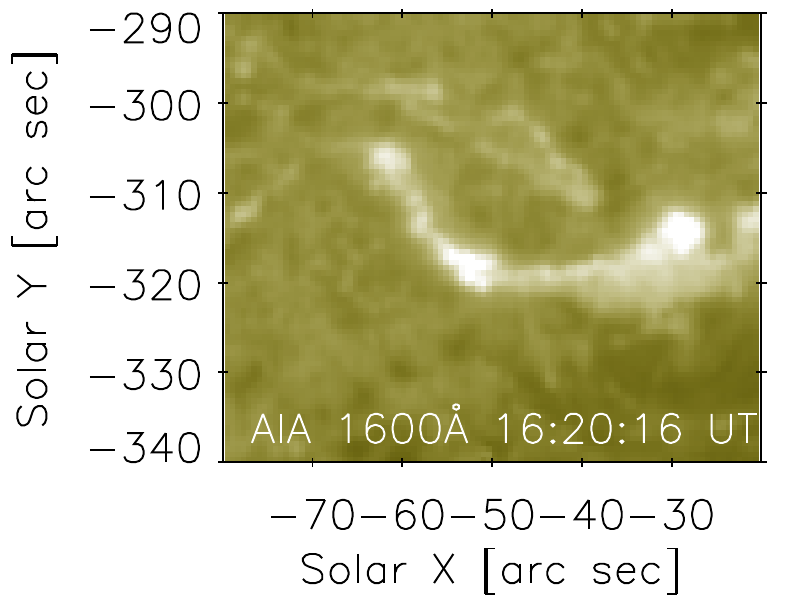}

       \caption{Example of slipping magnetic loops along the developing NR/NRH. Dark lines in the \textit{top left} panel show positions of the cuts used to construct $S$-$t$ plots (stackplots) shown in Fig. \ref{Fig:Slip3_stackplots}.The intensities are scaled logarithmically, except the AIA 131\AA~running difference in the \textit{middle column}. An animation of the AIA 131\AA~observations (\textit{left column}) is available as the online Movie 6. 
        }
       \label{Fig:Slip3}
   \end{figure*}
%
%
   \begin{figure}
    \centering
    \includegraphics[height=4.7cm,bb=0  0 187 249,clip]{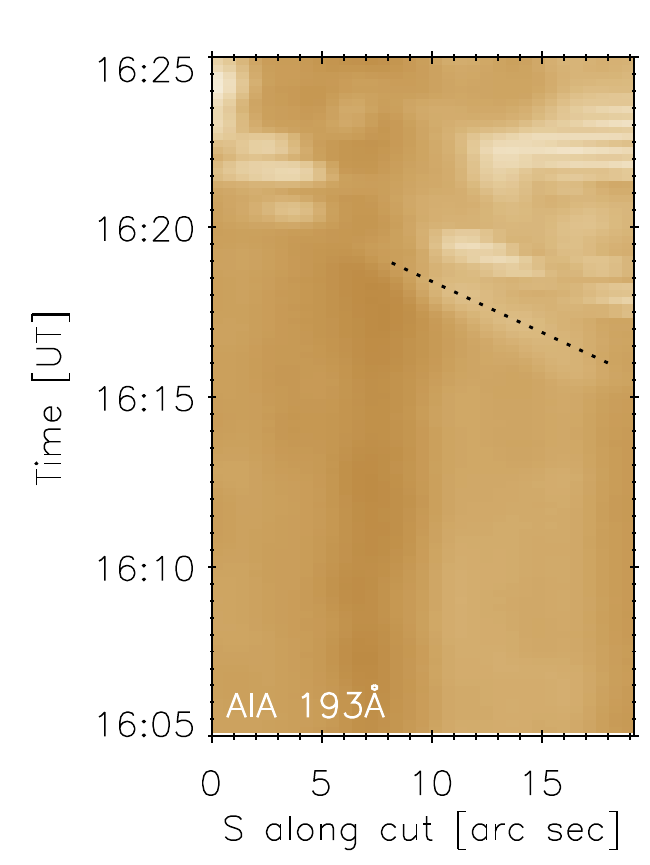}
    \includegraphics[height=4.7cm,bb=58 0 187 249,clip]{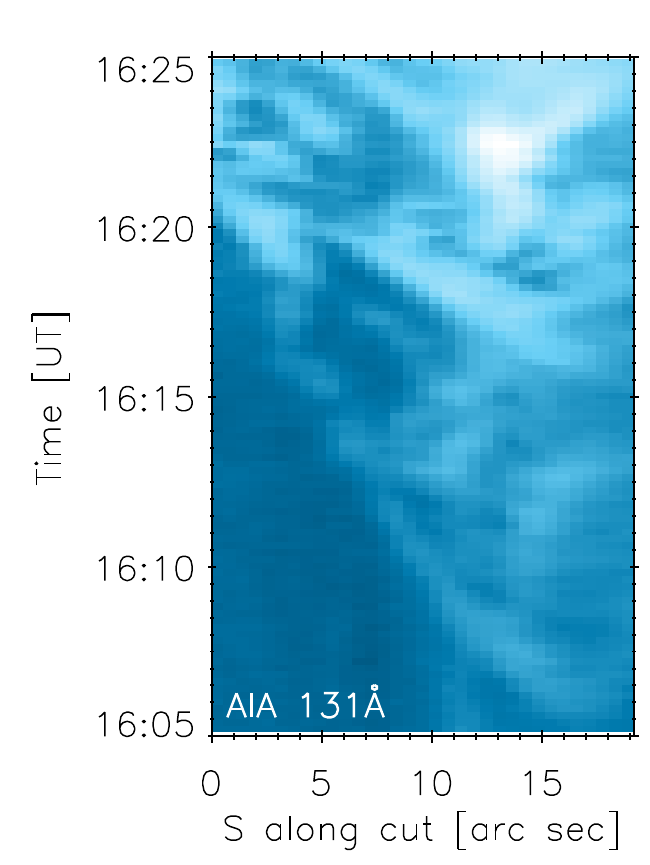}
    \includegraphics[height=4.7cm,bb=58 0 187 249,clip]{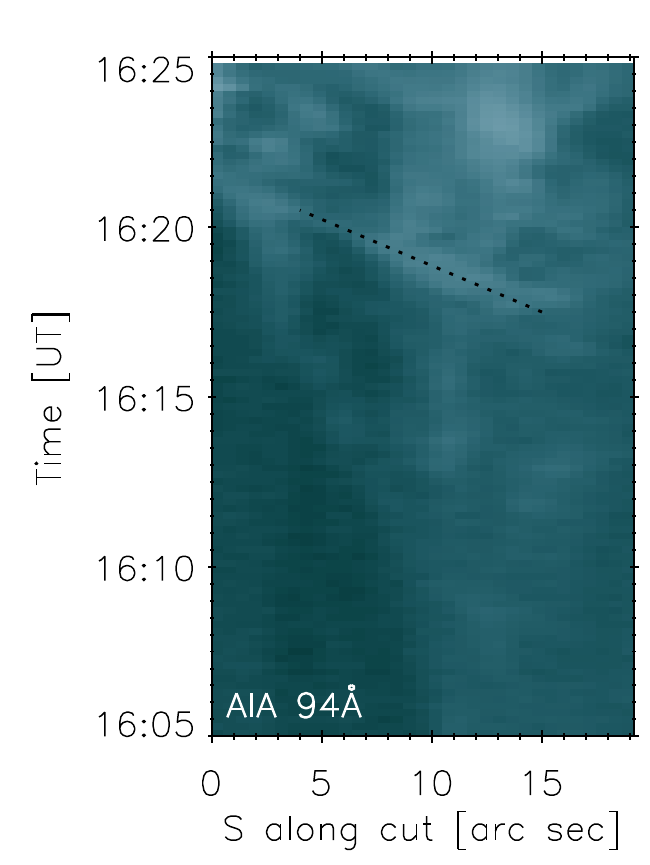}
    \caption{$S$-$t$ stackplots along the cut plotted in Fig. \ref{Fig:Slip3} showing slipping loops in the AIA 193\AA\,131\AA\,and 94\AA\,filters. The dotted black line on the 193\AA\,stackplot (\textit{left}) corresponds to the velocity of 40 km\,s$^{-1}$, while the dotted line on the 94\AA\, stackplot (\textit{right}) stands for 44\,km\,s$^{-1}$.
        }
       \label{Fig:Slip3_stackplots}
   \end{figure}
%
%
%
   \begin{figure*}[!ht]
       \centering
       \includegraphics[height=1.15cm, bb=0   0 220 56, clip]{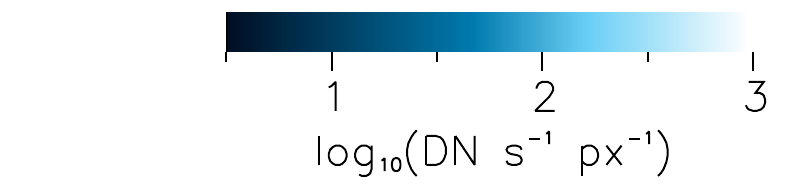}
       \includegraphics[height=1.15cm, bb=60  0 220 56, clip]{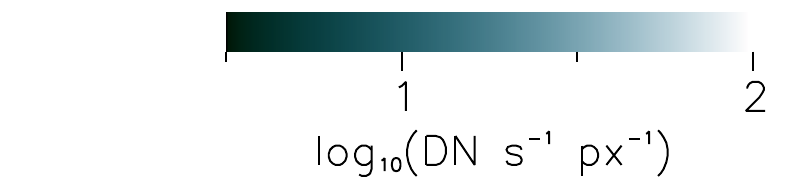}
       \includegraphics[height=1.15cm, bb=60  0 220 56, clip]{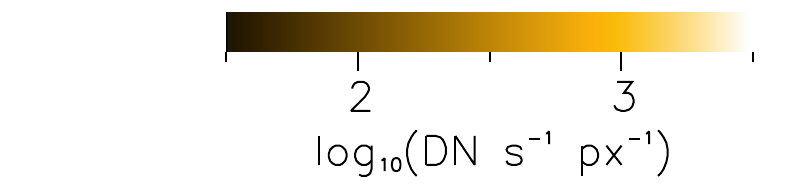}
       \includegraphics[height=1.15cm, bb=60  0 220 56, clip]{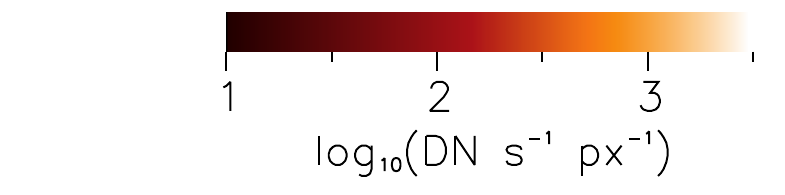}
       \includegraphics[height=1.15cm, bb=60  0 220 56, clip]{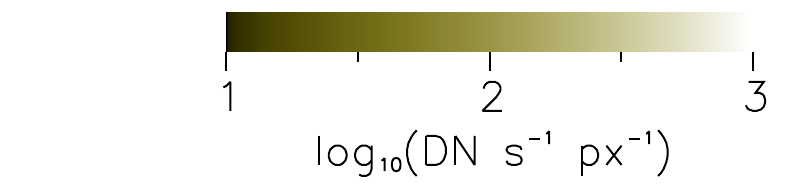}

       \includegraphics[height=1.70cm, bb=0  40 243 136, clip]{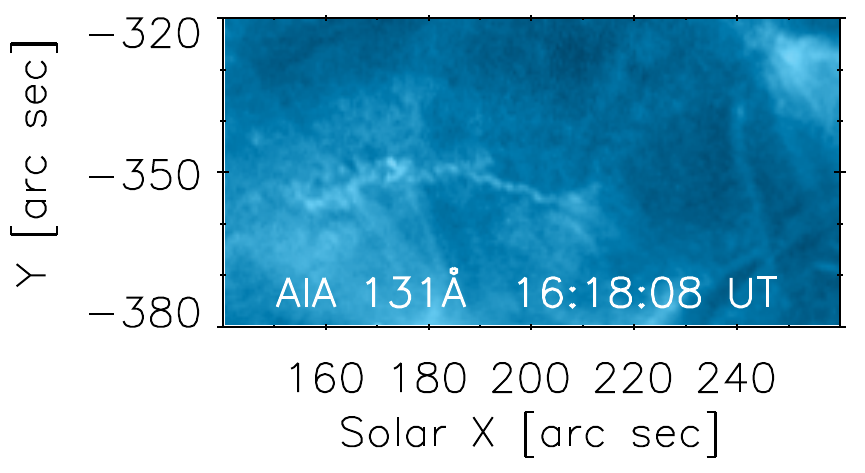}
       \includegraphics[height=1.70cm, bb=60 40 243 136, clip]{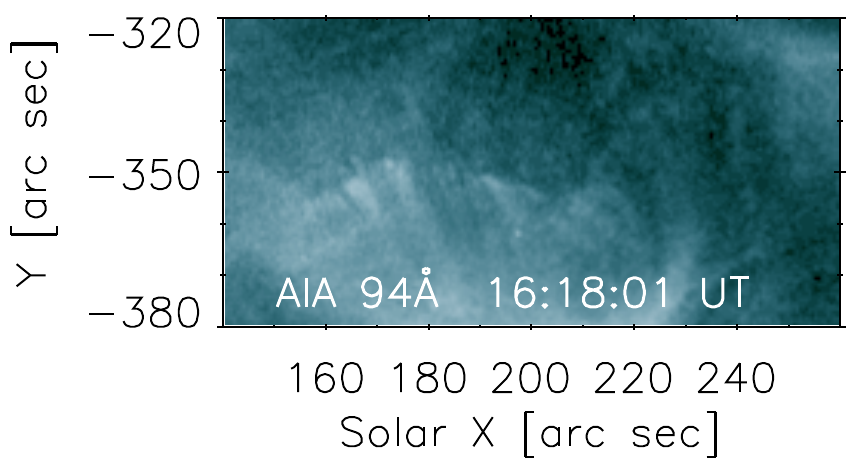}
       \includegraphics[height=1.70cm, bb=60 40 243 136, clip]{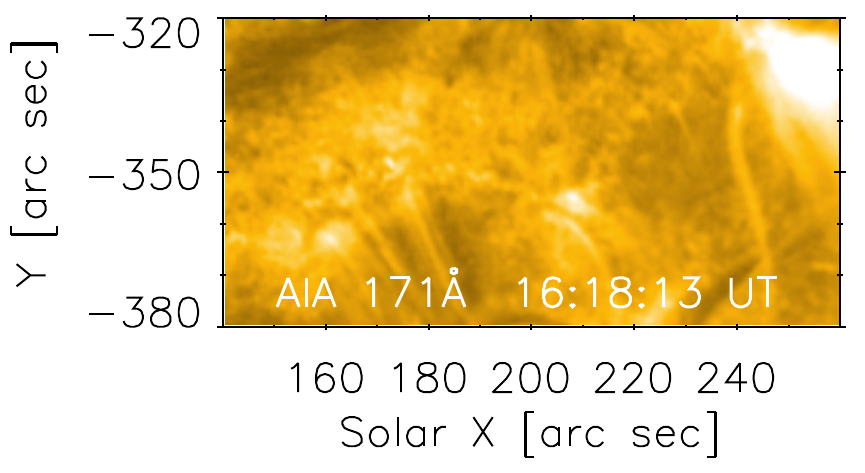}
       \includegraphics[height=1.70cm, bb=60 40 243 136, clip]{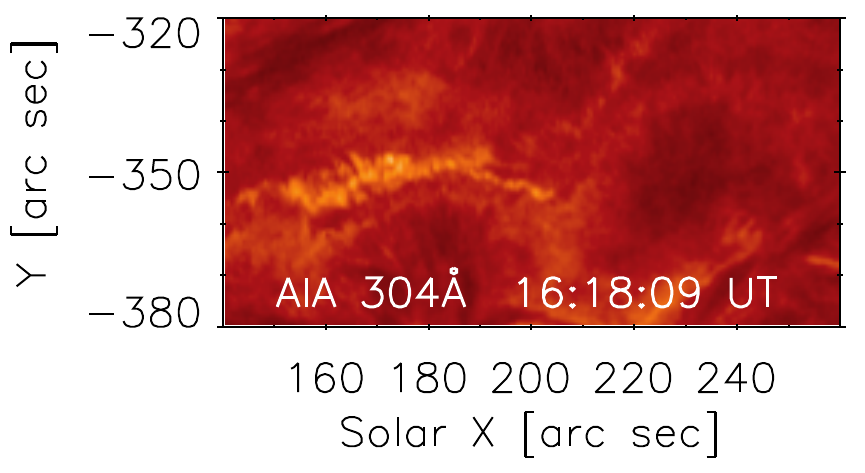}
       \includegraphics[height=1.70cm, bb=60 40 243 136, clip]{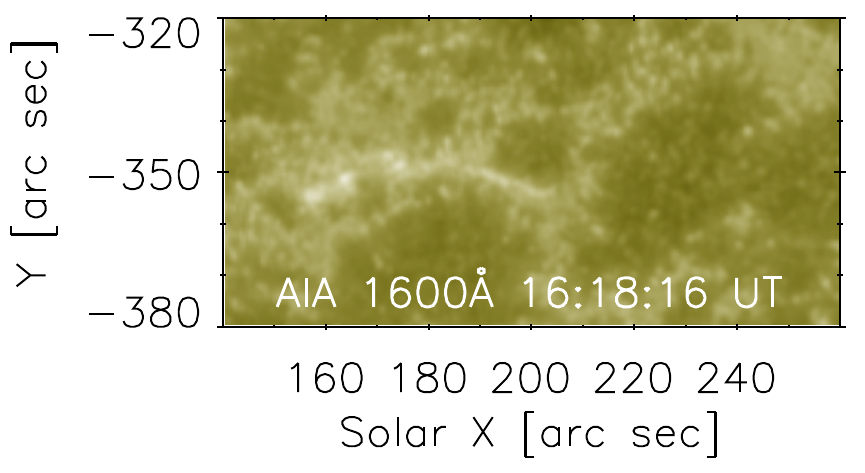}

       \includegraphics[height=1.70cm, bb=0  40 243 136, clip]{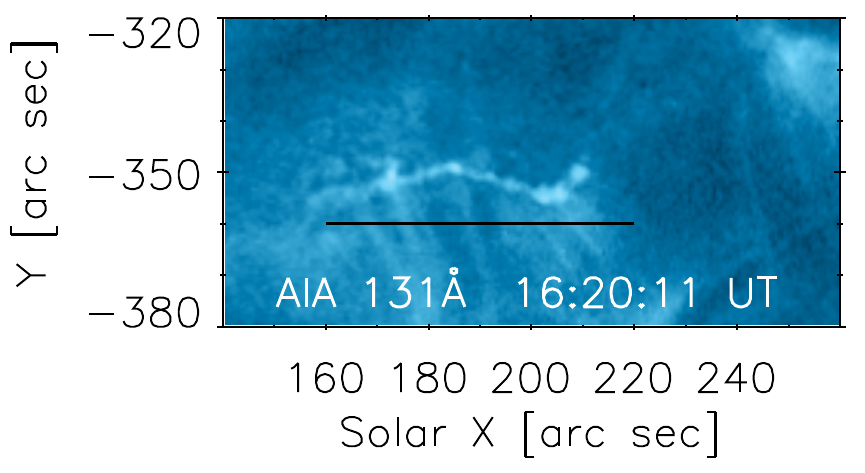}
       \includegraphics[height=1.70cm, bb=60 40 243 136, clip]{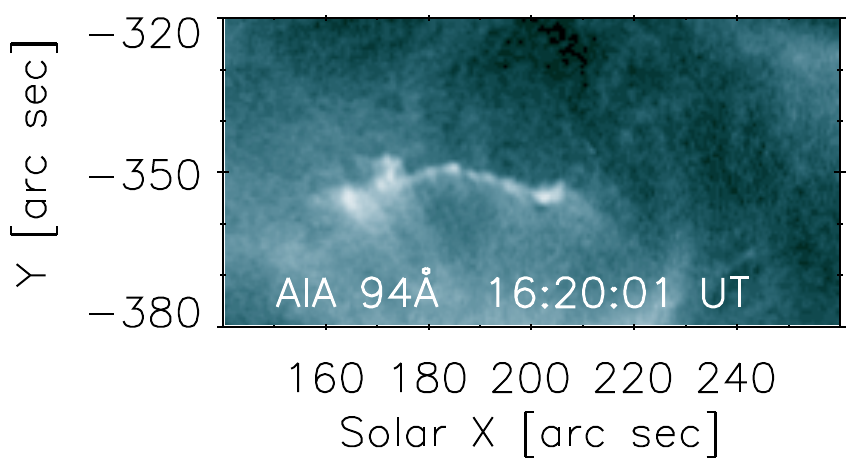}
       \includegraphics[height=1.70cm, bb=60 40 243 136, clip]{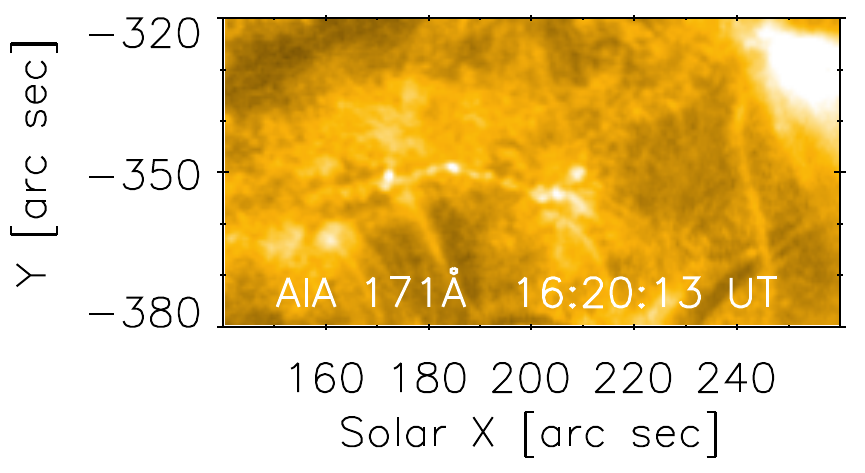}
       \includegraphics[height=1.70cm, bb=60 40 243 136, clip]{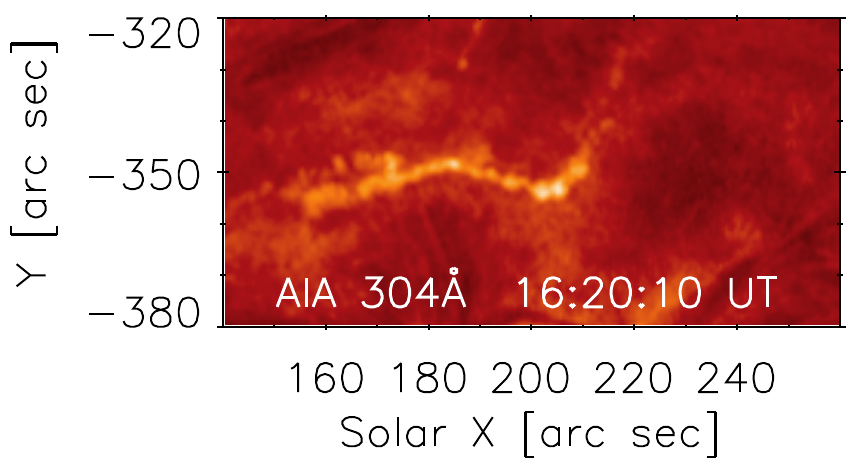}
       \includegraphics[height=1.70cm, bb=60 40 243 136, clip]{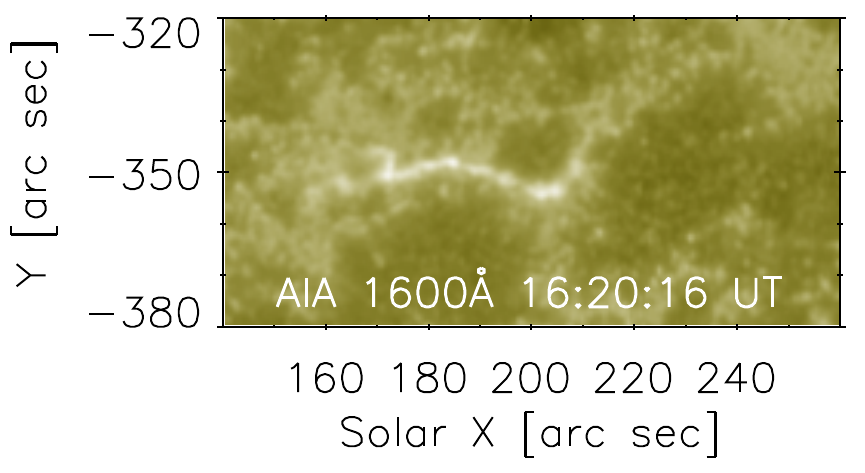}

       \includegraphics[height=1.70cm, bb=0  40 243 136, clip]{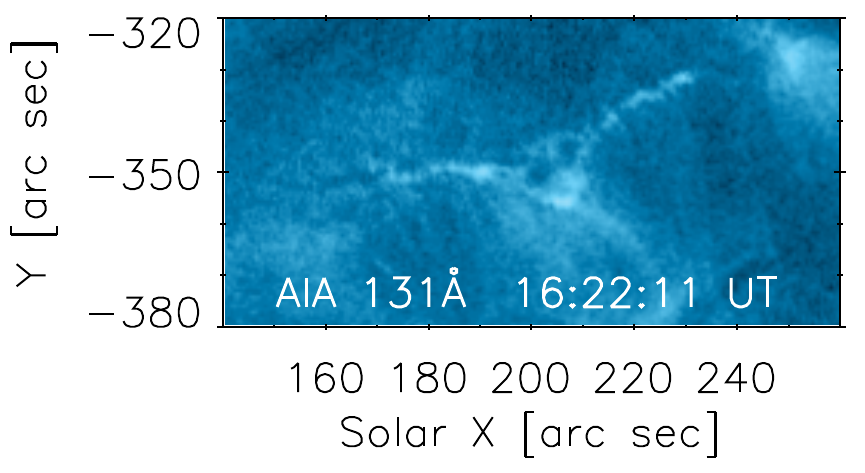}
       \includegraphics[height=1.70cm, bb=60 40 243 136, clip]{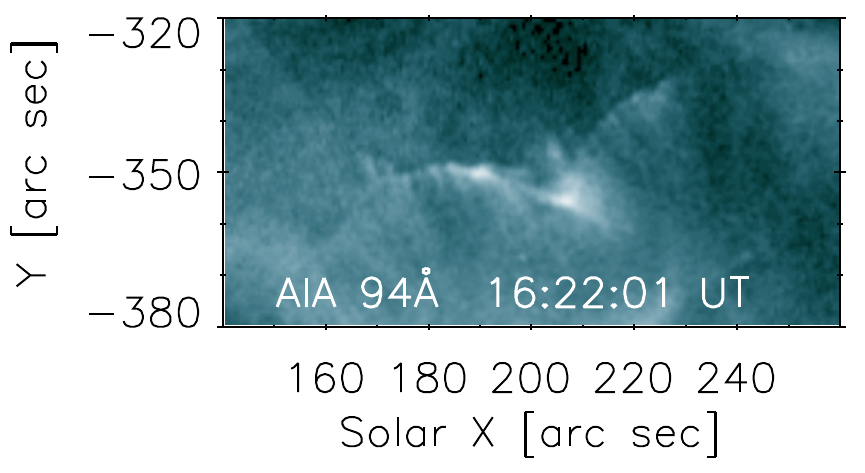}
       \includegraphics[height=1.70cm, bb=60 40 243 136, clip]{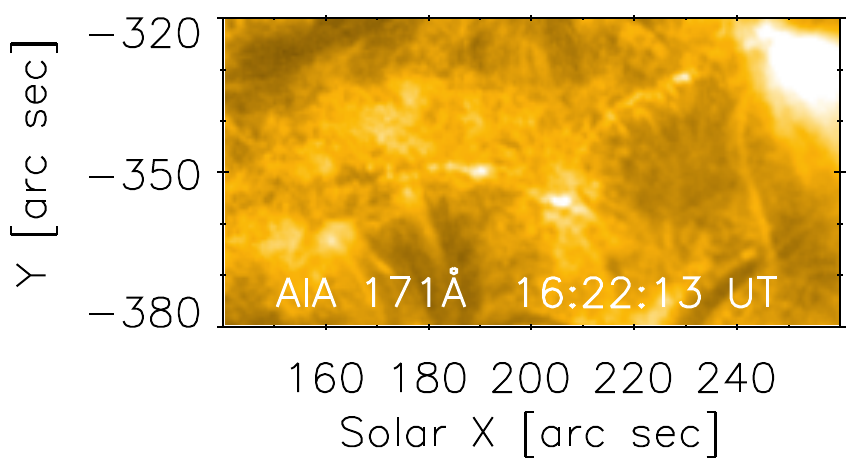}
       \includegraphics[height=1.70cm, bb=60 40 243 136, clip]{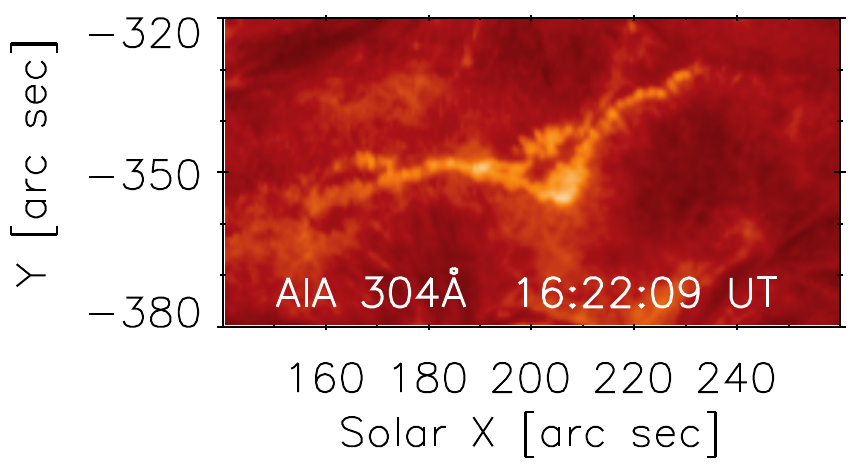}
       \includegraphics[height=1.70cm, bb=60 40 243 136, clip]{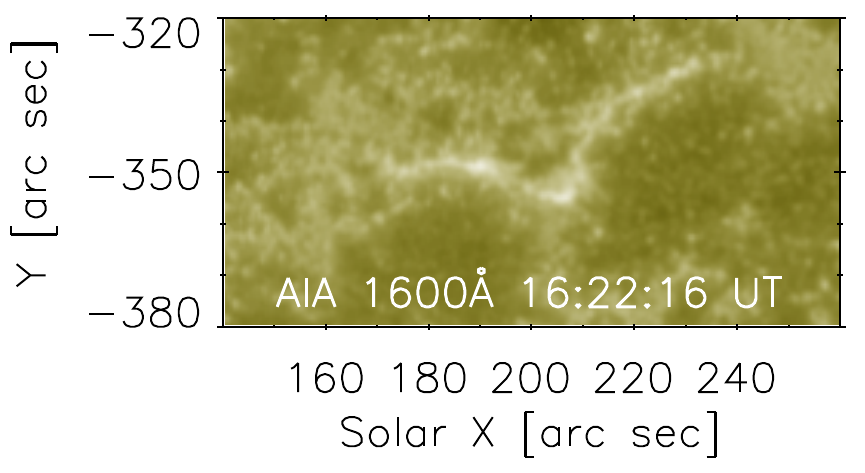}

       \includegraphics[height=1.70cm, bb=0  40 243 136, clip]{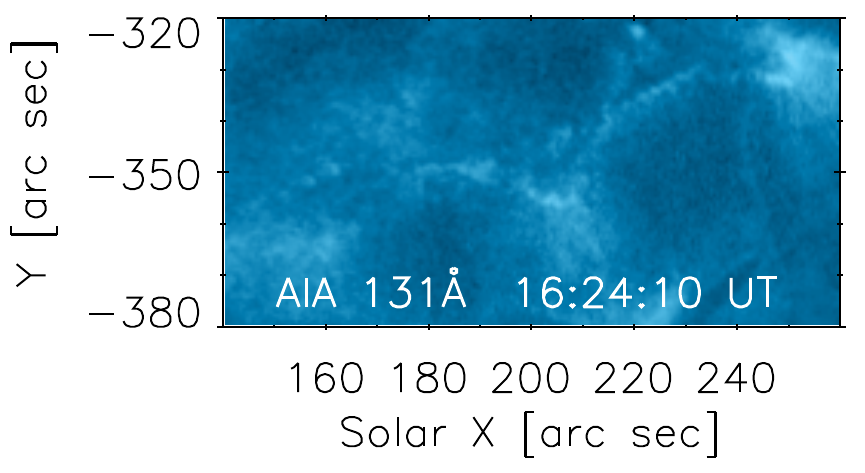}
       \includegraphics[height=1.70cm, bb=60 40 243 136, clip]{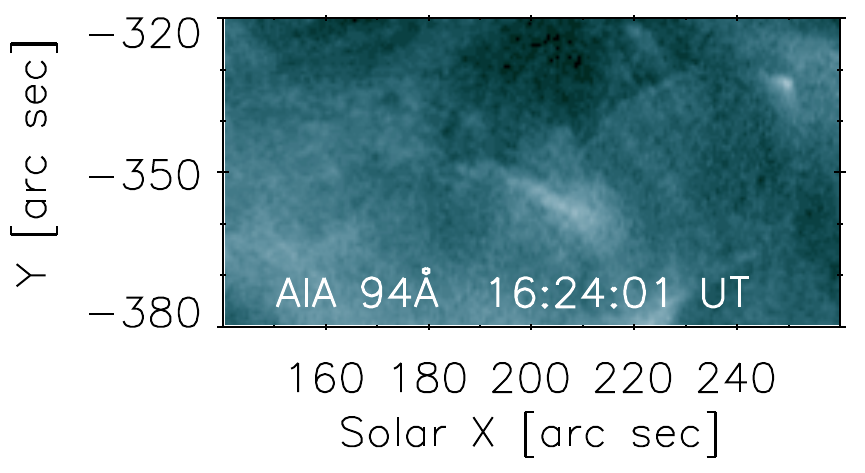}
       \includegraphics[height=1.70cm, bb=60 40 243 136, clip]{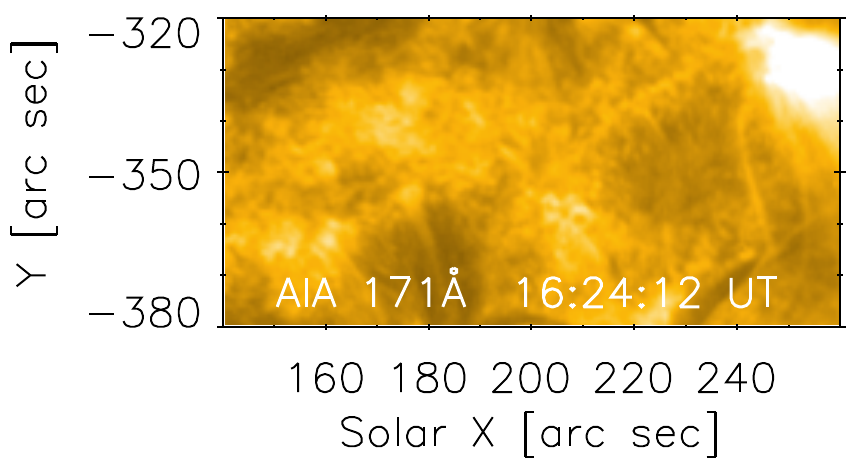}
       \includegraphics[height=1.70cm, bb=60 40 243 136, clip]{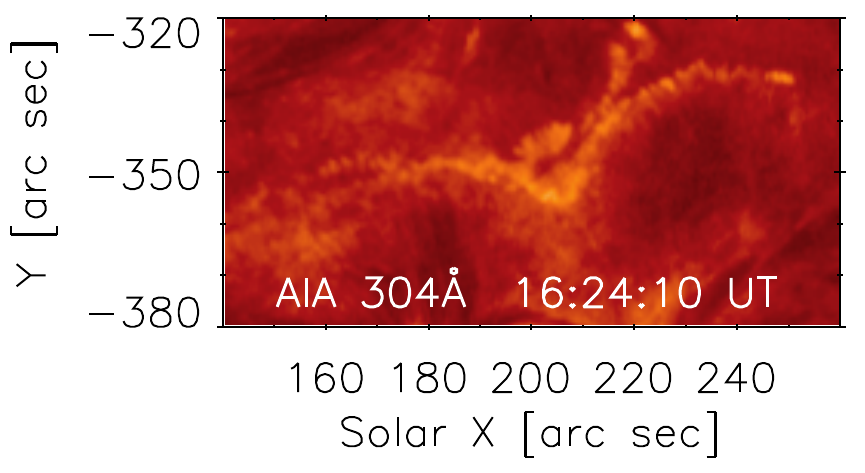}
       \includegraphics[height=1.70cm, bb=60 40 243 136, clip]{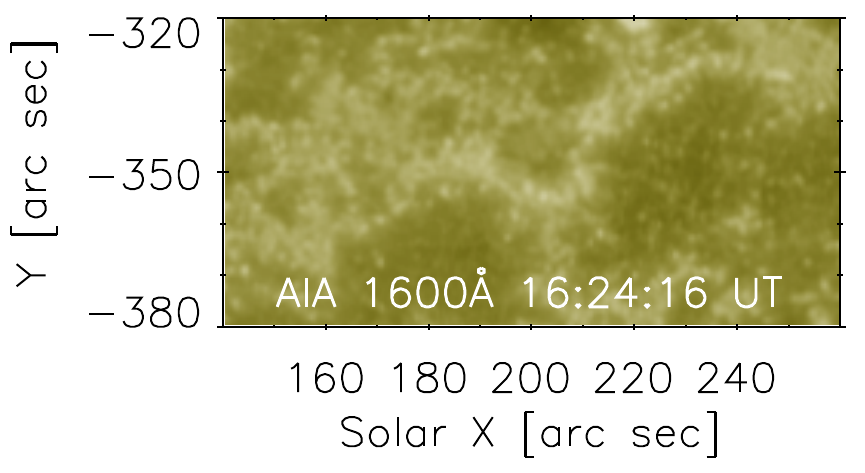}

       \includegraphics[height=2.41cm, bb=0   0 243 136, clip]{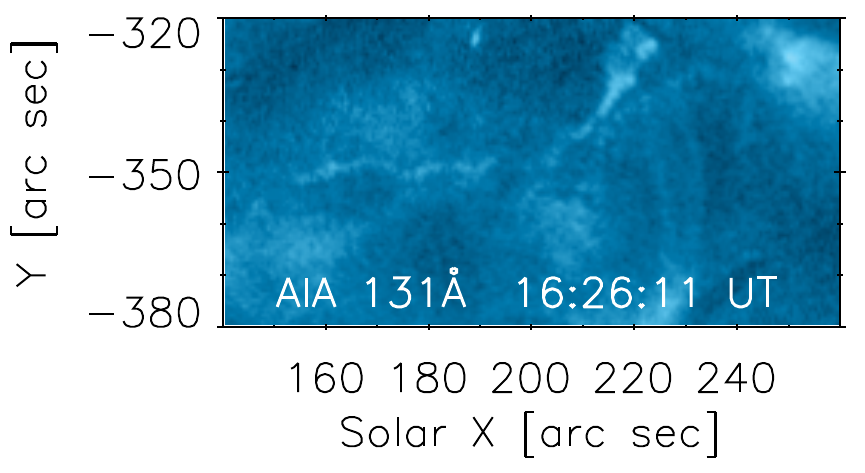}
       \includegraphics[height=2.41cm, bb=60  0 243 136, clip]{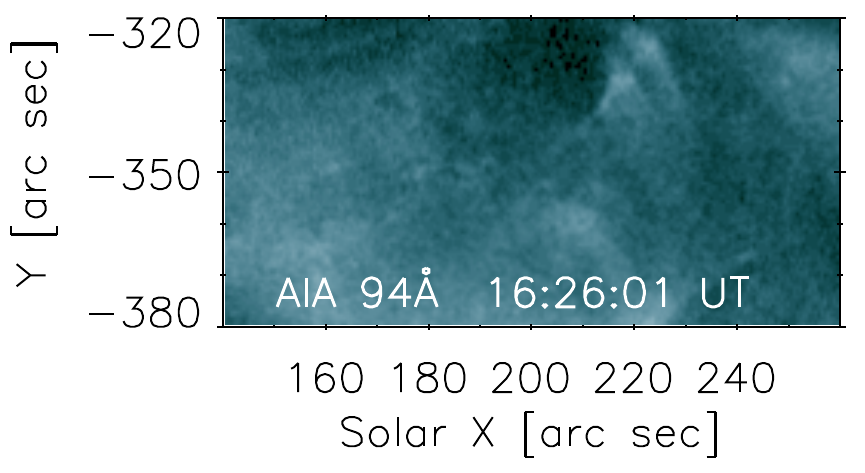}
       \includegraphics[height=2.41cm, bb=60  0 243 136, clip]{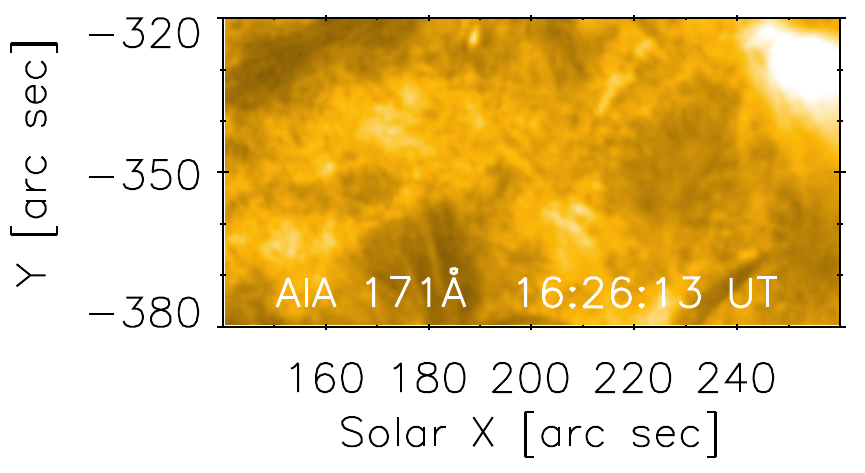}
       \includegraphics[height=2.41cm, bb=60  0 243 136, clip]{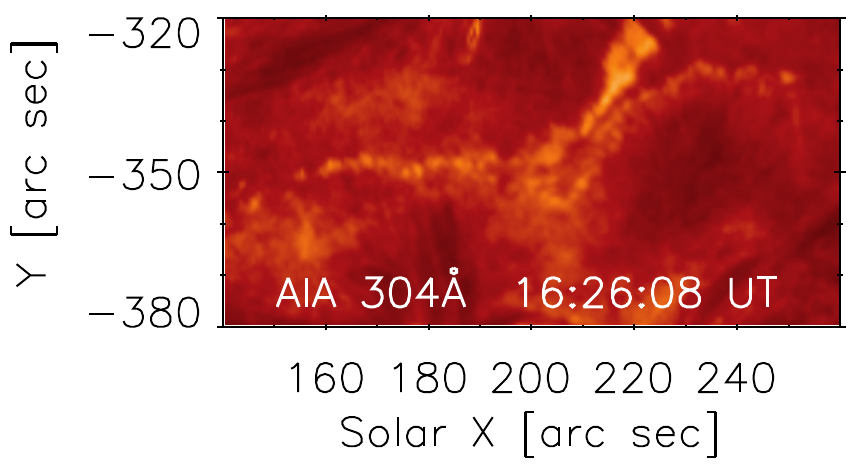}
       \includegraphics[height=2.41cm, bb=60  0 243 136, clip]{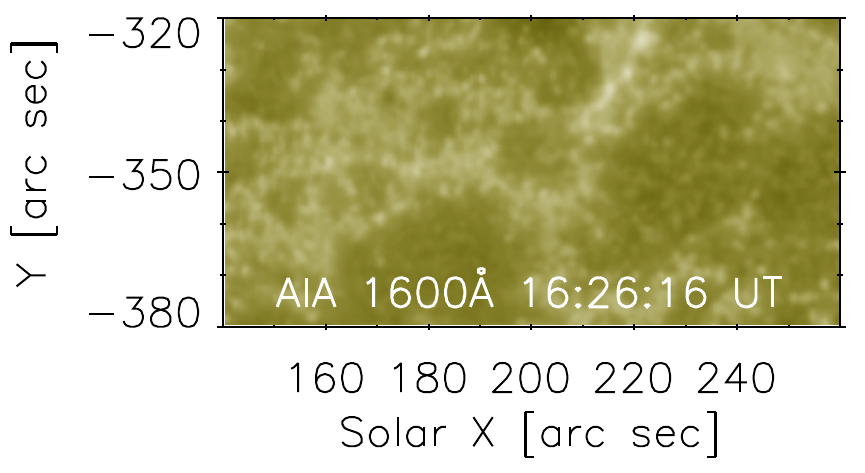}

       \caption{Fast evolution of the PRH during eruption. Dark line in the AIA 131\AA\,image at 16:20 UT shows  position of the cut used to construct $X$-$t$ plot shown in Fig. \ref{Fig:Slipe_stackplots}. The intensities are scaled logarithmically, with units of DN\,s$^{-1}$\,px$^{-1}$.
        }
       \label{Fig:Slipe}
   \end{figure*}
%
%
%
   \begin{figure}
    \centering
    \includegraphics[height=4.9cm,bb=0  0 249 249,clip]{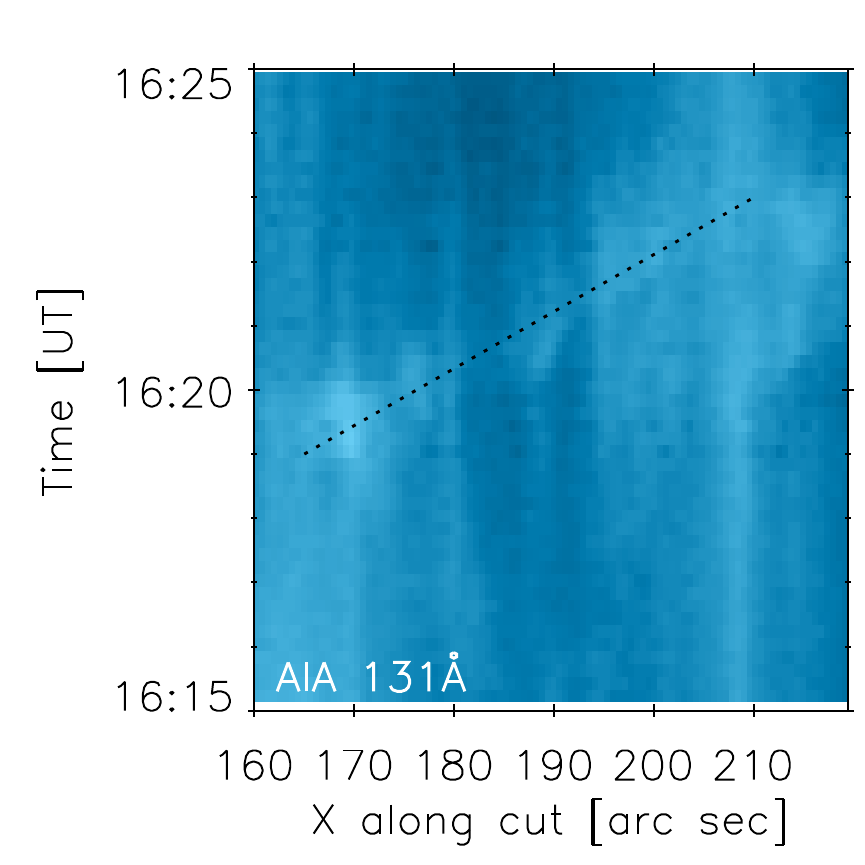}
    \includegraphics[height=4.9cm,bb=70 0 249 249,clip]{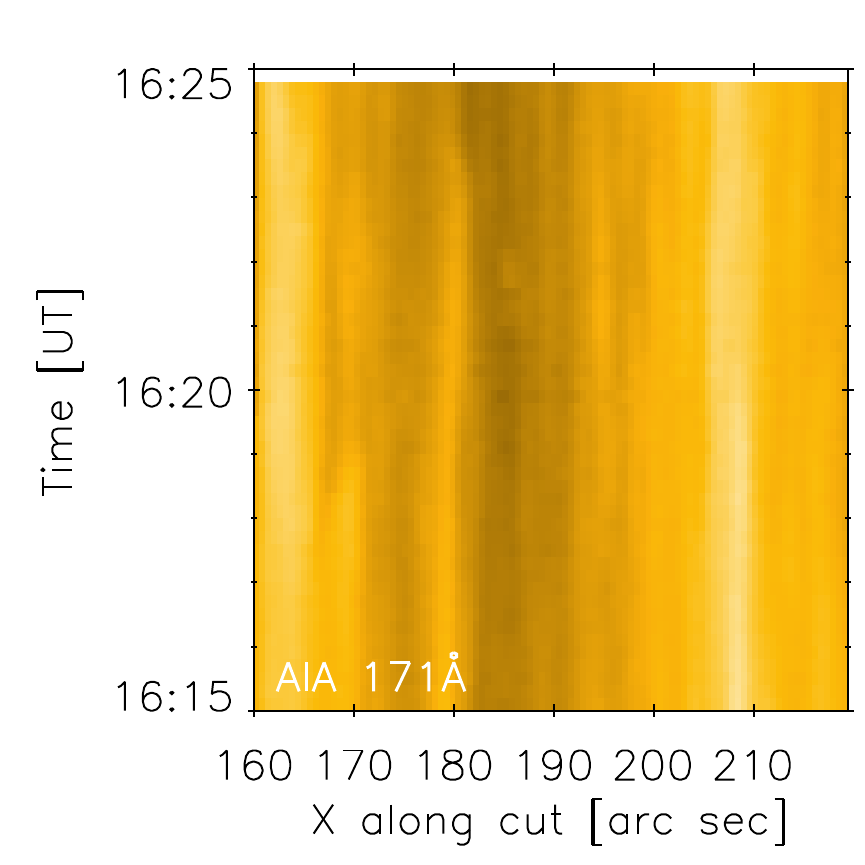}
    \caption{$X$-$t$ stackplot along the cut plotted in Fig. \ref{Fig:Slipe} showing moving structure observed in the AIA 131\AA, but not in AIA 171\AA. The dotted black line corresponds to the velocity of 136 km\,s$^{-1}$.
        }
       \label{Fig:Slipe_stackplots}
   \end{figure}
%
%
%
\section{SDO/AIA Observations and Data Analysis}
\label{Sect:2}
An X1.4 flare occurred on 2012 July 12 in the large complex of active regions NOAA 11519, 11520 and 11521. It was an eruptive, long-duration event with a peak in the GOES 1--8\AA\,flux at 16:49 UT (Fig. \ref{Fig:1500UT}), but with flaring activity starting as early as 15:00 UT. The bulk of the flare EUV and X-ray emission (Sect. \ref{Sect:2.1}) occurred in AR 11520 (Hale class $\beta\gamma\delta / \beta\gamma\delta$), with one ribbon extending to AR 11521 ($\beta\gamma/\beta$). The magnetogram for these active regions is shown in Fig. \ref{Fig:1500UT}. The small, old AR 11519 ($\alpha/\alpha$) located further 100$\arcsec$ westward was not involved in the flare. Altogether, the active region complex spanned nearly 40$^\circ$ in solar longitude.

%
\subsection{Overview}
\label{Sect:2.1}
The Atmospheric Imaging Assembly \citep[AIA,][]{Lemen12,Boerner12} on board the \textit{Solar Dynamics Observatory} (SDO) consists of 4 identical, normal-incidence two-channel telescopes providing multiple, near-simultaneous full-Sun images with both high temporal resolution and high spatial resolution (1.5$\arcsec$, pixel size 0.6$\arcsec$). AIA images of the Sun are taken in 10 filters, 7 of which are centered on EUV wavelengths (94\AA,\,131\AA,\,171\AA,\,193\AA,\,211\AA,\,304\AA,\,and 335\AA), and 3 on UV or visible wavelengths (1600\AA,\,1700\AA, and 4500\AA). The EUV filters are centered on some of the strongest lines in the solar EUV spectrum. Such multi-filter AIA observations allow a study of the thermal structure of the solar atmosphere. However, the presence of a variety of emission lines originating at different temperatures within one filter bandpass makes the temperature responses of the AIA EUV filters (Fig. \ref{Fig:AIA_resp}) highly multithermal in nature \citep[e.g.,][]{ODwyer10,DelZanna11c,Schmelz13}. This behaviour is also reflected in the responses of the 131\AA\,and 94\AA~filters used for flare observations. Their responses are double-peaked due to sensitivity to both flare and coronal temperatures. The response of the AIA 131\AA\,filter peaks at log$(T/$K)\,=\,5.75 and again at 7.05 due to contributions from \ion{Fe}{8} and \ion{Fe}{21}, respectively. The main contributors to the response of the 94\AA\,filter are \ion{Fe}{10} and \ion{Fe}{18}, producing peaks at log$(T/$K)\,=\,6.0 and 6.85, respectively. \ion{Fe}{14} contributes to this filter as well \citep{DelZanna13}. Nevertheless, the flare emission observed by AIA is now well understood \citep{Petkaki12,DelZanna13}. Together with its high temporal and spatial resolution, AIA is well-suited for studies of dynamical phenomena that occur on small scales, such as the evolution of individual features during a flare.

The flare evolution, as observed by SDO/AIA, is summarized in Fig. \ref{Fig:Overview} and in Table \ref{Table:1}, which should act as a reference guide for the online movies, as well as individual events during the flare discussed in this paper.

%
\subsubsection{Pre-flare state, brightening in AR 11521 and the large-scale magnetic topology}
\label{Sect:2.1.1}
The magnetic configuration of the AR 11520 before the flare is that of a forward S-shaped sigmoid visible only in AIA 94\AA~(Fig. \ref{Fig:1500UT}, \textit{second row}). This sigmoid overlies an active region filament F1 visible in AIA 304\AA~(Fig. \ref{Fig:Overview}, \textit{top right}).

In the neighbouring AR 11521, a brightening of several loop systems occurs at around 14:48, i.e., shortly before the flare in AR 11520. One of the loop systems involved in the brightening is rooted in the vicinity of the position $[X, Y] = [+250\arcsec, -330\arcsec]$ (arrow in Fig. \ref{Fig:1500UT}). This position corresponds to the leftmost extension of the positive-polarity ribbon (PR) during the flare, and also to an intersection of one of the strongest QSLs with the photosphere (Fig. \ref{Fig:1500UT}, \textit{bottom}).

We obtained these QSL footpoints from a potential extrapolation of a SDO/HMI \citep[Helioseismic Magnetic Imager,][]{Scherrer12} magnetogram using the method of \citet{Alissandrakis81} and \citet{Gary89}. The potential extrapolation is an approximative method which assumes that there are no electric currents in the region. Clearly, this assumption is not valid either during the flare or near the vicinity of a sigmoid. Nevertheless, it can be used to infer the number and approximate shape of the large-scale QSLs that separate the magnetic flux closed within the active region complex from other closed, or locally ``open'' magnetic field lines \citep[as also done in e.g.,][]{Chandra09}. We found that there are two strong, large-scale QSLs in the active region complex (Fig. \ref{Fig:1500UT}, \textit{bottom}): one semi-circular shaped on the left-hand side of the image in the negative polarities, and a second one at $Y \approx -350\arcsec$ in the positive polarities. A portion of the QSL in the negative polarities is shifted to the north with respect to the footpoint locations of the sigmoidal loops. Such mismatch can be expected because of the electric currents present in the sigmoid.

As already noted, the positive-polarity QSL at $Y \approx -350\arcsec$ corresponds well to the ribbon PR involved the flare (Fig. \ref{Fig:Overview}). This QSL provides the spatial connection between the flare in AR 11520 and the brightening in AR 11521 occuring immediately before the flare. We note that flaring, erupting or even radio events closely related in time have higher than a random occurrence \citep{Wheatland06}, which could be a result of a ``domino effect'' \citep{Chifor06}. Events related in time can have spatial connections through the magnetic field and its topology \citep[e.g.][]{Zuccarello09,Liu09,Jiang11,Torok11,Shen12,Meszarosova13,Schrijver13}. Therefore, the brightening in AR 11521 is the earliest signature of the flaring activity in the entire AR complex.

%
\subsubsection{Flare evolution}
\label{Sect:2.1.2}

The flare itself starts at approximately 15:00 UT, when the first flare loop can be clearly identified in the AIA 131\AA\,filter (Arrow 1 in Fig. \ref{Fig:Overview}). This is the first signature of high temperature flare emission in any of the AIA filters. At one end, the loop is rooted near $[+40\arcsec, -350\arcsec]$ in the QSL in the strongest positive-polarity spots in AR 11520. It then encircles the western spot counterclockwise from SW to NE and is rooted on the other end in the negative polarity. The loop is highly sheared and lies along the outer edge of the curved filament F1 (Fig. \ref{Fig:Overview}, \textit{top right}). The 131\AA~flare loop is easily identifiable with one of the pre-existing sigmoidal loops.

Over the next 40 minutes, the flare gradually develops into a rather compact, highly sheared bundle of flare loops. Some of the flare loops undergo expansion in the SW direction (Arrow 2 in Fig. \ref{Fig:Overview}). Their footpoints gradually move along the QSL which develops into the positive-polarity ribbon (PR) and its hook (PRH, Fig. \ref{Fig:Overview}, \textit{second to fifth row}). The apparent motion of the footpoints along the PR accelerates, and by around 16:25, the loops have erupted. This eruption is observed by the twin STEREO satellites as a CME (Fig. \ref{Fig:CME}) consisting of expanding concentric coronal loops in 171\AA. Unfortunately, the STEREO/EUVI instrument \citep{Wuelser04} does not contain ``hot'' flare filters, so the erupting loops cannot be directly identified in STEREO observations. The expansion of the 171\AA\,loops can be driven by the erupting hot loops, as reported for another event by \citet{Zhang12}.

The arcade of flare loops meanwhile continues to widen, with the flare ribbons brightening and growing in lateral directions (Fig. \ref{Fig:Overview}, \textit{second to fifth row}). These extensions and brightenings are often in the form of bright blobs in 304\AA\,or 1600\AA\,moving along the ribbons, associated with the apparent slippage of flare loops. This is discussed in detail in Sect. \ref{Sect:2.2}. Around 16:25 UT, the ribbons are the most prominent emitting structures in the 304\AA~observations (dominated by \ion{He}{2}). Both ribbons exhibit hooks on their ends. These hooks are typically not as bright as the rest of the ribbon (Fig. \ref{Fig:Overview}, \textit{fifth row}). The hooks show a quite rapid evolution with the eruption. For example, the hook of the negative-polarity ribbon (NRH) undergoes a deformation starting around 16:34 UT and subsequently extends by more than 100$\arcsec$ in the N-S direction grazing along the large-scale QSL (shown in Fig. \ref{Fig:1500UT}, \textit{bottom}). The width of both the NRH and PRH increases over time to a maximum of about 20--30$\arcsec$. Therefore, even if their width is increasing, they still stay rather narrow. Both of them also stay continuous and enclose narrow regions of coronal dimming observed in all AIA EUV filters.

This coronal dimming confirms that the erupting loops are rooted in both hooks. In fact, a bright loop arc located at the end of NRH can be identified around 16:02--16:17 UT (Arrow 3 in Fig. \ref{Fig:Overview}). This bright loop arc is a portion of a system of long loops with projected length of more than 250$\arcsec$ (Fig. \ref{Fig:Eruption}). These long loops are a part of the erupting structure. Because these loops are very long, for most of their length they are barely visible in the AIA 131\AA\,observations due to low signal caused in turn by decreasing density along the loops. However, the motion of this loop system can be discerned in Fig. \ref{Fig:Eruption} or in the corresponding online Movie 3, where the running-difference of the 131\AA~observations with time lag of 1 min is shown. To guide the eye, the discernible portions of two long loops are outlined in Fig. \ref{Fig:Eruption}, \textit{right} by yellow and dark red lines. The running-difference also shows a series of other moving loops following the outlined ones. We emphasize that these long loops have general S-shaped appearance, and so are presumably non-potential. Unfortunately, due to the overlay of many emitting structures, the footpoints of these loops in the PRH cannot be traced with confidence.

One of the interesting features of this flare is that the filament F1, as well as other filaments within the AR complex (e.g. F2, Fig. \ref{Fig:Overview}, \textit{top right}) do not erupt during the flare. Most importantly, F1 is still present and visible at 16:10--16:25\,UT, i.e., the time of eruption of the hot loops. This suggests that F1 is constituted by a portion of the magnetic field not participating in the eruption and/or the flare. We note that the intensity of F1 as observed in the 304\AA\,filter is highly variable with time, although there are little or no morphological changes. This is due to the fact that the \ion{He}{2} emission originates from scattering \citep[e.g.,][and references therein]{Andretta03,Labrosse10}. The scattering increases with the ribbon emission. An additional contribution comes from the diffraction pattern showing many secondary maxima \citep{Poduval13}.

The GOES 1-8\AA\,flux peaked at 16:49 UT, at which time the CME is in the interplanetary space and the arcade of flare loops is well developed and is cooling (Fig. \ref{Fig:Overview} \textit{bottom}).

%
%
\subsection{Individual slipping events}
\label{Sect:2.2}
We now focus on the apparent slipping motion of the flare loops observed in 131\AA. This motion is observed throughout the early stage of the flare. Four times, when the apparent slipping motion is most prominent, can be identified. They are listed in Table \ref{Table:1} and discussed in the next subsections. Note that the plasma velocity can be decoupled from the velocity of the apparent motion of the field lines \citep{Priest03}. Hereafter, we use the terms ``apparent slipping motion'' or ``slipping motion'' to describe the apparent motion of the flare loops as observed by AIA. The physical origin of the apparent slipping motion of flare loops is further discussed in Sect. \ref{Sect:4.3}.

\subsubsection{15:00--15:34 UT}
\label{Sect:2.2.1}
The first signature of apparent slipping motion of the flare loops is observed immediately at the start of the flare. The slipping is best visible in the negative-polarity footpoints of the first flare loops. Part of the time-sequence of this event is shown in Fig. \ref{Fig:Slip1} and in the online Movie 4. This figure is a zoom in the region NR indicated in Fig. \ref{Fig:Overview}. It shows AIA observations in filters 131\AA, 94\AA, 171\AA, 304\AA\,and 1600\AA, with a time step of 5\,min. The apparent slipping motions are best seen in 131\AA~and are weak in 94\AA. All other filters show only concentrated enhancements of emission consisting of several point-like features. This means that this emission originates in the rather compact transition region near the loop photospheric footpoints \citep[similarly as in][]{Graham11,Young13}. In other words, the loops emit primarily in \ion{Fe}{21} and not in \ion{Fe}{8} or \ion{Fe}{9}, meaning that their temperature is around 10\,MK. We examine this point in detail in Sect. \ref{Sect:3}. We emphasize here that the transition region emission seen in all AIA filters is the first signature of the developing ribbon NR. During these early stages of the flare, the ribbon emission is only due to the footpoints of these dense, hot flare loops.

We note that some of the footpoints are very close to F1, especially around 15:05--15:10 UT. The F1 does not exhibit any significant structural changes, suggesting that its magnetic field is stable even to perturbations of the surrounding field as close as 1--2$\arcsec$. Similarly, the sigmoid in 94\AA~remains largely unperturbed during this time (Fig. \ref{Fig:Slip1}, \textit{Column 2}), except for the widening of the arcade and apparent slippage of the flare loops.

To study the apparent slipping motion of the hot flare loops, we construct stackplots along artificial ``cuts'' inserted at $Y$\,=\,$-$315 and\,$-$320$\arcsec$ ($X$--$t$ plots). These cuts are shown in Fig. \ref{Fig:Slip1} as dark lines. The stackplots are shown in Fig. \ref{Fig:Slip1_stackplots}. Several intensity structures can be discerned, moving in the negative $X$ direction, in agreement with a visual inspection of the time-sequence in Fig. \ref{Fig:Slip1}. One of the brightest structures is moving with an apparent velocity of 8.7$\pm$0.3\,km\,s$^{-1}$ (dotted line in Fig. \ref{Fig:Slip1_stackplots}). The error in velocity is estimated as the error of the line slope. There are other structures exhibiting apparent motion in the same direction, but they show large and intermittent intensity variations. There is no distinguishable velocity component in the perpendicular ($Y$) direction.

The apparent slipping motion is also evident in the online Movies 1, 2 and 4, which have full temporal cadence (12\,s). We note that the slipping motion can easily be missed in the visual inspection of AIA movies with lower cadence (e.g., 1\,min). 

\subsubsection{15:43--16:07 UT}
\label{Sect:2.2.2}
After the first time interval discussed in the previous section, the apparent slipping motion of the flare loops becomes less evident or nearly invisible. Then, a series of apparently slipping loops reappear shortly after 15:43 UT. These loops appear brighter and more dynamical than during the time interval described in Sect. \ref{Sect:2.2.1}. Figure \ref{Fig:Slip2} shows a portion of the time-sequence with a cadence of 1 minute. The slipping is again predominantly in the negative $X$ direction. The intensity variations of the apparently slipping loops make it difficult to distinguish individual structures moving in the opposite direction. We again construct stackplots along an artificial cut placed at $Y$\,=\,$-$320$\arcsec$. The stackplots (Fig. \ref{Fig:Slip2_stackplots}) show a series of moving intensity features, with the one denoted by a long, dotted line having a velocity of 16.6$\pm$2\,km\,s$^{-1}$. The apparently moving loops are also clearly visible on the running-difference images (Fig. \ref{Fig:Slip2_stackplots}, \textit{bottom row}) with a time delay of 12\,s. At 15:52 UT, only one intense, apparently moving loop is visible, while at 15:57 UT, there are several, as shown by the stackplots. Note also that there are indications of somewhat weaker structures moving short distances in the opposite direction. The brightest one is outlined by the short dotted short line (Fig. \ref{Fig:Slip2_stackplots}), which corresponds to a velocity of 4.5\,km\,s$^{-1}$.

At 94\AA, only the most intense of the 131\AA~loops can be clearly distinguished (Fig. \ref{Fig:Slip2}). These can also be seen in the 94\AA~stackplot shown in Fig. \ref{Fig:Slip2_stackplots} \textit{right}. There are no discernible temporal shifts between the 131\AA~and 94\AA~bands, suggesting no temporally resolved cooling of the hot plasma.

The transition region emission near the footpoints of the apparently slipping loops is again visible in all AIA filters, with a clear one-to-one correspondence. The emission morphology is that of a chain of bright dots, resembling a pearl necklace in the AIA 304\AA\,and 1600\AA\,images at 15:57 UT. This transition region emission constitutes the elongating, developing NR.

%
\subsubsection{16:05--16:35 UT}
\label{Sect:2.2.3}

As the NR continues to develop, flare loops apparently move along it and along its hook (NRH). The most prominent example is shown in Fig. \ref{Fig:Slip3}, with stackplots along the cut shown in Fig. \ref{Fig:Slip3_stackplots}. We denote the coordinate along the cut as $S$, measured from left to right. The stackplots show multiple structures moving in both directions, but predominantly in the negative-$S$ direction. This is not surprising, since this is the local direction of the ribbon extension. The brightest apparently moving loops are seen at 131\AA~and 94\AA. However, one of the loops is seen at 193\AA~and 131\AA~rather than at 94\AA, apparently moving with a velocity of 44$\pm$5\,km\,s$^{-1}$ (Fig. \ref{Fig:Slip3_stackplots}, dotted line in the \textit{left} panel). Considering the temperature responses (Fig. \ref{Fig:AIA_resp}), this means that the loop emits in \ion{Fe}{24}. This occurence of a flare loop at 193\AA~is the only example we were able to find in this dataset. This is due to the fact that the 193\AA~channel is normally dominated by transition region and coronal emission from moss or warm coronal loops, with the secondary peak at log$(T/$K)\,=\,7.2 which is more than an order of magnitude lower than the primary one.

%
\subsubsection{16:14--16:27 UT}
\label{Sect:2.2.4}

The bulk of the brightest flare loops are rooted in a small portion of the PR oriented in the N-S direction, directly in the strongest positive-polarity sunspots. Since the magnetic field is strong here, the footpoints of the flare loops are concentrated and any apparent slippage here is not easily distinguished.

There is however one prominent event, exhibiting moving structures along the PR and its developing hook. This event is the eruption of a portion of the hot 131\AA~loops, associated with a travelling brightening along the PRH (Fig. \ref{Fig:Slipe}). This brightening is seen in all AIA filters simultaneously. We note especially that this brightening happens immediately after the eruption of the long loop (Fig. \ref{Fig:Eruption}) described in Sect. \ref{Sect:2.1}. We therefore interpret it as the apparently moving footpoints of the erupting loops.

We again construct a stackplot along the artificial cut inserted at $Y$\,=\,$-360\arcsec$ (black line in Fig. \ref{Fig:Slipe}). The stackplot (Fig. \ref{Fig:Slipe_stackplots}, \textit{left}) shows a single weak structure moving in the positive-$X$ direction with the velocity of 136$\pm$15\,km\,s$^{-1}$ (dotted line). The vertical stripes of enhanced intensity are stationary, warm coronal loops. They are seen in the 171\AA, both in Fig. \ref{Fig:Slipe} and on the 171\AA~stackplot (Fig. \ref{Fig:Slipe_stackplots}, \textit{right}). Where the moving 131\AA~feature temporarily overlies these warm coronal loops, the intensity becomes enhanced, since the emission is optically thin.

The brightening of PRH due to the apparently moving footpoints of erupting loops causes intensity variations of the warm coronal loops anchored in the PRH. As can be seen from Figs. \ref{Fig:Slipe} and \ref{Fig:Slipe_stackplots}, these loops fade or disappear after the eruption.

   \begin{figure*}[!ht]
    \centering
        \includegraphics[height=3.25cm,bb=0  40 331 198,clip]{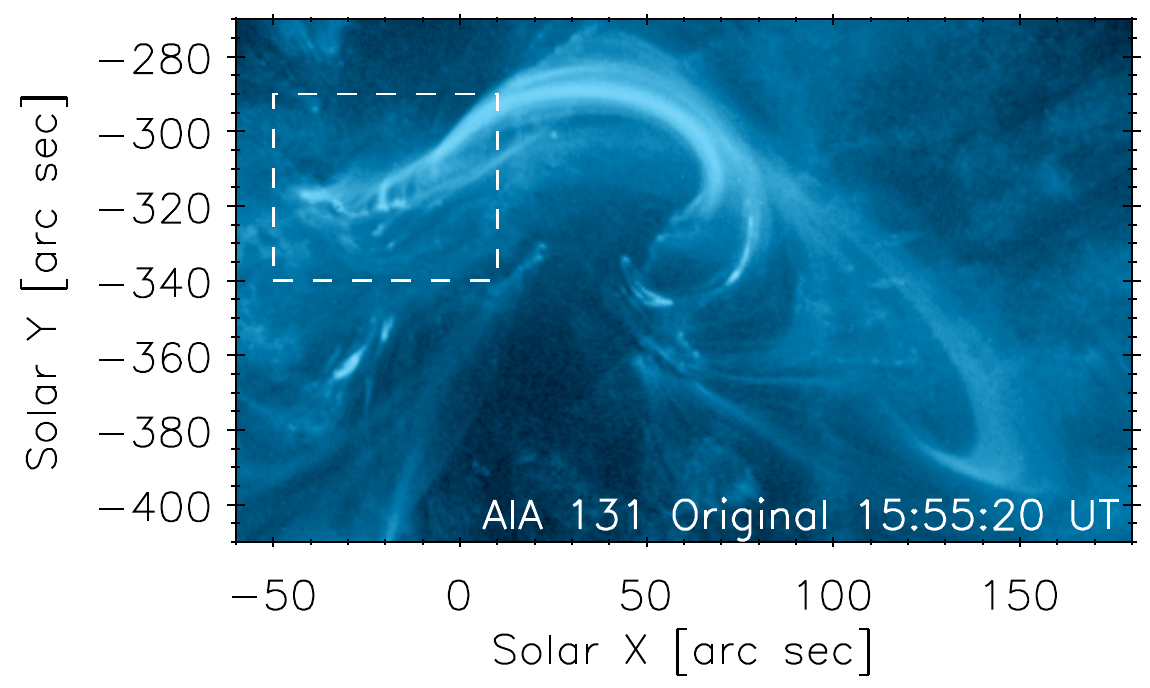}
        \includegraphics[height=3.25cm,bb=65 40 331 198,clip]{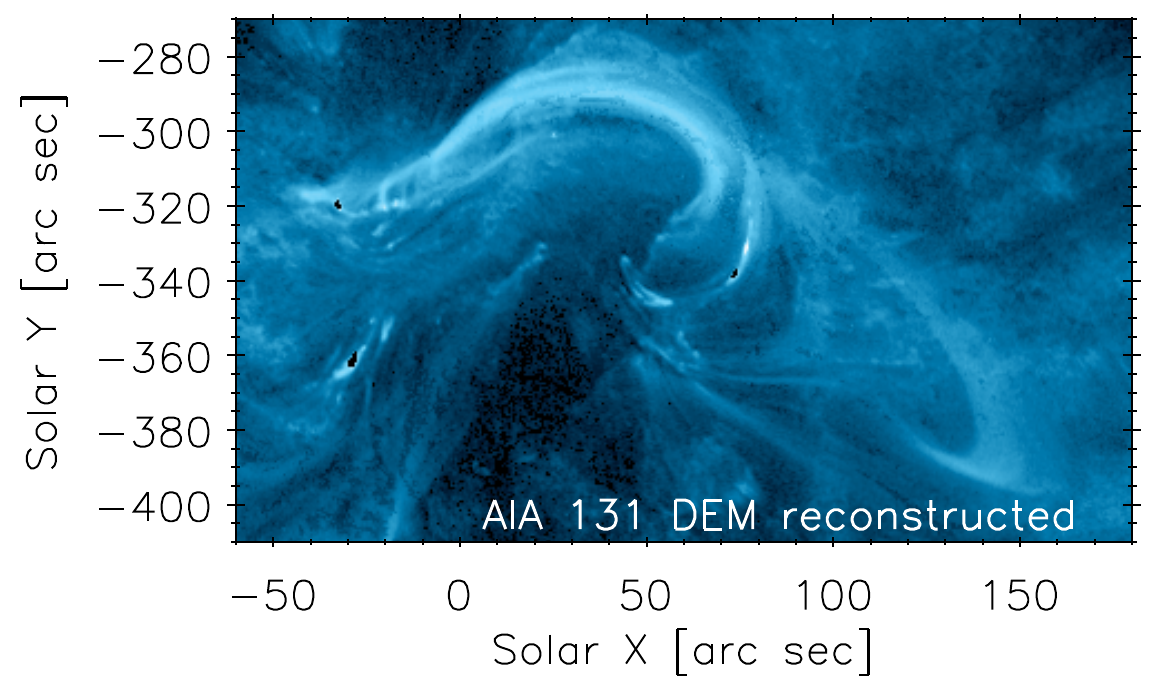}
        \includegraphics[height=3.25cm,bb=65 40 331 198,clip]{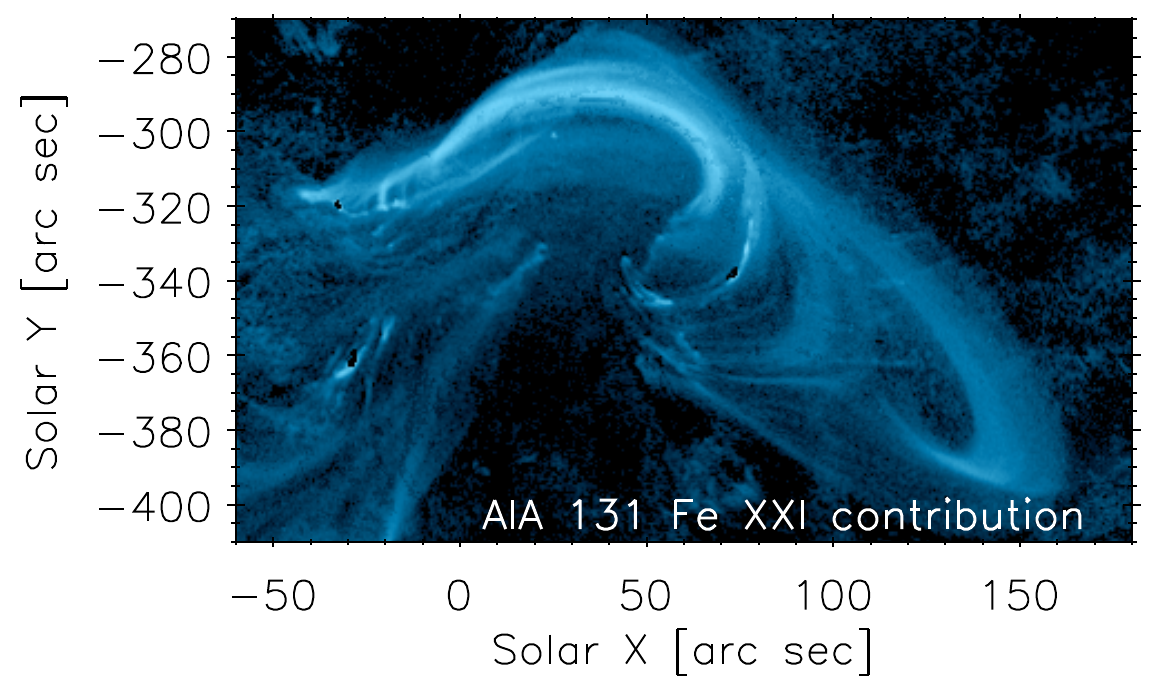}
        \includegraphics[height=4.08cm,bb=0   0 331 198,clip]{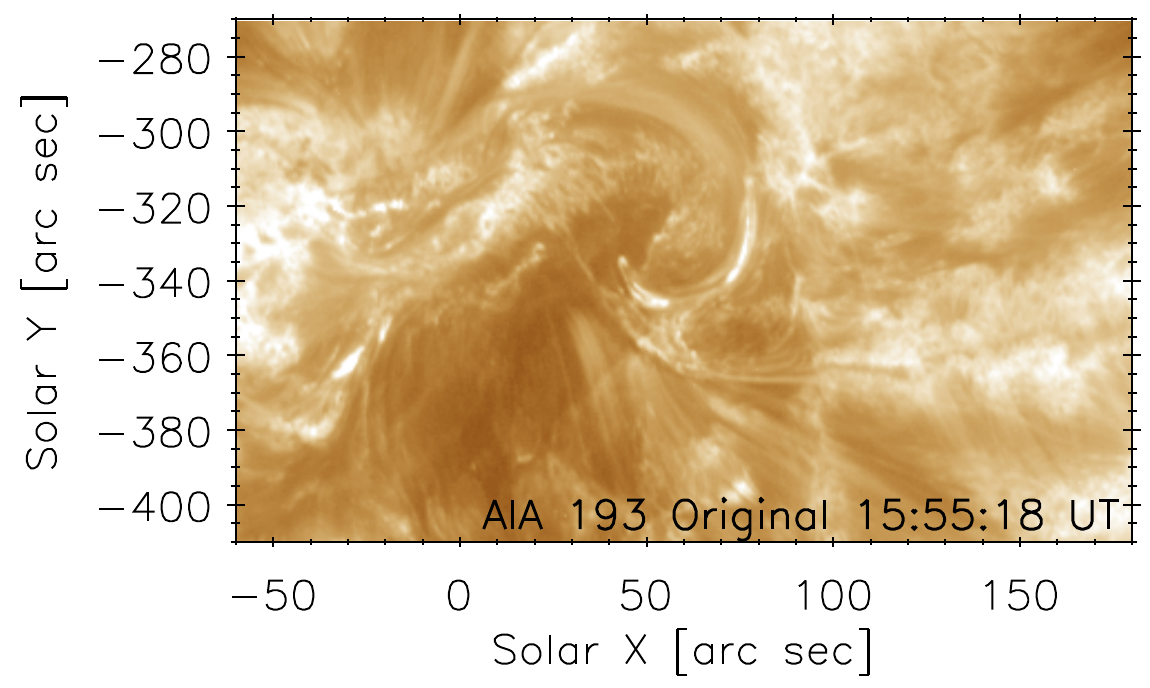}
        \includegraphics[height=4.08cm,bb=65  0 331 198,clip]{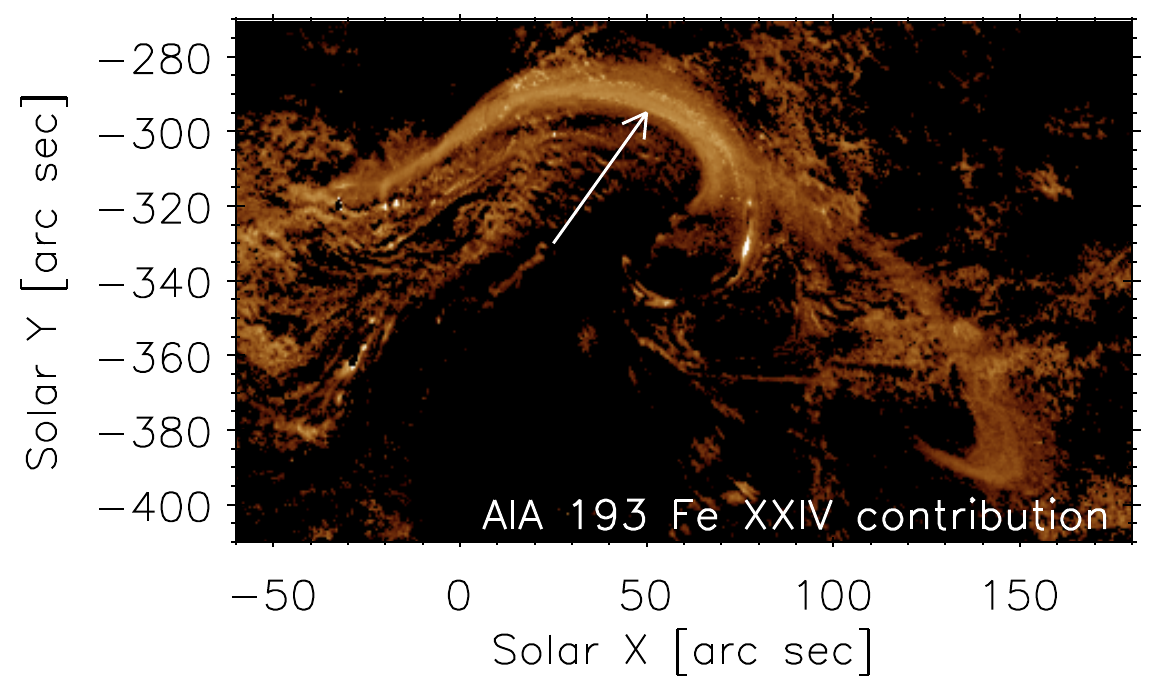}
        \includegraphics[height=4.08cm,bb=65  0 331 198,clip]{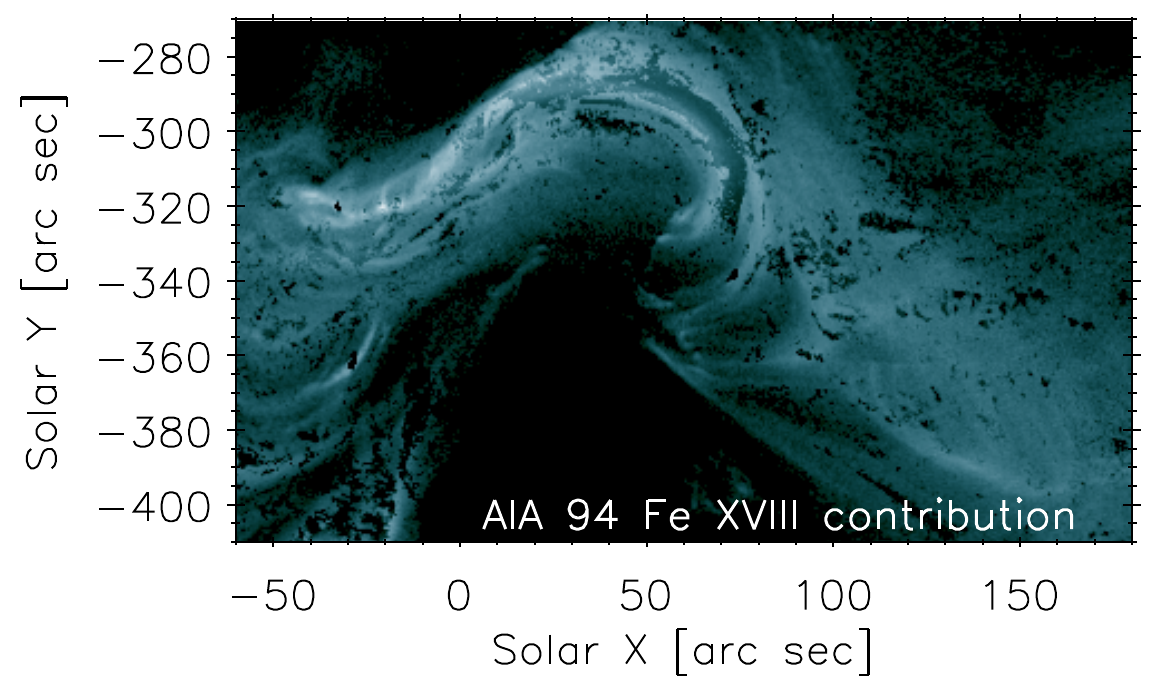}
        \includegraphics[height=3.58cm,bb=0   0 220 175,clip]{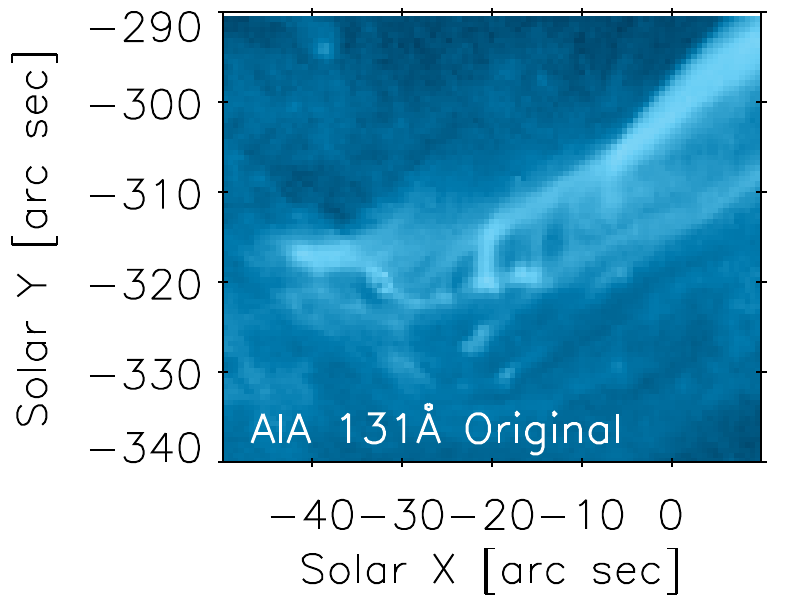}
        \includegraphics[height=3.58cm,bb=60  0 220 175,clip]{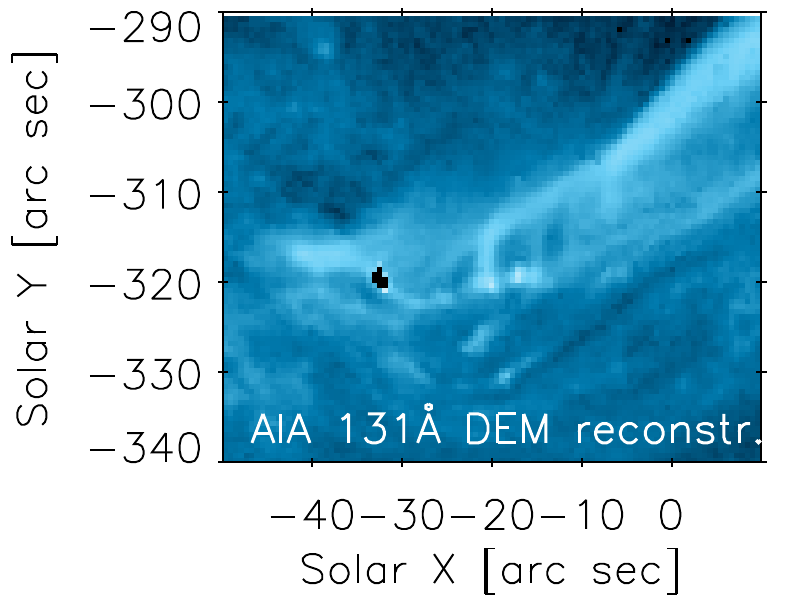}
        \includegraphics[height=3.58cm,bb=60  0 220 175,clip]{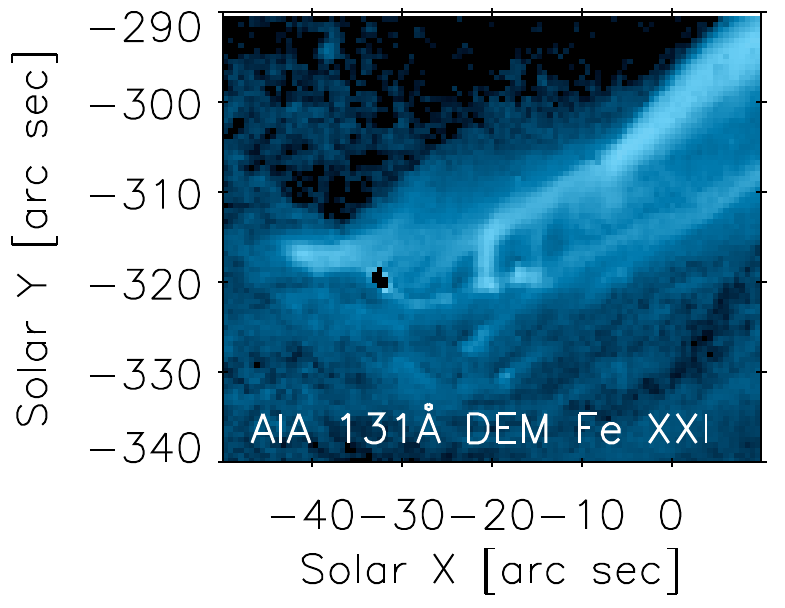}
        \includegraphics[height=3.58cm,bb=60  0 220 175,clip]{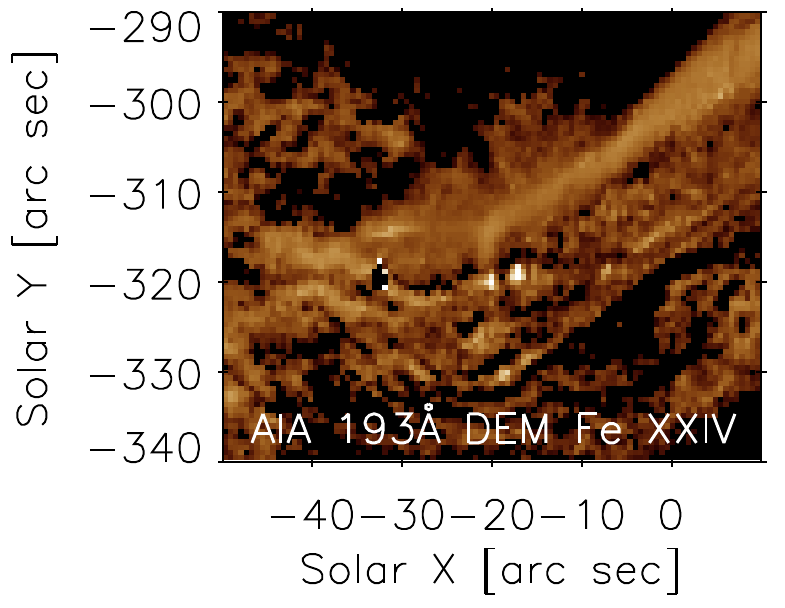}
        \includegraphics[height=3.58cm,bb=60  0 220 175,clip]{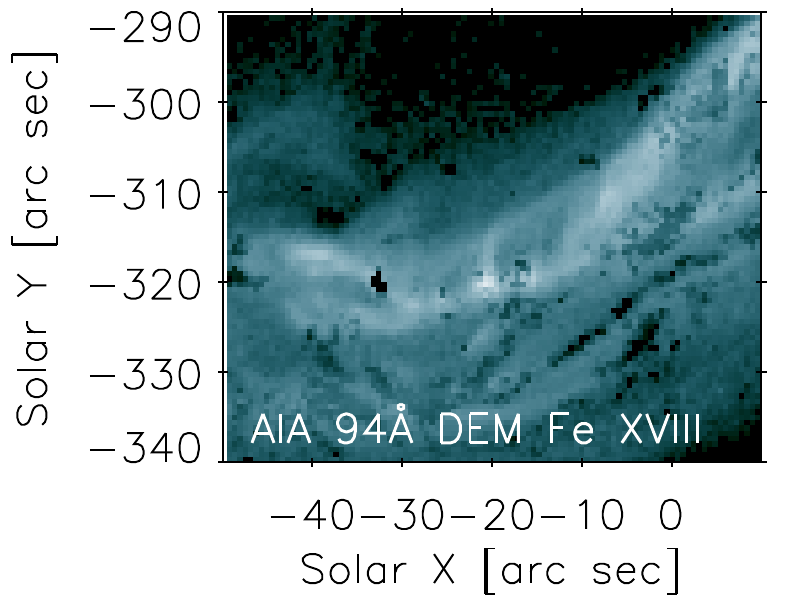}
    \caption{DEM reconstruction of the AIA data at 15:55\,UT and the main contributions to individual AIA channels. \textit{Top}, from \textit{left} to \textit{right}: AIA 131\AA~observed, AIA 131\AA~reconstructed from DEM, and \ion{Fe}{21} contribution to AIA 131\AA. \textit{Middle}, from \textit{left} to \textit{right}: AIA 193 observed, contribution from \ion{Fe}{24} to AIA 193\AA~channel, and contribution from \ion{Fe}{18} to AIA 94\AA~channel. The box in the \textit{top left} image indicates the zoom-in region shown in the \textit{bottom} row. White Arrow denotes flare loops emitting in \ion{Fe}{24}. \textit{Bottom: Close-up on the footpoints of the apparently slipping flare loops. The field of view corresponds to that of Fig.\,\ref{Fig:Slip2}. All intensities are scaled logarithmically, with units of DN\,s$^{-1}$\,px$^{-1}$. The scale is the same as in Fig.\,\ref{Fig:Slip2}, with AIA 193\AA~having the same scale as AIA 171\AA~in Fig.\,\ref{Fig:Slip2}} 
           }
    \label{Fig:DEMREG}
   \end{figure*}
%
%
%
\section{DEM Analysis}
\label{Sect:3}
We now investigate the temperature structure of the individual structures within the flare, in particular the slipping and erupting loops. To do this, we performed a differential emission measure (DEM) analysis on each pixel of a selected AIA frame, within the field of view given by the $X$\,=\,$[-60\arcsec, 190\arcsec]$ and $Y$\,=\,$[-410\arcsec, -270\arcsec]$. We selected the observations at 15:55 UT as representative of the flare (c.f. Fig. \ref{Fig:Overview}, \textit{third row}). At this time, both the slipping loops are visible (Sect. \ref{Sect:2.2.2}), together with the erupting loops moving along the PRH appearing as a single wide bundle (Arrow 2 in Fig. \ref{Fig:Overview} \textit{third row}, also Fig. \ref{Fig:DEMREG} \textit{top left}). We remind the reader that the presence of plasma at flare temperatures can already be discerned by visual inspection of the AIA images, in particular the 131\AA~and 171\AA~channels (Sect. \ref{Sect:2.2.1}). The aim of the DEM reconstruction is to estimate the contribution of the \ion{Fe}{18} 93.93\AA, \ion{Fe}{21} 128.75\AA~and \ion{Fe}{24} 192.02\AA~lines to the AIA flare bands 94\AA, 131\AA~and 193\AA, respectively. The contribution function of these lines peaks for log$(T_\mathrm{max}$/K)\,=\,6.85, 7.05 and 7.25, respectively. Therefore, these lines sample well the contribution of the hot plasma to the AIA observations. 

We note that in general, the DEM inversion problem is ill-posed and under-constrained \citep[e.g.,][]{Craig76,Craig86,Judge97}. Any solution found is not unique, as it typically contains additional constraints, such as some form of regularisation or smoothing of the solution, or an a-priori assumption on the functional form of the solution. \citep[e.g.,][]{DelZanna99,Aschwanden11,Hannah12}. The regularized inversion of \citet{Hannah12} has been specifically adapted for DEM reconstruction of the AIA data and used e.g. to recover DEMs at each AIA pixel in an observation of an eruptive off-limb event \citep{Hannah13}. We adopt this method for DEM analysis of the selected AIA images at 15:55 UT and use the implementation provided in the \textit{dn2dem\_map\_pos.pro} IDL routine. In the reconstruction, the photospheric abundances of \citet{Asplund09} are used similarly as in the flare modeling of \citet{Petkaki12}. Newest atomic data benchmarked against best available solar and laboratory spectra \citep[see][Sect. 3 therein for details]{DelZanna13} are also used together with the atomic data from the CHIANTI database, v7.1 \citep{Landi13,Dere97}. We verified that the assumption of abundances and atomic data has only a small effect on the shape of the reconstructed DEMs. This is because the AIA responses are dominated by Fe ions \citep{ODwyer10,DelZanna13} and the atomic data for flare lines are reliable \citep{Petkaki12}.

Once the DEM is recovered for each AIA pixel, predicted intensity maps for each AIA filter are calculated as follows. First, the DEM obtained for a given pixel is used to calculate a corresponding synthetic spectrum. To do this, CHIANTI v7.1 is used together with the newest atomic data available. The obtained synthetic spectrum is then multiplied by the spectral response of a given AIA filter, and finally integrated in the wavelength direction. Predicted contribution of a specific spectral line to a given AIA band is calculated in a similar manner. This is done for each pixel to obtain the predicted intensity map.

We determined that the AIA observations are best reproduced using 19 temperature bins of log$(T/$K)\,=\,5.5--7.3. This temperature interval is chosen to adequately cover the range of temperatures observed by the AIA instrument (Fig. \ref{Fig:AIA_resp}) and the many contributions to its bandpasses \citep{ODwyer10,DelZanna13}. The DEM reconstruction obtained for each AIA pixel results in good agreement between observed and predicted intensities for the 131\AA, 171\AA, 193\AA, and 211\AA~filters. An example is shown in Fig. \ref{Fig:DEMREG}, \textit{top left} and \textit{top center} for the AIA 131\AA. The 94\AA~and 335\AA~contain some areas, especially within the flare loops arcade, where the reconstructed intensities do not approximate the observed ones. At these locations, the DEM$(T)$ has large horizontal errors in the log$(T/$K)\,=\,6.6--6.8 temperature bins, affecting the \ion{Fe}{16} and \ion{Fe}{18} contributions to the 335\AA~and 94\AA~bands, respectively. As an example, the predicted intensity map for \ion{Fe}{18} is shown in Fig.\,\ref{Fig:DEMREG}, \textit{middle row, right}. The locations of poorly recovered DEMs correspond to dark spots and patches of darker areas, especially within the flare loops arcade. Note that these locations of poorly recovered DEMs at log$(T/$K)\,=\,6.6--6.8 correspond to locations where the LOS pierces the filament F1 and overlying flare loop arcade. The F1 shows dark as well as bright threads \citep[e.g., Figs. \ref{Fig:Overview} and \ref{Fig:Slip2}; see also][]{Alexander13}, while the overlying flare loops are visible in all of the AIA flare filters, namely 94\AA, 131\AA~and 193\AA~(Fig. \ref{Fig:DEMREG}). We therefore suspect that the true DEM$(T)$ structure in such locations is complicated, with more than two peaks, and cannot be adequately recovered by the \citep{Hannah12} method due to the enforced smoothness. Including more temperature bins at lower or higher log$(T$/K) improves the reconstruction only marginally. Nevertheless, outside of these areas, the DEM is recovered successfully, and the results confirm strong contribution of \ion{Fe}{18} to the AIA 94\AA~channel.

The results of the DEM reconstruction also confirm that the observed apparently slipping flare loops are indeed emitting strongly in \ion{Fe}{21} (Fig. \ref{Fig:DEMREG}, \textit{bottom}), which is the dominant contributor to the 131\AA~bandpass. In our case, \ion{Fe}{21} contributes up to $\approx$50--85\% of the observed flare emission in the 131\AA~channel. Weak \ion{Fe}{24} emission is also present in the 193\AA~bandpass (Fig. \ref{Fig:DEMREG}, \textit{middle row, center}.) This \ion{Fe}{24} emission contributes of about $\approx$45\% to the observed flare loops in the 193\AA~image (white arrow in Fig. \ref{Fig:DEMREG}). We note that these loops are weak in the 193\AA~image and can be discerned only outside of areas of strong moss emisson that dominate the observed 193\AA~morphology.

We note that there is some spurious contribution in the moss areas in the recovered \ion{Fe}{21} and \ion{Fe}{24} intensities. However, this emission is typically $\approx$30 times weaker compared to the observed signal in the 193\AA~channel. The recovered DEM at these locations exhibit a weak secondary peak at log$(T/$K)\,=\,7.2 characterized again by large errors, and therefore uncertain.

In summary, the DEM reconstruction confirms that the slipping loops consist dominantly of flare plasma with temperatures up to log$(T/$K)\,=\,7.1--7.3.

%
%
%
\section{Numerical simulation of a flux rope expansion and associated slipping reconnection}
\label{Sect:4}
The mechanisms of solar flares and the associated formation of magnetic structures can be well reproduced with numerical models. In the following, we exploit the 3D MHD simulation performed earlier by \citet{Aulanier12}, recreating the evolution of a flux rope expansion during an eruptive flare. This simulation exhibits naturally occurring slipping magnetic reconnection as a result of the evolution of the model. This slipping reconnection builds both the flare loop arcade and the erupting flux rope \citep{Aulanier12,Janvier13}, which, as we will show in this section, compares well with the observations in qualitative terms. Although the simulation was not designed to fit any specific event, it is well suited for this particular flare (described in Sect. \ref{Sect:2}), as the photospheric magnetic field contains flux asymmetry, and geometrical comparison with the observations can be easily achieved by appropriate rotation of the simulation box.

%
\subsection{Description of the numerical simulation}
\label{Sect:4.1}
The initial conditions of the model are dynamically built so that the whole magnetic structure is torus-unstable. The details of the physical ingredients needed to build such conditions are described in \citet{Aulanier10}. The development of the torus instability leads to the upward expansion of the flux rope core, as well as the formation of a thin current layer where reconnection takes place. The reconnected field lines are of two types: they either further add to the envelope of the flux rope or they form the flare loops, as described in \citet{Aulanier12} and \citet{Janvier13}. The numerical simulation is non-dimensionalized, and extends over a time period of $46\ t_{A}$, where $t_{A}$ represents the Alfv\'en time, i.e., the travel time for a distance $d=1$ at the Alfv\'en speed $c_A=1$.

The region modelled in the present numerical simulation has similar features as that of the AR 11520 where the X1.4 flare was observed (Sect. \ref{Sect:2}). First, the asymmetry of the magnetic polarities is reproduced, with a 27\% flux imbalance favouring the positive polarity in the photosphere. This asymmetry reflects the stronger, leading positive polarity and the weaker, trailing negative polarity of AR 11520 (Fig. \ref{Fig:1500UT}). The numerical simulation also reproduces well the shape of the sigmoid as seen in the 94\AA\ filter at the beginning of the flare (Fig. \ref{Fig:1500UT}). This sigmoid, present at $t$=15:00 UT, contains both $J$-shaped pre-reconnected loops and $S$-shaped post-reconnected loops as seen from the top, as is shown in the top right image of Fig.~\ref{Fig:FR}, where field lines are drawn at one time in the simulation.

Lack of null points and separatrices in the simulation implies that reconnection takes place in QSLs. Field lines then undergo a succession of reconnection processes as they cross the evolving QSLs, resulting in an apparent slipping motion. The details of such a mechanism have been thoroughly investigated in \citet{Janvier13}.

   \begin{figure*}[!ht]
    \centering
    \includegraphics[height=5.0cm,bb=120 220 520 580,clip]{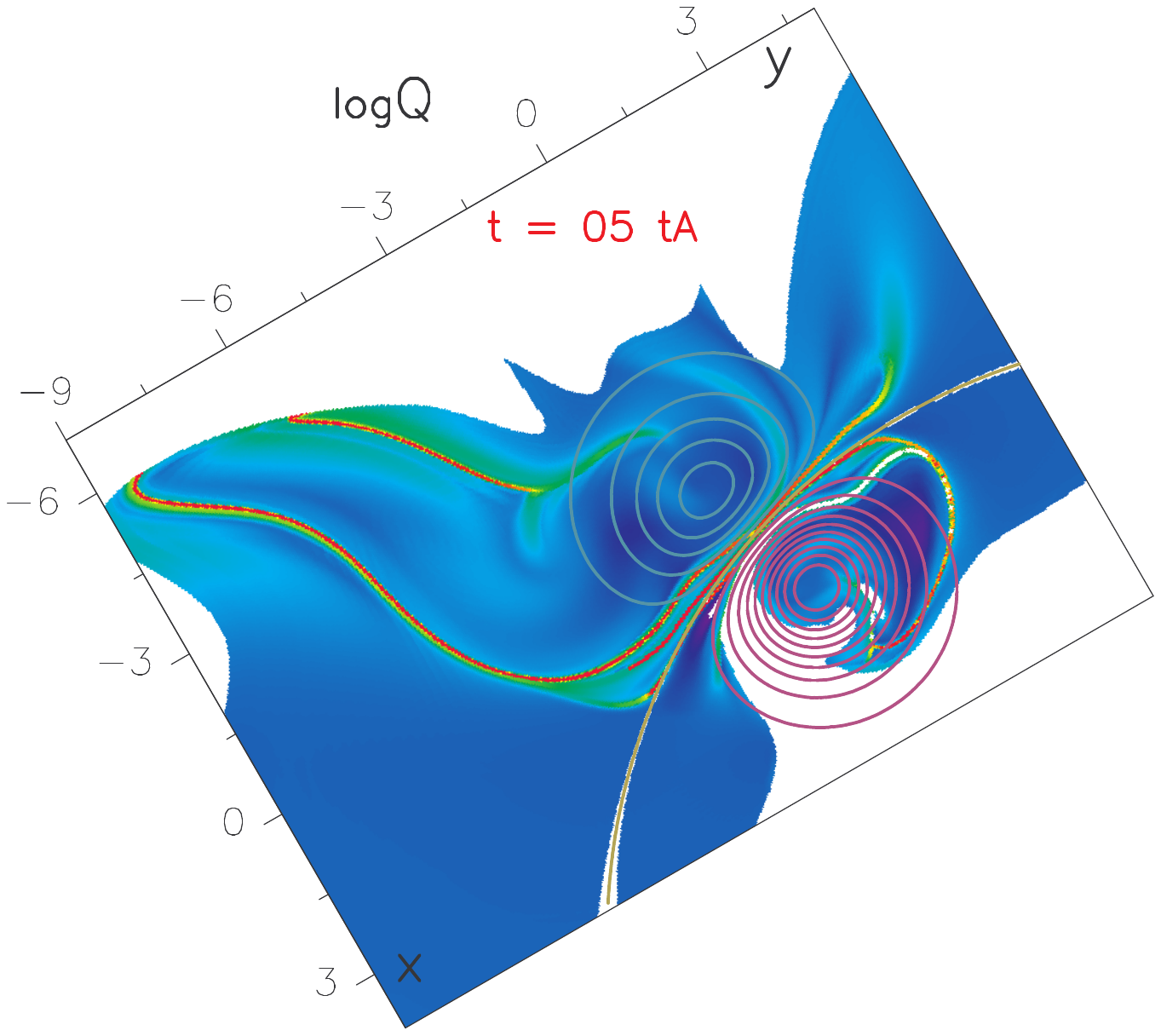}
    \includegraphics[height=5.0cm,bb=120 220 520 580,clip]{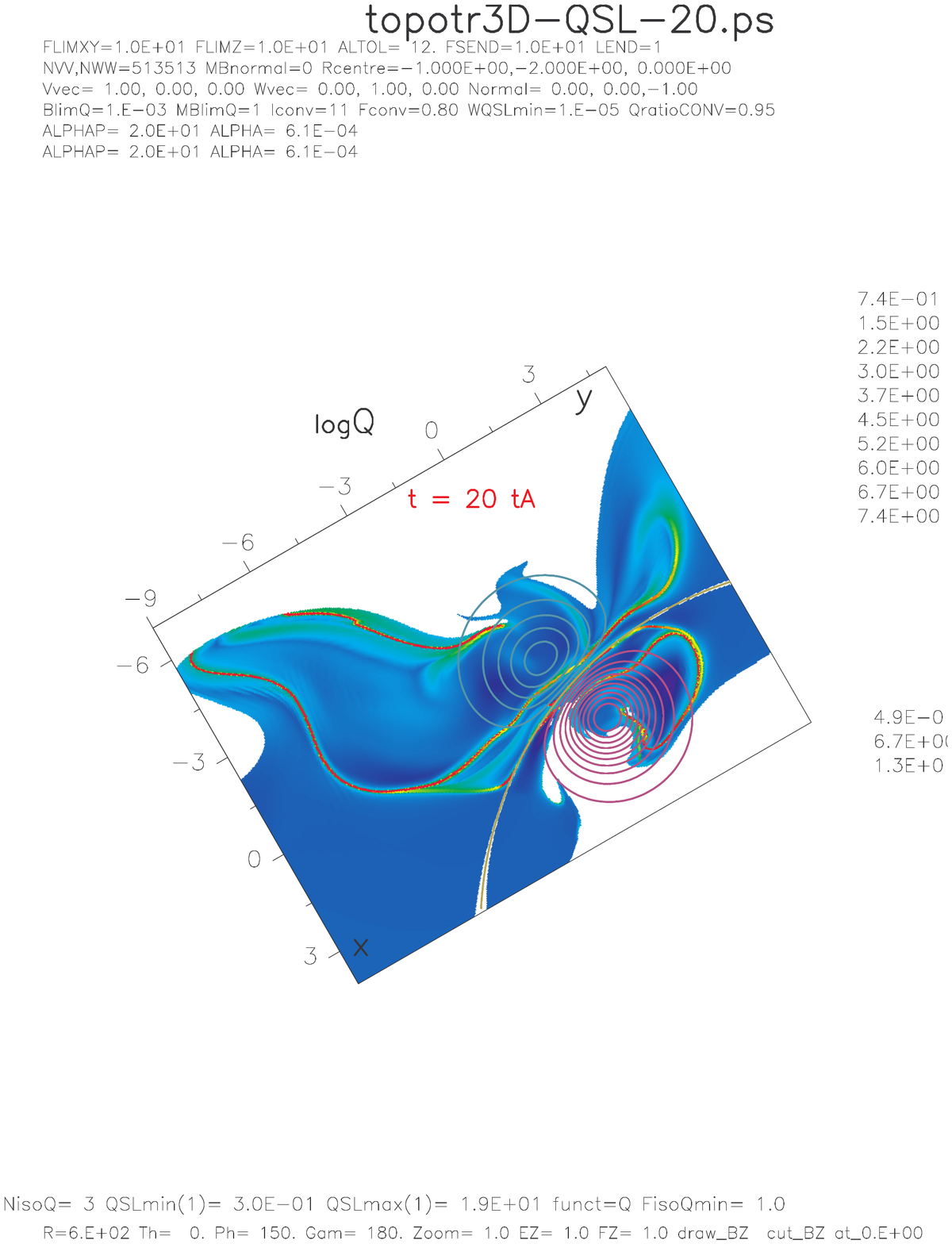}
    \includegraphics[height=5.0cm,bb=120 220 520 580,clip]{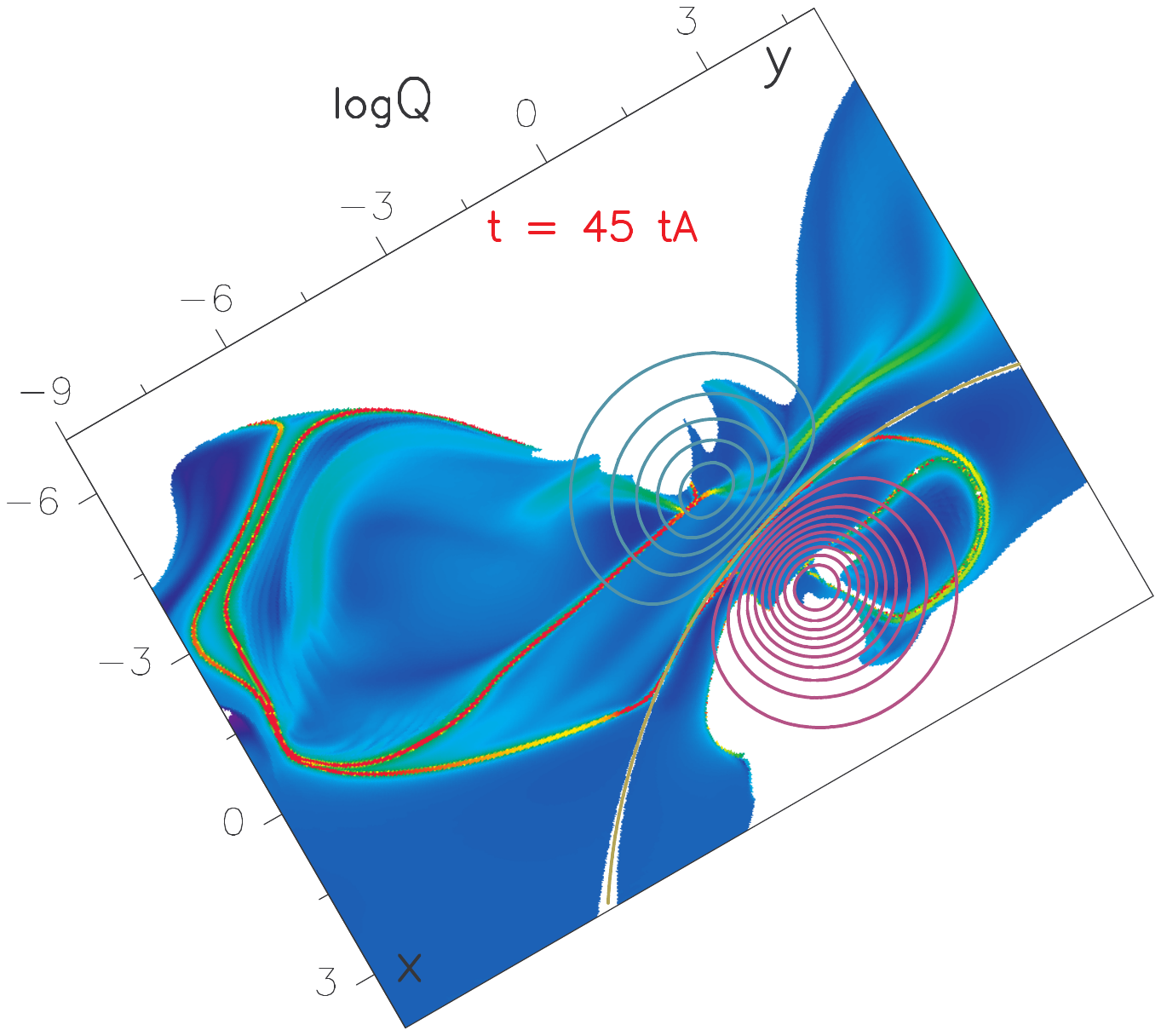}
    \includegraphics[height=5.0cm,bb= 35  30  90 300,clip]{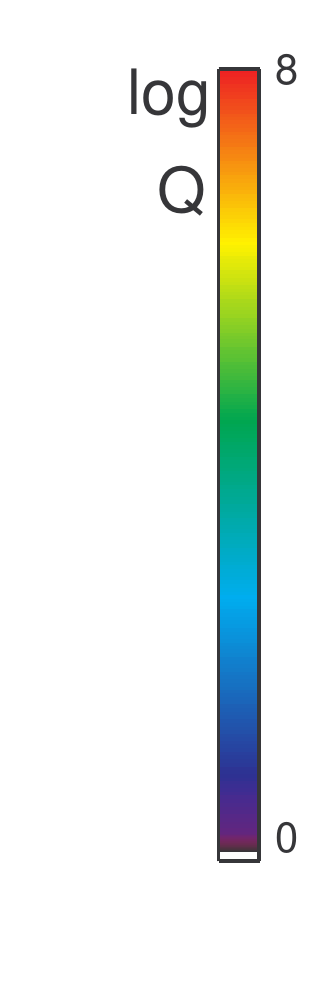}
   \caption{Evolution of the photospheric footpoints of the QSLs in the torus-unstable flux rope simulation with increasing time, showing the structural chages of the QSLs, as well as changes in their magnitude. Individual colors depict the value of log$(Q)$ of the squashing factor $Q$ at the photosphere. $Q$ is plotted only in areas where the magnetic field lines are locally closed within the computational box. Pink and cyan contours denote the positive and negative magnetic polarities in the model, with the positive one being stronger than the negative one.
        }
       \label{Fig:QSLs}
   \end{figure*}
%

   \begin{figure*}[!ht]
    \centering
    \includegraphics[width=5.6cm,bb=60 200 550 600,clip]{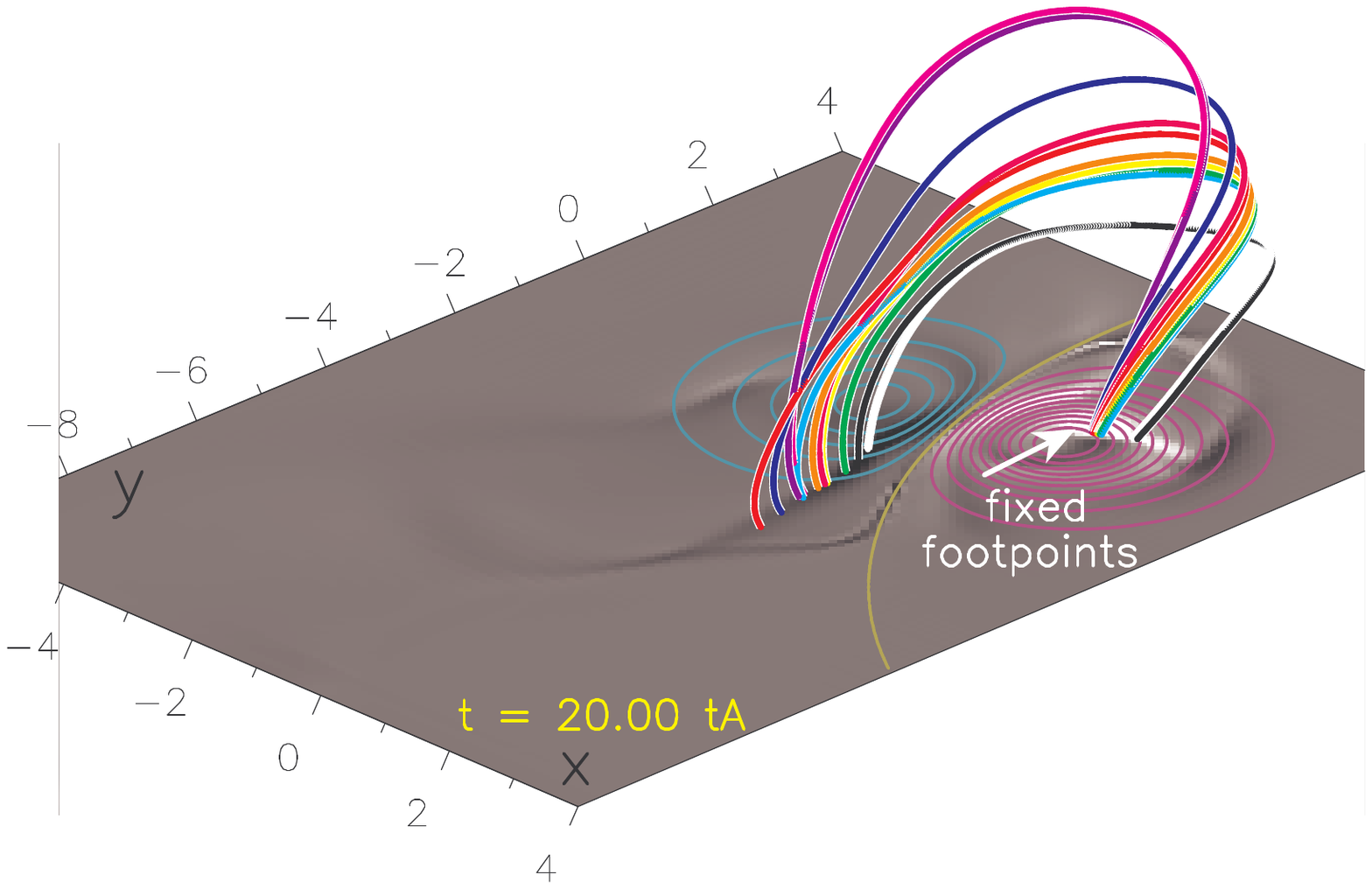}
    \includegraphics[width=5.6cm,bb=60 200 550 600,clip]{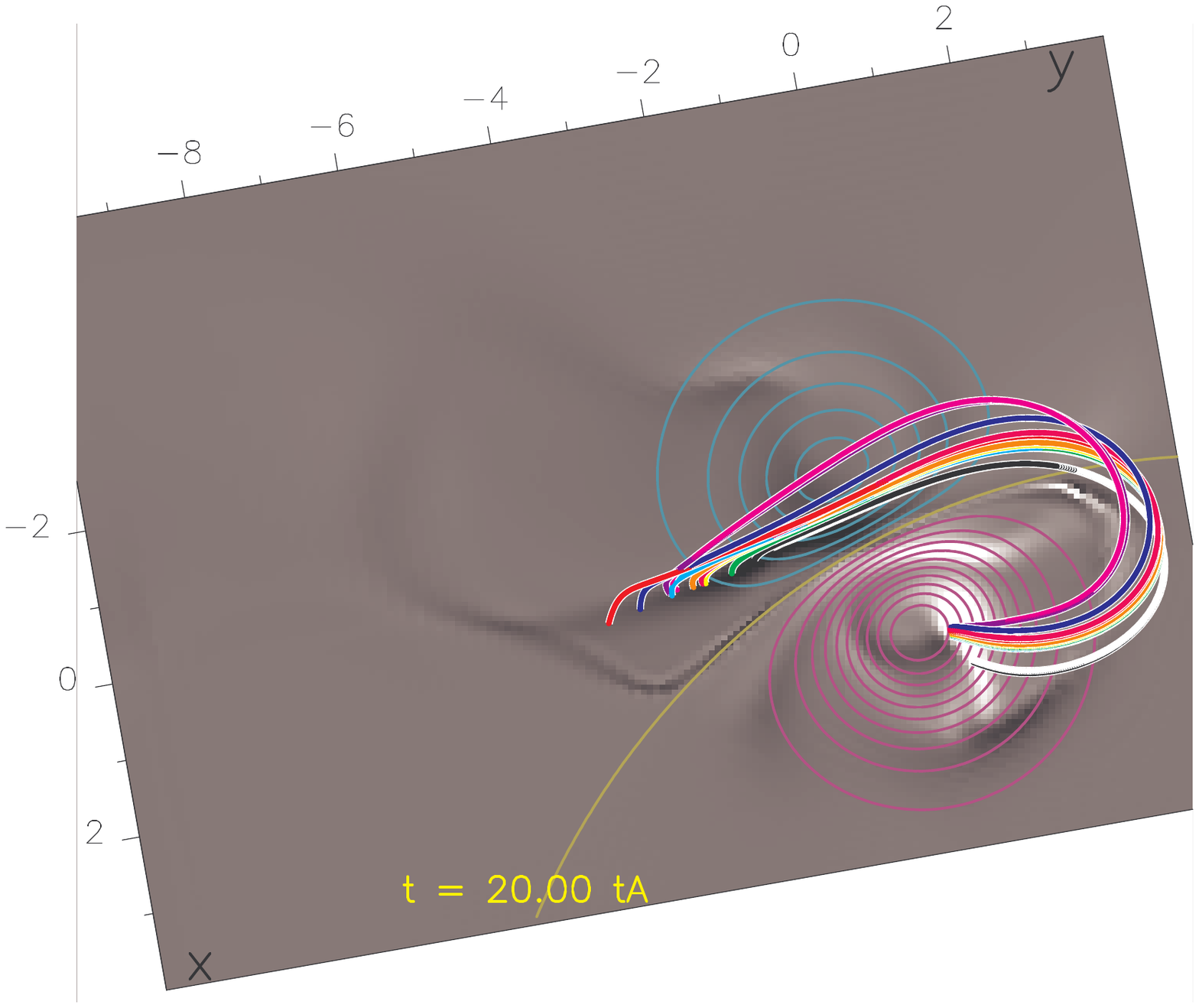} \\
    \includegraphics[width=5.6cm,bb=60 200 550 600,clip]{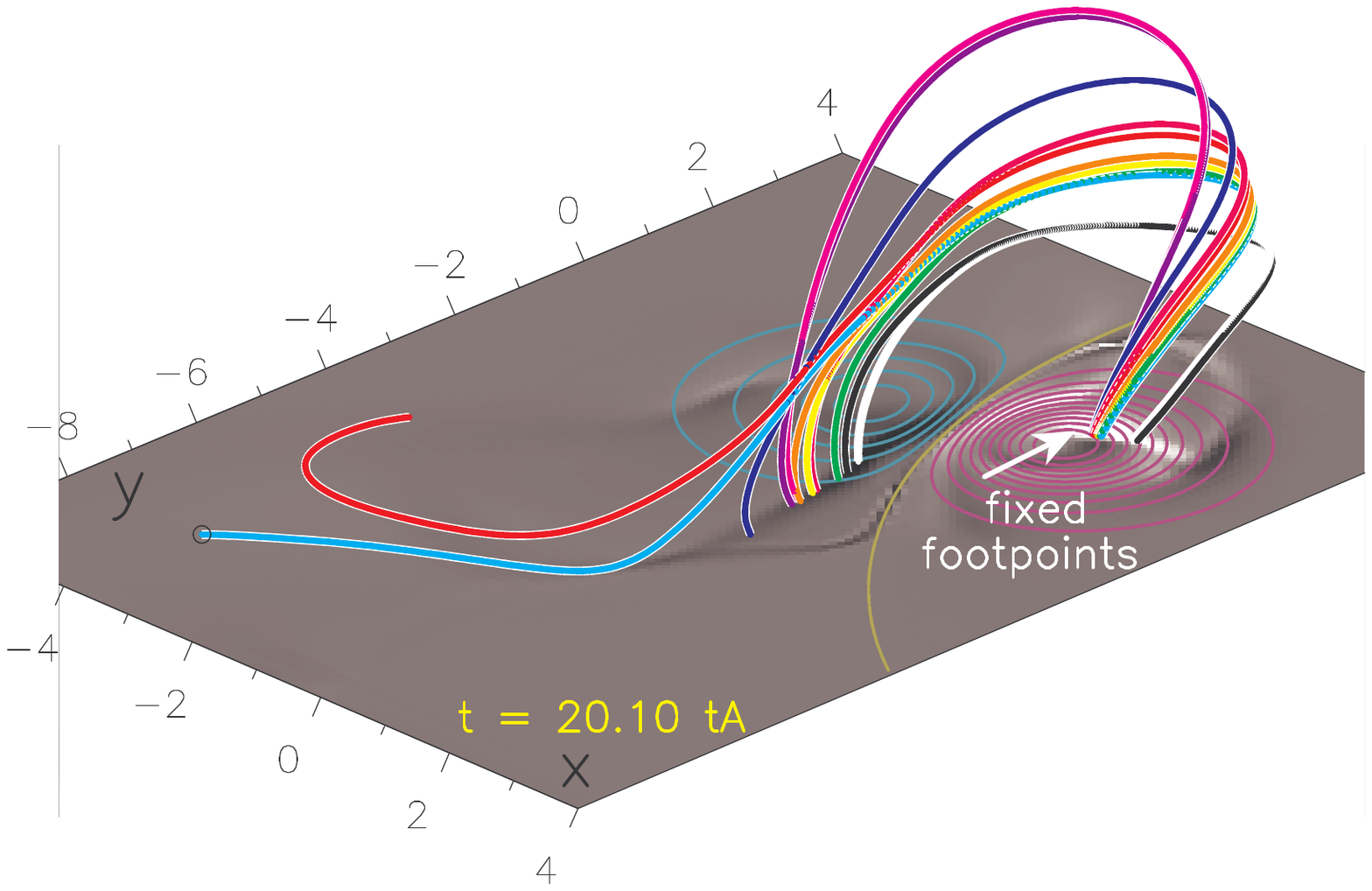}
    \includegraphics[width=5.6cm,bb=60 200 550 600,clip]{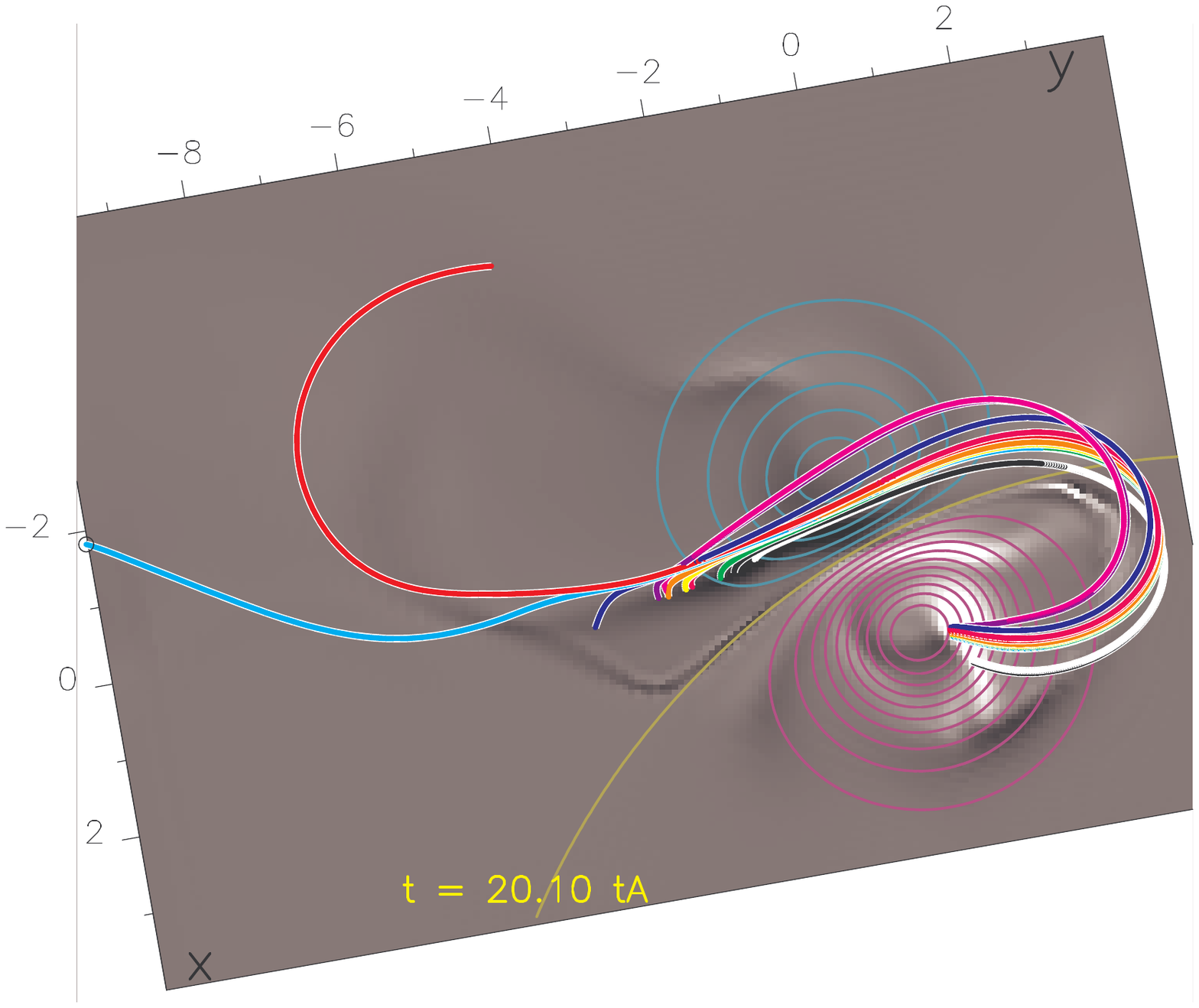} \\
    \includegraphics[width=5.6cm,bb=60 200 550 600,clip]{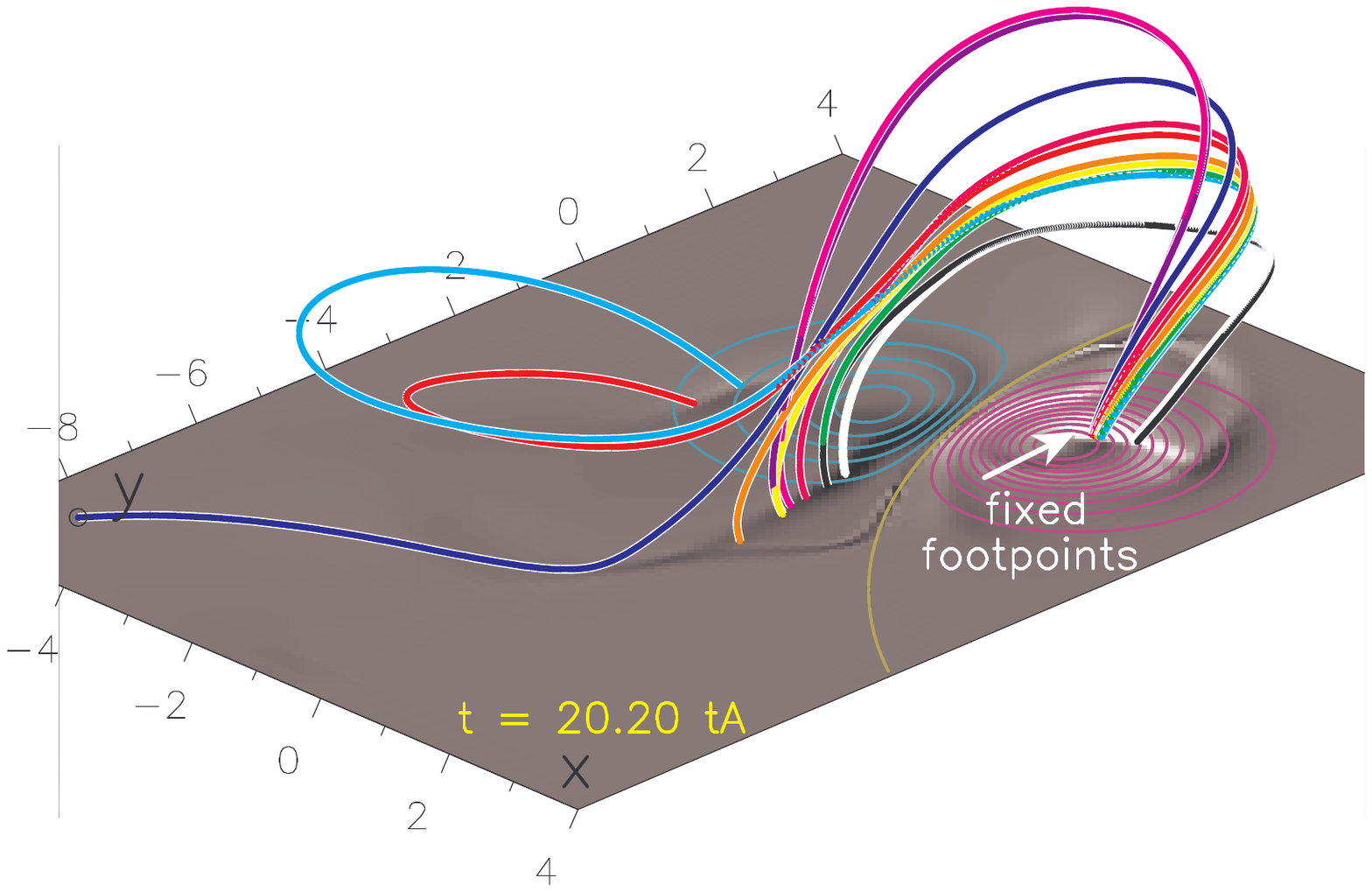}
    \includegraphics[width=5.6cm,bb=60 200 550 600,clip]{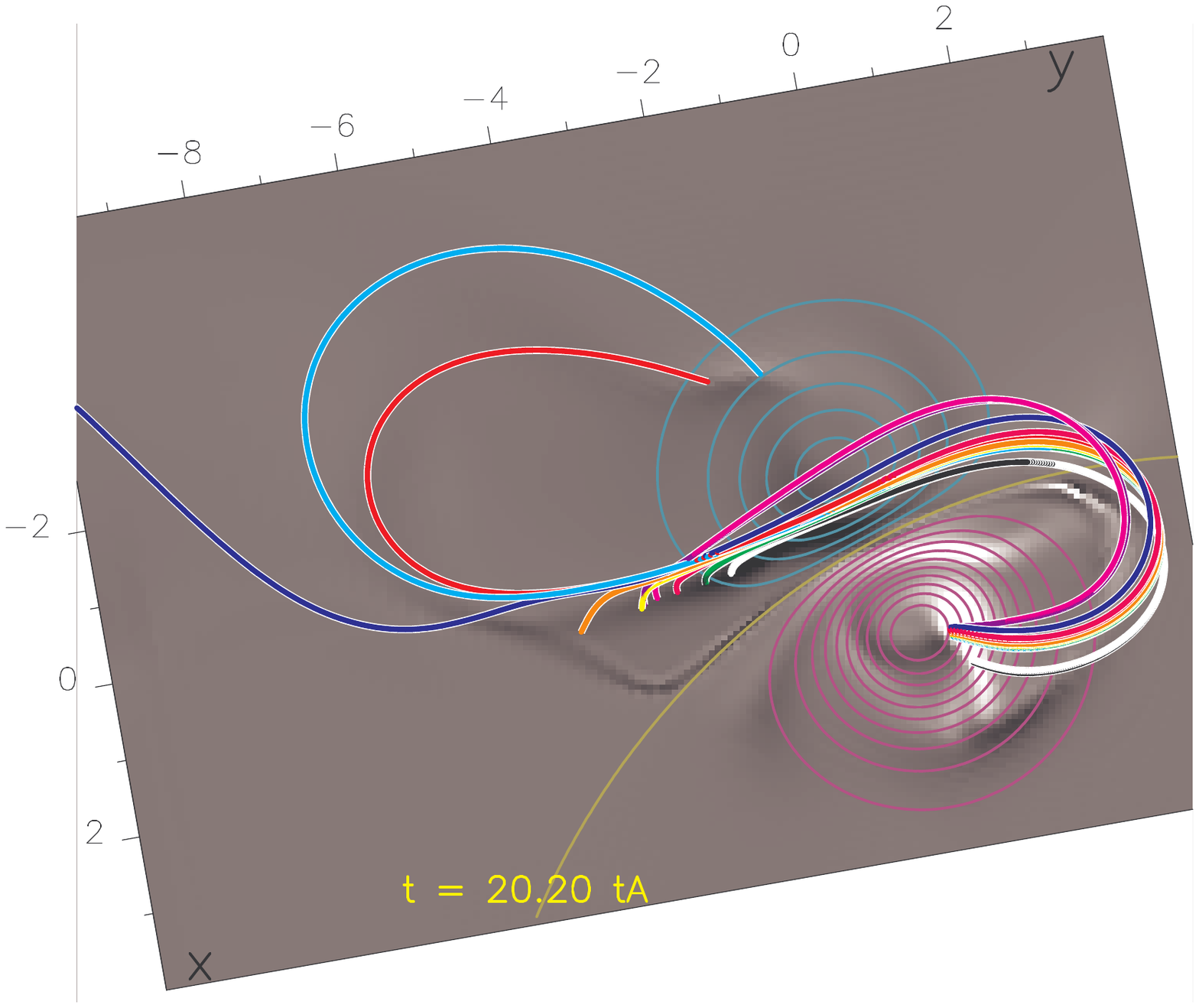} \\
    \includegraphics[width=5.6cm,bb=60 200 550 600,clip]{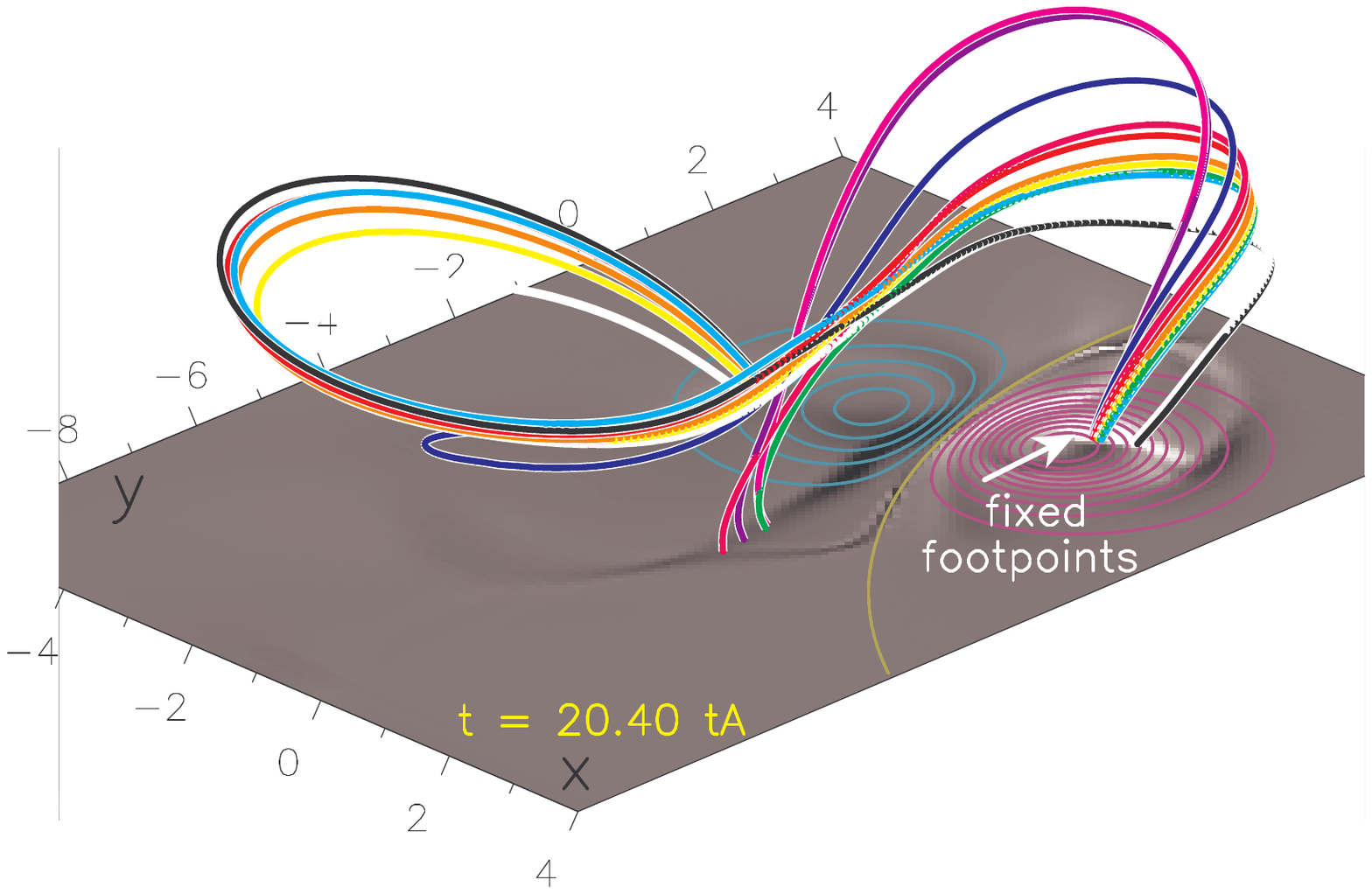}
    \includegraphics[width=5.6cm,bb=60 200 550 600,clip]{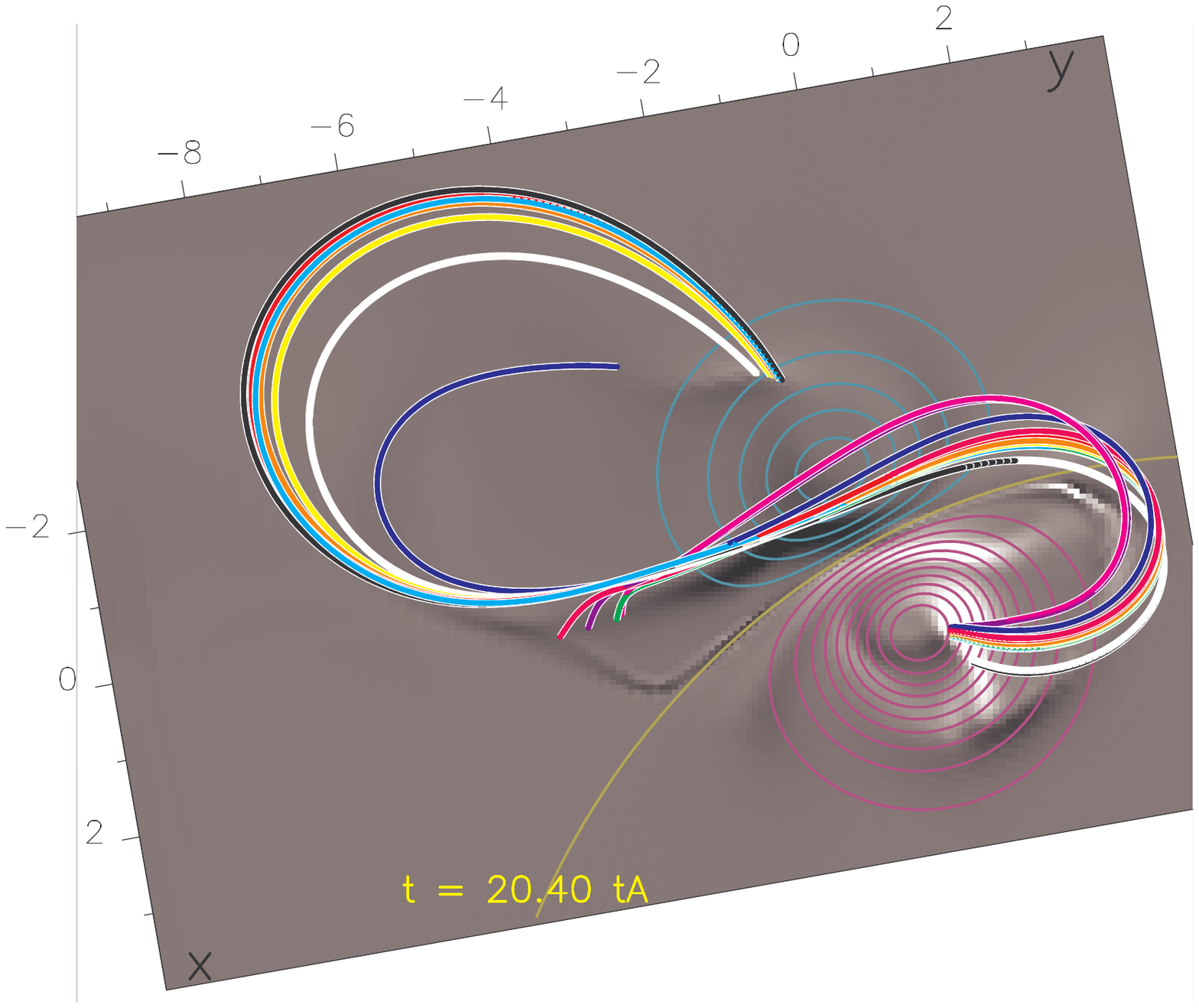} \\
    \includegraphics[width=5.6cm,bb=60 200 550 600,clip]{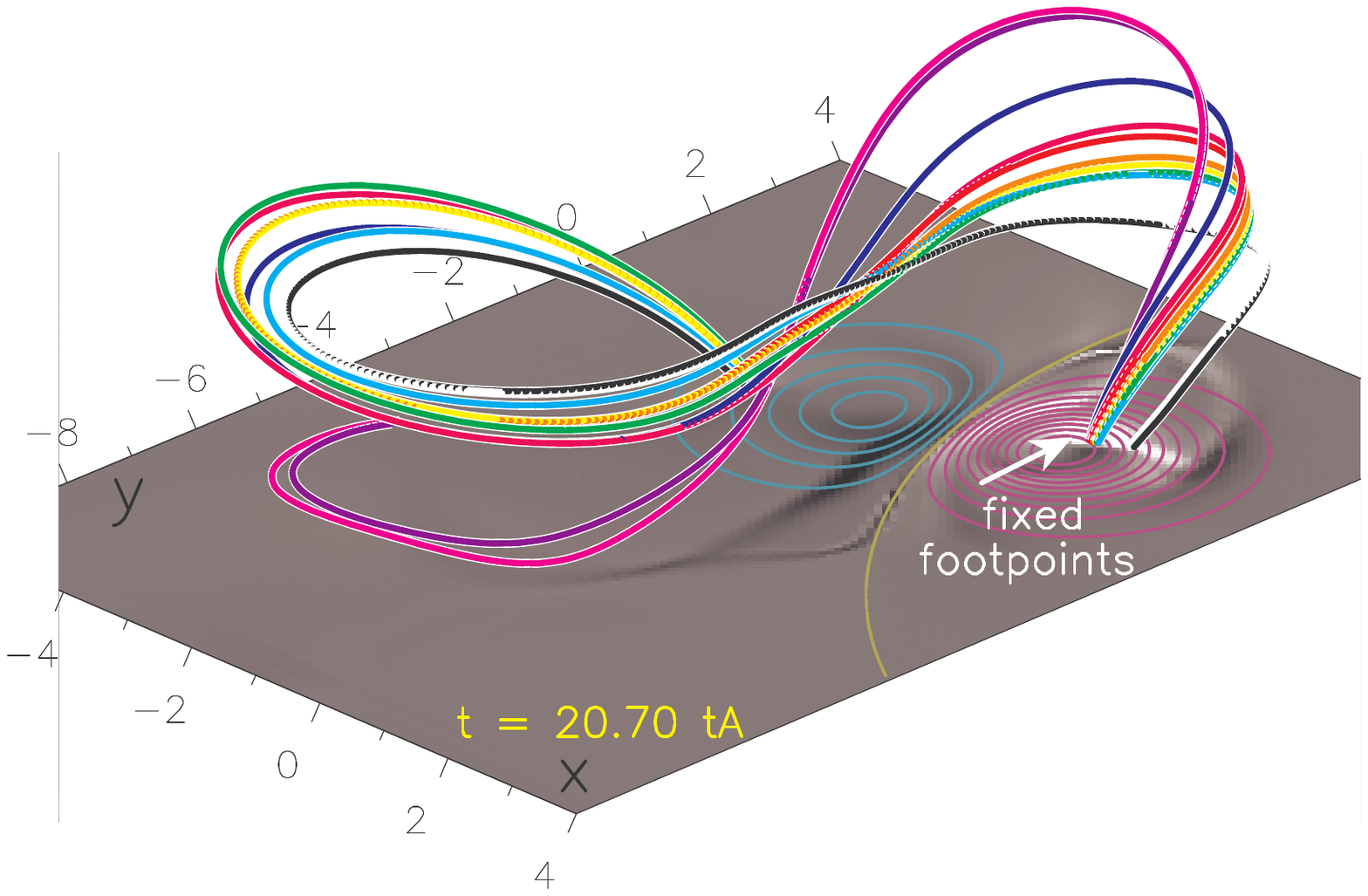}
    \includegraphics[width=5.6cm,bb=60 200 550 600,clip]{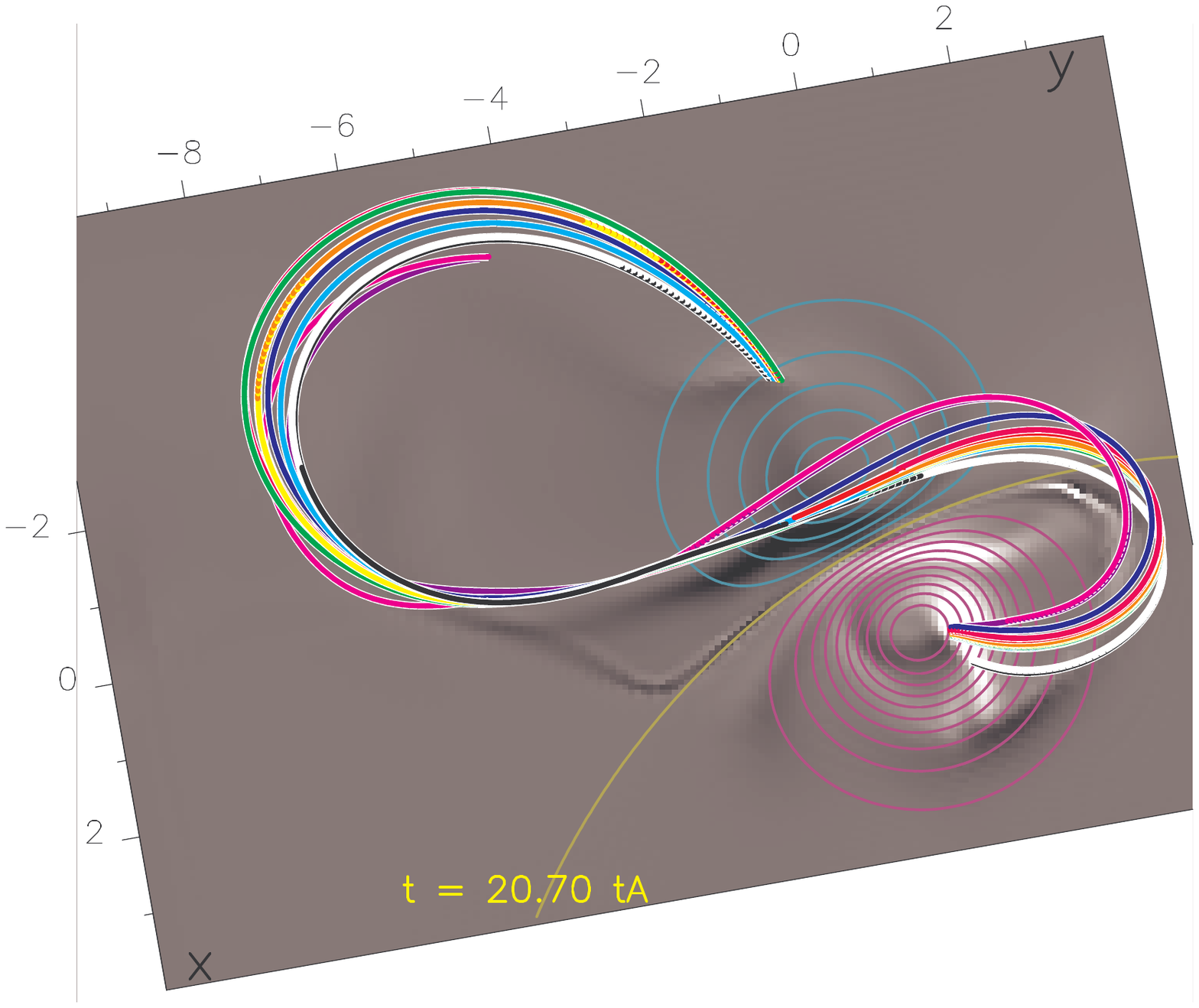}
    \caption{Images from different times of the flux rope simulation showing the evolution of the field lines undergoing slipping reconnection. \textit{Right}: top view; \textit{left}: side view. With increasing time, the field lines move from the central part of the J-shaped current region (dark area along the PIL) lying along the QSL to the tip of the hook, and finally become a part of the flux rope envelope.
            }
       \label{Fig:Slipping}
   \end{figure*}
%

   \begin{figure*}[!ht]
    \centering
    \includegraphics[width=5.9cm,bb=100 220 540 600,clip]{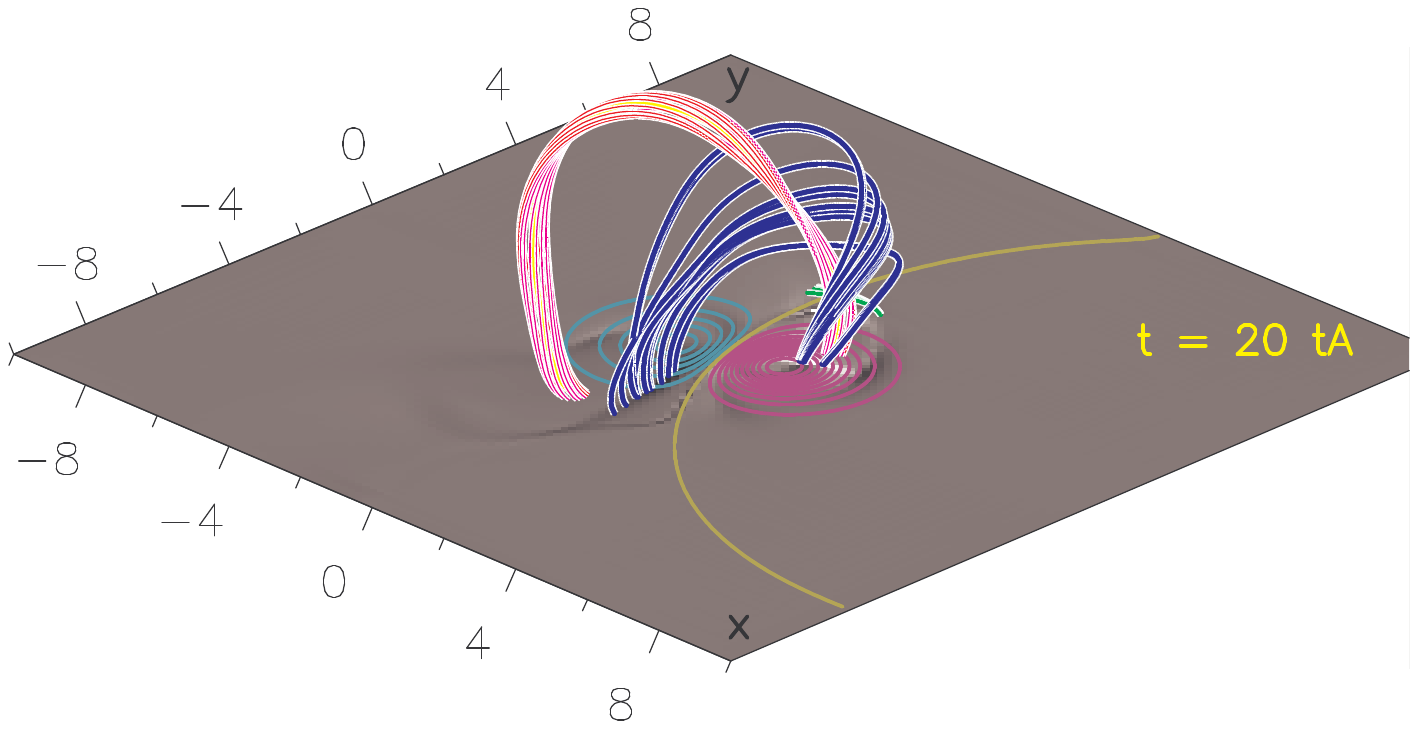}
    \includegraphics[width=5.9cm,bb=60  175 550 640,clip]{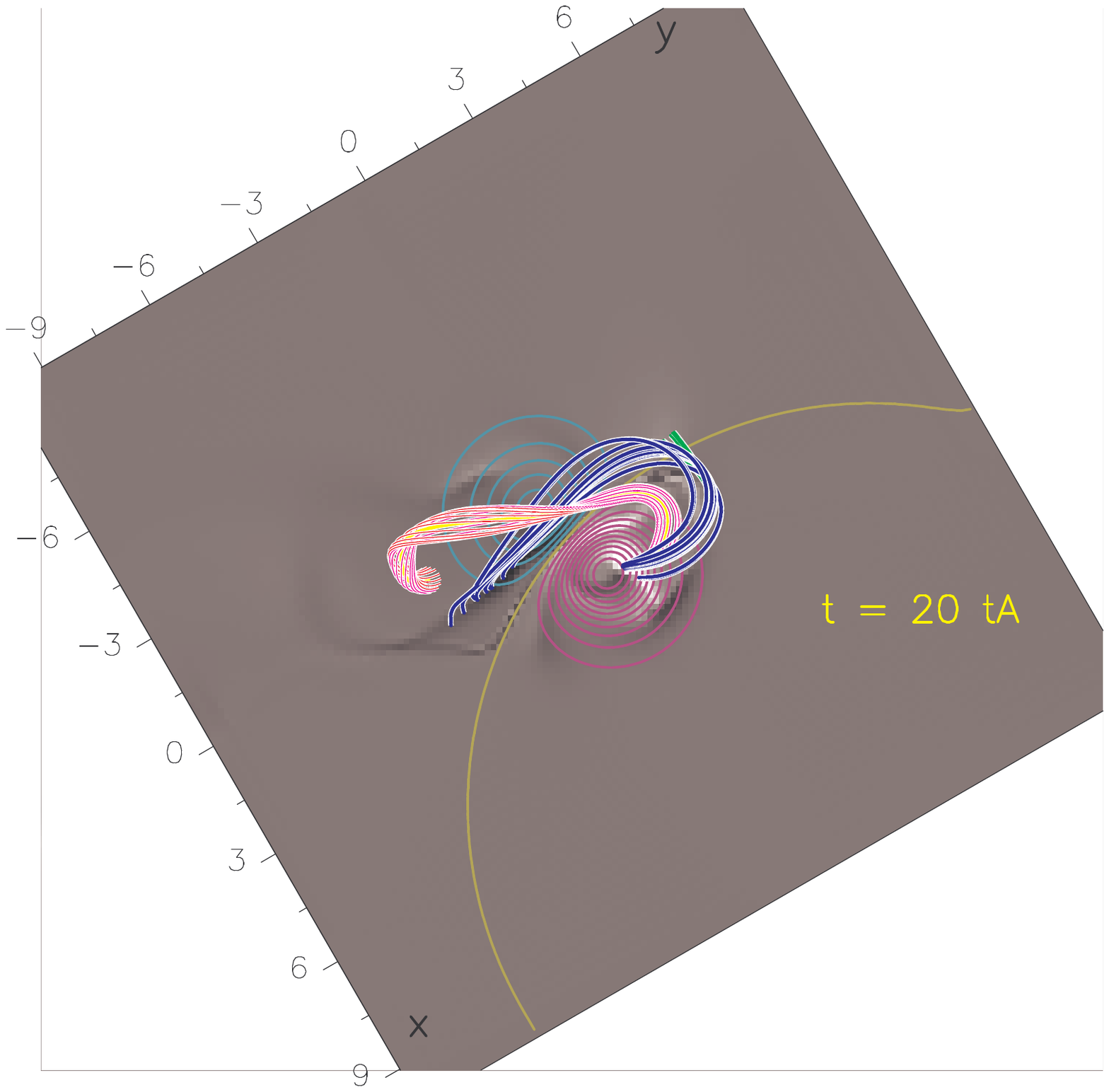} \\
    \includegraphics[width=5.9cm,bb=100 220 540 600,clip]{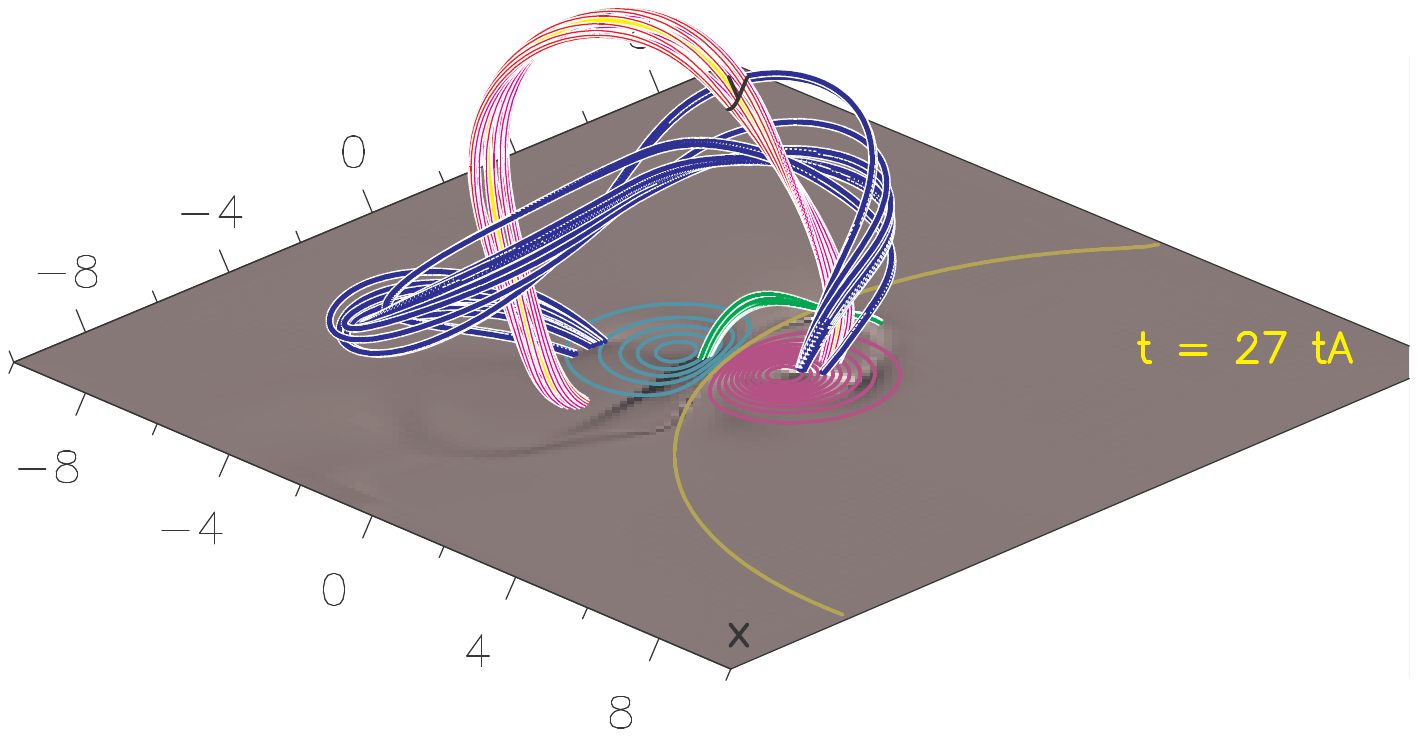}
    \includegraphics[width=5.9cm,bb=60  175 550 640,clip]{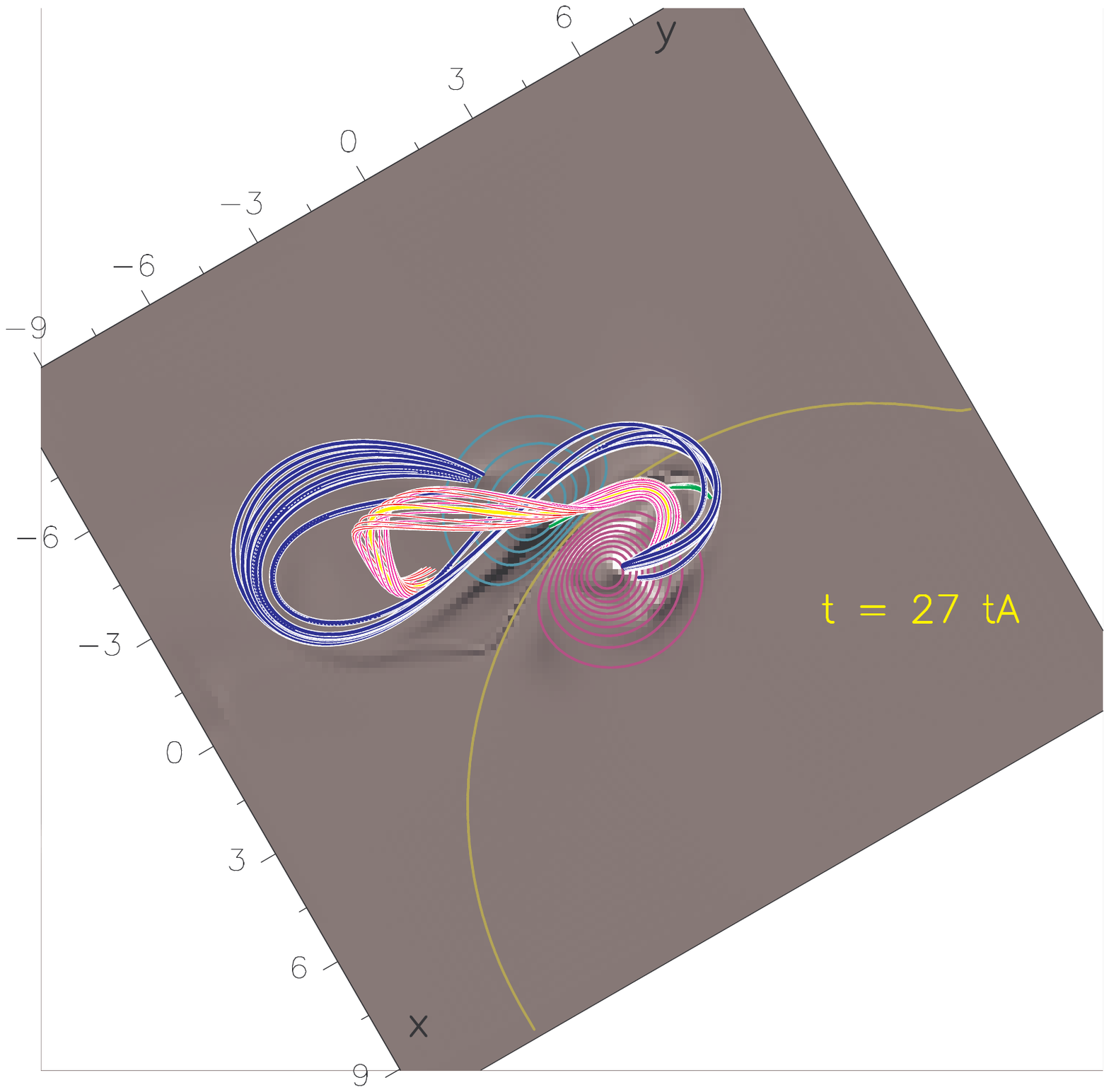} \\
    \includegraphics[width=5.9cm,bb=100 220 540 600,clip]{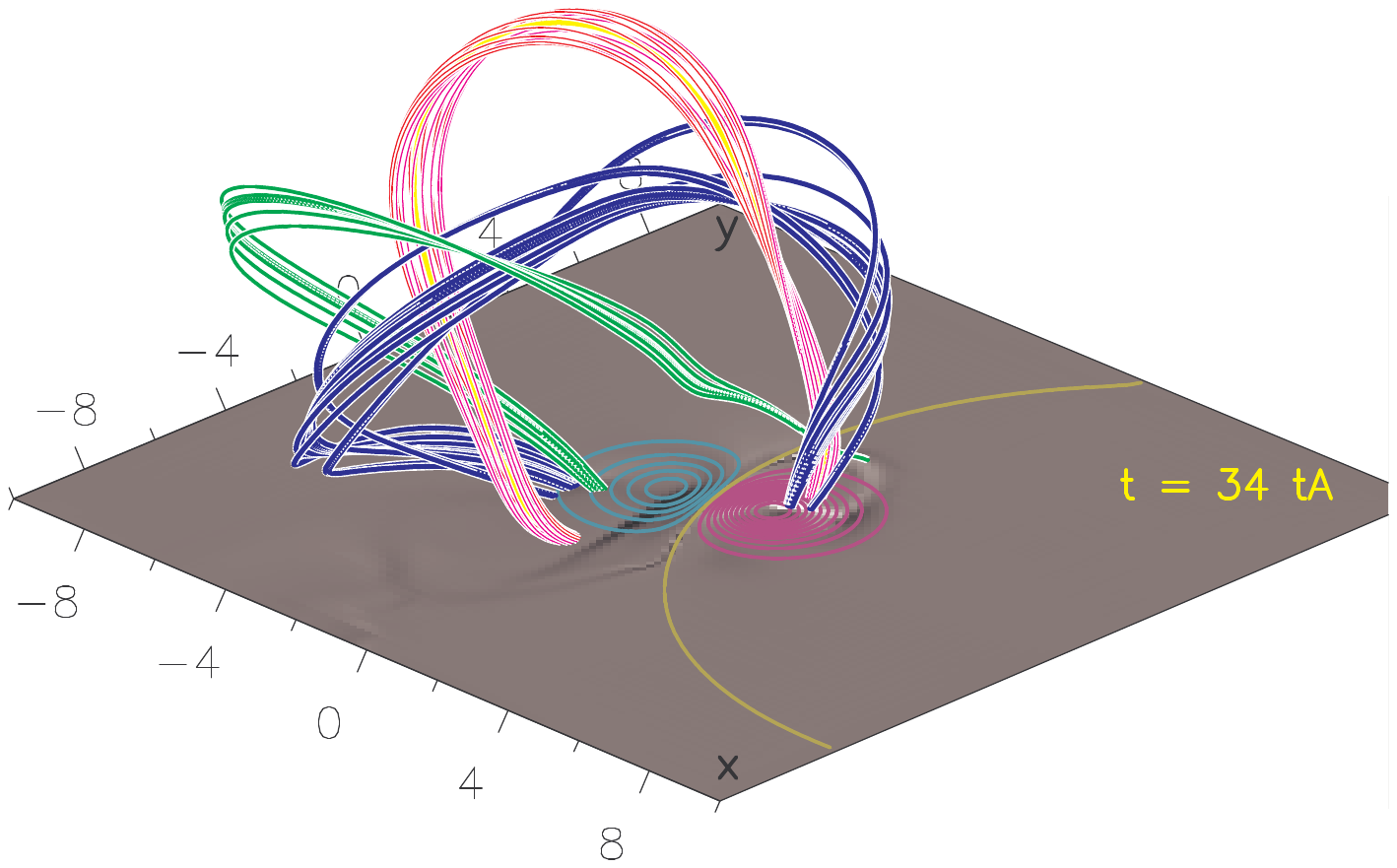}
    \includegraphics[width=5.9cm,bb=60  175 550 640,clip]{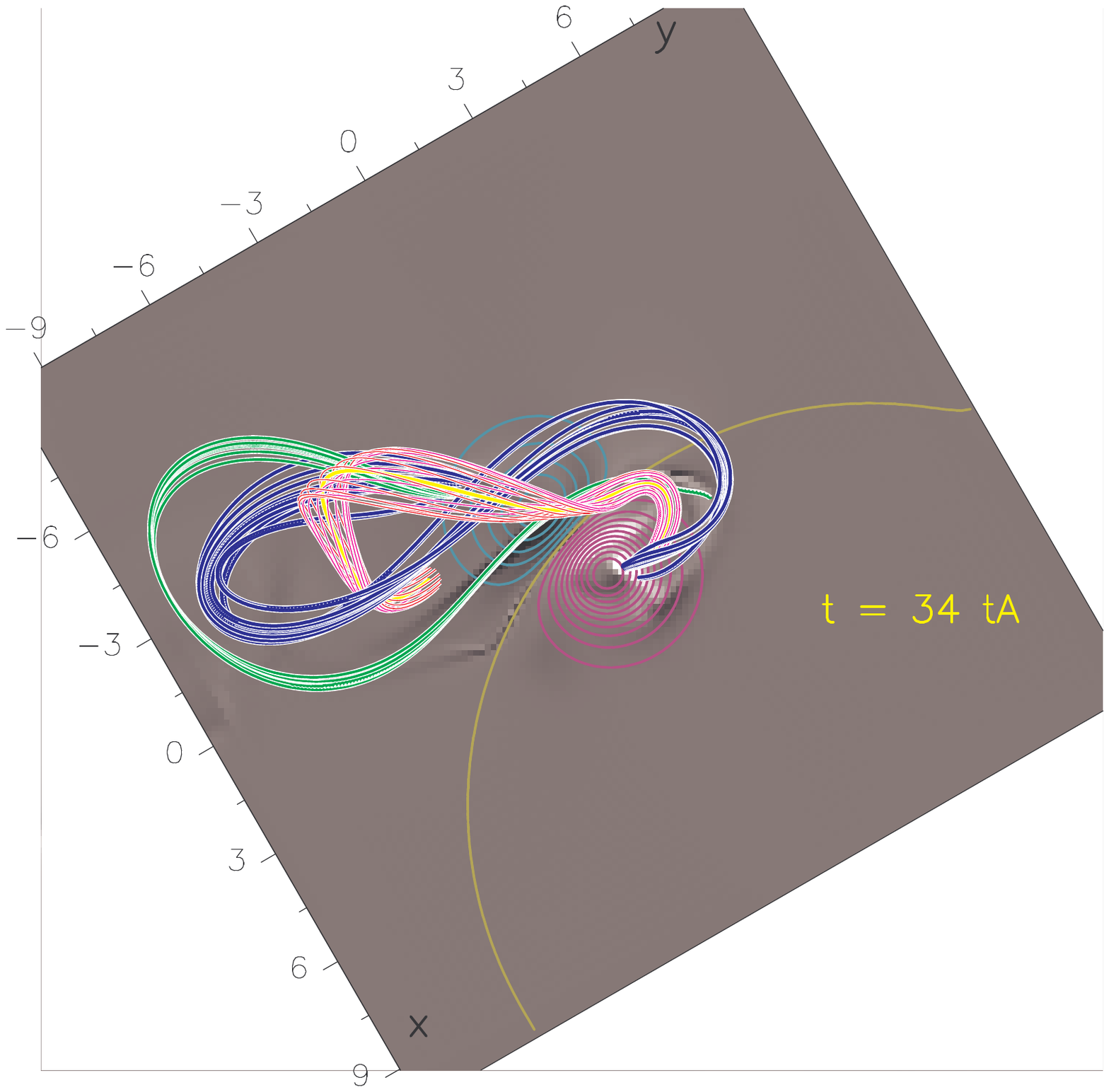} \\
    \includegraphics[width=5.9cm,bb=100 220 540 600,clip]{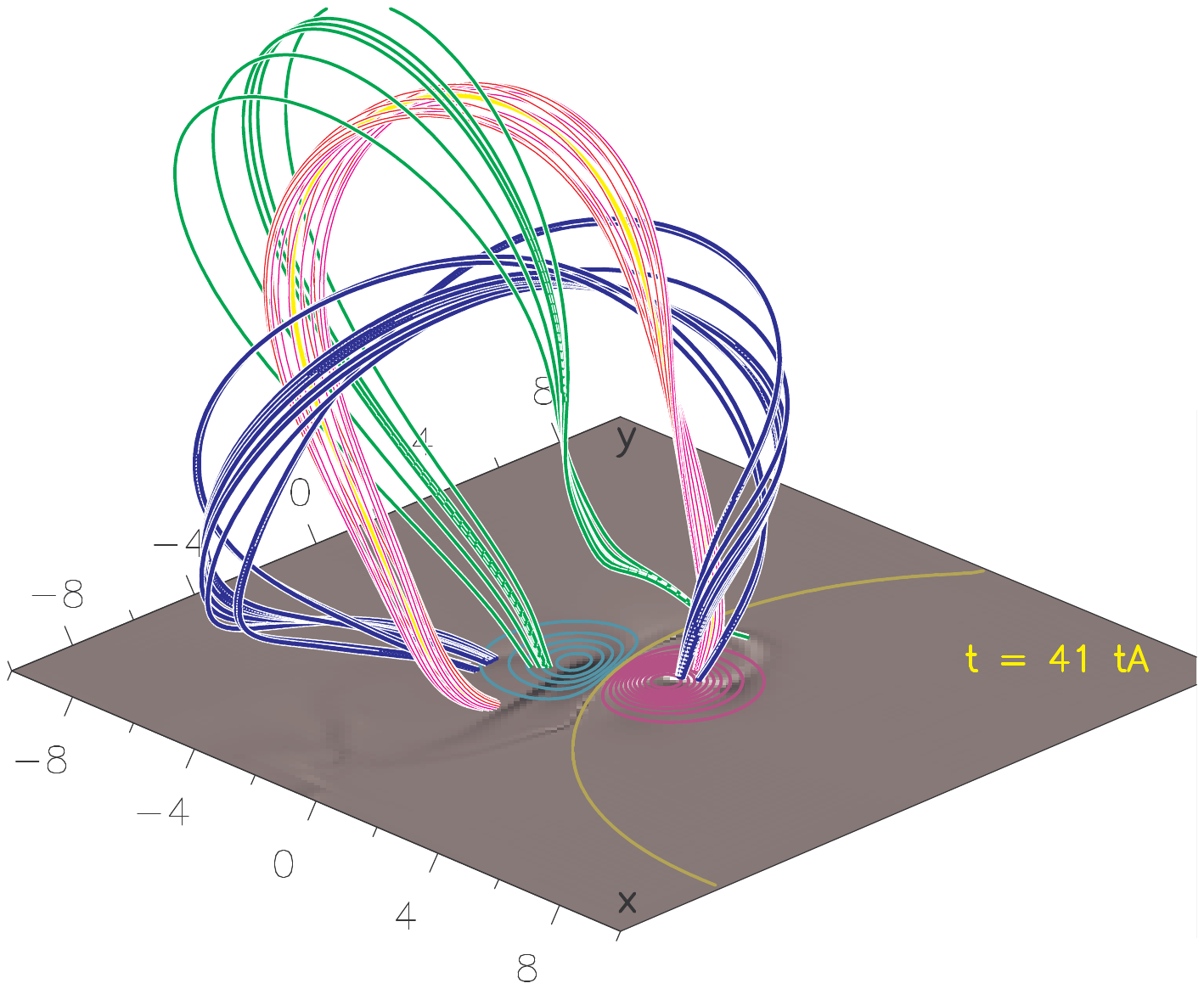}
    \includegraphics[width=5.9cm,bb=60  175 550 640,clip]{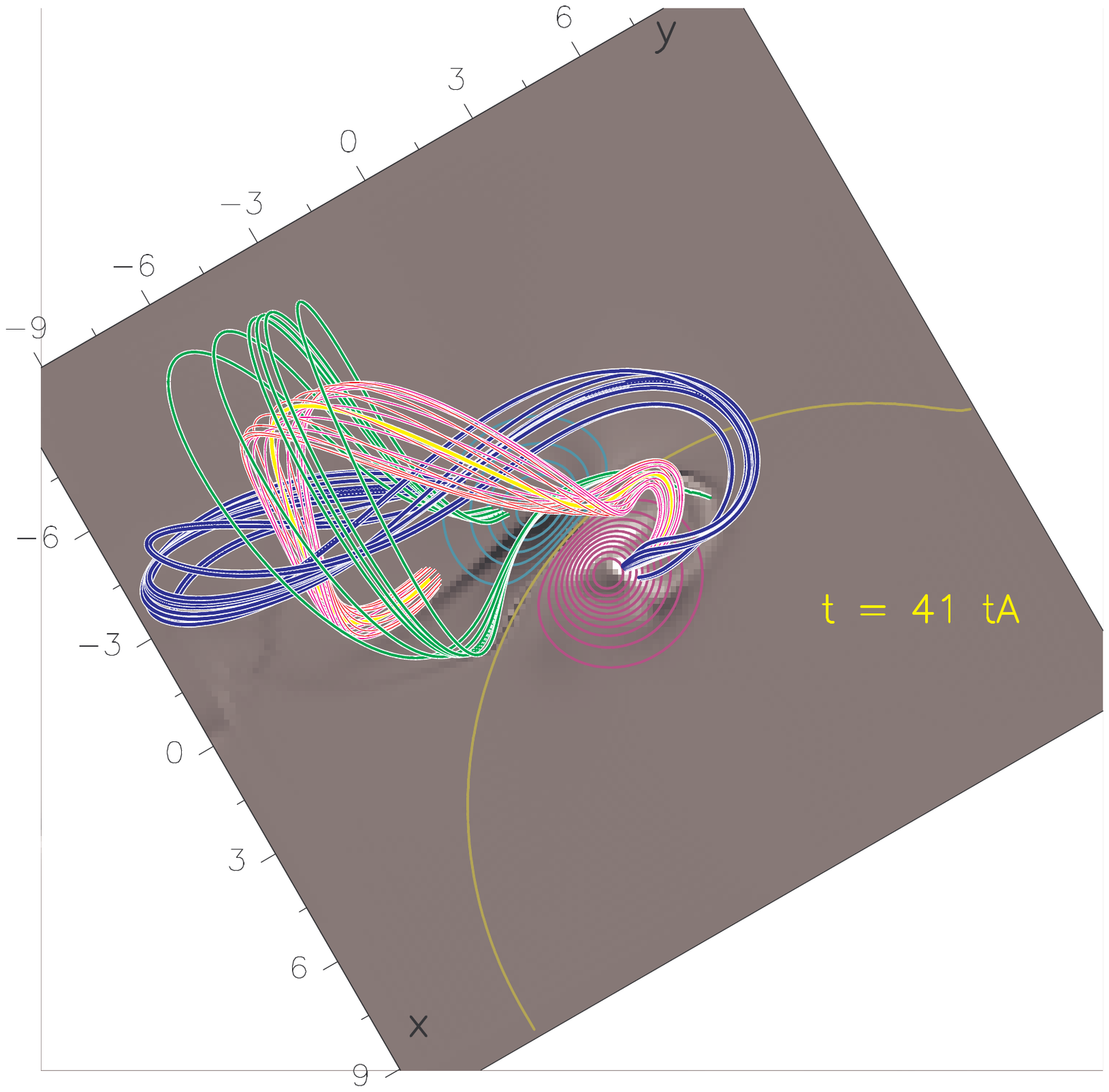}
    \caption{Evolution of the flux rope with increasing time. \textit{Right}: top view; \textit{left}: side view. The initial flux rope is indicated in pink, while successively reconnecting field lines are indicated in blue (reconnection from $t$\,=\,20\,$t_\mathrm{A}$) and green (reconnection from $t$\,=\,27\,$t_\mathrm{A}$). The final image shows the expansion of the whole structure and the envelope successively formed by magnetic lines reconnecting at different times.
            }
       \label{Fig:FR}
   \end{figure*}
%
%
\subsection{QSL structural evolution}
\label{Sect:4.2}

Quasi-separatrix layers, that correspond to a strong distortion of the magnetic field, are for the present simulation calculated with the TOPOTR routine \citep{Demoulin96a} that integrates all the field lines of the volume from a fixed position in one polarity to its counterpart in the other polarity. TOPOTR then measures the squashing degree $Q$, that is a quantification of the magnetic field distortion \citep{Pariat12}, and that defines QSLs for $Q>2$. 2D maps of QSLs can then be drawn anywhere in the volume, representing e.g. the QSL footpoints with a cut at the photosphere, or the hyperbolic flux tube (HFT) with a vertical cut into the volume which corresponds to the central part of the QSL volume.

An example of these QSL footpoints at the photosphere ($z=0$) can be found in Fig. \ref{Fig:QSLs}, where we have plotted the contour plot of log$(Q)$ at three different times during the flux rope expansion of the present numerical simulation. The colour table indicates the strongest magnetic field gradients corresponding to log$(Q) \geqq 8$ in red, and weaker gradients with log$(Q) \to 0$ in blue. The regions in white correspond to areas where magnetic field lines are open, making the calculation of $Q$ impossible \citep[see also][]{Janvier13}. The magnetic polarities in the photosphere are indicated by contour plots, with purple and cyan for the positive and negative polarities, respectively.

We note that since QSLs correspond to regions of high magnetic field distortion, they are expected to be formed near strong current density locations, as was demonstrated in \citet{Savcheva12a}, \citet{Gekelman12} and \citet{Janvier13}, although there is not necessarily a one-to-one correspondence \citep{Wilmot09}. Then, the evolution of QSLs is very similar to that of flare ribbons, as is described in Sec.2 of \citet{Janvier13}.

At $t=5\ t_A$, two very thin QSL structures exist (Fig. \ref{Fig:QSLs}, \textit{left}, red color). Both are QSLs consisting of a straight part and a hook. Note that similarly to the magnetic polarity asymmetry, there is an asymmetry in the shape of those QSLs, most pronounced for the hooks. The QSL in the stronger positive polarity has a rounder hook, and is localized very near the center of the polarity. In contrast to this, the hook of the QSL in the weaker negative polarity extends toward negative values and is much broader, while the tip of the hook remains located near the centre of the polarity. The straight part of both QSLs is close to the polarity inversion line (PIL) indicated in yellow colour.

As time advances, two evolutions can be seen. First, the QSLs move away from each other, i.e., their straight parts move away from the PIL. This motion resembles the flare ribbon motion, as was suggested by \citet{Janvier13} and as can be seen here by comparison with observations. Indeed, the distance between flare ribbons NR and PR in 304\AA\ filter (Fig.\ref{Fig:Overview}) increases over time, suggesting the separation of flare ribbons as is commonly observed during eruptive flare events.

Secondly, both QSL hooks become rounder with time. This is straightforward with the QSL in the negative polarity, as it extends toward negative $y$-direction in the simulation (Fig. \ref{Fig:QSLs}) and becomes rounder. The orientation of this extension corresponds to the east direction on the surface of the Sun, and is similar to the evolution of the NRH in AIA 304\AA, (Fig.\ref{Fig:Overview}, \textit{right}) between 16:25 and 17:00 UT. However, the observed NRH does not become rounder, instead it elongates in the south direction and remains narrow, as was described in Sect. \ref{Sect:2.1}. This is because the large-scale magnetic configuration of the AR 11520 and 11521 is much more complex than in the simulation. The presence of surrounding magnetic fields on the Sun implies the existence of numerous large-scale QSLs that prevent the extension of the flare-associated QSLs \citep{Chen12}. The broadening QSL is then squeezed by the surrounding structures, therefore explaining the unidirectional elongation in the observations instead of the isotropic broadening seen in the simulation.

In the simulation, the QSL hook in the positive polarity does not extend much and while it becomes rounder, it remains located near the center of the polarity. It is however difficult to compare the evolution of this QSL hook in the positive polarity with the observations. In Fig.\ref{Fig:Overview}, the PRH is in fact an elongated ribbon, whose structure could be more complicated than in the simulation due to presence of other magnetic polarities, i.e., AR 11521.

For completeness, we note that at $t=45\ t_A$ another QSL is present in the negative polarity in the simulation. Its straight part lies very close to the PIL, and the hook very close to the previously described QSL. This QSL is associated with a bald patch corresponding to field lines tangent to the surface. The bald patch does not play any role in the reconnection of field lines in the simulation. Its description will then be omitted in the further paragraphs.

The evolution of the QSLs is associated with the reconnection process leading to the formation of new pairs of reconnected field lines, including the flux rope envelope. This process is detailed in the following.

%
\subsection{Slipping motion and associated kernel brightening}
\label{Sect:4.3}

In the simulation, the upward expansion of the flux rope during its ejection creates a very thin current layer where reconnection occurs. This current layer is associated with a HFT, as described in Fig.\,2 in \citet{Janvier13}. The field lines passing through the HFT undergo multiple reconnections, and this effect is seen as an apparent slippage of the field lines. In order to investigate this fast apparent motion in detail, the numerical data have been output from the simulation at a time cadence of less than one Alfv\'en time.

Figure \ref{Fig:Slipping} shows a set of selected field lines that are evolving from $t=20$ to $t=20.7\ t_A$. We picked these times since the magnetic field lines resemble the shape of the flare loops (Fig. \ref{Fig:Overview}). The photospheric surface at $z=0$ is shaded according to the photospheric current density $j_z$, with the magnetic polarities indicated by pink and cyan contour plots. The field lines are integrated from their fixed footpoints anchored in the positive polarity close to the tip of the QSL hook. This hook is apparent in both the QSL trace (Fig.\ref{Fig:QSLs}, \textit{middle}) and the $j_z$ structures. With advancing time, the connectivity of the field lines changes. For example, the blue field line is anchored in the negative polarity near the PIL at $t=20\ t_A$, while at $t=20.2\ t_A$ it extends beyond the field of view of the simulation box, and at $t=20.4\ t_A$ is connected to the tip of the hook of strong $j_z$, similar in shape to the QSL hook in the negative polarity. This motion occurs similarly for the other field lines so that at $t=20.7\ t_A$, they are all connected in the negative polarity near the tip of the QSL or $j_z$ hook. The continuity of this motion can be seen in the online Movie 7.

This motion can in principle be observed with SDO/AIA imaging by two means. First, the loops themselves can be seen to be moving, as is the case for their NR footpoints discussed in Sects. \ref{Sect:2.2.1}--\ref{Sect:2.2.3} (also Figs. \ref{Fig:Slip1}, \ref{Fig:Slip2}, \ref{Fig:Slip3}). Second, the field line footpoints can be seen to be moving as are the ribbon brightenings. These correspond to chromospheric or transition region emission due to the impact of energetic particles accelerated in the reconnection site. In practice, both are observed at the same time (Sect. \ref{Sect:2.2}).

We note that in the simulation, the fast motion of field lines is \textit{only an apparent motion} due to the diffusion of magnetic field in the current layer, not necessarily associated with motion of plasma dragged by the moving field lines. This is because the time scale needed to fill the field lines by chromospheric evaporation is generally much larger than the time scale for multiple reconnections to take place in the current sheet. However, if reconnection were to happen on a time scale that is comparable with evaporation-field line filling, then the slipping motion of field lines could be identical with that of the flare loops. Note that the AIA instrument is able to observe only the portions of the field lines filled with high-density plasma at temperatures given by its temperature response (Fig. \ref{Fig:AIA_resp}). Therefore, the coherent, apparently moving loops, reported in Sect. \ref{Sect:2.2}, do not neccessarily lie on a single, co-moving (slipping) magnetic field line. Rather, the apparent slipping motion of these loops is an illusion created by the apparent slipping motion of the magnetic field itself, coupled with the plasma thermal response.

We also note that the speed of the slipping motion along the QSL is not uniform. This can be inferred also from Fig.\ref{Fig:Slipping}. At $t=20\ t_A$ and $t=20.4\ t_A$, field lines are only slightly moving in the straight part of the QSL, while their motion becomes much faster in the hook, as can be seen by comparing $t=20.4\ t_A$ with $t=20.7\ t_A$. This slipping velocity profile has been investigated in detail in Sect.\,3 of \citet{Janvier13}, where a peak in the velocity profile was found near the hook of the QSL. Smaller velocities were found at the beginning and at the end of the slipping motion, corresponding to the straight part of the $J$-shape and tip part of the hook. The peak in the velocity profile can be explained by the fact that the reconnecting field lines are passing through the HFT, where $Q$ is the highest, and the changes in magnetic connectivity the most drastic.

This change of slipping motion speed can also be seen in the observations (Sect. 2). As seen in Figs.~\ref{Fig:Slip1}, \ref{Fig:Slip2} and \ref{Fig:Slip3}, one can see different sets of loops slipping at different times during the flare, yet their footpoints are mostly located in the straight part of the NR (Fig. \ref{Fig:Overview}). These slipping loops can be seen in this location because the field line slipping motion is slow, leaving enough time for filling the field lines via chromospheric evaporation. It is then possible to measure the velocity of this motion with stackplots along different cuts (Figs. \ref{Fig:Slip1_stackplots}, \ref{Fig:Slip2_stackplots}, and \ref{Fig:Slip3_stackplots}). Similarly, a bright loop appears at the tip of the NRH at 16:10:08 UT (Arrow 3 in Fig.\ref{Fig:Overview}). This loop can be associated with a slipping set of field lines found at 15:55:08 UT.

Contrary to flare loops, kernel brightenings are much more visible moving along the ribbons, as can be seen in the NRH in the online Movie 2 and also in the Fig. \ref{Fig:Overview} (\textit{right column}). However, this kernel motion is not continuous but rather scattered, although the associated reconnection process should be continuous. This can be explained in terms of energy deposition along the QSL footpoints. The energy from the reconnection process is deposited at the chromosphere thanks to the energetic particles accelerated at the reconnection site. When the field lines are moving slowly, the energy is deposited in a small area of the QSL footpoint, so that the resulting chromospheric or EUV emission due to evaporation is large. In contrast, when the field lines move fast, the energy deposited per area is small, so that the kernel brightening will be weaker or even practically invisible. This is illustrated in Fig.\ref{Fig:Slipping} where the footpoints of the field lines are seen to jump from the straight part of the QSL to the tip of the hook. This jump indicates that the change of connectivity is much more important in the hook. Therefore, the deposition of energy in the hook should be very low. At 16:25 UT, the intensity of the NRH is indeed slightly lower than the straight part of the NR, or the NRH hook (Fig. \ref{Fig:Overview}, \textit{fifth column}).

The direction of the propagation of slipping loops during the observed flare is also consistent with the simulation. In the numerical model, the footpoints of the reconnecting field lines move along the QSL footpoint (Fig.\ref{Fig:Slipping}) toward the tip of the hook. This propagation direction is clearly seen in Figs. \ref{Fig:Slip1}, \ref{Fig:Slip2}, and \ref{Fig:Slip3}.

%
\subsection{Evolution of the expanding flux rope}
\label{Sect:4.4}

The slipping motion of field lines, as discussed above, occurs on a very short time scale, typically within one Alfv\'en time. When the magnetic field lines reconnect, they either add to flare arcade or they contribute to the flux rope envelope. We are now concerned with the latter process.

As time passes, the reconnection process feeds the flux rope with twisted field lines surrounding its core, leading to the continuous growth of the expanding flux rope \citep[e.g., Fig.5 in][]{Aulanier12}. This process, reproduced in the present 3D simulation, is similar to that depicted in the CSHKP model where twisted field lines construct the envelope of the flux rope. This growth of the flux rope is related with the motion of QSLs within the volume. Similar to separatrices in a quadripolar configuration, QSLs move away from the reconnection region with increasing time (Sect. \ref{Sect:4.2}). The moving QSLs swipe the field lines that subsequently reconnect as they are passing in the QSLs.

Therefore, the footpoints of the field lines constituting the flux rope, that have therefore already reconnected, are located inside the two hooks of the QSL footpoints in the negative and positive polarities, as shown in Fig.\ref{Fig:QSLs}. Pre-reconnected field lines that will become a part of the flux rope are in contrast situated on the periphery, i.e., outside the hook. The shape of the hooks therefore gives a good indication on the localization of the flux rope footpoints as well as its growth with time: the rounder and bigger the hooks become indicates how ``big'' the flux rope becomes as well. In summary, the envelope of the flux rope increases from the peripheral region by feeding the flux rope as time goes by.

The evolution of the flux rope envelope is shown in Fig.\ref{Fig:FR}. There, we have represented a set of field lines in pink representing the unstable flux rope core present from the beginning of the simulation. Another set of field lines, reconnecting between $t=20\ t_A$ and $t=27\ t_A$, is shown in blue. These lines are actually the same set of field lines that are slipping from $t=20$ to $t=20.7\ t_A$ in Fig.\ref{Fig:Slipping}. Finally, green represent field lines reconnecting at a later time, between $t=27$ to $t=34\ t_A$. We note that field lines reconnecting at later times surround the field lines already reconnected, i.e., the green lines are winding around both the pink and blue ones. Note also that as the flux rope expands, the whole structure becomes increasingly stretched, leading to almost vertical field lines near their footpoints at the photosphere, as is apparent from the side view in Fig. \ref{Fig:FR} (\textit{left column}) at $t=41\ t_A$.

The top view shown in the right panel of Fig.\ref{Fig:FR} can be compared directly with observations of the flare (Fig.\ref{Fig:Overview}, arrows 2 and 3; Fig. \ref{Fig:Eruption}). These long, hot loops appearing on the sides of the active region are added to the whole flux rope structure by the periphery and constitute its envelope (Fig. \ref{Fig:Eruption}). The flux rope core itself is not visible.

The mechanism behind these erupting loops can be understood when looking at the top view (right panel) of Fig.\ref{Fig:FR}. The flux rope tends to erupt in a privileged direction (green lines in Fig. \ref{Fig:FR}). In the simulation, this privileged direction is north-east. In the observations, the erupting loops also have a preferential direction, although this is south-west (Arrow 2 in Fig. \ref{Fig:Overview}). The existence of a privileged direction arises due to the asymmetry of the flux rope expansion, but can be constrained by the surrounding magnetic field, as observed here and described in Sect. \ref{Sect:2}. Moreover, the green field lines are seen to erupt in the model from $t=34$ to $t=41\ t_A$. This shows that only parts of the flux rope erupt at different times, explaining the appearance of erupting loops on the west (arrow 2) at $t=$ 15:40:08 UT before those on the east (arrow 3) at $t=$ 16:10:08 UT as shown in Fig.\ref{Fig:Overview}. Therefore, the classical approach to flux rope ejection, as described in \citet{McKenzie08} is not complete: the flux rope is continuously fed by on-going reconnection, leading to different sets of field lines constituting the flux rope envelope having their own ejection dynamics.

%
   \begin{figure}[!ht]
       \centering
       \includegraphics[width=8.4cm,bb=25 10 495 340, clip]{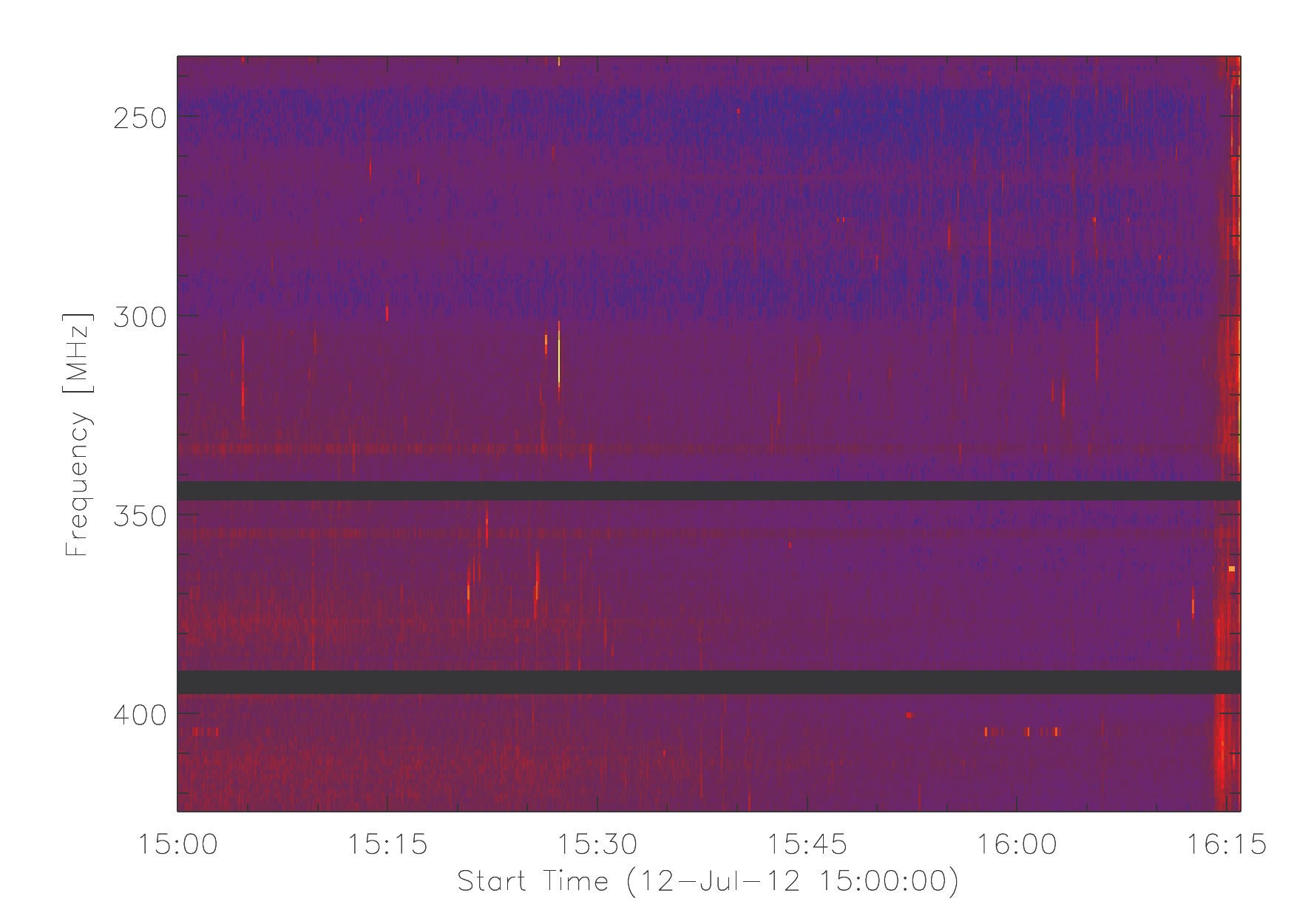}
       \includegraphics[width=8.4cm,bb=25 10 495 340, clip]{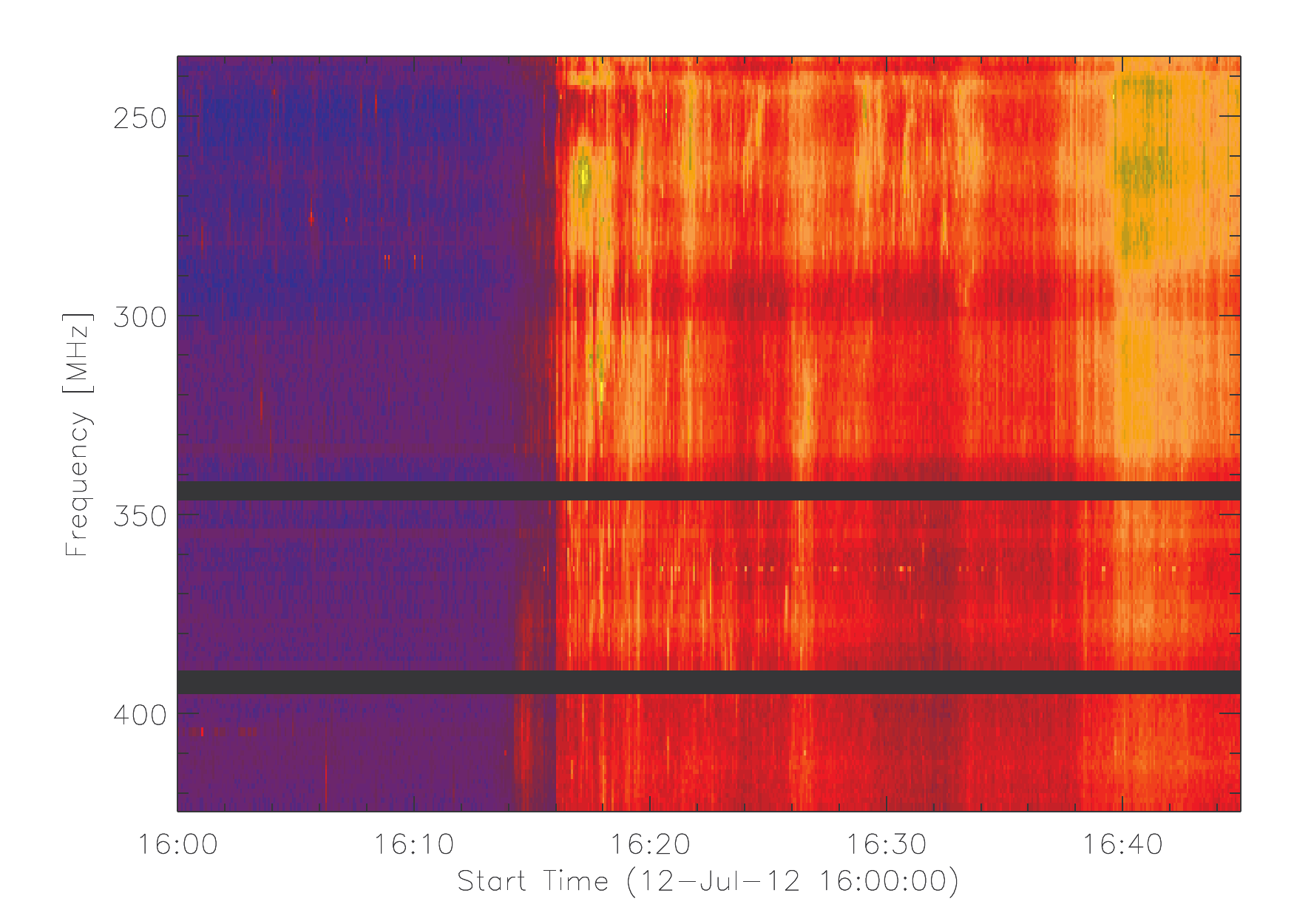}
       \includegraphics[width=8.4cm,bb=25 10 495 340, clip]{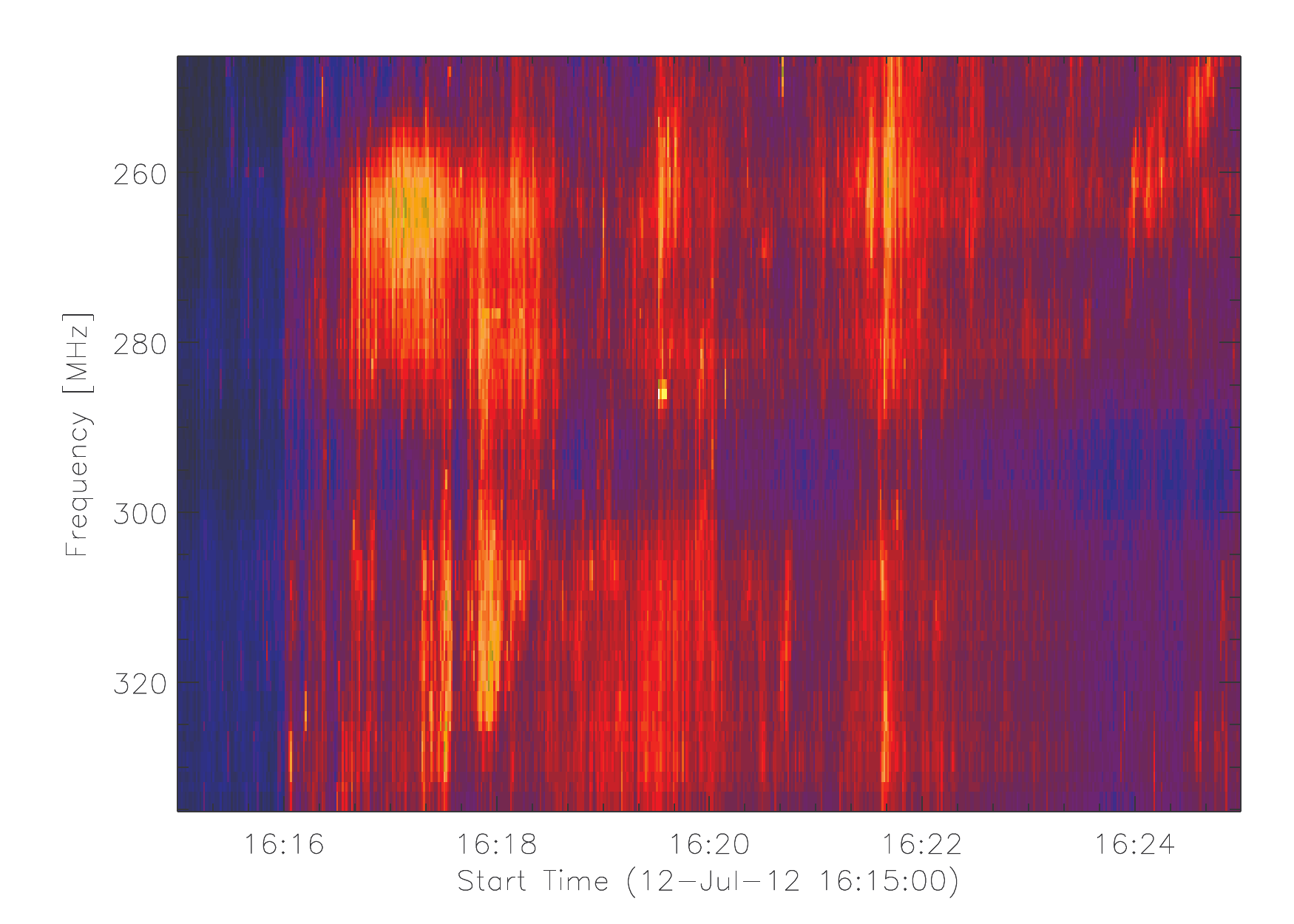}
       \includegraphics[width=8.0cm]{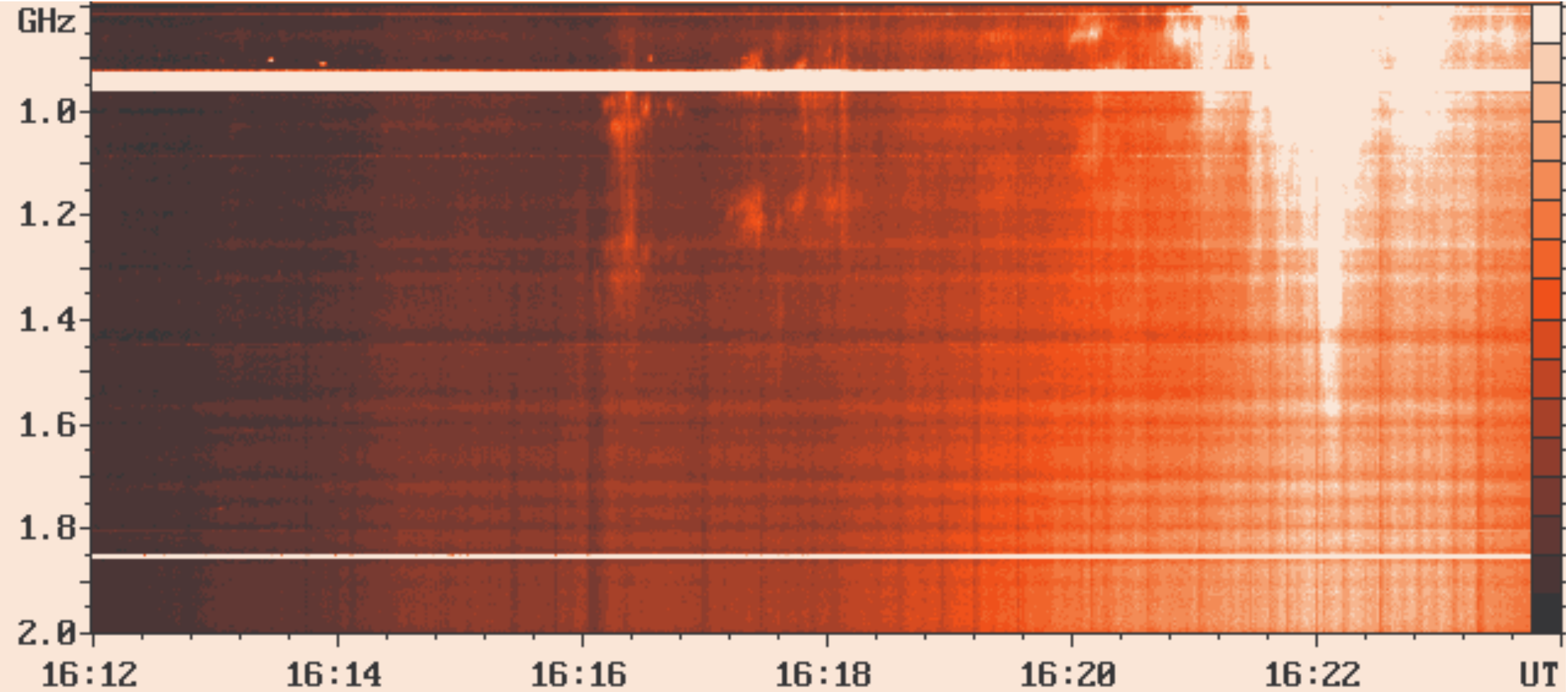}
    \caption{Radio spectra of the flare observed by the Callisto instrument in Trieste (\textit{first to third row}) and by the radio spectrograph at the Ond\v{r}ejov observatory (\textit{bottom}). \textit{First and second rows}: Spectra from 15:00 UT to 16:49 UT, capturing the early phase characterized by the noise storm, and the impulsive phase beginning at 16:16 UT. \textit{Third and bottom rows}: Detail of the DPSs at the start of the impulsive phase around 16:16 UT.}
       \label{Fig:Radio}
   \end{figure}
%
%
%
%
\section{Radio Observations and Interpretation}
\label{Sect:5}

To gain more insight into the nature of the slipping reconnection process, we analysed the available radio data for the flare. These also complement the EUV observations from SDO/AIA. 

At about 15:00 UT, radio burst activity started as a noise storm in the 200--500 MHz range (Fig. \ref{Fig:Radio}, \textit{first row}). The noise storm was observed by the Callisto instrument \citep{Benz09,Monstein13} in Trieste and consisted of a group of narrowband bursts resembling narrowband type III bursts. Some of the individual bursts exhibited drifts towards lower and some to higher frequencies. The typical frequency drifts of these bursts, which are about 20\,MHz\,s$^{-1}$, are more than an order of magnitude smaller than the typical mean frequency drift of the type III bursts, which is about 360\,MHz\,s$^{-1}$ \citep{Alvarez73}. This means that the velocity of the electron beams generating the noise storm bursts is more than an order of magnitude slower than for type III radio bursts, which is about $c$/3, where $c$ is the speed of light.

The noise storm lasted until 16:16 UT, when the strong radio flare started. The beginning of the strong radio flare is marked with two drifting pulsation structures \citep[DPSs, e.g.,][]{Karlicky02,Karlicky04} occurring during 16:16--16:21 UT: one starting at 1.3\,GHz and the other one at 1.0\,GHz (Fig. \ref{Fig:Radio}, \textit{bottom row}). Since these frequencies are not in harmonic relationship, this indicates spatially separated plasmoids. Both these DPSs drifted towards lower frequencies with the frequency drift d$f$/d$t$\,=\,$-$0.8\,MHz\,s$^{-1}$. Based on the timing, we identify them with the eruption of the series of long S-shaped loops observed by AIA (Fig. \ref{Fig:Eruption}). Note also that at about 16:16:30 UT, several DPSs also appeared in the 250--300\,MHz range (Fig.\,\ref{Fig:Radio}, \textit{third row}), c.f. \citet{Karlicky04}, indicating a range of erupting structures with different densities.

At 16:16:30 UT, broadband radio continuum was registered in the range of 1--5\,GHz simultaneously with the DPSs. The 1--5\,GHz radio emission peaked at about 16:22 UT. After a short decrease at about of 16:24\,UT the radio emission increased to its main maximum at 16:26 UT. At the lower frequencies (200--500\,MHz), the main maximum occurred at about 16:40\,UT.

The bursty nature of the radio emission in both the noise storm and the DPSs shows that the reconnection during the flare is intermittent and that the energy release leading to plasma heating is not uniformly distributed in time or space. This is already hinted at by the EUV observations from AIA (Sect. \ref{Sect:2}), which showed discrete, apparently moving features in all filters, rather than a continuous, near-uniform heating over the spatial locations (ribbons) where the slipping motion of the field line footpoints is occurring. Furthermore, presence of the DPSs during the impulsive phase indicates (1) presence of a current sheet and its fragmentation, (2) enhancement of the reconnection rate \citep{Kliem00,Barta08,Barta11a} at 16:16 UT, and that (3) the fast reconnection is not of the \citet{Petschek64} type, but turbulent with plasmoids in 3D \citep{Daughton11b,Daughton11a}. Note that the 3D MHD model of \citet{Aulanier12} does not have sufficient spatial resolution to resolve the individual small-scale processes within the current sheet. Furthermore, since the model is pressureless, it does not include spatial and temporal distributions of plasma heating and particle acceleration. Therefore, the radio data give an important complementary information to the flare physics treated by the 3D MHD model.
%
%
%
\section{Summary and Discussion}
\label{Sect:6}
We have presented observations of an eruptive X1.4 long-duration flare that occurred on July 12, 2012 in active region 11520, which is a part of an active region complex. The observations were compared to the 3D MHD ``standard solar flare model'' of \citet{Aulanier12}. The model qualitatively allows for an explanation of the observed apparent slipping motion of both the flare and erupting loops in terms of a torus-unstable erupting flux rope that is fed continually by the slipping magnetic reconnection.

The flare itself is preceded by a brightening of several loop systems located in active region 11521. Using magnetic field extrapolations, we found that one of these loop systems shares the footpoint of a quasi-separatrix layer corresponding to the positive-polarity flare ribbon PR.

The flare starts with the appearance of a highly sheared flare loop in AIA 131\AA, outlying an active region filament F1 seen in 304\AA. This flare loop appears in a pre-existing coronal sigmoid. It subsequently develops into an arcade of flare loops, with individual loops exhibiting apparent slipping motion. In the early stages of the flare, there are several episodes when the slipping motion is clearly visible. This apparent slipping motion is of the order of several tens of km\,s$^{-1}$ and is most pronounced in the flare loop footpoints located in the ribbon in the trailing negative polarities. Transition region emission from the loop footpoints is clearly identifiable in 171\AA, 304\AA~and 1600\AA~from the beginning of the flare. These footpoints represent the first signature of the ribbon and subsequent local brightenings along the developing ribbon.

A number of the flare loops observed in 131\AA~expand in the SW direction and subsequently erupt. These loops connect both flare ribbons and contain a series of faint, S-shaped non-potential loops more than 250$\arcsec$ long. The footpoints of the erupting loops are seen to slip along the extended hook of the positive-polarity ribbon with a velocity of approximately 136 km\,s$^{-1}$. A CME is subsequently observed by the STEREO spacecrafts.

The DEM analysis method of \citet{Hannah12} was applied to each AIA pixel in a limited field of view. It confirms that both the apparently slipping and erupting flare loops emit strongly in \ion{Fe}{21}, originating around 10\,MK. This emission is seen in the 131\AA~channel. Portions of these loops are also visible in the 94\AA\,or 193\AA~channels, dominated by contributions from \ion{Fe}{18}, \ion{Ca}{17} and \ion{Fe}{24}. A DEM analysis leads to an estimation of these contributions. We show that the flare loops do indeed emit in \ion{Fe}{24}. This emission is, except for a portion of the flare loops above the inversion line, obscured by much stronger \ion{Fe}{12} emission coming from the moss and warm coronal loops.

The observations have been qualitatively explained by a 3D pressureless MHD simulation of \citet{Aulanier12} and \citet{Janvier13}. In this simulation, a torus-unstable flux rope is located within an active region exhibiting flux imbalance similarly to our observations. The simulation does not contain null points or separatrices, but the presence of quasi-separatrix layers leads to a slipping reconnection regime. That is, the field lines with one fixed footpoint in a QSL exhibit an apparent slipping motion of the footpoint in the conjugate QSL. The direction of the slipping predicted by the model is consistent with the observations. The difference between the simulated and observed velocities may be caused by the difference between the timescales of the slipping motion and chromospheric evaporation, as the flare loops must first be filled with heated plasma in order to be observable. Alternatively, the slipping reconnection during the flare can happen in thicker QSLs, involving sub-Alfv{\' e}nic flare loop slipping motion, as suggested in \citet{Aulanier07}, instead of apparent super-Alfv{\'e}nic motion in thin QSLs as in \citet{Janvier13}.

The QSL footpoints in the simulation are in the shape of a hook, with a straight portion in the strong photospheric magnetic field, and a curved hook portion located further away. The simulated slipping motion is faster in the hook, in agreement with the observations. During the simulation, the hook evolves and becomes rounder. The presence of a hook and its evolution is reflected in the observations, with the exception that large-scale magnetic field constricts the negative-polarity hook and does not allow it to become round. Instead, the hook is deformed because of the presence of a large-scale QSL and extends more than 100$\arcsec$ to the south. The EUV intensity of the ribbon hook is lower than that of the straight portion of the ribbon, which can be understood in terms of energy deposition. The slipping motion is faster in the hook than in the straight part of the QSL in both the observations and the simulation, resulting in a lower amount of energy per unit time and area available for chromospheric evaporation.

The flux rope in the simulation is unstable and expands, which leads to its eruption. As the reconnection proceeds, the flux rope is fed with newly reconnected field lines that participate in the eruption. The flux rope envelope is observed by the AIA as long, S-shaped erupting hot loops. The erupting flux rope has a preferential direction in both the model and observations. In the observations this direction is modified by the large-scale magnetic field not present in the model.

An interesting feature of the flare is that despite the presence of an erupting flux rope and a flare arcade, the filament F1 remains unperturbed during the entire flare. This suggests that the real flare configurations may be complicated by the presence of another, filament-related flux rope that does not evolve with the rest of the magnetic configuration, in particular the overlying sigmoid. We note that although there are magnetic dips close to the photosphere in the simulation \citep[][Fig. 6 therein]{Aulanier12}, these are part of the sigmoid and not of any additional flux rope. This additional flux rope must then be connected to the topological complexity of real solar magnetic fields, as opposed to the simplified modelled ones. We also note that such tightly-packed flux-ropes constituting active-region filaments have indeed been recently reported by \citet{Kuckein12} and \citet{Yelles12}.

The apparent motion of EUV loops in early stages of the flare, interpreted by the slipping reconnection, was associated with the noise storm in the metric radio range. The noise storm ended at 16:16 UT with the appearance of the dm-drifting pulsation structures, indicating plasmoid formation within the flare current sheet and their subsequent ejection. This marks the enhancement of the reconnection rate and the impulsive phase of the flare, as evidenced by radio bursts observed in very broad range of radio frequencies. Note that at this instant the long S-shaped loop erupted and the GOES X-ray flux rapidly increased.

In summary, we have shown that the apparent slipping motion as a result of slipping reconnection, is indeed occurring during eruptive flares. This motion is a typical feature of the ``standard solar flare model in 3D'', which allows for a consistent explanation of many of the individual magnetically-controlled phenomena during the eruptive flares. It also shows that null-points and true separatrices are not required for the eruptive flares to occur. Radio data indicate that the slipping reconnection is also associated with intermittent particle acceleration and plasmoid formation.

\begin{acknowledgements}
AIA data are courtesy of NASA/SDO and the AIA science team. JD acknowledges support from the Royal Society via the Newton Fellowship Programme. This work was supported by Scientific Grant Agency, VEGA, Slovakia, Grant No. 1/0240/11. GDZ acknowledges support from STFC (UK) via the Advanced Fellowships Programme. HEM also acknowledges support from STFC. The work of MK was supported by the Grant No. 209/12/0103 of the Grant Agency of the Czech Republic. CHIANTI is a collaborative project involving the NRL (USA), RAL (UK), MSSL (UK), the Universities of Florence (Italy) and Cambridge (UK), and George Mason University (USA). 
\end{acknowledgements}

\bibliographystyle{apj}
\bibliography{Xflare}

\begin{thebibliography}{125}
\expandafter\ifx\csname natexlab\endcsname\relax\def\natexlab#1{#1}\fi

\bibitem[{{Alexander} {et~al.}(2013){Alexander}, {Walsh}, {R{\'e}gnier},
  {Cirtain}, {Winebarger}, {Golub}, {Kobayashi}, {Platt}, {Mitchell},
  {Korreck}, {DePontieu}, {DeForest}, {Weber}, {Title}, \&
  {Kuzin}}]{Alexander13}
{Alexander}, C.~E., {Walsh}, R.~W., {R{\'e}gnier}, S., {et~al.} 2013, \apjl,
  775, L32

\bibitem[{{Alissandrakis}(1981)}]{Alissandrakis81}
{Alissandrakis}, C.~E. 1981, \aap, 100, 197

\bibitem[{{Alvarez} \& {Haddock}(1973)}]{Alvarez73}
{Alvarez}, H., \& {Haddock}, F.~T. 1973, \solphys, 29, 197

\bibitem[{{Amari} {et~al.}(2000){Amari}, {Luciani}, {Mikic}, \&
  {Linker}}]{Amari00}
{Amari}, T., {Luciani}, J.~F., {Mikic}, Z., \& {Linker}, J. 2000, \apjl, 529,
  L49

\bibitem[{{Andretta} {et~al.}(2003){Andretta}, {Del Zanna}, \&
  {Jordan}}]{Andretta03}
{Andretta}, V., {Del Zanna}, G., \& {Jordan}, S.~D. 2003, \aap, 400, 737

\bibitem[{{Antiochos} {et~al.}(1999){Antiochos}, {DeVore}, \&
  {Klimchuk}}]{Antiochos99}
{Antiochos}, S.~K., {DeVore}, C.~R., \& {Klimchuk}, J.~A. 1999, \apj, 510, 485

\bibitem[{{Aschwanden} \& {Boerner}(2011)}]{Aschwanden11}
{Aschwanden}, M.~J., \& {Boerner}, P. 2011, \apj, 732, 81

\bibitem[{{Asplund} {et~al.}(2009){Asplund}, {Grevesse}, {Sauval}, \&
  {Scott}}]{Asplund09}
{Asplund}, M., {Grevesse}, N., {Sauval}, A.~J., \& {Scott}, P. 2009, \araa, 47,
  481

\bibitem[{{Aulanier} {et~al.}(2005){Aulanier}, {D{\'e}moulin}, \&
  {Grappin}}]{Aulanier05}
{Aulanier}, G., {D{\'e}moulin}, P., \& {Grappin}, R. 2005, \aap, 430, 1067

\bibitem[{{Aulanier} {et~al.}(2012){Aulanier}, {Janvier}, \&
  {Schmieder}}]{Aulanier12}
{Aulanier}, G., {Janvier}, M., \& {Schmieder}, B. 2012, \aap, 543, A110

\bibitem[{{Aulanier} {et~al.}(2006){Aulanier}, {Pariat}, {D{\'e}moulin}, \&
  {DeVore}}]{Aulanier06}
{Aulanier}, G., {Pariat}, E., {D{\'e}moulin}, P., \& {DeVore}, C.~R. 2006,
  \solphys, 238, 347

\bibitem[{{Aulanier} {et~al.}(2010){Aulanier}, {T{\"o}r{\"o}k}, {D{\'e}moulin},
  \& {DeLuca}}]{Aulanier10}
{Aulanier}, G., {T{\"o}r{\"o}k}, T., {D{\'e}moulin}, P., \& {DeLuca}, E.~E.
  2010, \apj, 708, 314

\bibitem[{{Aulanier} {et~al.}(2007){Aulanier}, {Golub}, {DeLuca}, {Cirtain},
  {Kano}, {Lundquist}, {Narukage}, {Sakao}, \& {Weber}}]{Aulanier07}
{Aulanier}, G., {Golub}, L., {DeLuca}, E.~E., {et~al.} 2007, Science, 318, 1588

\bibitem[{{Aurass} {et~al.}(2011){Aurass}, {Mann}, {Zlobec}, \&
  {Karlick{\'y}}}]{Aurass11}
{Aurass}, H., {Mann}, G., {Zlobec}, P., \& {Karlick{\'y}}, M. 2011, \apj, 730,
  57

\bibitem[{{B{\'a}rta} {et~al.}(2011){B{\'a}rta}, {B{\"u}chner}, {Karlick{\'y}},
  \& {Sk{\'a}la}}]{Barta11a}
{B{\'a}rta}, M., {B{\"u}chner}, J., {Karlick{\'y}}, M., \& {Sk{\'a}la}, J.
  2011, \apj, 737, 24

\bibitem[{{B{\'a}rta} {et~al.}(2008){B{\'a}rta}, {Vr{\v s}nak}, \&
  {Karlick{\'y}}}]{Barta08}
{B{\'a}rta}, M., {Vr{\v s}nak}, B., \& {Karlick{\'y}}, M. 2008, \aap, 477, 649

\bibitem[{{Benz} {et~al.}(2009){Benz}, {Monstein}, {Meyer}, {Manoharan},
  {Ramesh}, {Altyntsev}, {Lara}, {Paez}, \& {Cho}}]{Benz09}
{Benz}, A.~O., {Monstein}, C., {Meyer}, H., {et~al.} 2009, Earth Moon and
  Planets, 104, 277

\bibitem[{{Boerner} {et~al.}(2012){Boerner}, {Edwards}, {Lemen}, {Rausch},
  {Schrijver}, {Shine}, {Shing}, {Stern}, {Tarbell}, {Title}, {Wolfson},
  {Soufli}, {Spiller}, {Gullikson}, {McKenzie}, {Windt}, {Golub}, {Podgorski},
  {Testa}, \& {Weber}}]{Boerner12}
{Boerner}, P., {Edwards}, C., {Lemen}, J., {et~al.} 2012, \solphys, 275, 41

\bibitem[{{Brosius}(2013)}]{Brosius13}
{Brosius}, J.~W. 2013, \apj, 762, 133

\bibitem[{{Brosius} \& {Holman}(2010)}]{Brosius10}
{Brosius}, J.~W., \& {Holman}, G.~D. 2010, \apj, 720, 1472

\bibitem[{{Carmichael}(1964)}]{Carmichael64}
{Carmichael}, H. 1964, NASA Special Publication, 50, 451

\bibitem[{{Chandra} {et~al.}(2009){Chandra}, {Schmieder}, {Aulanier}, \&
  {Malherbe}}]{Chandra09}
{Chandra}, R., {Schmieder}, B., {Aulanier}, G., \& {Malherbe}, J.~M. 2009,
  \solphys, 258, 53

\bibitem[{{Chen} {et~al.}(2012){Chen}, {Su}, {Guo}, \& {Deng}}]{Chen12}
{Chen}, P.~F., {Su}, J.~T., {Guo}, Y., \& {Deng}, Y.~Y. 2012, Chi. Sci. Bull.,
  57, 1393

\bibitem[{{Cheng} {et~al.}(2013){Cheng}, {Zhang}, {Ding}, {Liu}, \&
  {Poomvises}}]{Cheng13}
{Cheng}, X., {Zhang}, J., {Ding}, M.~D., {Liu}, Y., \& {Poomvises}, W. 2013,
  \apj, 763, 43

\bibitem[{{Chifor} {et~al.}(2006){Chifor}, {Mason}, {Tripathi}, {Isobe}, \&
  {Asai}}]{Chifor06}
{Chifor}, C., {Mason}, H.~E., {Tripathi}, D., {Isobe}, H., \& {Asai}, A. 2006,
  \aap, 458, 965

\bibitem[{{Craig} \& {Brown}(1976)}]{Craig76}
{Craig}, I.~J.~D., \& {Brown}, J.~C. 1976, \aap, 49, 239

\bibitem[{{Craig} \& {Brown}(1986)}]{Craig86}
{Craig}, I. J.~D., \& {Brown}, J.~C. 1986, Inverse problems in astronomy: A
  guide to inversion strategies for remotely sensed data (Adam Hilger, Ltd.,
  159 p.)

\bibitem[{{Daughton} {et~al.}(2011{\natexlab{a}}){Daughton}, {Roytershteyn}, \&
  {Karimabadi}}]{Daughton11b}
{Daughton}, W., {Roytershteyn}, V., \& {Karimabadi}, H. 2011{\natexlab{a}}, in
  APS Meeting Abstracts, 9016P

\bibitem[{{Daughton} {et~al.}(2011{\natexlab{b}}){Daughton}, {Roytershteyn},
  {Karimabadi}, {Yin}, {Albright}, {Gary}, \& {Bowers}}]{Daughton11a}
{Daughton}, W., {Roytershteyn}, V., {Karimabadi}, H., {et~al.}
  2011{\natexlab{b}}, in American Institute of Physics Conference Series, Vol.
  1320, American Institute of Physics Conference Series, ed. D.~{Vassiliadis},
  S.~F. {Fung}, X.~{Shao}, I.~A. {Daglis}, \& J.~D. {Huba}, 144--159

\bibitem[{{del Zanna}(1999)}]{DelZanna99}
{del Zanna}, G. 1999, PhD thesis, , Univ.~of Central Lancashire, (1999)

\bibitem[{{Del Zanna}(2013)}]{DelZanna13}
{Del Zanna}, G. 2013, \aap, 558, A73

\bibitem[{{Del Zanna} {et~al.}(2011{\natexlab{a}}){Del Zanna}, {Aulanier},
  {Klein}, \& {T{\"o}r{\"o}k}}]{DelZanna11b}
{Del Zanna}, G., {Aulanier}, G., {Klein}, K.-L., \& {T{\"o}r{\"o}k}, T.
  2011{\natexlab{a}}, \aap, 526, A137

\bibitem[{{Del Zanna} {et~al.}(2011{\natexlab{b}}){Del Zanna}, {Mitra-Kraev},
  {Bradshaw}, {Mason}, \& {Asai}}]{DelZanna11a}
{Del Zanna}, G., {Mitra-Kraev}, U., {Bradshaw}, S.~J., {Mason}, H.~E., \&
  {Asai}, A. 2011{\natexlab{b}}, \aap, 526, A1

\bibitem[{{Del Zanna} {et~al.}(2011{\natexlab{c}}){Del Zanna}, {O'Dwyer}, \&
  {Mason}}]{DelZanna11c}
{Del Zanna}, G., {O'Dwyer}, B., \& {Mason}, H.~E. 2011{\natexlab{c}}, \aap,
  535, A46

\bibitem[{{D{\'{e}}moulin} {et~al.}(1997){D{\'{e}}moulin}, {Bagala},
  {Mandrini}, {H{\'{e}}noux}, \& {Rovira}}]{Demoulin97}
{D{\'{e}}moulin}, P., {Bagala}, L.~G., {Mandrini}, C.~H., {H{\'{e}}noux},
  J.~C., \& {Rovira}, M.~G. 1997, \aap, 325, 305

\bibitem[{{D{\'{e}}moulin} {et~al.}(1996){D{\'{e}}moulin}, {H{\'{e}}noux},
  {Priest}, \& {Mandrini}}]{Demoulin96a}
{D{\'{e}}moulin}, P., {H{\'{e}}noux}, J.~C., {Priest}, E.~R., \& {Mandrini},
  C.~H. 1996, \aap, 308, 643

\bibitem[{{D{\'e}moulin} {et~al.}(1996){D{\'e}moulin}, {Priest}, \&
  {Lonie}}]{Demoulin96b}
{D{\'e}moulin}, P., {Priest}, E.~R., \& {Lonie}, D.~P. 1996, \jgr, 101, 7631

\bibitem[{{Dere} {et~al.}(1999){Dere}, {Brueckner}, {Howard}, {Michels}, \&
  {Delaboudiniere}}]{Dere99}
{Dere}, K.~P., {Brueckner}, G.~E., {Howard}, R.~A., {Michels}, D.~J., \&
  {Delaboudiniere}, J.~P. 1999, \apj, 516, 465

\bibitem[{{Dere} {et~al.}(1997){Dere}, {Landi}, {Mason}, {Monsignori Fossi}, \&
  {Young}}]{Dere97}
{Dere}, K.~P., {Landi}, E., {Mason}, H.~E., {Monsignori Fossi}, B.~C., \&
  {Young}, P.~R. 1997, \aaps, 125, 149

\bibitem[{{Doschek} {et~al.}(2013){Doschek}, {Warren}, \& {Young}}]{Doschek13}
{Doschek}, G.~A., {Warren}, H.~P., \& {Young}, P.~R. 2013, \apj, 767, 55

\bibitem[{{Elgaroy}(1977)}]{Elgaroy77}
{Elgaroy}, O. 1977, {Solar noise storms}, 43

\bibitem[{{Fan}(2012)}]{Fan12}
{Fan}, Y. 2012, \apj, 758, 60

\bibitem[{{Fletcher} {et~al.}(2011){Fletcher}, {Dennis}, {Hudson}, {Krucker},
  {Phillips}, {Veronig}, {Battaglia}, {Bone}, {Caspi}, {Chen}, {Gallagher},
  {Grigis}, {Ji}, {Liu}, {Milligan}, \& {Temmer}}]{Fletcher11}
{Fletcher}, L., {Dennis}, B.~R., {Hudson}, H.~S., {et~al.} 2011, \ssr, 159, 19

\bibitem[{{Gary}(1989)}]{Gary89}
{Gary}, G.~A. 1989, \apjs, 69, 323

\bibitem[{{Gekelman} {et~al.}(2012){Gekelman}, {Lawrence}, \& {Van
  Compernolle}}]{Gekelman12}
{Gekelman}, W., {Lawrence}, E., \& {Van Compernolle}, B. 2012, \apj, 753, 131

\bibitem[{{Graham} {et~al.}(2011){Graham}, {Fletcher}, \& {Hannah}}]{Graham11}
{Graham}, D.~R., {Fletcher}, L., \& {Hannah}, I.~G. 2011, \aap, 532, A27

\bibitem[{{Green} \& {Kliem}(2009)}]{Green09}
{Green}, L.~M., \& {Kliem}, B. 2009, \apjl, 700, L83

\bibitem[{{Green} {et~al.}(2011){Green}, {Kliem}, \& {Wallace}}]{Green11}
{Green}, L.~M., {Kliem}, B., \& {Wallace}, A.~J. 2011, \aap, 526, A2

\bibitem[{{Hannah} \& {Kontar}(2012)}]{Hannah12}
{Hannah}, I.~G., \& {Kontar}, E.~P. 2012, \aap, 539, A146

\bibitem[{{Hannah} \& {Kontar}(2013)}]{Hannah13}
---. 2013, \aap, 553, A10

\bibitem[{{Hirayama}(1974)}]{Hirayama74}
{Hirayama}, T. 1974, \solphys, 34, 323

\bibitem[{{Inglis} \& {Gilbert}(2013)}]{Inglis13}
{Inglis}, A.~R., \& {Gilbert}, H.~R. 2013, ArXiv e-prints

\bibitem[{{Janvier} {et~al.}(2013){Janvier}, {Aulanier}, {Pariat}, \&
  {D{\'e}moulin}}]{Janvier13}
{Janvier}, M., {Aulanier}, G., {Pariat}, E., \& {D{\'e}moulin}, P. 2013, \aap,
  555, A77

\bibitem[{{Jiang} {et~al.}(2011){Jiang}, {Yang}, {Hong}, {Bi}, \&
  {Zheng}}]{Jiang11}
{Jiang}, Y., {Yang}, J., {Hong}, J., {Bi}, Y., \& {Zheng}, R. 2011, \apj, 738,
  179

\bibitem[{{Judge} {et~al.}(1997){Judge}, {Hubeny}, \& {Brown}}]{Judge97}
{Judge}, P.~G., {Hubeny}, V., \& {Brown}, J.~C. 1997, \apj, 475, 275

\bibitem[{{Kane}(1974)}]{Kane74}
{Kane}, S.~R. 1974, in IAU Symposium, Vol.~57, Coronal Disturbances, ed. G.~A.
  {Newkirk}, 105--141

\bibitem[{{Karlick{\'y}}(2004)}]{Karlicky04}
{Karlick{\'y}}, M. 2004, \aap, 417, 325

\bibitem[{{Karlick{\'y}} \& {B{\'a}rta}(2007)}]{KarlickyBarta07}
{Karlick{\'y}}, M., \& {B{\'a}rta}, M. 2007, \aap, 464, 735

\bibitem[{{Karlick{\'y}} \& {B{\'a}rta}(2011)}]{KarlickyBarta11}
---. 2011, \apj, 733, 107

\bibitem[{{Karlick{\'y}} {et~al.}(2012){Karlick{\'y}}, {B{\'a}rta}, \&
  {Nickeler}}]{Karlicky12}
{Karlick{\'y}}, M., {B{\'a}rta}, M., \& {Nickeler}, D. 2012, \aap, 541, A86

\bibitem[{{Karlick{\'y}} {et~al.}(2010){Karlick{\'y}}, {B{\'a}rta}, \&
  {Ryb{\'a}k}}]{Karlicky10}
{Karlick{\'y}}, M., {B{\'a}rta}, M., \& {Ryb{\'a}k}, J. 2010, \aap, 514, A28

\bibitem[{{Karlick{\'y}} {et~al.}(2002){Karlick{\'y}}, {F{\'a}rn{\'{\i}}k}, \&
  {M{\'e}sz{\'a}rosov{\'a}}}]{Karlicky02}
{Karlick{\'y}}, M., {F{\'a}rn{\'{\i}}k}, F., \& {M{\'e}sz{\'a}rosov{\'a}}, H.
  2002, \aap, 395, 677

\bibitem[{{Kliem} {et~al.}(2000){Kliem}, {Karlick{\'y}}, \& {Benz}}]{Kliem00}
{Kliem}, B., {Karlick{\'y}}, M., \& {Benz}, A.~O. 2000, \aap, 360, 715

\bibitem[{{Kliem} {et~al.}(2010){Kliem}, {Linton}, {T{\"o}r{\"o}k}, \&
  {Karlick{\'y}}}]{Kliem10}
{Kliem}, B., {Linton}, M.~G., {T{\"o}r{\"o}k}, T., \& {Karlick{\'y}}, M. 2010,
  \solphys, 266, 91

\bibitem[{{Ko{\l}oma{\'n}ski} \& {Karlick{\'y}}(2007)}]{Kolomanski07}
{Ko{\l}oma{\'n}ski}, S., \& {Karlick{\'y}}, M. 2007, \aap, 475, 685

\bibitem[{{Kopp} \& {Pneuman}(1976)}]{Kopp76}
{Kopp}, R.~A., \& {Pneuman}, G.~W. 1976, \solphys, 50, 85

\bibitem[{{Kuckein} {et~al.}(2012){Kuckein}, {Mart{\'{\i}}nez Pillet}, \&
  {Centeno}}]{Kuckein12}
{Kuckein}, C., {Mart{\'{\i}}nez Pillet}, V., \& {Centeno}, R. 2012, \aap, 539,
  A131

\bibitem[{{Labrosse} {et~al.}(2010){Labrosse}, {Heinzel}, {Vial}, {Kucera},
  {Parenti}, {Gun{\'a}r}, {Schmieder}, \& {Kilper}}]{Labrosse10}
{Labrosse}, N., {Heinzel}, P., {Vial}, J.-C., {et~al.} 2010, \ssr, 151, 243

\bibitem[{{Landi} {et~al.}(2013){Landi}, {Young}, {Dere}, {Del Zanna}, \&
  {Mason}}]{Landi13}
{Landi}, E., {Young}, P.~R., {Dere}, K.~P., {Del Zanna}, G., \& {Mason}, H.~E.
  2013, \apj, 763, 86

\bibitem[{{Lemen} {et~al.}(2012){Lemen}, {Title}, {Akin}, {Boerner}, {Chou},
  {Drake}, {Duncan}, {Edwards}, {Friedlaender}, {Heyman}, {Hurlburt}, {Katz},
  {Kushner}, {Levay}, {Lindgren}, {Mathur}, {McFeaters}, {Mitchell}, {Rehse},
  {Schrijver}, {Springer}, {Stern}, {Tarbell}, {Wuelser}, {Wolfson}, {Yanari},
  {Bookbinder}, {Cheimets}, {Caldwell}, {Deluca}, {Gates}, {Golub}, {Park},
  {Podgorski}, {Bush}, {Scherrer}, {Gummin}, {Smith}, {Auker}, {Jerram},
  {Pool}, {Soufli}, {Windt}, {Beardsley}, {Clapp}, {Lang}, \&
  {Waltham}}]{Lemen12}
{Lemen}, J.~R., {Title}, A.~M., {Akin}, D.~J., {et~al.} 2012, \solphys, 275, 17

\bibitem[{{Lin} \& {Forbes}(2000)}]{Lin00}
{Lin}, J., \& {Forbes}, T.~G. 2000, \jgr, 105, 2375

\bibitem[{{Liu} {et~al.}(2009){Liu}, {Lee}, {Karlick{\'y}}, {Prasad Choudhary},
  {Deng}, \& {Wang}}]{Liu09}
{Liu}, C., {Lee}, J., {Karlick{\'y}}, M., {et~al.} 2009, \apj, 703, 757

\bibitem[{{Loureiro} {et~al.}(2012){Loureiro}, {Samtaney}, {Schekochihin}, \&
  {Uzdensky}}]{Loureiro12}
{Loureiro}, N.~F., {Samtaney}, R., {Schekochihin}, A.~A., \& {Uzdensky}, D.~A.
  2012, Physics of Plasmas, 19, 042303

\bibitem[{{Lynch} {et~al.}(2008){Lynch}, {Antiochos}, {DeVore}, {Luhmann}, \&
  {Zurbuchen}}]{Lynch08}
{Lynch}, B.~J., {Antiochos}, S.~K., {DeVore}, C.~R., {Luhmann}, J.~G., \&
  {Zurbuchen}, T.~H. 2008, \apj, 683, 1192

\bibitem[{{Magara} {et~al.}(1996){Magara}, {Mineshige}, {Yokoyama}, \&
  {Shibata}}]{Magara96}
{Magara}, T., {Mineshige}, S., {Yokoyama}, T., \& {Shibata}, K. 1996, \apj,
  466, 1054

\bibitem[{{Masson} {et~al.}(2012){Masson}, {Aulanier}, {Pariat}, \&
  {Klein}}]{Masson12}
{Masson}, S., {Aulanier}, G., {Pariat}, E., \& {Klein}, K.-L. 2012, \solphys,
  276, 199

\bibitem[{{Masson} {et~al.}(2009){Masson}, {Pariat}, {Aulanier}, \&
  {Schrijver}}]{Masson09}
{Masson}, S., {Pariat}, E., {Aulanier}, G., \& {Schrijver}, C.~J. 2009, \apj,
  700, 559

\bibitem[{{McKenzie} \& {Canfield}(2008)}]{McKenzie08}
{McKenzie}, D.~E., \& {Canfield}, R.~C. 2008, \aap, 481, L65

\bibitem[{{M{\'e}sz{\'a}rosov{\'a}} {et~al.}(2013){M{\'e}sz{\'a}rosov{\'a}},
  {Dud{\'{\i}}k}, {Karlick{\'y}}, {Madsen}, \& {Sawant}}]{Meszarosova13}
{M{\'e}sz{\'a}rosov{\'a}}, H., {Dud{\'{\i}}k}, J., {Karlick{\'y}}, M.,
  {Madsen}, F.~R.~H., \& {Sawant}, H.~S. 2013, \solphys, 283, 473

\bibitem[{{Milligan} \& {Dennis}(2009)}]{Milligan09}
{Milligan}, R.~O., \& {Dennis}, B.~R. 2009, \apj, 699, 968

\bibitem[{{Milligan} {et~al.}(2010){Milligan}, {McAteer}, {Dennis}, \&
  {Young}}]{Milligan10}
{Milligan}, R.~O., {McAteer}, R.~T.~J., {Dennis}, B.~R., \& {Young}, C.~A.
  2010, \apj, 713, 1292

\bibitem[{{Monstein}(2013)}]{Monstein13}
{Monstein}, C. 2013, in EGU General Assembly Conference Abstracts, Vol.~15, EGU
  General Assembly Conference Abstracts, 2027

\bibitem[{{Moore} {et~al.}(1997){Moore}, {Schmieder}, {Hathaway}, \&
  {Tarbell}}]{Moore97}
{Moore}, R.~L., {Schmieder}, B., {Hathaway}, D.~H., \& {Tarbell}, T.~D. 1997,
  \solphys, 176, 153

\bibitem[{{Moore} {et~al.}(2001){Moore}, {Sterling}, {Hudson}, \&
  {Lemen}}]{Moore01}
{Moore}, R.~L., {Sterling}, A.~C., {Hudson}, H.~S., \& {Lemen}, J.~R. 2001,
  \apj, 552, 833

\bibitem[{{Neupert}(1968)}]{Neupert68}
{Neupert}, W.~M. 1968, \apjl, 153, L59

\bibitem[{{Ning}(2011)}]{Ning11}
{Ning}, Z.~J. 2011, in Astronomical Society of India Conference Series, Vol.~2,
  Astronomical Society of India Conference Series, 279

\bibitem[{{O'Dwyer} {et~al.}(2010){O'Dwyer}, {Del Zanna}, {Mason}, {Weber}, \&
  {Tripathi}}]{ODwyer10}
{O'Dwyer}, B., {Del Zanna}, G., {Mason}, H.~E., {Weber}, M.~A., \& {Tripathi},
  D. 2010, \aap, 521, A21

\bibitem[{{Ohyama} \& {Shibata}(1998)}]{Ohyama98}
{Ohyama}, M., \& {Shibata}, K. 1998, \apj, 499, 934

\bibitem[{{Pariat} \& {D{\'e}moulin}(2012)}]{Pariat12}
{Pariat}, E., \& {D{\'e}moulin}, P. 2012, \aap, 541, A78

\bibitem[{{Parker}(1957)}]{Parker57}
{Parker}, E.~N. 1957, \jgr, 62, 509

\bibitem[{{Patsourakos} {et~al.}(2013){Patsourakos}, {Vourlidas}, \&
  {Stenborg}}]{Patsourakos13}
{Patsourakos}, S., {Vourlidas}, A., \& {Stenborg}, G. 2013, \apj, 764, 125

\bibitem[{{Petkaki} {et~al.}(2012){Petkaki}, {Del Zanna}, {Mason}, \&
  {Bradshaw}}]{Petkaki12}
{Petkaki}, P., {Del Zanna}, G., {Mason}, H.~E., \& {Bradshaw}, S.~J. 2012,
  \aap, 547, A25

\bibitem[{{Petschek}(1964)}]{Petschek64}
{Petschek}, H.~E. 1964, NASA Special Publication, 50, 425

\bibitem[{{Poduval} {et~al.}(2013){Poduval}, {DeForest}, {Schmelz}, \&
  {Pathak}}]{Poduval13}
{Poduval}, B., {DeForest}, C.~E., {Schmelz}, J.~T., \& {Pathak}, S. 2013, \apj,
  765, 144

\bibitem[{{Priest} \& {Forbes}(2000)}]{Priest00}
{Priest}, E., \& {Forbes}, T. 2000, {Magnetic Reconnection}

\bibitem[{{Priest} \& {D{\'e}moulin}(1995)}]{Priest95}
{Priest}, E.~R., \& {D{\'e}moulin}, P. 1995, \jgr, 1002, 23443

\bibitem[{{Priest} {et~al.}(2003){Priest}, {Hornig}, \& {Pontin}}]{Priest03}
{Priest}, E.~R., {Hornig}, G., \& {Pontin}, D.~I. 2003, Journal of Geophysical
  Research (Space Physics), 108, 1285

\bibitem[{{Raftery} {et~al.}(2009){Raftery}, {Gallagher}, {Milligan}, \&
  {Klimchuk}}]{Raftery09}
{Raftery}, C.~L., {Gallagher}, P.~T., {Milligan}, R.~O., \& {Klimchuk}, J.~A.
  2009, \aap, 494, 1127

\bibitem[{{Reid} {et~al.}(2012){Reid}, {Vilmer}, {Aulanier}, \&
  {Pariat}}]{Reid12}
{Reid}, H.~A.~S., {Vilmer}, N., {Aulanier}, G., \& {Pariat}, E. 2012, \aap,
  547, A52

\bibitem[{{Savcheva} {et~al.}(2012){Savcheva}, {Pariat}, {van Ballegooijen},
  {Aulanier}, \& {DeLuca}}]{Savcheva12a}
{Savcheva}, A., {Pariat}, E., {van Ballegooijen}, A., {Aulanier}, G., \&
  {DeLuca}, E. 2012, \apj, 750, 15

\bibitem[{{Scherrer} {et~al.}(2012){Scherrer}, {Schou}, {Bush}, {Kosovichev},
  {Bogart}, {Hoeksema}, {Liu}, {Duvall}, {Zhao}, {Title}, {Schrijver},
  {Tarbell}, \& {Tomczyk}}]{Scherrer12}
{Scherrer}, P.~H., {Schou}, J., {Bush}, R.~I., {et~al.} 2012, \solphys, 275,
  207

\bibitem[{{Schmelz} {et~al.}(2013){Schmelz}, {Jenkins}, \&
  {Pathak}}]{Schmelz13}
{Schmelz}, J.~T., {Jenkins}, B.~S., \& {Pathak}, S. 2013, \apj, 770, 14

\bibitem[{{Schmelz} {et~al.}(2012){Schmelz}, {Reames}, {von Steiger}, \&
  {Basu}}]{Schmelz12}
{Schmelz}, J.~T., {Reames}, D.~V., {von Steiger}, R., \& {Basu}, S. 2012, \apj,
  755, 33

\bibitem[{{Schmieder} {et~al.}(1996){Schmieder}, {Heinzel}, {van
  Driel-Gesztelyi}, \& {Lemen}}]{Schmieder96}
{Schmieder}, B., {Heinzel}, P., {van Driel-Gesztelyi}, L., \& {Lemen}, J.~R.
  1996, \solphys, 165, 303

\bibitem[{{Schrijver} {et~al.}(2013){Schrijver}, {Title}, {Yeates}, \&
  {DeRosa}}]{Schrijver13}
{Schrijver}, C.~J., {Title}, A.~M., {Yeates}, A.~R., \& {DeRosa}, M.~L. 2013,
  ArXiv e-prints

\bibitem[{{Shen} {et~al.}(2012){Shen}, {Liu}, \& {Su}}]{Shen12}
{Shen}, Y., {Liu}, Y., \& {Su}, J. 2012, \apj, 750, 12

\bibitem[{{Shibata} \& {Tanuma}(2001)}]{Shibata01}
{Shibata}, K., \& {Tanuma}, S. 2001, Earth, Planets, and Space, 53, 473

\bibitem[{{Sturrock}(1966)}]{Sturrock66}
{Sturrock}, P.~A. 1966, \nat, 211, 695

\bibitem[{{Sweet}(1958)}]{Sweet58}
{Sweet}, P.~A. 1958, in IAU Symposium, Vol.~6, Electromagnetic Phenomena in
  Cosmical Physics, ed. B.~{Lehnert}, 123

\bibitem[{{Testa} {et~al.}(2013){Testa}, {De Pontieu},
  {Mart{\'{\i}}nez-Sykora}, {DeLuca}, {Hansteen}, {Cirtain}, {Winebarger},
  {Golub}, {Kobayashi}, {Korreck}, {Kuzin}, {Walsh}, {DeForest}, {Title}, \&
  {Weber}}]{Testa13}
{Testa}, P., {De Pontieu}, B., {Mart{\'{\i}}nez-Sykora}, J., {et~al.} 2013,
  \apjl, 770, L1

\bibitem[{{Titov} {et~al.}(2002){Titov}, {Hornig}, \& {D{\'e}moulin}}]{Titov02}
{Titov}, V.~S., {Hornig}, G., \& {D{\'e}moulin}, P. 2002, Journal of
  Geophysical Research (Space Physics), 107, 1164

\bibitem[{{T{\"o}r{\"o}k} {et~al.}(2004){T{\"o}r{\"o}k}, {Kliem}, \&
  {Titov}}]{Torok04}
{T{\"o}r{\"o}k}, T., {Kliem}, B., \& {Titov}, V.~S. 2004, \aap, 413, L27

\bibitem[{{T{\"o}r{\"o}k} {et~al.}(2011){T{\"o}r{\"o}k}, {Panasenco}, {Titov},
  {Miki{\'c}}, {Reeves}, {Velli}, {Linker}, \& {De Toma}}]{Torok11}
{T{\"o}r{\"o}k}, T., {Panasenco}, O., {Titov}, V.~S., {et~al.} 2011, \apjl,
  739, L63

\bibitem[{{Uzdensky} {et~al.}(2010){Uzdensky}, {Loureiro}, \&
  {Schekochihin}}]{Uzdensky10}
{Uzdensky}, D.~A., {Loureiro}, N.~F., \& {Schekochihin}, A.~A. 2010, Physical
  Review Letters, 105, 235002

\bibitem[{{van Ballegooijen} \& {Martens}(1989)}]{vanB89}
{van Ballegooijen}, A.~A., \& {Martens}, P.~C.~H. 1989, \apj, 343, 971

\bibitem[{{Warren} \& {Warshall}(2001)}]{Warren01}
{Warren}, H.~P., \& {Warshall}, A.~D. 2001, \apjl, 560, L87

\bibitem[{{Wheatland}(2006)}]{Wheatland06}
{Wheatland}, M.~S. 2006, \solphys, 236, 313

\bibitem[{{White} {et~al.}(2011){White}, {Benz}, {Christe},
  {F{\'a}rn{\'{\i}}k}, {Kundu}, {Mann}, {Ning}, {Raulin}, {Silva-V{\'a}lio},
  {Saint-Hilaire}, {Vilmer}, \& {Warmuth}}]{White11}
{White}, S.~M., {Benz}, A.~O., {Christe}, S., {et~al.} 2011, \ssr, 159, 225

\bibitem[{{Wilmot-Smith} {et~al.}(2009){Wilmot-Smith}, {Hornig}, \&
  {Pontin}}]{Wilmot09}
{Wilmot-Smith}, A.~L., {Hornig}, G., \& {Pontin}, D.~I. 2009, \apj, 704, 1288

\bibitem[{{Wuelser} {et~al.}(2004){Wuelser}, {Lemen}, {Tarbell}, {Wolfson},
  {Cannon}, {Carpenter}, {Duncan}, {Gradwohl}, {Meyer}, {Moore}, {Navarro},
  {Pearson}, {Rossi}, {Springer}, {Howard}, {Moses}, {Newmark},
  {Delaboudiniere}, {Artzner}, {Auchere}, {Bougnet}, {Bouyries}, {Bridou},
  {Clotaire}, {Colas}, {Delmotte}, {Jerome}, {Lamare}, {Mercier}, {Mullot},
  {Ravet}, {Song}, {Bothmer}, \& {Deutsch}}]{Wuelser04}
{Wuelser}, J.-P., {Lemen}, J.~R., {Tarbell}, T.~D., {et~al.} 2004, in Society
  of Photo-Optical Instrumentation Engineers (SPIE) Conference Series, Vol.
  5171, Society of Photo-Optical Instrumentation Engineers (SPIE) Conference
  Series, ed. S.~{Fineschi} \& M.~A. {Gummin}, 111--122

\bibitem[{{Yelles Chaouche} {et~al.}(2012){Yelles Chaouche}, {Kuckein},
  {Mart{\'{\i}}nez Pillet}, \& {Moreno-Insertis}}]{Yelles12}
{Yelles Chaouche}, L., {Kuckein}, C., {Mart{\'{\i}}nez Pillet}, V., \&
  {Moreno-Insertis}, F. 2012, \apj, 748, 23

\bibitem[{{Young} {et~al.}(2013){Young}, {Doschek}, {Warren}, \&
  {Hara}}]{Young13}
{Young}, P.~R., {Doschek}, G.~A., {Warren}, H.~P., \& {Hara}, H. 2013, \apj,
  766, 127

\bibitem[{{Zhang} {et~al.}(2012){Zhang}, {Cheng}, \& {Ding}}]{Zhang12}
{Zhang}, J., {Cheng}, X., \& {Ding}, M.-D. 2012, Nature Communications, 3

\bibitem[{{Zuccarello} {et~al.}(2009){Zuccarello}, {Romano}, {Farnik},
  {Karlicky}, {Contarino}, {Battiato}, {Guglielmino}, {Comparato}, \&
  {Ugarte-Urra}}]{Zuccarello09}
{Zuccarello}, F., {Romano}, P., {Farnik}, F., {et~al.} 2009, \aap, 493, 629

\bibitem[{{Zweibel} \& {Yamada}(2009)}]{Zweibel09}
{Zweibel}, E.~G., \& {Yamada}, M. 2009, \araa, 47, 291

\end{thebibliography}

\end{document}